%% file: paper_cls.tex
\documentclass[final,authoryear,5p,times,twocolumn]{elsarticle}
\usepackage{amsmath}
\usepackage[usenames,dvipsnames]{color}
\usepackage{booktabs}
\journal{Journal of Quantitative Spectroscopy \& Radiative Transfer}

\newcommand{\tripoli}{{\sc Tripoli-4}\textsuperscript{ \textregistered}}

\begin{document}

\begin{frontmatter}

\title{Monte Carlo Chord Length Sampling for $d$-dimensional Markov binary mixtures}


\author[label1]{Coline Larmier}
\address[label1]{Den-Service d'Etudes des R\'eacteurs et de Math\'ematiques Appliqu\'ees (SERMA), CEA, Universit\'e Paris-Saclay, 91191 Gif-sur-Yvette, FRANCE.}
\author[label2]{Adam Lam}
\address[label2]{School of Nuclear Science \& Engineering, Oregon State University, Corvallis, Oregon, USA.}
\author[label3]{Patrick Brantley}
\address[label3]{Lawrence Livermore National Laboratory, P.O. Box 808, Livermore, CA 94551.}
\author[label1]{Fausto Malvagi}
\author[label2]{Todd Palmer}
\author[label1]{Andrea Zoia\corref{cor1}}
\cortext[cor1]{Corresponding author. Tel. +33 (0)1 69 08 79 76}
\ead{andrea.zoia@cea.fr}





\begin{abstract}
The Chord Length Sampling (CLS) algorithm is a powerful Monte Carlo method that models the effects of stochastic media on particle transport by generating on-the-fly the material interfaces seen by the random walkers during their trajectories. This annealed disorder approach, which formally consists of solving the approximate Levermore-Pomraning equations for linear particle transport, enables a considerable speed-up with respect to transport in quenched disorder, where ensemble-averaging of the Boltzmann equation with respect to all possible realizations is needed. However, CLS intrinsically neglects the correlations induced by the spatial disorder, so that the accuracy of the solutions obtained by using this algorithm must be carefully verified with respect to reference solutions based on quenched disorder realizations. When the disorder is described by Markov mixing statistics, such comparisons have been attempted so far only for one-dimensional geometries, of the rod or slab type. In this work we extend these results to Markov media in two-dimensional (extruded) and three-dimensional geometries, by revisiting the classical set of benchmark configurations originally proposed by Adams, Larsen and Pomraning~\cite{benchmark_adams} and extended by Brantley~\cite{brantley_benchmark}. In particular, we examine the discrepancies between CLS and reference solutions for scalar particle flux and transmission/reflection coefficients as a function of the material properties of the benchmark specifications and of the system dimensionality.
\end{abstract}


\end{frontmatter}

%\linenumbers

\section{Introduction}

Several applications in nuclear science and engineering involve linear particle transport theory in stochastic media. Examples include neutron diffusion in pebble-bed reactors or randomly mixed water-vapor phases in boiling water reactors~\cite{pomraning, larsen, levermore, sanchez, wong}, and inertial confinement fusion~\cite{zimmerman, zimmerman_adams, haran}. Particle propagation in random media emerges more broadly in material and life sciences and in radiative transport~\cite{torquato, NatureOptical, davis, kostinski, clouds, tuchin, brantley_jeju}. Assuming that particles undergo single-speed transport with isotropic scattering, the angular particle flux $\varphi({\bf r}, {\boldsymbol \omega}) $ for each physical realization of the system obeys the linear Boltzmann equation
\begin{equation}
{\boldsymbol \omega} \cdot \nabla  \varphi + \Sigma({\bf r}) \varphi = \frac{\Sigma_s({\bf r})}{\Omega_d} \int\varphi({\bf r}, {\boldsymbol \omega}') d{\boldsymbol \omega}' + S.
\label{boltzmann}
\end{equation}
Here ${\bf r}$ and ${\boldsymbol \omega}$ denote the position and direction variables, respectively, $\Sigma({\bf r})$ being the total cross section and $S=S({\bf r}, {\boldsymbol \omega})$ the source term. The quantity $\Omega_d = 2\pi^{d/2}/\Gamma(d/2)$ is the surface area of the unit sphere in dimension $d$, $\Gamma(a)$ being the Gamma function. The quantities $\Sigma({\bf r})$, $\Sigma_s({\bf r})$ and $S({\bf r}, {\boldsymbol \omega})$ are in principle random variables, since the materials composing the traversed medium are assumed to be possibly distributed according to some statistical law. The physical observable of interest is typically the ensemble-averaged angular particle flux $\langle \varphi({\bf r}, {\boldsymbol \omega}) \rangle$, or more generally some ensemble-averaged functional $\langle F[ \varphi] \rangle$ of the particle flux, namely,
\begin{equation}
\langle \varphi({\bf r}, {\boldsymbol \omega}) \rangle = \int {\cal P}(q) \varphi^{(q)}({\bf r}, {\boldsymbol \omega}) dq,
\end{equation}
where $\varphi^{(q)}({\bf r}, {\boldsymbol \omega})$ is the solution of the Boltzmann equation~\eqref{boltzmann} corresponding to a single realization $q$, and ${\cal P}(q)$ is the stationary probability of observing the state $q$ for the functions $\Sigma^{(q)}({\bf r})$, $\Sigma^{(q)}_s({\bf r} )$ and $S^{(q)}({\bf r}, {\boldsymbol \omega})$~\cite{pomraning, renewal}.

Exact solutions for $\langle F[ \varphi] \rangle$ can be in principle obtained in the following way: first, a realization of the medium is sampled from the underlying mixing statistics; then, the linear transport equation~\eqref{boltzmann} corresponding to this realization is solved by either deterministic or Monte Carlo methods, and the physical observables of interest $F[ \varphi]$ are determined; a sufficiently large collection of realizations is produced; and ensemble averages are finally taken for the physical observables.

Reference solutions are very demanding in terms of computational resources, especially if transport is to be solved by Monte Carlo methods in order to preserve the highest possible accuracy in solving the Boltzmann equation. In principle, it would be thus desirable to directly derive a single equation for the ensemble-averaged flux $\langle \varphi \rangle$. A widely adopted model of random media is the so-called binary stochastic mixing, where only two immiscible materials (say $\alpha$ and $\beta$) are present~\cite{pomraning}. Then, by averaging Eq.~\eqref{boltzmann} over realizations having material $\alpha$ at ${\bf r}$, we obtain the following equation for $\langle \varphi_\alpha({\bf r}, {\boldsymbol \omega}) \rangle$
\begin{align}
& \left[ {\boldsymbol \omega} \cdot \nabla  + \Sigma_\alpha \right] p_\alpha \langle \varphi_\alpha \rangle = \frac{p_\alpha \Sigma_{s,\alpha} }{\Omega_d} \int \langle \varphi_\alpha({\bf r}, {\boldsymbol \omega}') \rangle d{\boldsymbol \omega}'  \nonumber \\
&+ p_{\beta,\alpha} \langle \varphi_{\beta, \alpha} \rangle - p_{\alpha,\beta} \langle \varphi_{\alpha, \beta} \rangle +p_\alpha S_\alpha
\label{boltzmann_ave}
\end{align}
where $p_i({\bf r}) $ is the probability of finding the material of index $i$ at position ${\bf r}$. Here $p_{i,j}=p_{i,j}({\bf r}, {\boldsymbol \omega})$ represents the probability per unit length of crossing the interface from material $i$ to material $j$ for a particle located at ${\bf r}$ and travelling in direction ${\boldsymbol \omega}$. The quantity $\langle \varphi_{i, j} \rangle$ denotes the angular flux averaged over those realizations where there is a transition from material $i$ to material $j$ for a particle located at ${\bf r}$ and travelling in direction ${\boldsymbol \omega}$. The cross sections $\Sigma_\alpha$ and $\Sigma_{s,\alpha}$ are those of material $\alpha$. The equation for $\langle \varphi_\beta({\bf r}, {\boldsymbol \omega}) \rangle$ is immediately obtained from Eq.~\eqref{boltzmann_ave} by permuting the indices $\alpha$ and $\beta$. Excluding the special case of particle transport in the absence of scattering, we are thus led to an infinite hierarchy for $\langle \varphi_\alpha \rangle$ in Eqs.~\eqref{boltzmann_ave}.

In order to explicitly derive the ensemble-averaged flux $\langle \varphi_\alpha \rangle$, it is therefore necessary to introduce a closure formula, which will in general depend on the underlying mixing statistics~\cite{pomraning, renewal, su}. The celebrated Levermore-Pomraning model assumes for instance $\langle \varphi_{\alpha, \beta} \rangle = \langle \varphi_{\alpha} \rangle$ for homogeneous Markov mixing statistics, with
\begin{equation}
p_{i,j}({\bf r}, {\boldsymbol \omega}) = \frac{p_{i}}{\Lambda_i({\boldsymbol \omega})},
\label{closure_lp_boltzmann}
\end{equation}
where $\Lambda_i({\boldsymbol \omega})$ is the mean chord length for trajectories crossing material $i$ in direction ${\boldsymbol \omega}$~\cite{pomraning}. Several generalisations of this model have been later proposed, including higher-order closure schemes~\cite{pomraning, su}.

In parallel, a family of Monte Carlo algorithms have been conceived in order to approximate the ensemble-averaged solutions to various degrees of accuracy~\cite{zimmerman_adams, sutton, donovan}. Their common feature is that they allow a simpler treatment of transport in stochastic mixtures (typically by neglecting the correlations on particle trajectories induced by the spatial disorder), which might be convenient in practical applications. In this context, a prominent role is played by the so-called Chord Length Sampling (CLS) algorithm, which is supposed to solve the Levermore-Pomraning model for Markovian binary mixing~\cite{zimmerman_adams, sahni1, sahni2}. The basic idea behind CLS is that the interfaces between the constituents of the stochastic medium are sampled on-the-fly during the particle displacements by drawing the distances to the following material boundaries from a distribution depending on the mixing statistics. The free parameters of the CLS model are the average chord length $\Lambda_i$ through each material and the volume fraction $p_i$. Since the spatial configuration seen by each particle is regenerated at each particle flight, the CLS corresponds to an annealed disorder model, as opposed to the quenched disorder of the reference solutions, where the spatial configuration is frozen for all the traversing particles. Generalization of these Monte Carlo algorithms including partial memory effects due to correlations for particles crossing back and forth the same materials have been also proposed~\cite{zimmerman_adams}.

In order to quantify the accuracy of the various approximate models, comparisons with respect to reference solutions are mandatory. For instance, although originally formulated for Markov statistics, CLS has been extensively applied also to randomly dispersed spherical inclusions into background matrices, with application to pebble-bed and very high temperature gas-cooled reactors~\cite{sutton, donovan}, and several benchmark problems have been examined in two and three dimensions~\cite{sutton, donovan, brantley_martos, brantley_lp}. Some methods to mitigate the errors between CLS and the reference solutions have been presented in the context of eigenvalue calculations, e.g., in~\cite{ji_brown}. For Markov mixing, a number of benchmark problems comparing CLS and reference solutions have been proposed in the literature so far~\cite{benchmark_adams, brantley_benchmark, renewal, brantley_conf, brantley_conf_2}, with focus exclusively on $1d$ geometries, either of the rod or slab type. Flat two-dimensional configurations have received less attention~\cite{haran}.

In a series of recent papers, some of the authors have provided reference solutions for particle transport in extruded two-dimensional and full three-dimensional random media with Markov statistics~\cite{larmier_benchmark, larmier_models}, where the spatial disorder has been generated by means of homogeneous and isotropic $d$-dimensional Poisson tessellations~\cite{larmier}. In this work, we will compare the CLS simulation results to the reference solutions for the classical benchmark problem proposed by Adams, Larsen and Pomraning for transport in stochastic media~\cite{benchmark_adams} and revisited by Brantley~\cite{brantley_benchmark}. The case of $1d$ slab disorder has been considered previously in the literature~\cite{benchmark_adams, brantley_benchmark, renewal, brantley_conf, brantley_conf_2} and will be reported here for the sake of completeness. In addition, we will also consider $2d$ extruded and full $3d$ Markov mixing configurations. The physical observables of interest will be the particle flux $\langle \varphi \rangle$, the transmission coefficient $\langle T \rangle$ and the reflection coefficient $\langle R \rangle$: we will examine the discrepancies between reference and CLS simulation results as a function of the benchmark configurations and of the system dimensionality $d$. In order to verify the consistency of the proposed results, the CLS calculations will be performed by using two independent Monte Carlo implementations of the CLS algorithm, in the \tripoli{} code~\cite{T4} and in the Mercury code~\cite{mercury2016, BrantleyEtAl2017}, respectively.

This paper is organized as follows: in Sec.~\ref{benchmark_definition} we will recall the benchmark specifications that will be used in this work. In Secs.~\ref{quenched_approach} and~\ref{cls_approach} we will detail the methods and the algorithms that we have adopted in order to produce reference and CLS results, respectively. Simulation findings will be illustrated and discussed in Sec.~\ref{simulation_results}. Conclusions will be finally drawn in Sec.~\ref{conclusions}.

\section{Benchmark specifications}
\label{benchmark_definition}

In order for the paper to be self-contained, we start by recalling the benchmark specifications that have been selected for this work, which are essentially taken from those originally proposed in~\cite{benchmark_adams} and~\cite{renewal}, and later extended in~\cite{brantley_benchmark, brantley_conf, brantley_conf_2}.

We consider single-speed linear particle transport through a stochastic binary medium with homogeneous Markov mixing. The medium is non-multiplying, with isotropic scattering. The geometry consists of a cubic box of side $L=10$ (in arbitrary units), with reflective boundary conditions on all sides of the box except two opposite faces (say those perpendicular to the $x$ axis), where leakage boundary conditions are imposed~\footnote{In~\cite{benchmark_adams} and~\cite{renewal}, system sizes $L=0.1$ and $L=1$ were also considered, but in this work we will focus on the case $L=10$, which leads to more physically relevant configurations.}. Two kinds of non-stochastic sources will be considered: either an imposed normalized incident angular flux on the leakage surface at $x=0$ (with zero interior sources), or a distributed homogeneous and isotropic normalized interior source (with zero incident angular flux on the leakage surfaces). Following the notation in~\cite{brantley_benchmark}, the benchmark configurations pertaining to the former kind of source will be called {\em suite} I, whereas those pertaining to the latter will be called {\em suite} II. The material properties for the Markov mixing are entirely defined by assigning the average chord length for each material $i = \alpha, \beta$, namely $\Lambda_i$, which in turn allows deriving the homogeneous probability $p_i$ of finding material $i$ at an arbitrary location within the box, namely
\begin{equation}
p_{i}= \frac{\Lambda_i}{\Lambda_i + \Lambda_j}.
\end{equation}
By definition, the material probability $p_i$ yields the volume fraction for material $i$. The cross sections for each material will be denoted as customary $\Sigma_i$ for the total cross section and $\Sigma_{s,i}$ for the scattering cross section. The average number of particles surviving a collision in material $i$ will be denoted by $c_i = \Sigma_{s,i} / \Sigma_i \le 1$. The physical parameters for the benchmark configurations are recalled in Tabs.~\ref{tab_param1} and~\ref{tab_param2}: the benchmark specifications include three cases (numbered $1$, $2$ and $3$, corresponding to different materials), and three sub-cases (noted $a$, $b$ and $c$, corresponding to different $c_i $ for a given material) for each case~\cite{benchmark_adams}. The so-called atomic mix limit~\cite{pomraning}, where one assumes that the statistical disorder can be approximated by simply taking a full homogenization of the physical properties based on the ensemble-averaged cross sections, has been examined, e.g.,  in~\cite{brantley_benchmark} for $d=1$ and in~\cite{larmier_benchmark} for $d=2$ and $d=3$ and will not be considered here.

\begin{table}[!ht]
\begin{center}
\begin{tabular}{lcccc}
\toprule
Case & $\Sigma_\alpha$ & $\Lambda_\alpha$ & $\Sigma_\beta$ & $\Lambda_\beta$ \\
\midrule
1 & 10/99 & 99/100 & 100/11 & 11/100 \\
2 & 10/99 & 99/10 & 100/11 & 11/10 \\
3 & 2/101 & 101/20 & 200/101 & 101/20 \\
\bottomrule
\end{tabular}
\end{center}
\caption{Material parameters for the three cases of the benchmark configurations.}
\label{tab_param1}
\end{table}

\begin{table}[!ht]
\begin{center}
\begin{tabular}{lccc}
\toprule
Sub-case & a & b & c \\
\midrule
$c_\alpha$ & 0 & 1 & 0.9 \\
$c_\beta$ & 1 & 0 & 0.9 \\
\bottomrule
\end{tabular}
\end{center}
\caption{Scattering probabilities for the three sub-cases of the benchmark configurations.}
\label{tab_param2}
\end{table}

The physical observables of interest for the proposed benchmark will be the ensemble-averaged outgoing particle currents $\langle J \rangle $ on the two surfaces with leakage boundary conditions, the ensemble-averaged scalar particle flux $\langle \varphi(x) \rangle= \langle \int \int \int\varphi({\bf r}, {\boldsymbol \omega}) d {\boldsymbol \omega} dy dz \rangle$ along $0 \le x \le L$, and the total scalar flux $\langle \varphi \rangle = \langle \int \varphi(x) dx \rangle$. For the {\em suite} I configurations, the outgoing particle current on the side opposite to the imposed current source will represent the ensemble-averaged transmission coefficient, namely, $\langle T \rangle = \langle J_{x=L} \rangle $, whereas the outgoing particle current on the side of the current source will represent the ensemble-averaged reflection coefficient, namely, $\langle R \rangle = \langle J_{x=0} \rangle $. For the {\em suite} II configurations, the outgoing currents on opposite faces are expected to be equal (within statistical fluctuations), for symmetry reasons. In this case, we also introduce the average leakage current $\langle L \rangle = \langle (T+R)/2 \rangle$.

\section{Reference solutions}
\label{quenched_approach}

For particle transport in the presence of quenched disorder with $d$-dimensional Markov mixing, reference solutions for the ensemble-averaged scalar particle flux $\langle \varphi(x) \rangle$ and the currents $\langle R \rangle$ and $\langle T \rangle$ have been first obtained and thoroughly described in~\cite{larmier_benchmark}. For this work, CEA has produced a new set of reference solutions with lower statistical uncertainty~\footnote{In order to reduce computer time for highly fragmented geometries including hundreds of thousands of polyhedra, in~\cite{larmier_benchmark} we used thin empty boxes adjacent to the two free surfaces in order to sample the source particles for {\em suite} I configurations and to record the weights of the particles escaping from the viable domain. In the new simulations, we have also decreased the thickness of these boxes in order to improve the overall accuracy.}. Here we will briefly detail the methods and the simulation parameters that have been used and recall the changes introduced with respect to~\cite{larmier_benchmark}.

\subsection{Poisson tessellations}

Random tessellations are stochastic aggregates of disjoint and space-filling cells obeying a given distribution~\cite{santalo}. Poisson tessellations are obtained by partitioning a domain of a $d$-dimensional space by sampling $(d-1)$-dimensional hyper-planes from an auxiliary Poisson process~\cite{santalo, miles1964, miles1972}. An explicit construction amenable to Monte Carlo realizations for two-dimensional homogeneous and isotropic Poisson geometries of finite size has been established in~\cite{switzer}. A generalization of this algorithm to $d$-dimensional domains has been recently proposed~\cite{mikhailov}. The construction of Poisson stochastic geometries depends on a single free parameter $\rho$, which takes the name of tessellation density, and is such that an arbitrary segment of length $s$ will have on average $\rho s$ intersections with the random hyper-planes.

The algorithm for the $1d$ slab tessellations is recalled in~\cite{benchmark_adams}, based on the Poisson process on the line. For the $2d$ extruded tessellations, we begin by creating an isotropic Poisson tessellation of a square of side $L$, according to the algorithm detailed in~\cite{lepage}. The full geometrical description for the cube is simply achieved by extruding the random polygons of the plane along the orthogonal (say $z$) axis. The algorithm for $3d$ tessellations of a cube of side $L$ by drawing random planes has been detailed in~\cite{larmier}.

Isotropic Poisson geometries satisfy a Markov property: for domains of infinite size, arbitrary drawn lines will be cut by the $(d-1)$-surfaces of the $d$-polyhedra into segments whose lengths $\ell$ are exponentially distributed, with average chord length $\langle \ell \rangle = 1/\rho$~\cite{santalo}. The quantity $\Lambda = 1/\rho$ intuitively defines the correlation length of the Poisson geometry, i.e, the typical linear size of a volume composing the random tessellation.

\subsection{Colored stochastic geometries}

Binary Markov mixtures required for the benchmark specifications are obtained as follows: first, a $d$-dimensional Poisson tessellation is constructed as described above. Then, each polyhedron of the geometry is assigned a material composition by formally attributing a distinct `color', say $\alpha$ or $\beta$, with associated complementary probabilities $p_\alpha$ and $p_\beta = 1-p_\alpha$~\cite{pomraning}. This gives rise to (generally) non-convex $\alpha$ and $\beta$ clusters, each composed of a random number of convex polyhedra. It can be shown that the average chord length $\Lambda_\alpha$ through clusters with composition $\alpha$ is related to the correlation length $\Lambda$ of the geometry via $\Lambda = (1-p_\alpha) \Lambda_\alpha$, and for $\Lambda_\beta$ we similarly have $\Lambda =  p_\alpha \Lambda_\beta$. This yields $1 / \Lambda_\alpha + 1 / \Lambda_\beta =  1 / \Lambda$, and we recover
\begin{equation}
p_\alpha = \frac{\Lambda}{\Lambda_\beta} = \frac{\Lambda_\alpha}{\Lambda_\alpha + \Lambda_\beta}.
\end{equation}
Based on the formulas above, and using $\rho = 1/\Lambda$, the parameters of the colored Poisson geometries corresponding to the benchmark specifications provided in Tab.~\ref{tab_param1} are easily derived.

\subsection{Particle transport and ensemble averages}

For each benchmark case and sub-case, a large number $M$ of geometries has been generated, and the material properties have been attributed to each volume as described above. Then, for each realization $k$ of the ensemble, linear particle transport has been simulated by using the production Monte Carlo code \tripoli{}, developed at CEA~\cite{T4}. \tripoli{} is a general-purpose stochastic transport code capable of simulating the propagation of neutral and charged particles with continuous-energy cross sections in arbitrary geometries. In order to comply with the benchmark specifications, constant cross sections adapted to mono-energetic transport and isotropic angular distributions have been prepared. The number of simulated particle histories per configuration is $10^6$. For a given physical observable ${\cal O}$, the benchmark solution is obtained as the ensemble average
\begin{equation}
\langle {\cal O} \rangle = \frac{1}{M} \sum_{k=1}^M {\cal O}_k,
\end{equation}
where ${\cal O}_k$ is the Monte Carlo estimate for the observable ${\cal O}$ obtained for the $k$-th realization. Specifically, currents $R_k$ and $T_k$ at a given surface are estimated by summing the statistical weights of the particles crossing that surface. Scalar fluxes $\varphi_k(x)$ have been tallied using the standard track length estimator over a pre-defined spatial grid containing $10^2$ uniformly spaced meshes along the $x$ axis.

The error affecting the average observable $\langle {\cal O} \rangle$ results from two separate contributions, the dispersion
\begin{equation}
\sigma^2_G = \frac{1}{M} \sum_{k=1}^M {{\cal O}_k}^2 - {\langle {\cal O} \rangle}^2
\end{equation}
of the observables exclusively due to the stochastic nature of the geometries and of the material compositions, and 
\begin{equation}
\sigma^2_{{\cal O}}=\frac{1}{M} \sum_{k=1}^M \sigma_{{\cal O}_k}^2,
\end{equation}
which is an estimate of the variance due to the stochastic nature of the Monte Carlo method for particle transport, $\sigma_{{\cal O}_k}^2$ being the dispersion of a single calculation~\cite{donovan, sutton}. The statistical error on $\langle {\cal O} \rangle$ is then estimated as
\begin{equation}
\sigma[ \langle {\cal O}\rangle ] = \sqrt{\frac{\sigma^2_G}{M}+\sigma^2_{{\cal O}}}.
\end{equation}

In order to reduce the dispersion of the observables due to the statistical nature of the geometries, for the new set of reference solutions computed for this work we have increased the number of realizations for the benchmark configurations displaying larger correlation lengths (i.e., larger material chunks).

For $1d$ slab tessellations, we have taken $M=4\times 10^4$ for sub-cases $1b$; $M=5 \times 10^4$ for sub-cases $2b$; and $M=10^5$ for sub-cases $3b$. Otherwise, we have used the same number of realizations as in~\cite{larmier_benchmark}, namely, $M=5 \times 10^4$ for sub-case $2a$ of the \textit{suite} II, and $M=10^4$ for the remaining cases and sub-cases.

For the $2d$ extruded tessellations, we have taken $M=2 \times 10^4$ for the sub-cases $2b$ and $M=5 \times 10^4$ for the sub-cases $3b$. Otherwise, we have used the same number of realizations as in~\cite{larmier_benchmark}, namely, $M=4 \times 10^3$.

Finally, for the $3d$ tessellations we have taken $M=2 \times 10^4$ for the sub-case $2b$ of the \textit{suite} II; $M=5 \times 10^3$ for all the other sub-cases of case $2$; $M=5 \times 10^4$ for the sub-case $3b$ of the \textit{suite} II; and $M=10^4$ for all the other sub-cases of case $3$. For all remaining cases and sub-cases, we have used the same number of realizations as in~\cite{larmier_benchmark}, namely, $M=10^3$.

Actually, increasing the dimension $d$ implies a better statistical mixing (in other words, a single realization is more representative of the average behaviour), at the expense of increasing the computational burden (each realization takes longer both for generation and for Monte Carlo transport).

Transport calculations have been run on a cluster based at CEA, with Intel Xeon E5-2680 V2 2.8 GHz processors. The average computer time globally increases as a function of dimension, but depends also on the correlation lengths, volume fractions, and material properties such as cross sections and scattering probabilities. For the simulations discussed here we have largely benefited from a feature implemented in the code \tripoli{}, namely the possibility of reading pre-computed connectivity maps for the volumes composing the geometry. During the generation of the Poisson tessellations, care has been taken so as to store the indices of the neighbouring volumes for each realization, which means that during the geometrical tracking a particle will have to find the following crossed volume in a list that might be considerably smaller than the total number of random volumes composing the box (depending on the features of the random geometry).

\section{The Chord Length Sampling approach}
\label{cls_approach}

Reference solutions based on the quenched disorder approach are computationally expensive, so that intensive research efforts have been devoted to the development of Monte Carlo-based annealed disorder models capable of approximating the ensemble observables on-the-fly, i.e., with a single particle transport simulation. The pioneering work by Zimmerman and Adams~\cite{zimmerman, zimmerman_adams} has led to a family of algorithms that go now under the name of Chord Length Sampling methods. In particular, it has been shown that the standard form of the CLS (Algorithm A in~\cite{zimmerman_adams}) formally solves the Levermore-Pomraning equations, i.e., Eq.~\eqref{boltzmann_ave} with the closure formula~\eqref{closure_lp_boltzmann}, corresponding to Markov mixing with the approximation that memory of the crossed material interfaces is lost at each particle flight~\cite{sahni1, sahni2}.

Algorithm A proceeds as follows~\cite{zimmerman_adams}: each particle history begins by sampling position, angle and velocity from the specified source, as customary. Moreover, the particle is assigned a supplementary attribute, the material label, which is sampled from the probability $p_i$. Then we need to compute three distances, denoted respectively $\ell_b$, $\ell_c$, and $\ell_i$. The quantity $\ell_b$ is the distance to the next physical boundary, along the current direction of the particle. The quantity $\ell_c$ is the distance to the next collision, which is determined by using the material cross section that has been chosen at the previous step: if the particle has a material $\alpha$, e.g., then $\ell_c$ will be drawn from an exponential distribution of parameter $1/\Sigma_\alpha$. Finally, the quantity $\ell_i$ is the distance to the next material interface, which is sampled from an exponential distribution with parameter $\Lambda_\alpha$, i.e., the average chord length of material $\alpha$, if the particle has a material label $\alpha$ (whence the name of CLS).

Then, the minimum distance among $\ell_b, \ell_c$ and $\ell_i$ must be selected: if the minimum is $\ell_b$, the particle is moved along a straight line until it hits the external boundary; if the minimum is $\ell_c$, the particle is moved to the collision point, and the outgoing particle features are selected according to the collision kernel pertaining to the current material label. If the minimum is $\ell_i$, the particle is moved to the interface between the two materials, and the material label is switched. If the particle is not absorbed, a new set of distances $\ell_b, \ell_c$ and $\ell_i$ are determined. During the time spent within the random medium, the particle will be thus either colliding within a random chunk, or crossing the interface between two chunks; the particle will ultimately get absorbed in one of the chunks, or escape out of the boundaries of the random medium. The Monte Carlo estimators for the scalar flux and the currents are the same as those for the reference solutions described above.

As observed above, Algorithm A assumes that the particle has no memory of its past history, and in particular the crossed interfaces are immediately forgotten (which is coherent with the closure formula of the Levermore-Pomraning model). In this respect, CLS is an approximation of the exact treatment of disorder-induced spatial correlations (actually, it can be shown that CLS is exact only for pure absorbers). As a result, Algorithm A is expected to be less accurate in the presence of strong scatterers with optically thick mean material chunk length. A thorough discussion of the shortcomings of the CLS approach for $d=1$ can be found, e.g., in~\cite{ji_cls}.

\subsection{Slab geometries}

For mono-energetic particle transport in slab geometries with isotropic scattering, the Boltzmann equation~\eqref{boltzmann} yields
\begin{equation}
\mu \frac{\partial}{\partial x} \varphi+ \Sigma(x) \varphi = \frac{\Sigma_s(x)}{2} \int_{-1}^{1} d\mu' \varphi(x,\mu'),
\label{eq_lp_1d}
\end{equation}
where $\varphi=\varphi(x,\mu)$ is the angular particle flux for particles at position $x$ with a direction cosine $\mu = \cos(\theta)$ with respect to the $x$ axis. The source and the boundary conditions depend on the benchmark specifications.

Correspondingly, the CLS algorithm that formally solves the Levermore-Pomraning model as applied to Eq.~\ref{eq_lp_1d} is the following. For {\em suite} I, the source particle position is set to $x=0$, and the direction cosine is sampled from a cosine distribution, namely,
\begin{equation}
\mu = \sqrt{\xi},
\end{equation}
where $\xi$ is a uniform random number in $[0,1)$, in order to ensure the isotropic incident flux condition. For {\em suite} II, the starting position $x$ is sampled uniformly in $[0,L]$, and the direction cosine is sampled uniformly in $[-1,1]$ in order to ensure the uniform and isotropic source condition. According to the Levermore-Pomraning prescription, the distance to material interfaces for a particle in material $\alpha$ is sampled from an exponential distribution as follows:
\begin{equation}
d_i = - \frac{\Lambda_\alpha}{|\mu|} \ln(1-\xi), 
\end{equation}
where the factor $1/|\mu|$ accounts for the projection of the distance along the $x$ axis. The distance to the next collision is sampled from the exponential distribution of parameter $1/\Sigma_\alpha(x)$, and the distance to the boundary is computed as customary. For isotropic scattering, the cosine direction after collision is sampled uniformly in $[-1,1]$.

\subsection{Two-dimensional extruded geometries}

Assuming again mono-energetic particle transport with isotropic scattering, the Boltzmann equation for two-dimensional geometries extruded in the $z$ axis direction yields
\begin{align}
& \sqrt{1-\mu^2}\cos(\phi)\frac{\partial}{\partial x}\varphi + \sqrt{1-\mu^2}\sin(\phi)\frac{\partial}{\partial y}\varphi  =\nonumber \\
&\Sigma(x,y) \varphi + \frac{\Sigma_s(x,y)}{4 \pi} \int_{-1}^{1} d\mu' \int_{0}^{2\pi} d\phi' \varphi(x,y,\mu',\phi') ,
\label{eq_lp_2d}
\end{align}
where $\varphi = \varphi(x,y,\mu,\phi)$ is the angular particle flux for particles being at position $x,y$ with a direction cosine $\mu=\cos(\theta)$ with respect to the $z$ axis and a polar angle $\phi$ with respect to the $x$ axis.

The CLS algorithm that formally corresponds to solving the Levermore-Pomraning model as applied to Eq.~\ref{eq_lp_2d} is the following. For {\em suite} I, the source particle positions are set to $x=0$ and $y$ taken uniformly in $[0,L]$. Then we sample a direction cosine $\mu'$ (with respect to the $x$ axis) from
\begin{equation}
\mu' = \sqrt{\xi}
\end{equation}
where $\xi$ is taken in $[0,1)$, and a polar angle $\phi'$ (with respect to the $y$ axis) uniform in $[0,2 \pi]$. The initial particle direction is
\begin{equation}
{\boldsymbol \omega}_0 = \left\lbrace  \frac{\mu'}{Q} ,\frac{ \sqrt{1-\mu'^2}\cos(\phi') }{Q}  \right\rbrace,
\end{equation}
with
\begin{equation}
Q = \sqrt{\mu'^2+(1-\mu'^2)\cos^2(\phi')},
\end{equation}
in order to ensure the isotropic incident flux condition, and the initial direction cosine $\mu_0$ is defined by
\begin{equation}
\mu_0 = \sqrt{1-\mu'^2}\sin(\phi').
\end{equation}
For {\em suite} II, the starting positions $x,y$ are sampled uniformly in $[0,L]\times[0,L]$, the direction cosine $\mu$ is sampled uniformly in $[-1,1]$ and the polar angle $\phi$ is sampled uniformly in $[0,2\pi]$ in order to ensure the uniform and isotropic source condition, which yields the initial particle direction
\begin{equation}
{\boldsymbol \omega}_0  = \left\lbrace \cos(\phi) ,\sin(\phi)  \right\rbrace.
\end{equation}
According to the Levermore-Pomraning prescription, the distance to material interfaces for a particle in material $\alpha$ is sampled from an exponential distribution as follows:
\begin{equation}
d_i = - \frac{\Lambda_\alpha}{\sqrt{1-\mu^2}} \ln(1-\xi), 
\end{equation}
where the factor $1/\sqrt{1-\mu^2}$ again accounts for the projection of the distance on the $x-y$ plane. The distance to the next collision is sampled from the exponential distribution of parameter $1/\Sigma_\alpha(x,y)$, and the distance to the boundary is computed as customary. For isotropic scattering, the cosine direction $\mu$ after collision is sampled uniformly in $[-1,1]$, and the polar angle $\phi$ is sampled uniformly in $[0,2\pi]$; the particle direction is then given by
\begin{equation}
{\boldsymbol \omega} = \left\lbrace \cos(\phi) ,\sin(\phi)  \right\rbrace.
\end{equation}

\subsection{Three-dimensional geometries}

The Boltzmann equation for mono-energetic transport with isotropic scattering in three-dimensional geometries yields
\begin{align}
&\left(  \sqrt{1-\mu^2}\cos(\phi)\frac{\partial}{\partial x} + \sqrt{1-\mu^2}\sin(\phi)\frac{\partial}{\partial y} + \mu \frac{\partial}{\partial z} \right) \varphi =\nonumber \\
&\Sigma(x,y,z) \varphi + \frac{\Sigma_s(x,y,z)}{4 \pi} \int_{-1}^{1} d\mu' \int_{0}^{2\pi} d\phi' \varphi(x,y,z,\mu',\phi'),
\label{eq_lp_3d}
\end{align}
where $\varphi = \varphi(x,y,z,\mu,\phi)$ is the angular particle flux for particles being at position $x,y,z$ with a direction cosine $\mu=\cos(\theta)$ with respect to the $z$ axis and a polar angle $\phi$ with respect to the $x$ axis.

The CLS algorithm that formally corresponds to solving the Levermore-Pomraning model as applied to Eq.~\ref{eq_lp_3d} is the following. For {\em suite} I, the source particle positions are set to $x=0$ and $y,z$ taken uniformly in $[0,L]\times[0,L]$. Then we sample a direction cosine $\mu'$ (with respect to the $x$ axis) from
\begin{equation}
\mu' = \sqrt{\xi}
\end{equation}
where $\xi$ is taken in $[0,1)$, and a polar angle $\phi'$ (with repect to the $y$ axis) uniform in $[0,2 \pi]$. The initial particle direction is
\begin{equation}
{\boldsymbol \omega}_0 = \left\lbrace  \mu' , \sqrt{1-\mu'^2}\cos(\phi') , \sqrt{1-\mu'^2}\sin(\phi') \right\rbrace
\end{equation}
in order to ensure the isotropic incident flux condition. For {\em suite} II, the starting positions $x,y,z$ are sampled uniformly in $[0,L]\times[0,L]\times[0,L]$, the direction cosine $\mu$ is sampled uniformly in $[-1,1]$ and the polar angle is sampled uniformly in $[0,2\pi]$ in order to ensure the uniform and isotropic source condition, which yields the initial particle direction
\begin{equation}
{\boldsymbol \omega}_0 = \left\lbrace \sqrt{1-\mu^2}\cos(\phi) , \sqrt{1-\mu^2}\sin(\phi) , \mu \right\rbrace.
\end{equation}
According to the Levermore-Pomraning prescription, the distance to material interfaces for a particle in material $\alpha$ is sampled from an exponential distribution as follows:
\begin{equation}
d_i = - \Lambda_\alpha \ln(1-\xi).
\end{equation}
The distance to the next collision is sampled from the exponential distribution of parameter $1/\Sigma_\alpha(x,y,z)$, and the distance to the boundary is computed as customary. For isotropic scattering, the cosine direction $\mu$ after collision is sampled uniformly in $[-1,1]$, and the polar angle $\phi$ is sampled uniformly in $[0,2\pi]$; the particle direction is then given by
\begin{equation}
{\boldsymbol \omega} = \left\lbrace \sqrt{1-\mu^2}\cos(\phi) , \sqrt{1-\mu^2}\sin(\phi) , \mu \right\rbrace.
\end{equation}

\section{Simulation results}
\label{simulation_results}

The simulation results for the total scalar flux $\langle \varphi \rangle$, the transmission coefficient $\langle T \rangle$ and the reflection coefficient $\langle R \rangle$ are provided in Tabs.~\ref{tab_suite1_case1} to \ref{tab_suite1_case3} for the benchmark cases corresponding to {\em suite} I, and in Tabs.~\ref{tab_suite2_case1} to \ref{tab_suite2_case3} for the benchmark cases corresponding to {\em suite} II, respectively. The reference solutions have been computed by following the procedure detailed in Sec.~\ref{quenched_approach}, based on~\cite{larmier_benchmark}.

The CLS results have been obtained with both \tripoli{} and Mercury Monte Carlo codes by following the procedure described in Sec.~\ref{cls_approach}. We will denote by $\sigma_\text{CLS}[ {\cal O}]$ the resulting statistical uncertainty associated to each physical observable ${\cal O}$. For the \tripoli{} CLS simulations of the $d$-dimensional benchmark configurations we have used 10$^9$ particle histories. Mercury is a Monte Carlo particle transport code being developed at Lawrence Livermore National Laboratory~\cite{mercury2016, BrantleyEtAl2017}. The Monte Carlo Levermore-Pomraning CLS algorithm was previously implemented in Mercury~\cite{brantley_lp} in a manner consistent with the algorithmic descriptions in~\cite{zimmerman_adams, brantley_benchmark} and Sec.~\ref{cls_approach}. The Mercury Levermore-Pomraning implementation has been demonstrated~\cite{brantley_lp} to accurately reproduce the independent one-dimensional slab geometry Monte Carlo Levermore-Pomraning results in~\cite{brantley_benchmark}. We modelled the three-dimensional benchmark {\em suites} I and II using the Mercury Levermore-Pomraning CLS implementation with 10$^{9}$ particle histories. We obtained results that are generally statistically equivalent to the \tripoli{} CLS results to typically within three standard deviations for the reflection and transmission coefficients and the scalar flux distributions (agreement to typically four to five digits).  For this paper, we will present only the \tripoli{} simulation results. Computer times for the reference and CLS solutions are also provided in the same tables: not surprisingly, the CLS approach is much faster than the reference method, since a single transport simulation is needed.

As a general remark, the accuracy of CLS with respect to reference solutions increases with increasing system dimensionality $d$. This is expected on physical grounds, since the higher $d$ and the smaller is the impact of the spatial correlations: a particle undergoing back-scattering is less likely to cross exactly the same material interface as the one crossed during the previous flight. In other words, the approximations introduced in the CLS algorithm by neglecting spatial correlations will have a weaker effect on particle transport. Nonetheless, simulation results show a few exceptions among the examined configurations. Moreover, the accuracy of CLS also generally improves when increasing the tessellation density, i.e., decreasing the average chord length: configurations pertaining to case $1$ globally show a better agreement than those of case $2$, and those of case $2$ show a better agreement than those of case $3$.

The effects of system dimensionality on the discrepancies between CLS and exact solutions are stronger for configurations with smaller average chord lengths. This behaviour is again consistent with the fact that increasing the chord length induces larger chunks of materials, and for chunks that span a large fraction of the entire geometry the impact of dimensionality must be rather weak: in this regime, particle transport is mostly influenced by the material volume fractions (i.e., the coloring probability).

The behaviour of {\em suite} II configurations is quite similar to that of {\em suite} I configurations, and no specific trend due to the source and/or initial conditions can be easily detected.

\begin{figure*}
\begin{center}
\,\,\,\, Case 1a \,\,\,\,\\
\scalebox{0.7}{\input{flux_1A_I}}\,\,\,\,
\scalebox{0.7}{\input{flux_1A_II}}\\
\,\,\,\, Case 1b \,\,\,\,\\
\scalebox{0.7}{\input{flux_1B_I}}\,\,\,\,
\scalebox{0.7}{\input{flux_1B_II}}\\
\,\,\,\, Case 1c \,\,\,\,\\
\scalebox{0.7}{\input{flux_1C_I}}\,\,\,\,
\scalebox{0.7}{\input{flux_1C_II}}\\
\end{center}
\caption{Ensemble-averaged spatial scalar flux for the benchmark configurations: Case 1. Left column: {\em suite} I configurations; right column: {\em suite} II configurations. Blue lines correspond to $d=1$, red lines to $d=2$ and green lines to $d=3$. Solid lines represent the benchmark solutions (quenched disorder approach), dotted or dashed lines represent the solutions from the Chord Length Sampling algorithm (annealed disorder approach).\label{fig_space_1}}
\end{figure*}
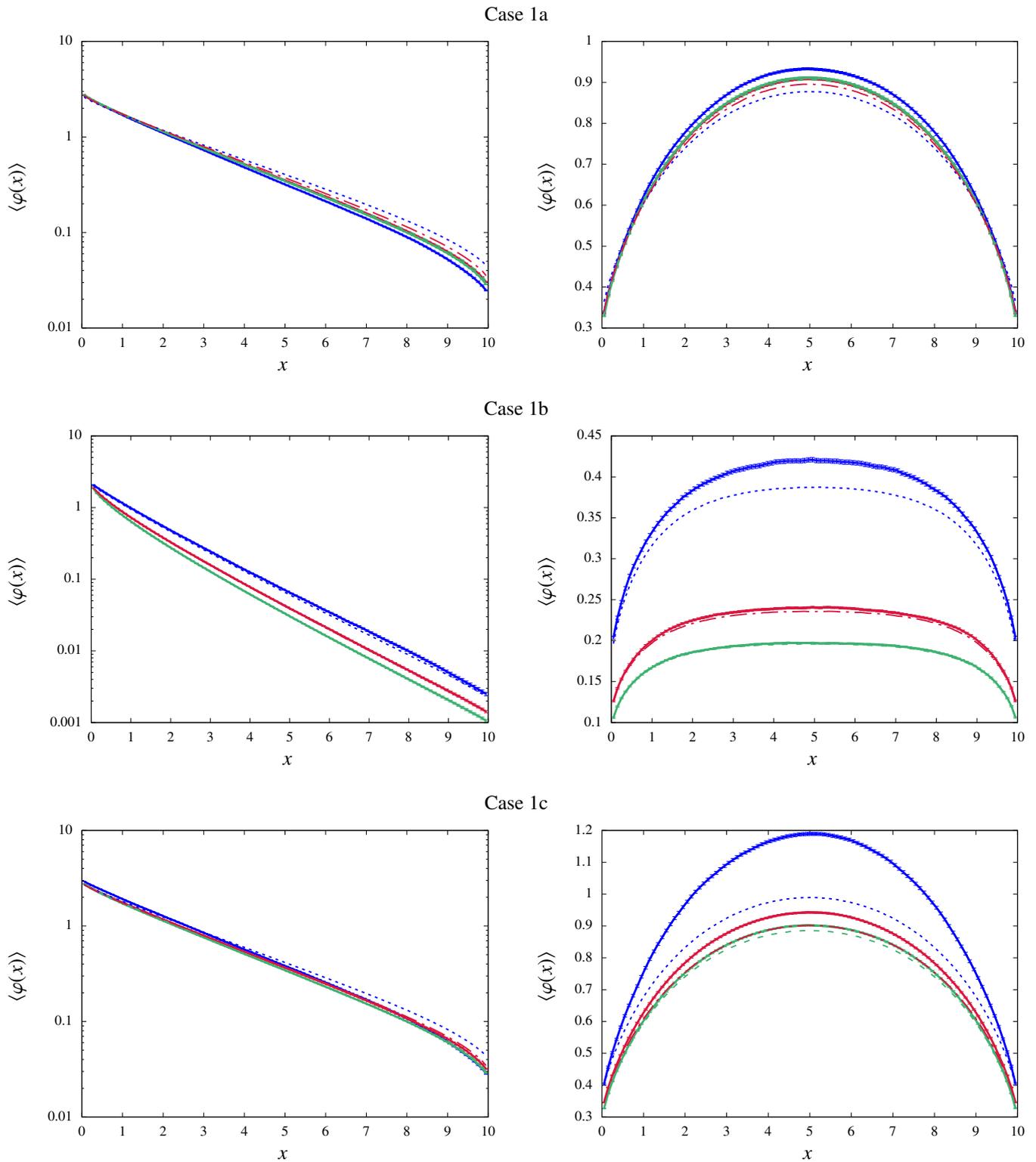

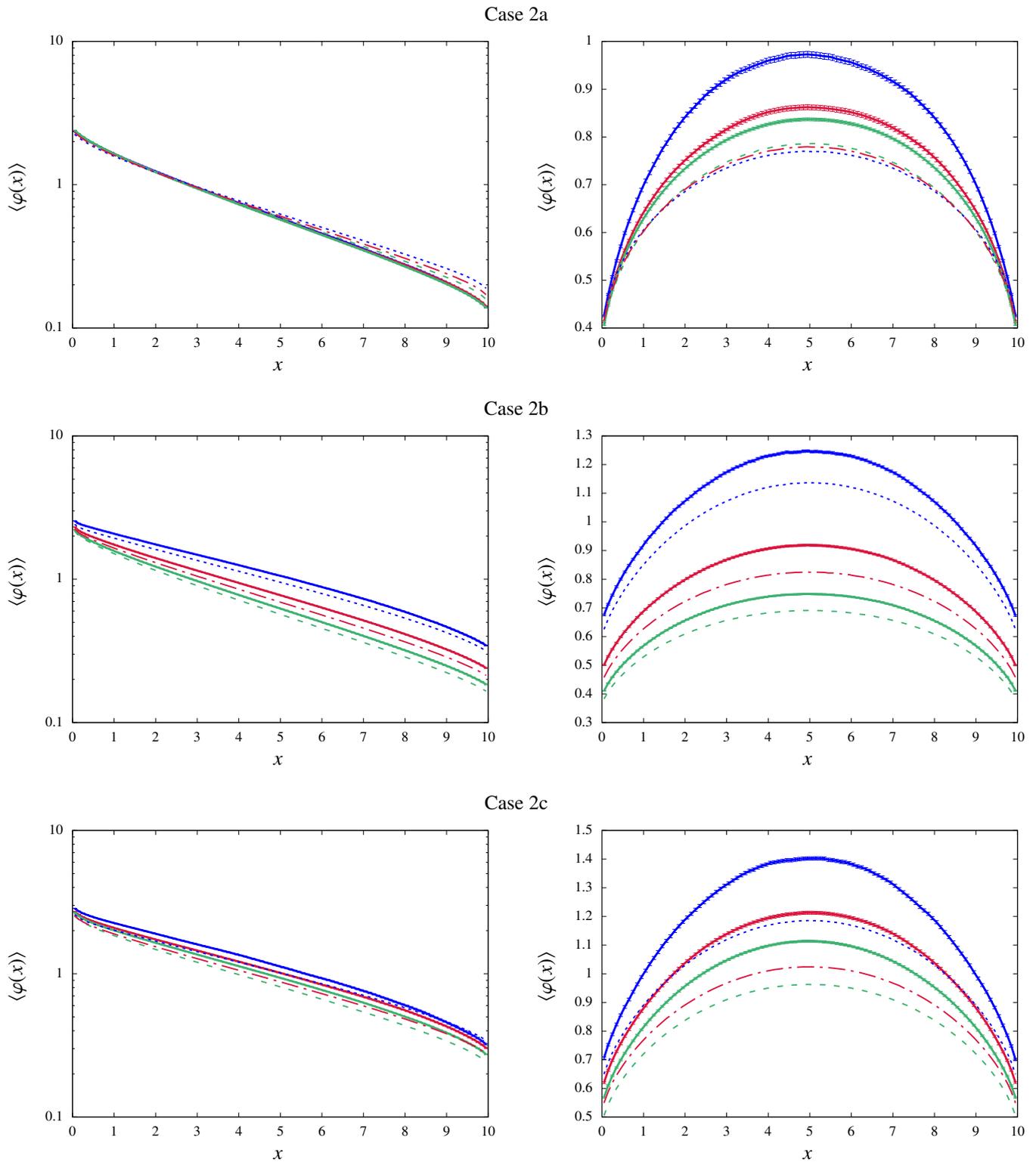
\begin{figure*}
\begin{center}
\,\,\,\, Case 2a \,\,\,\,\\
\scalebox{0.7}{\input{flux_2A_I}}\,\,\,\,
\scalebox{0.7}{\input{flux_2A_II}}\\
\,\,\,\, Case 2b \,\,\,\,\\
\scalebox{0.7}{\input{flux_2B_I}}\,\,\,\,
\scalebox{0.7}{\input{flux_2B_II}}\\
\,\,\,\, Case 2c \,\,\,\,\\
\scalebox{0.7}{\input{flux_2C_I}}\,\,\,\,
\scalebox{0.7}{\input{flux_2C_II}}\\
\end{center}
\caption{Ensemble-averaged spatial scalar flux for the benchmark configurations: Case 2. Left column: {\em suite} I configurations; right column: {\em suite} II configurations. Blue lines correspond to $d=1$, red lines to $d=2$ and green lines to $d=3$. Solid lines represent the benchmark solutions (quenched disorder approach), dotted or dashed lines represent the solutions from the Chord Length Sampling algorithm (annealed disorder approach).\label{fig_space_2}}
\end{figure*}

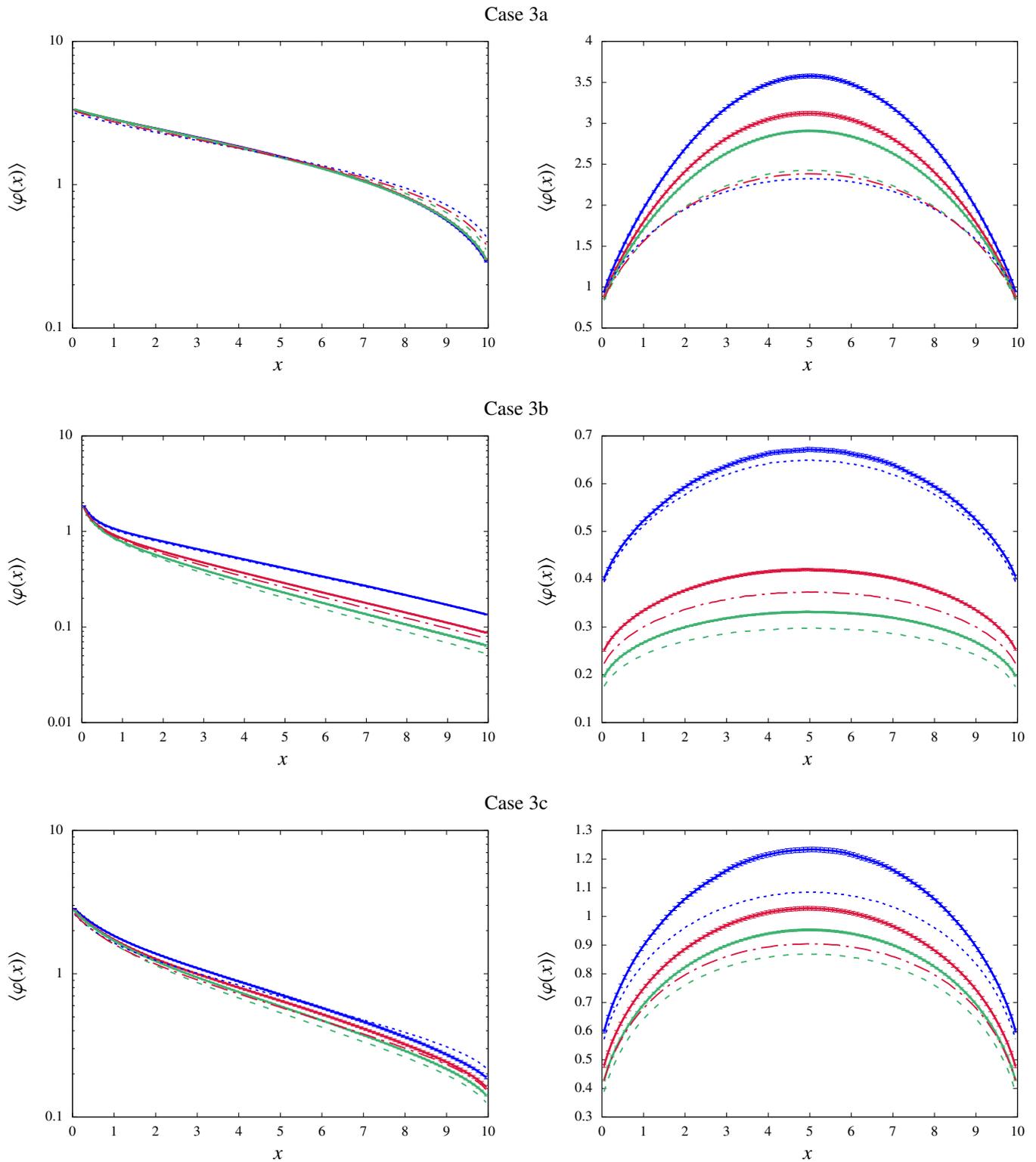
\begin{figure*}
\begin{center}
\,\,\,\, Case 3a \,\,\,\,\\
\scalebox{0.7}{\input{flux_3A_I}}\,\,\,\,
\scalebox{0.7}{\input{flux_3A_II}}\\
\,\,\,\, Case 3b \,\,\,\,\\
\scalebox{0.7}{\input{flux_3B_I}}\,\,\,\,
\scalebox{0.7}{\input{flux_3B_II}}\\
\,\,\,\, Case 3c \,\,\,\,\\
\scalebox{0.7}{\input{flux_3C_I}}\,\,\,\,
\scalebox{0.7}{\input{flux_3C_II}}\\
\end{center}
\caption{Ensemble-averaged spatial scalar flux for the benchmark configurations: Case 3. Left column: {\em suite} I configurations; right column: {\em suite} II configurations.  Blue lines correspond to $d=1$, red lines to $d=2$ and green lines to $d=3$. Solid lines represent the results from the benchmark (quenched disorder approach), dotted or dashed lines represent the results from the Chord Length Sampling algorithm (annealed disorder approach).\label{fig_space_3}}
\end{figure*}

\begin{figure*}
\begin{center}
\,\,\,\, Case 1a \,\,\,\,\\
\scalebox{0.7}{\input{flux_1A_I_ecart}}\,\,\,\,
\scalebox{0.7}{\input{flux_1A_II_ecart}}\\
\,\,\,\, Case 1b \,\,\,\,\\
\scalebox{0.7}{\input{flux_1B_I_ecart}}\,\,\,\,
\scalebox{0.7}{\input{flux_1B_II_ecart}}\\
\,\,\,\, Case 1c \,\,\,\,\\
\scalebox{0.7}{\input{flux_1C_I_ecart}}\,\,\,\,
\scalebox{0.7}{\input{flux_1C_II_ecart}}\\
\end{center}
\caption{Discrepancy $\Delta[\langle\varphi(x)\rangle]$ between the ensemble-averaged spatial flux $\langle\varphi(x)\rangle$ obtained with Poisson tessellations (quenched disorder approach) and that obtained with the Chord Length Sampling algorithm (annealed disorder approach) for the benchmark configurations: Case 1. Left column: {\em suite} I configurations; right column: {\em suite} II configurations. Blue lines correspond to $d=1$, red lines to $d=2$ and green lines to $d=3$. Error bars are computed as in Eq.~\eqref{combined_var}.\label{fig_space_1_diff}}
\end{figure*}
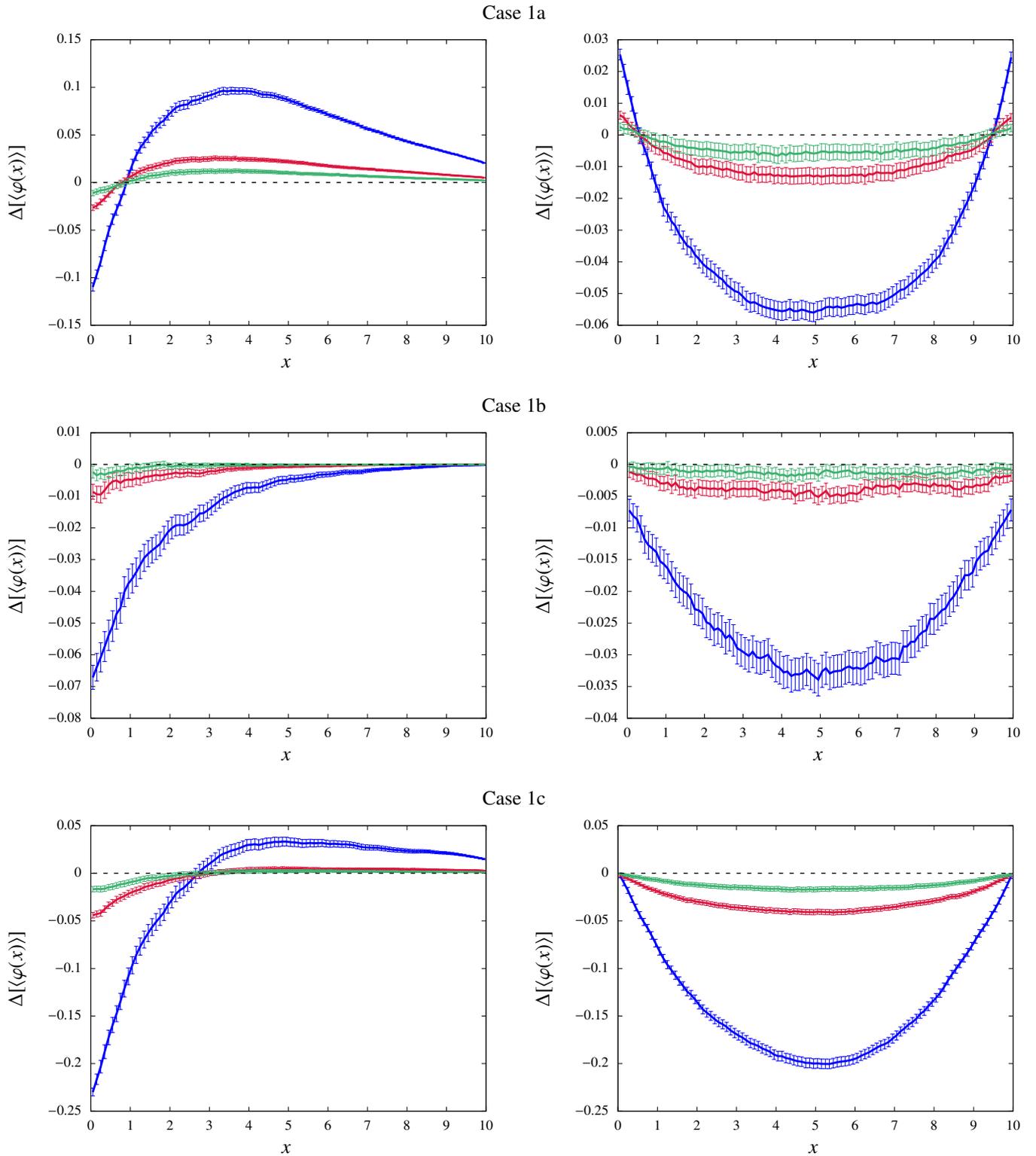

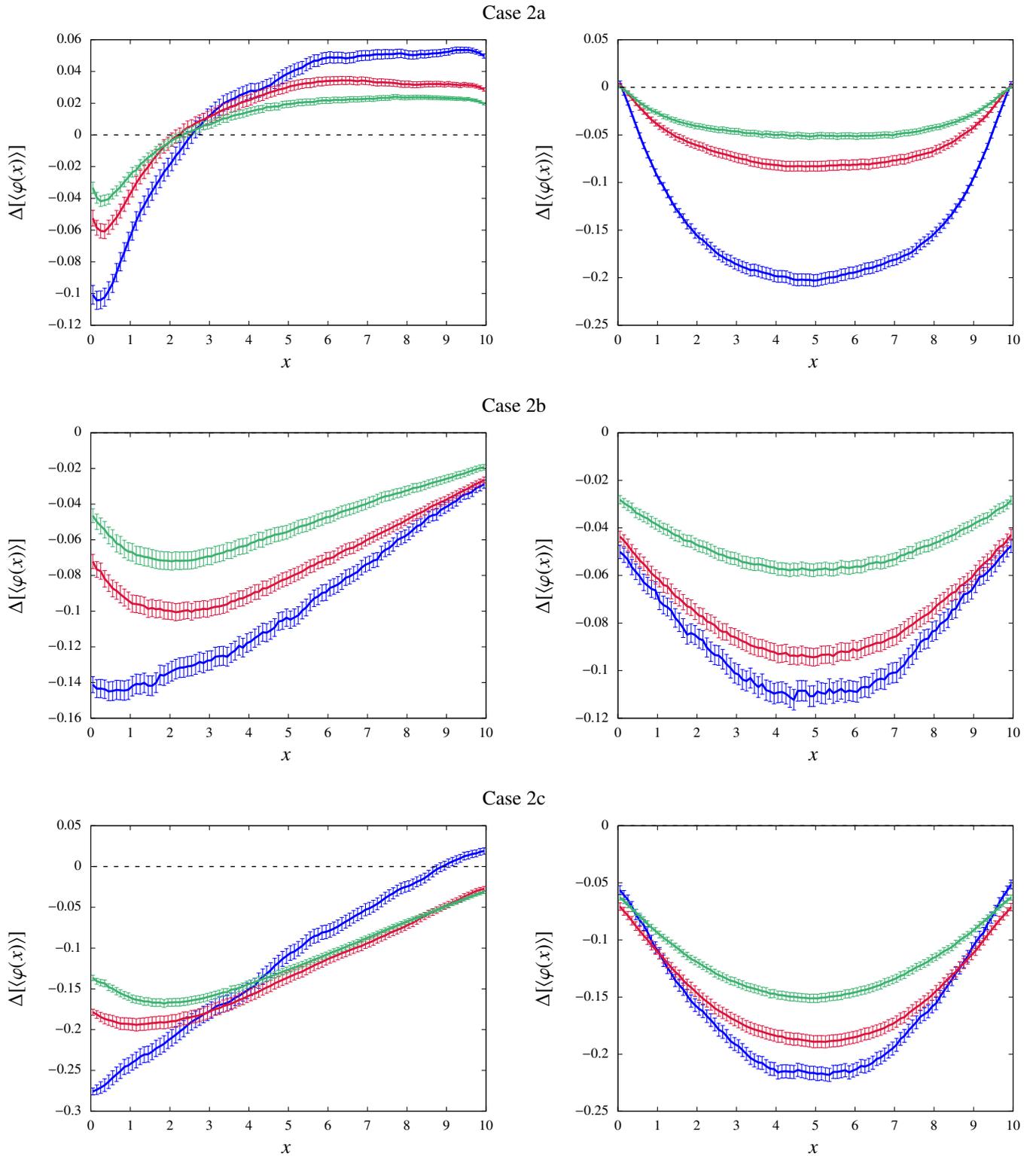
\begin{figure*}
\begin{center}
\,\,\,\, Case 2a \,\,\,\,\\
\scalebox{0.7}{\input{flux_2A_I_ecart}}\,\,\,\,
\scalebox{0.7}{\input{flux_2A_II_ecart}}\\
\,\,\,\, Case 2b \,\,\,\,\\
\scalebox{0.7}{\input{flux_2B_I_ecart}}\,\,\,\,
\scalebox{0.7}{\input{flux_2B_II_ecart}}\\
\,\,\,\, Case 2c \,\,\,\,\\
\scalebox{0.7}{\input{flux_2C_I_ecart}}\,\,\,\,
\scalebox{0.7}{\input{flux_2C_II_ecart}}\\
\end{center}
\caption{Discrepancy $\Delta[\langle\varphi(x)\rangle]$ between the ensemble-averaged spatial flux $\langle\varphi(x)\rangle$ obtained with Poisson tessellations (quenched disorder approach) and that obtained with the Chord Length Sampling algorithm (annealed disorder approach) for the benchmark configurations: Case 2. Left column: {\em suite} I configurations; right column: {\em suite} II configurations. Blue lines correspond to $d=1$, red lines to $d=2$ and green lines to $d=3$. Error bars are computed as in Eq.~\eqref{combined_var}.\label{fig_space_2_diff}} 
\end{figure*}

\begin{figure*}
\begin{center}
\,\,\,\, Case 3a \,\,\,\,\\
\scalebox{0.7}{\input{flux_3A_I_ecart}}\,\,\,\,
\scalebox{0.7}{\input{flux_3A_II_ecart}}\\
\,\,\,\, Case 3b \,\,\,\,\\
\scalebox{0.7}{\input{flux_3B_I_ecart}}\,\,\,\,
\scalebox{0.7}{\input{flux_3B_II_ecart}}\\
\,\,\,\, Case 3c \,\,\,\,\\
\scalebox{0.7}{\input{flux_3C_I_ecart}}\,\,\,\,
\scalebox{0.7}{\input{flux_3C_II_ecart}}\\
\end{center}
\caption{Discrepancy $\Delta[\langle\varphi(x)\rangle]$ between the ensemble-averaged spatial flux $\langle\varphi(x)\rangle$ obtained with Poisson tessellations (quenched disorder approach) and that obtained with the Chord Length Sampling algorithm (annealed disorder approach) for the benchmark configurations: Case 3. Left column: {\em suite} I configurations; right column: {\em suite} II configurations. Blue lines correspond to $d=1$, red lines to $d=2$ and green lines to $d=3$. Error bars are computed as in Eq.~\eqref{combined_var}.\label{fig_space_3_diff}} 
\end{figure*}
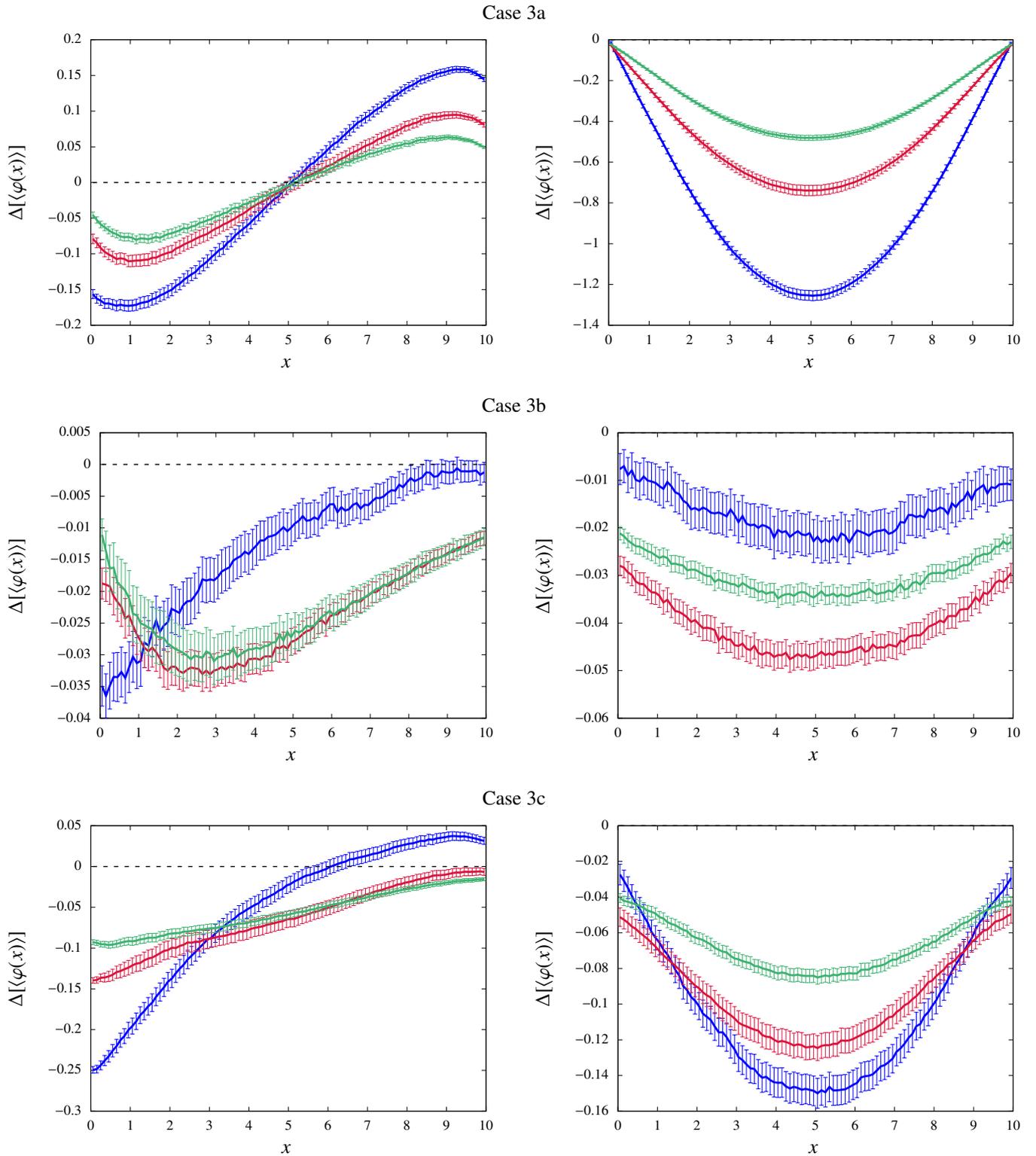

The spatial scalar flux $\langle \varphi \rangle$ within the box is illustrated in Figs.~\ref{fig_space_1} to \ref{fig_space_3} for case $1$ to case $3$, respectively. The discrepancies between CLS and reference solutions for this observable have the same behaviour as for the scalar quantities described above. The discrepancy decreases with increasing system dimensionality and with decreasing average chord length. For dense geometries (case $1$) the effects of dimensionality on the discrepancy are rather strong, and become less appreciable for less dense geometries. The kind of source and/or initial conditions plays again a minor role. This analysis is confirmed by plotting the differences $\Delta[\langle\varphi(x)\rangle]$ between reference and CLS solutions (see Figs.~\ref{fig_space_1_diff} to \ref{fig_space_3_diff} for case $1$ to case $3$, respectively). Since both reference and CLS solutions are affected by a statistical uncertainty, the error bars on $\Delta[\langle\varphi(x)\rangle]$ have been computed by taking the combined variance
\begin{equation}
\sigma[\Delta[ {\cal O}]] = \sqrt{\sigma^2[ \langle {\cal O}\rangle ] + \sigma^2_\text{CLS}[ {\cal O}]}
\label{combined_var}
\end{equation}
for each observable ${\cal O}$.

\section{Conclusions}
\label{conclusions}

The Chord Length Sampling algorithm efficiently provides approximate ensemble-averaged observables corresponding to the Levermore-Pomraning model for Markovian binary mixing. The interfaces between the constituents of the random medium are sampled on-the-fly during the particle displacements by drawing the distances to the following material boundaries from a distribution depending on the mixing statistics: the correlations on particle trajectories induced by the spatial disorder are thus neglected. Comparisons of CLS solutions with respect to reference results are mandatory in order to quantify the degree of approximations introduced in these models. For Markov mixing, a number of benchmark problems have been proposed in the literature for this purpose, but so far analyses have been conducted in one-dimensional media of the rod or slab type.

In this work we have contrasted CLS simulation results to the reference solutions for the classical benchmark problem proposed by Adams, Larsen and Pomraning, and recently revisited by Brantley, for particle propagation in stochastic media with binary Markov mixing. In particular, we have examined the evolution of the particle flux, the transmission coefficient and the reflection coefficient as a function of the benchmark configurations and of the system dimension $d$.

Two main trends have been detected: the accuracy of CLS algorithm with respect to reference solutions generally increases with increasing system dimensionality. Moreover, the accuracy of the CLS algorithm increases for decreasing average chord length, i.e., for denser stochastic tessellations. The impact of dimensionality is particularly relevant for case $1$ configurations (which have smaller chord lengths), and progressively diminishes for configurations having larger material chunks. The considerations presented in this paper, although derived strictly speaking for the Adams, Larsen and Pomraning benchmark considered here, seem to be quite general.

This work represents a first step towards extensive comparisons between CLS and reference solutions for Markov mixing statistics in higher dimensions. Furthermore, extension of these comparisons to reference solutions for other types of $d$-dimensional mixing statistics based on spatial tessellations (such as the Poisson-Voronoi model presented in~\cite{larmier_models}) would be interesting topics for future research.

\section*{Acknowledgements}
TRIPOLI-4\textsuperscript{ \textregistered} is a registered trademark of CEA. C.~Larmier, A.~Zoia and F.~Malvagi wish to thank \'Electricit\'e de France (EDF) for partial financial support and A.~Mazzolo and E.~Dumonteil for fruitful discussions. Work of P.~Brantley performed under the auspices of the U.~S. Department of Energy by Lawrence Livermore National Laboratory under Contract DE-AC52-07NA27344.

\clearpage

\begin{table*}
\footnotesize
\begin{center}
\begin{tabular}{ccccccccccc}
\toprule
Case & d & Algorithm & $\langle R \rangle$ & $\langle T \rangle$ & $\langle \varphi \rangle$ & $t_{\mathrm{tot}}$ [s] \\
\midrule
 &  & Ref & $0.437 \pm 0.002$ & $0.0148 \pm 2\times 10^{-4}$ & $6.10 \pm 0.01$ & $1.1 \times 10^{6}$\\
 & $1$ & CLS & $0.37814 \pm 2\times 10^{-5}$ & $0.026403	\pm 5\times 10^{-6}$ & $6.6288 \pm 2\times 10^{-4}$ & $2.6 \times 10^{3}$\\
& & Err [$\%$] & $-13.49 \pm	0.33$ & $78.43 \pm	2.35$ & $8.64 \pm	0.26$\\
  \cmidrule(lr){2-7}
 &  & Ref & $0.4060 \pm 6\times 10^{-4}$ & $0.0173 \pm 10^{-4}$ & $6.365 \pm 0.008$ & $6.5 \times 10^{5}$\\
1a &  $2$ & CLS & $0.39001 \pm 2\times 10^{-5}$ & $0.020100 \pm 5\times 10^{-6}$ & $6.5056 \pm 2\times 10^{-4}$  & $4.2 \times 10^{3}$\\
 & & Err [$\%$] & $-3.93 \pm	0.15$ &	$16.03 \pm	0.91$ & $2.21 \pm	0.13$\\
    \cmidrule(lr){2-7}
&  & Ref & $0.4091 \pm 5\times 10^{-4}$ & $0.0163 \pm 10^{-4}$ & $6.328 \pm 0.007$ & $3.9 \times 10^{6}$\\
 & $3$  & CLS & $0.40176 \pm 2\times 10^{-5}$ &	$0.017491 \pm 4\times 10^{-6}$ & $6.3933 \pm 2\times 10^{-4}$ & $4.6 \times 10^{3}$\\
 & & Err [$\%$] & $-1.79	\pm 0.13$ & $7.53 \pm	0.86$ &	$1.03 \pm	0.12$ \\
\midrule
 &  & Ref & $0.0845 \pm 4\times 10^{-4}$ & $0.00164 \pm 7\times 10^{-5}$ & $2.90 \pm 0.01$ & $2.6 \times 10^{6}$\\
 &  $1$ & CLS & $0.058641 \pm 8\times 10^{-6}$ & $0.001545 \pm 10^{-6}$ & $2.7738 \pm 2\times 10^{-4}$  & $6.2 \times 10^{2}$ \\
 & & Err [$\%$] & $-30.59 \pm	0.36$ & $-5.79 \pm	4.23$ & $-4.26 \pm	0.40$\\
   \cmidrule(lr){2-7}
 &  & Ref & $0.0454 \pm 2\times 10^{-4}$ & $0.00108 \pm 3\times 10^{-5}$ &  $2.163 \pm 0.005$ & $2.9 \times 10^{5}$\\
1b & $2$ & CLS & $0.042346 \pm 6\times 10^{-6}$ &	$0.001067\pm 10^{-6}$ & $2.1467 \pm 2\times 10^{-4}$ & $9.6 \times 10^{2}$ \\
 & & Err [$\%$] & $-6.70 \pm	0.49$ & $-1.04 \pm 3.10$ &	$-0.78 \pm	0.23$\\
    \cmidrule(lr){2-7}
& & Ref & $0.0377 \pm 2\times 10^{-4}$ & $0.00085 \pm 3\times 10^{-5}$ & $1.918 \pm 0.003$ & $1.8 \times 10^{6}$ \\
& $3$ & CLS & $0.036714 \pm 6\times 10^{-6}$	& $0.0008413 \pm 9\times 10^{-7}$ & $1.91440 \pm 6\times 10^{-5}$ & $1.0 \times 10^{3}$ \\
& & Err [$\%$] & $-2.52 \pm	0.52$ &	$-1.03 \pm	3.46$ &	$-0.20 \pm	0.17$ \\
\midrule
 &  & Ref & $0.4767 \pm 5\times 10^{-4}$ & $0.0159 \pm 3\times 10^{-4}$ & $6.97 \pm 0.03$ & $1.1 \times 10^{6}$\\
 &  $1$ & CLS & $0.36953 \pm 10^{-5}$ & $0.023765 \pm 3\times 10^{-6}$ & $6.9137 \pm 2\times 10^{-4}$ & $5.6 \times 10^{3}$\\
 & & Err [$\%$] & $-22.48 \pm	0.08$ &	$49.14 \pm 3.21$ & $-0.82 \pm	0.49$ \\
     \cmidrule(lr){2-7}
 &  & Ref & $0.4078 \pm 5\times 10^{-4}$ & $0.0179 \pm 10^{-4}$ & $6.52 \pm 0.01$ & $6.6 \times 10^{5}$\\
1c &  $2$ & CLS & $0.38557 \pm 10^{-5}$ &	$0.019478 \pm 3\times 10^{-6}$ & $6.4952 \pm 2\times 10^{-4}$ & $9.8 \times 10^{3}$\\
 & & Err [$\%$] & $-5.45 \pm	0.12$	& $8.59 \pm	0.90$ &	$-0.35 \pm	0.17$\\
     \cmidrule(lr){2-7}
&  & Ref & $0.4059 \pm 5\times 10^{-4}$ & $0.0164 \pm 10^{-4}$ & $6.303 \pm 0.008$ & $4.4 \times 10^{6}$\\
& $3$  & CLS & $0.39619 \pm 10^{-5}$ &	$0.016992 \pm 2\times 10^{-6}$ & $6.2957	\pm 10^{-4}$ & $1.1 \times 10^{4}$ \\
& & Err [$\%$] & $-2.40 \pm	0.12$ & $3.62 \pm	0.84$ & $-0.12 \pm	0.13$\\
\bottomrule
\end{tabular}
\end{center}
\caption{Ensemble-averaged observables and computer time $t_{\mathrm{tot}}$ for the benchmark configurations: {\em suite} I - case $1$.\label{tab_suite1_case1}}
\end{table*}

\clearpage

\begin{table*}
\footnotesize
\begin{center}
\begin{tabular}{ccccccccccc}
\toprule
Case & d & Algorithm & $\langle R \rangle$ & $\langle T \rangle$ & $\langle \varphi \rangle$ & $t_{\mathrm{tot}}$ [s] \\
\midrule
&  & Ref & $0.239 \pm 0.003$ & $0.0973 \pm 9\times 10^{-4}$ & $7.64 \pm 0.02$ & $9.1 \times 10^{5}$\\
 &  $1$ & CLS &  $0.18051 \pm 10^{-5}$ & $0.12841 \pm 10^{-5}$ & $7.8140 \pm 10^{-4}$ & $2.1 \times 10^{3}$\\
 & & Err [$\%$] & $-24.46 \pm	0.91$ &	$32.01 \pm	1.26$ &	$2.21 \pm	0.28$ \\
  \cmidrule(lr){2-7}
 &  & Ref & $0.226 \pm 0.002$ & $0.0969 \pm 7\times 10^{-4}$ & $7.59 \pm 0.02$ & $3.5 \times 10^{5}$ \\
2a &  $2$ & CLS &  $0.18972	\pm 10^{-5}$ & $0.11403 \pm 10^{-5}$ & $7.7288	\pm 10^{-4}$ & $3.0 \times 10^{3}$\\
 & & Err [$\%$] & $-16.03 \pm	0.80$ & $17.70 \pm	0.86$ & $1.83 \pm	0.20$ \\
   \cmidrule(lr){2-7}
&  & Ref & $0.225 \pm 0.001$ & $0.0937 \pm 4\times 10^{-4}$ & $7.57 \pm 0.01$ & $4.4 \times 10^{5}$\\
 & $3$ & CLS &  $0.20043 \pm 10^{-5}$ & $0.105624	\pm 9\times 10^{-6}$ & $7.6615 \pm 2\times 10^{-4}$ & $3.1 \times 10^{3}$\\
  & & Err [$\%$] & $-11.08 \pm	0.45$ & $12.74 \pm	0.54$ & $1.22 \pm	0.13$\\
\midrule
 &  & Ref & $0.2866 \pm 8\times 10^{-4}$ & $0.194 \pm 0.001$ & $11.69 \pm 0.04$ & $3.0 \times 10^{6}$\\
 &  $1$ & CLS & $0.21827 \pm 10^{-5}$ & $0.17938 \pm 10^{-5}$	& $10.7138 \pm 5\times 10^{-4}$ & $5.4 \times 10^{2}$\\
 & & Err [$\%$] & $-23.84 \pm	0.22$	& $-7.45 \pm	0.56$ & $-8.33 \pm	0.28$ \\
  \cmidrule(lr){2-7}
 &  & Ref & $0.1980 \pm 8\times 10^{-4}$ & $0.1465 \pm 9\times 10^{-4}$ & $9.11 \pm 0.03$ & $1.0 \times 10^{6}$\\
2b &  $2$ & CLS & $0.16674 \pm 10^{-5}$ & $0.13377 \pm 10^{-5}$ & $8.3763 \pm	4\times 10^{-4}$ & $8.8 \times 10^{2}$\\
 & & Err [$\%$] & $-15.79 \pm	0.33$ & $-8.68 \pm 0.54$ & $-8.06 \pm	0.30$ \\
   \cmidrule(lr){2-7}
&  & Ref & $0.1616 \pm 8\times 10^{-4}$ & $0.1194 \pm 9\times 10^{-4}$ & $7.77 \pm 0.03$ & $3.4 \times 10^{5}$\\
 & $3$ & CLS & $0.14223 \pm 10^{-5}$ & $0.10996 \pm 10^{-5}$ & $7.2609 \pm 2\times 10^{-4}$ & $9.3 \times 10^{2}$\\
 & & Err [$\%$] & $-11.99 \pm	0.44$ & $-7.91 \pm	0.68$ & $-6.50 \pm	0.37$\\
\midrule
 &  & Ref & $0.4334 \pm 8\times 10^{-4}$ & $0.184 \pm 0.002$ & $12.51 \pm 0.06$ & $7.1 \times 10^{5}$\\
 &  $1$ & CLS & $0.28962 \pm 10^{-5}$ & $0.19497	\pm 10^{-5}$ & $11.3443 \pm 4\times 10^{-4}$ & $3.3 \times 10^{3}$\\
 & & Err [$\%$] & $-33.17 \pm	0.12$ &	$5.83 \pm	1.24$	& $-9.35 \pm	0.46$ \\
  \cmidrule(lr){2-7}
 &  & Ref &  $0.3677 \pm 6\times 10^{-4}$ & $0.179 \pm 0.002$ & $11.46 \pm 0.05$ & $4.1 \times 10^{5}$\\
2c & $2$  & CLS & $0.27853 \pm 10^{-5}$ & $0.16713	\pm 10^{-5}$	& $10.1679	\pm 3\times 10^{-4}$ & $5.6 \times 10^{3}$ \\
 & & Err [$\%$] & $-24.25 \pm	0.12$ & $-6.74 \pm 0.82$ & $-11.25 \pm	0.39$\\
   \cmidrule(lr){2-7}
&   & Ref & $0.3457 \pm 5\times 10^{-4}$ & $0.1651 \pm 9\times 10^{-4}$ & $10.76 \pm 0.03$ & $4.8 \times 10^{5}$\\
 & $3$ & CLS & $0.27693 \pm 10^{-5}$ &	$0.15031 \pm 10^{-5}$	& $9.6048 \pm 2\times 10^{-4}$ & $8.9 \times 10^{3}$\\
 & & Err [$\%$] & $-19.89 \pm	0.12$ & $-8.98 \pm	0.49$ & $-10.73 \pm	0.23$\\
\bottomrule
\end{tabular}
\end{center}
\caption{Ensemble-averaged observables and computer time $t_{\mathrm{tot}}$ for the benchmark configurations: {\em suite} I - case $2$.\label{tab_suite1_case2}}
\end{table*}

\clearpage

\begin{table*}
\footnotesize
\begin{center}
\begin{tabular}{ccccccccccc}
\toprule
Case & d & Algorithm & $\langle R \rangle$ & $\langle T \rangle$ & $\langle \varphi \rangle$ & $t_{\mathrm{tot}}$ [s] \\
\midrule
 &  & Ref & $0.692	\pm 0.003$ & $0.163	\pm 0.002$ & $16.44 \pm 0.05$ & $1.3 \times 10^{6}$\\
 & $1$ & CLS & $0.60758 \pm 2\times 10^{-5}$ & $0.24037	\pm 10^{-5}$	& $16.3738 \pm	7\times 10^{-4}$ & $3.8 \times 10^{3}$ \\
 & & Err [$\%$] & $-12.14 \pm	0.34$ & $47.09 \pm	1.60$ & $-0.43 \pm	0.30$  \\
  \cmidrule(lr){2-7}
 &  & Ref & $0.680 \pm 0.003$ & $0.168 \pm 0.002$ & $16.46 \pm 0.05$ & $5.4 \times 10^{5}$\\
3a &  $2$ & CLS &  $0.62678 \pm 2\times 10^{-5}$ & $0.21473	\pm 10^{-5}$ & $16.3866 \pm 7\times 10^{-4}$ & $6.0 \times 10^{3}$\\
 & & Err [$\%$] & $-7.77 \pm	0.40$ & $27.99 \pm 1.45$ & $-0.44 \pm	0.33$ \\
   \cmidrule(lr){2-7}
&  & Ref & $0.675 \pm 0.001$ & $0.1692 \pm 9\times 10^{-4}$ & $16.38 \pm 0.03$ & $1.4 \times 10^{6}$\\
 & $3$ & CLS & $0.64107 \pm 2\times 10^{-5}$	& $0.19957	\pm 10^{-5}$ & $16.3231 \pm 6\times 10^{-4}$ & $9.1 \times 10^{3}$\\
 & & Err [$\%$] & $-5.06 \pm	0.20$ & $17.96 \pm	0.65$ & $-0.36 \pm	0.19$ \\
\midrule
&  & Ref & $0.0361 \pm 2\times 10^{-4}$ & $0.0760 \pm 7\times 10^{-4}$ & $5.16 \pm 0.02$ & $4.2 \times 10^{6}$\\
 & $1$  & CLS & $0.024013 \pm 5\times 10^{-6}$ & $0.075671 \pm 8\times 10^{-6}$	& $5.0313	\pm 5\times 10^{-4}$ & $3.5 \times 10^{2}$\\
 & & Err [$\%$] & $-33.50 \pm	0.44$ & $-0.37 \pm	0.95$ & $-2.48 \pm	0.40$\\
  \cmidrule(lr){2-7}
 &   & Ref & $0.0217 \pm 2\times 10^{-4}$ & $0.0568 \pm 6\times 10^{-4}$ & $4.00 \pm 0.02$ & $2.8 \times 10^{6}$\\
3b &  $2$ & CLS & $0.015501 \pm 4\times 10^{-6}$ & $0.052503	\pm 7\times 10^{-6}$ & $3.7582 \pm 4\times 10^{-4}$ & $5.7 \times 10^{2}$ \\
  & & Err [$\%$] & $-28.55 \pm	0.61$ & $-7.51 \pm 0.97$ & $-6.03 \pm	0.41$\\
   \cmidrule(lr){2-7}
&  & Ref & $0.0165 \pm 2\times 10^{-4}$ & $0.0457 \pm 9\times 10^{-4}$ & $3.47 \pm 0.03$ & $5.0 \times 10^{5}$\\
 & $3$ & CLS & $0.012454 \pm 3\times 10^{-6}$ & $0.040345 \pm 6\times 10^{-6}$ & $3.2382 \pm 10^{-4}$ & $8.0 \times 10^{2}$\\
  & & Err [$\%$] & $-24.48 \pm 0.97$ & $-11.80	\pm 1.68$ & $-6.55 \pm	0.70$ \\
\midrule
&   & Ref & $0.445	\pm 0.001$ & $0.104	\pm 0.002$ & $9.00	\pm 0.07$ & $6.6 \times 10^{5}$\\
 &  $1$ & CLS & $0.32613	\pm 10^{-5}$ & $0.119665 \pm 9\times 10^{-6}$ & $8.4702 \pm 6\times 10^{-4}$ & $3.4 \times 10^{3}$\\
  & & Err [$\%$] & $-26.71	\pm 0.17$	& $15.11 \pm	2.54$ & $-5.91	\pm 0.75$ \\  
  \cmidrule(lr){2-7}
 &   & Ref & $0.411 \pm 0.001$ & $0.094 \pm 0.002$ & $8.30 \pm 0.07$ & $2.7 \times 10^{5}$\\
3c & $2$ & CLS & $0.33767	\pm 10^{-5}$ & $0.094998 \pm 9\times 10^{-6}$ & $7.6579 \pm 5\times 10^{-4}$ & $5.6 \times 10^{3}$\\
  & & Err [$\%$] & $-17.92 \pm	0.22$ & $0.72 \pm	2.54$ & $-7.72 \pm	0.83$\\
   \cmidrule(lr){2-7}
&   & Ref & $0.3979 \pm 7\times 10^{-4}$ & $0.086 \pm 0.001$ & $7.89 \pm 0.03$ & $7.0 \times 10^{5}$\\
 & $3$ & CLS & $0.34652 \pm 10^{-5}$ &	$0.080613 \pm 7\times 10^{-6}$ & $7.3217	\pm 2\times 10^{-4}$ & $8.8 \times 10^{3}$ \\
  & & Err [$\%$] & $-12.92 \pm	0.15$ & $-6.16 \pm	1.19$ & $-7.17 \pm 0.40$\\
\bottomrule
\end{tabular}
\end{center}
\caption{Ensemble-averaged observables and computer time $t_{\mathrm{tot}}$ for the benchmark configurations: {\em suite} I - case $3$.\label{tab_suite1_case3}}
\end{table*}

\clearpage

\begin{table*}
\footnotesize
\begin{center}
\begin{tabular}{ccccccccccc}
\toprule
Case & d & Algorithm & $\langle L \rangle$ & $\langle \varphi \rangle$ & $t_{\mathrm{tot}}$ [s] \\
\midrule
&  & Ref & $0.1525 \pm 3\times 10^{-4}$ & $7.70 \pm 0.01$ & $9.9 \times 10^{5}$\\
 &  $1$ & CLS & $0.165716 \pm 8\times 10^{-6}$  &	$7.3449 \pm 10^{-4} $ & $2.6 \times 10^{3}$\\
 & & Err [$\%$] & $8.69 \pm 0.24$ & $-4.65 \pm	0.14$ \\
  \cmidrule(lr){2-6}
 &  & Ref & $0.1592 \pm 3\times 10^{-4}$ & $7.512 \pm 0.008$ & $2.1 \times 10^{6}$\\
1a &  $2$ & CLS & $0.162634 \pm	8\times 10^{-6}$ & $7.4287 \pm 10^{-4}$ & $4.6 \times 10^{3}$\\
 & & Err [$\%$] &  $2.17 \pm	0.17$ & $-1.11 \pm	0.10$ \\
   \cmidrule(lr){2-6}
&  & Ref & $0.1583 \pm 3\times 10^{-4} $ & $7.530 \pm 0.008$ & $7.9 \times 10^{7}$\\
 & $3$ & CLS & $0.159828  \pm 8\times 10^{-6}$ & $7.4924 \pm 2\times 10^{-4}$ & $5.3 \times 10^{3}$\\
  & & Err [$\%$] &  $0.98 \pm	0.17$ & $-0.49 \pm	0.10$ \\
\midrule
&   & Ref & $0.0724 \pm 3\times 10^{-4}$ &  $3.735 \pm 0.009$ & $1.5 \times 10^{6}$\\
 &  $1$& CLS &  $0.069346	\pm 6\times 10^{-6}$ & $3.4898	\pm 2\times 10^{-4}$ & $5.8 \times 10^{2}$\\
  & & Err [$\%$] & $-4.28 \pm	0.36$ & $-6.55 \pm	0.22$\\
  \cmidrule(lr){2-6}
 &   & Ref & $0.0542 \pm 2\times 10^{-4}$ & $2.182 \pm 0.003$ & $1.8 \times 10^{6}$\\
1b &  $2$ & CLS & $0.053662 \pm 5\times 10^{-6}$ & $2.1468	\pm 2\times 10^{-4}$ & $8.9 \times 10^{2}$\\
  & & Err [$\%$] & $-0.92 \pm	0.33$ & $-1.63 \pm	0.16$ \\
   \cmidrule(lr){2-6}
&  & Ref & $0.0481 \pm 2\times 10^{-4}$ & $1.808 \pm 0.003$ & $7.4 \times 10^{7}$\\
 & $3$  & CLS & $0.047859 \pm 5\times 10^{-6}$ & $1.79609 \pm 6\times 10^{-5}$ & $1.0 \times 10^{3}$\\
  & & Err [$\%$] & $-0.42 \pm	0.33$ & $-0.63 \pm	0.14$\\
\midrule
 &   & Ref & $0.1742 \pm 7\times 10^{-4}$ & $9.62 \pm 0.03$ & $1.0 \times 10^{6}$ \\
 &  $1$ & CLS &  $0.172845 \pm 7\times 10^{-6}$ & $8.2618	\pm 3\times 10^{-4}$ & $6.7 \times 10^{3}$\\
  & & Err [$\%$] & $-0.76 \pm	0.38$ & $-14.11 \pm	0.22$ \\
  \cmidrule(lr){2-6}
 &   & Ref & $0.1630 \pm 3\times 10^{-4}$ & $7.77 \pm 0.01$ & $2.1 \times 10^{6}$\\
1c & $2$ & CLS & $0.162379 \pm 6\times 10^{-6}$ & $7.4824 \pm 2\times 10^{-4}$ & $1.2 \times 10^{4}$\\
  & & Err [$\%$] & $-0.38 \pm 0.18$ &	$-3.76 \pm	0.12$ \\
   \cmidrule(lr){2-6}
&  & Ref & $0.1577 \pm 3\times 10^{-4}$ & $7.455 \pm 0.008$ & $7.7 \times 10^{7}$\\
 &  $3$ & CLS & $0.157383 \pm 6\times 10^{-6}$ & $7.3335	\pm 10^{-4} $ & $1.4 \times 10^{4}$\\
  & & Err [$\%$] & $-0.19 \pm	0.17$ & $-1.63 \pm	0.10$\\
\bottomrule
\end{tabular}
\end{center}
\caption{Ensemble-averaged observables and computer time $t_{\mathrm{tot}}$ for the benchmark configurations: {\em suite} II - case $1$.\label{tab_suite2_case1}}
\end{table*}

\clearpage

\begin{table*}
\footnotesize
\begin{center}
\begin{tabular}{ccccccccccc}
\toprule
Case & d & Algorithm & $\langle L \rangle$ & $\langle \varphi \rangle$  & $t_{\mathrm{tot}}$ [s] \\
\midrule
& & Ref & $0.1904 \pm 3\times 10^{-4}$ & $8.29 \pm 0.03$ & $5.3 \times 10^{6}$\\
 &  $1$ & CLS & $0.195346 \pm 9\times 10^{-6}$ & $6.8189	\pm 2\times 10^{-4}$ & $1.7 \times 10^{3}$\\
   & & Err [$\%$] &  $2.57 \pm	0.16$ & $-17.70 \pm	0.30$\\
  \cmidrule(lr){2-6}
 &   & Ref & $0.1898 \pm 3\times 10^{-4}$ & $7.46 \pm 0.03$ & $3.2 \times 10^{5}$\\
2a & $2$ & CLS & $0.193217 \pm 9\times 10^{-6}$ & $6.8517	\pm 2\times 10^{-4}$ & $2.8 \times 10^{3}$\\
  & & Err [$\%$] &  $1.82 \pm	0.18$ & $-8.16 \pm	0.33$ \\
   \cmidrule(lr){2-6}
&   & Ref &  $0.1892 \pm 3\times 10^{-4}$ & $7.27 \pm 0.01$ & $5.8 \times 10^{5}$\\
 & $3$ & CLS & $0.191527\pm  9\times 10^{-6}$ & $6.8774	\pm 2\times 10^{-4}$ & $3.0 \times 10^{3}$\\
   & & Err [$\%$] & $1.21 \pm	0.16$ & $-5.36 \pm 0.18$\\
\midrule
&   & Ref & $0.2918 \pm 8\times 10^{-4}$ & $10.75 \pm 0.02$ & $1.5 \times 10^{6}$\\
 &  $1$ & CLS & $0.26783 \pm 10^{-5}$ & $9.8684	\pm 5\times 10^{-4}$ & $4.9 \times 10^{2}$\\
   & & Err [$\%$] & $-8.21 \pm	0.24$ & $-8.20 \pm	0.20$ \\
  \cmidrule(lr){2-6}
  &  & Ref &  $0.2274 \pm 6\times 10^{-4}$ & $7.97 \pm 0.02$ & $6.1 \times 10^{5}$\\
2b  &  $2$ & CLS & $0.209414 \pm 9\times 10^{-6}$ & $7.2072	\pm 4\times 10^{-4}$ & $7.8 \times 10^{2}$\\
  & & Err [$\%$] & $-7.91 \pm	0.26$ & $-9.60 \pm	0.24$ \\
   \cmidrule(lr){2-6}
&  & Ref & $0.1931 \pm 4\times 10^{-4}$ & $6.54 \pm 0.01$ & $1.7 \times 10^{6}$\\
 & $3$ & CLS & $0.181518	\pm 9\times 10^{-6}$ & $6.0577 \pm	2\times 10^{-4}$ & $8.6 \times 10^{2}$\\
   & & Err [$\%$] & $-6.01 \pm	0.21$ & $-7.31 \pm	0.19$ \\
\midrule
 & & Ref & $0.312 \pm 0.001$ & $11.92 \pm 0.03$ & $4.1 \times 10^{5}$\\
 &   $1$ & CLS & $0.283614 \pm 9\times 10^{-6}$ & $10.3022	\pm 4\times 10^{-4}$ & $2.8 \times 10^{3}$ \\
   & & Err [$\%$] & $-9.09 \pm	0.33$ & $-13.56 \pm 0.25$ \\
  \cmidrule(lr){2-6}
 &  & Ref & $0.286 \pm 0.001$ & $10.39 \pm 0.03$ & $2.1 \times 10^{5}$\\
2c & $2$  & CLS & $0.254187 \pm 8\times 10^{-6}$ & $8.8967	\pm 3\times 10^{-4}$ & $5.2 \times 10^{3}$\\
  & & Err [$\%$] & $-11.26 \pm	0.31$ & $-14.35 \pm	0.27$ \\
   \cmidrule(lr){2-6}
&   & Ref & $0.2688 \pm 6\times 10^{-4}$ & $9.55 \pm 0.02$ & $4.9 \times 10^{5}$ \\
 & $3$ & CLS &  $0.240117	\pm 8\times 10^{-6}$ & $8.3498 \pm	2\times 10^{-4}$ & $8.4 \times 10^{3}$\\
   & & Err [$\%$] & $-10.69 \pm	0.20$ & $-12.58 \pm	0.18$ \\
\bottomrule
\end{tabular}
\end{center}
\caption{Ensemble-averaged observables and computer time $t_{\mathrm{tot}}$ for the benchmark configurations: {\em suite} II - case $2$.\label{tab_suite2_case2}}
\end{table*}

\clearpage

\begin{table*}
\footnotesize
\begin{center}
\begin{tabular}{ccccccccccc}
\toprule
Case & d & Algorithm & $\langle L \rangle$ & $\langle \varphi \rangle$  & $t_{\mathrm{tot}}$ [s] \\
\midrule
 &   & Ref & $0.4112 \pm 6\times 10^{-4}$ & $27.3 \pm 0.2$ & $1.9 \times 10^{6}$\\
 &  $1$ & CLS & $0.40935	\pm 10^{-5}$ & $19.3460 \pm 8\times 10^{-4}$ & $4.3 \times 10^{3}$\\
    & & Err [$\%$] & $-0.45 \pm	0.15$ &	$-29.24 \pm	0.39$ \\
  \cmidrule(lr){2-6}
 &  & Ref & $0.4115 \pm 6\times 10^{-4}$ & $24.3 \pm 0.2$ & $6.8 \times 10^{5}$\\
3a & $2$  & CLS & $0.40967	\pm 10^{-5}$ & $19.5145	\pm 7\times 10^{-4}$ & $7.4 \times 10^{3}$\\
   & & Err [$\%$] & $-0.44 \pm	0.15$ & $-19.64 \pm	0.53$ \\
   \cmidrule(lr){2-6}
&   & Ref &  $0.4098 \pm 4\times 10^{-4}$ & $22.82 \pm 0.07$ & $1.7 \times 10^{6}$\\
 & $3$ & CLS &  $0.40807 \pm 10^{-5}$ & $19.7173 \pm	6\times 10^{-4}$  & $1.1 \times 10^{4}$\\
    & & Err [$\%$] & $-0.42 \pm	0.10$ & $-13.61 \pm	0.28$\\
\midrule
 &  & Ref & $0.1294	\pm 5\times 10^{-4}$  & $5.93 \pm 0.02$ & $1.4 \times 10^{6}$\\
 &  $1$  & CLS & $0.125785	\pm 7\times 10^{-6}$ & $5.7673	\pm 6\times 10^{-4}$ & $3.1 \times 10^{2}$\\
    & & Err [$\%$] &  $-2.80 \pm	0.35$ & $-2.82 \pm	0.34$\\
  \cmidrule(lr){2-6}
 &  & Ref & $0.1003 \pm 4\times 10^{-4}$  & $3.75 \pm 0.02$ & $6.7 \times 10^{5}$ \\
3b & $2$  & CLS & $0.093978 \pm	7\times 10^{-6}$ & $3.3419	\pm 4\times 10^{-4}$ & $5.2 \times 10^{2}$\\
   & & Err [$\%$] & $-6.26 \pm	0.38$ & $-10.88 \pm	0.36$ \\
   \cmidrule(lr){2-6}
&  & Ref &  $0.0868 \pm 3\times 10^{-4}$ & $2.98 \pm 0.01$ & $8.7 \times 10^{5}$ \\
 & $3$ & CLS &  $0.080949 \pm 6\times 10^{-6}$ & $2.6747 \pm 10^{-4}$ & $7.8 \times 10^{2}$\\
    & & Err [$\%$] & $-6.78 \pm	0.34$ & $-10.12 \pm	0.30$ \\
\midrule
 &  & Ref & $0.225	\pm 0.001$	& $10.56 \pm	0.05$ & $4.1 \times 10^{5}$\\
 & $1$  & CLS & $0.211761	\pm 8\times 10^{-6}$ & $9.5120	\pm 6\times 10^{-4}$ & $3.8 \times 10^{3}$\\
    & & Err [$\%$] &  $-6.02	\pm 0.55$ &	$-9.92 \pm	0.46$\\
  \cmidrule(lr){2-6}
 &  & Ref & $0.207 \pm 0.001$ & $8.78 \pm 0.05$ & $1.7 \times 10^{5}$\\
3c & $2$  & CLS & $0.191469	\pm 7\times 10^{-6}$ & $7.8470 \pm	5\times 10^{-4}$ & $6.7 \times 10^{3}$\\
   & & Err [$\%$] &  $-7.54 \pm	0.64$ & $-10.68 \pm	0.51$\\
   \cmidrule(lr){2-6}
&  & Ref & $0.1974 \pm 7\times 10^{-4}$ & $8.15 \pm 0.02$ & $5.0 \times 10^{4}$\\
 & $3$  & CLS &  $0.183044 \pm 7\times 10^{-6}$ & $7.4839 \pm 10^{-4} $ & $1.1 \times 10^{4}$\\
    & & Err [$\%$] &  $-7.26 \pm	0.33$ & $-8.20 \pm	0.25$\\
\bottomrule
\end{tabular}
\end{center}
\caption{Ensemble-averaged observables and computer time $t_{\mathrm{tot}}$ for the benchmark configurations: {\em suite} II - case $3$.\label{tab_suite2_case3}}
\end{table*}

\clearpage

\end{document}

%% file: flux_1A_I.tex
% GNUPLOT: LaTeX picture with Postscript
\begingroup
  \makeatletter
  \providecommand\color[2][]{%
    \GenericError{(gnuplot) \space\space\space\@spaces}{%
      Package color not loaded in conjunction with
      terminal option `colourtext'%
    }{See the gnuplot documentation for explanation.%
    }{Either use 'blacktext' in gnuplot or load the package
      color.sty in LaTeX.}%
    \renewcommand\color[2][]{}%
  }%
  \providecommand\includegraphics[2][]{%
    \GenericError{(gnuplot) \space\space\space\@spaces}{%
      Package graphicx or graphics not loaded%
    }{See the gnuplot documentation for explanation.%
    }{The gnuplot epslatex terminal needs graphicx.sty or graphics.sty.}%
    \renewcommand\includegraphics[2][]{}%
  }%
  \providecommand\rotatebox[2]{#2}%
  \@ifundefined{ifGPcolor}{%
    \newif\ifGPcolor
    \GPcolorfalse
  }{}%
  \@ifundefined{ifGPblacktext}{%
    \newif\ifGPblacktext
    \GPblacktexttrue
  }{}%
  % define a \g@addto@macro without @ in the name:
  \let\gplgaddtomacro\g@addto@macro
  % define empty templates for all commands taking text:
  \gdef\gplbacktext{}%
  \gdef\gplfronttext{}%
  \makeatother
  \ifGPblacktext
    % no textcolor at all
    \def\colorrgb#1{}%
    \def\colorgray#1{}%
  \else
    % gray or color?
    \ifGPcolor
      \def\colorrgb#1{\color[rgb]{#1}}%
      \def\colorgray#1{\color[gray]{#1}}%
      \expandafter\def\csname LTw\endcsname{\color{white}}%
      \expandafter\def\csname LTb\endcsname{\color{black}}%
      \expandafter\def\csname LTa\endcsname{\color{black}}%
      \expandafter\def\csname LT0\endcsname{\color[rgb]{1,0,0}}%
      \expandafter\def\csname LT1\endcsname{\color[rgb]{0,1,0}}%
      \expandafter\def\csname LT2\endcsname{\color[rgb]{0,0,1}}%
      \expandafter\def\csname LT3\endcsname{\color[rgb]{1,0,1}}%
      \expandafter\def\csname LT4\endcsname{\color[rgb]{0,1,1}}%
      \expandafter\def\csname LT5\endcsname{\color[rgb]{1,1,0}}%
      \expandafter\def\csname LT6\endcsname{\color[rgb]{0,0,0}}%
      \expandafter\def\csname LT7\endcsname{\color[rgb]{1,0.3,0}}%
      \expandafter\def\csname LT8\endcsname{\color[rgb]{0.5,0.5,0.5}}%
    \else
      % gray
      \def\colorrgb#1{\color{black}}%
      \def\colorgray#1{\color[gray]{#1}}%
      \expandafter\def\csname LTw\endcsname{\color{white}}%
      \expandafter\def\csname LTb\endcsname{\color{black}}%
      \expandafter\def\csname LTa\endcsname{\color{black}}%
      \expandafter\def\csname LT0\endcsname{\color{black}}%
      \expandafter\def\csname LT1\endcsname{\color{black}}%
      \expandafter\def\csname LT2\endcsname{\color{black}}%
      \expandafter\def\csname LT3\endcsname{\color{black}}%
      \expandafter\def\csname LT4\endcsname{\color{black}}%
      \expandafter\def\csname LT5\endcsname{\color{black}}%
      \expandafter\def\csname LT6\endcsname{\color{black}}%
      \expandafter\def\csname LT7\endcsname{\color{black}}%
      \expandafter\def\csname LT8\endcsname{\color{black}}%
    \fi
  \fi
    \setlength{\unitlength}{0.0500bp}%
    \ifx\gptboxheight\undefined%
      \newlength{\gptboxheight}%
      \newlength{\gptboxwidth}%
      \newsavebox{\gptboxtext}%
    \fi%
    \setlength{\fboxrule}{0.5pt}%
    \setlength{\fboxsep}{1pt}%
\begin{picture}(7200.00,5040.00)%
    \gplgaddtomacro\gplbacktext{%
      \csname LTb\endcsname%
      \put(946,704){\makebox(0,0)[r]{\strut{}$0.01$}}%
      \put(946,2061){\makebox(0,0)[r]{\strut{}$0.1$}}%
      \put(946,3418){\makebox(0,0)[r]{\strut{}$1$}}%
      \put(946,4775){\makebox(0,0)[r]{\strut{}$10$}}%
      \put(1078,484){\makebox(0,0){\strut{}$0$}}%
      \put(1651,484){\makebox(0,0){\strut{}$1$}}%
      \put(2223,484){\makebox(0,0){\strut{}$2$}}%
      \put(2796,484){\makebox(0,0){\strut{}$3$}}%
      \put(3368,484){\makebox(0,0){\strut{}$4$}}%
      \put(3941,484){\makebox(0,0){\strut{}$5$}}%
      \put(4513,484){\makebox(0,0){\strut{}$6$}}%
      \put(5086,484){\makebox(0,0){\strut{}$7$}}%
      \put(5658,484){\makebox(0,0){\strut{}$8$}}%
      \put(6231,484){\makebox(0,0){\strut{}$9$}}%
      \put(6803,484){\makebox(0,0){\strut{}$10$}}%
    }%
    \gplgaddtomacro\gplfronttext{%
      \csname LTb\endcsname%
      \put(176,2739){\rotatebox{-270}{\makebox(0,0){\strut{}\Large $\langle \varphi(x) \rangle$ \normalsize}}}%
      \put(3940,154){\makebox(0,0){\strut{}\Large $x$ \normalsize}}%
    }%
    \gplbacktext
    \put(0,0){\includegraphics{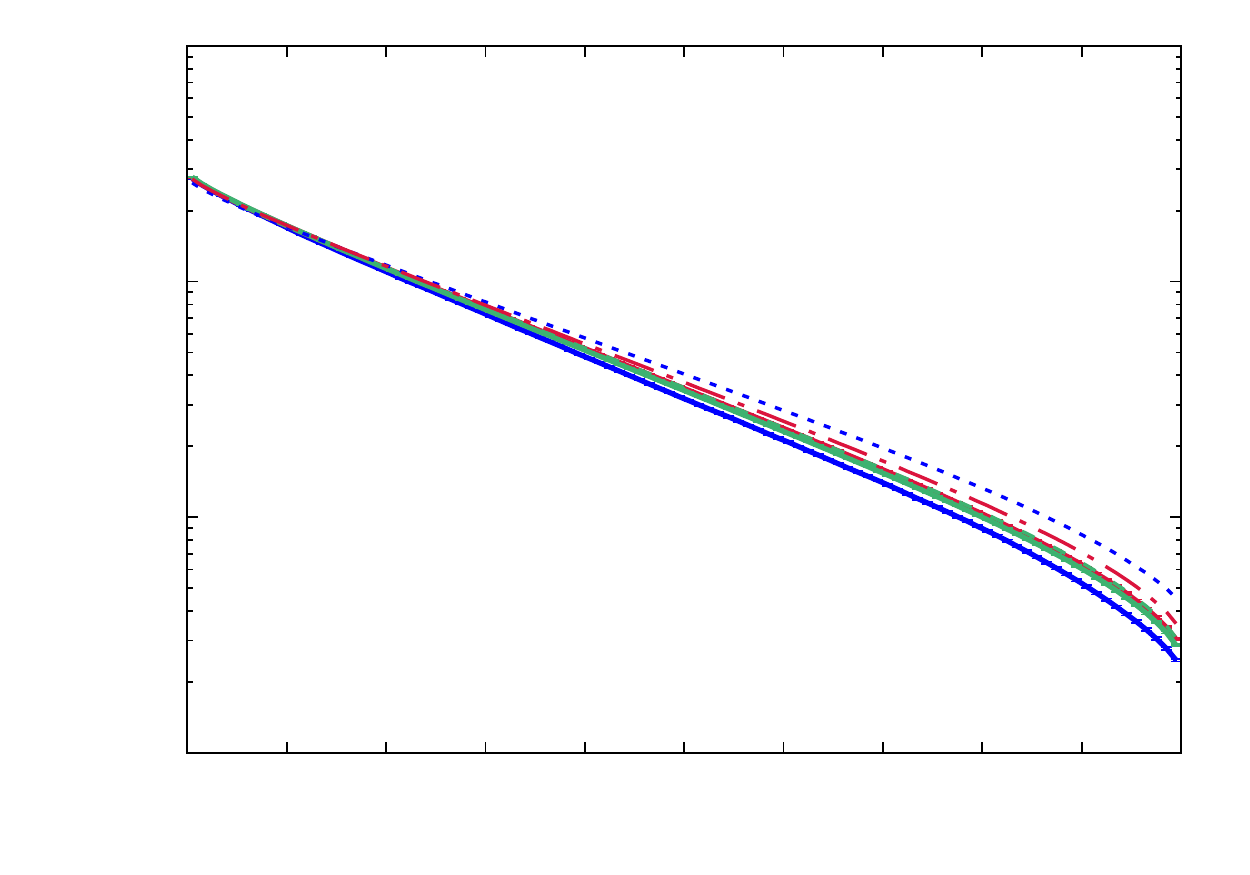}}%
    \gplfronttext
  \end{picture}%
\endgroup

%% file: flux_1A_II.tex
% GNUPLOT: LaTeX picture with Postscript
\begingroup
  \makeatletter
  \providecommand\color[2][]{%
    \GenericError{(gnuplot) \space\space\space\@spaces}{%
      Package color not loaded in conjunction with
      terminal option `colourtext'%
    }{See the gnuplot documentation for explanation.%
    }{Either use 'blacktext' in gnuplot or load the package
      color.sty in LaTeX.}%
    \renewcommand\color[2][]{}%
  }%
  \providecommand\includegraphics[2][]{%
    \GenericError{(gnuplot) \space\space\space\@spaces}{%
      Package graphicx or graphics not loaded%
    }{See the gnuplot documentation for explanation.%
    }{The gnuplot epslatex terminal needs graphicx.sty or graphics.sty.}%
    \renewcommand\includegraphics[2][]{}%
  }%
  \providecommand\rotatebox[2]{#2}%
  \@ifundefined{ifGPcolor}{%
    \newif\ifGPcolor
    \GPcolorfalse
  }{}%
  \@ifundefined{ifGPblacktext}{%
    \newif\ifGPblacktext
    \GPblacktexttrue
  }{}%
  % define a \g@addto@macro without @ in the name:
  \let\gplgaddtomacro\g@addto@macro
  % define empty templates for all commands taking text:
  \gdef\gplbacktext{}%
  \gdef\gplfronttext{}%
  \makeatother
  \ifGPblacktext
    % no textcolor at all
    \def\colorrgb#1{}%
    \def\colorgray#1{}%
  \else
    % gray or color?
    \ifGPcolor
      \def\colorrgb#1{\color[rgb]{#1}}%
      \def\colorgray#1{\color[gray]{#1}}%
      \expandafter\def\csname LTw\endcsname{\color{white}}%
      \expandafter\def\csname LTb\endcsname{\color{black}}%
      \expandafter\def\csname LTa\endcsname{\color{black}}%
      \expandafter\def\csname LT0\endcsname{\color[rgb]{1,0,0}}%
      \expandafter\def\csname LT1\endcsname{\color[rgb]{0,1,0}}%
      \expandafter\def\csname LT2\endcsname{\color[rgb]{0,0,1}}%
      \expandafter\def\csname LT3\endcsname{\color[rgb]{1,0,1}}%
      \expandafter\def\csname LT4\endcsname{\color[rgb]{0,1,1}}%
      \expandafter\def\csname LT5\endcsname{\color[rgb]{1,1,0}}%
      \expandafter\def\csname LT6\endcsname{\color[rgb]{0,0,0}}%
      \expandafter\def\csname LT7\endcsname{\color[rgb]{1,0.3,0}}%
      \expandafter\def\csname LT8\endcsname{\color[rgb]{0.5,0.5,0.5}}%
    \else
      % gray
      \def\colorrgb#1{\color{black}}%
      \def\colorgray#1{\color[gray]{#1}}%
      \expandafter\def\csname LTw\endcsname{\color{white}}%
      \expandafter\def\csname LTb\endcsname{\color{black}}%
      \expandafter\def\csname LTa\endcsname{\color{black}}%
      \expandafter\def\csname LT0\endcsname{\color{black}}%
      \expandafter\def\csname LT1\endcsname{\color{black}}%
      \expandafter\def\csname LT2\endcsname{\color{black}}%
      \expandafter\def\csname LT3\endcsname{\color{black}}%
      \expandafter\def\csname LT4\endcsname{\color{black}}%
      \expandafter\def\csname LT5\endcsname{\color{black}}%
      \expandafter\def\csname LT6\endcsname{\color{black}}%
      \expandafter\def\csname LT7\endcsname{\color{black}}%
      \expandafter\def\csname LT8\endcsname{\color{black}}%
    \fi
  \fi
    \setlength{\unitlength}{0.0500bp}%
    \ifx\gptboxheight\undefined%
      \newlength{\gptboxheight}%
      \newlength{\gptboxwidth}%
      \newsavebox{\gptboxtext}%
    \fi%
    \setlength{\fboxrule}{0.5pt}%
    \setlength{\fboxsep}{1pt}%
\begin{picture}(7200.00,5040.00)%
    \gplgaddtomacro\gplbacktext{%
      \csname LTb\endcsname%
      \put(814,704){\makebox(0,0)[r]{\strut{}$0.3$}}%
      \put(814,1286){\makebox(0,0)[r]{\strut{}$0.4$}}%
      \put(814,1867){\makebox(0,0)[r]{\strut{}$0.5$}}%
      \put(814,2449){\makebox(0,0)[r]{\strut{}$0.6$}}%
      \put(814,3030){\makebox(0,0)[r]{\strut{}$0.7$}}%
      \put(814,3612){\makebox(0,0)[r]{\strut{}$0.8$}}%
      \put(814,4193){\makebox(0,0)[r]{\strut{}$0.9$}}%
      \put(814,4775){\makebox(0,0)[r]{\strut{}$1$}}%
      \put(946,484){\makebox(0,0){\strut{}$0$}}%
      \put(1532,484){\makebox(0,0){\strut{}$1$}}%
      \put(2117,484){\makebox(0,0){\strut{}$2$}}%
      \put(2703,484){\makebox(0,0){\strut{}$3$}}%
      \put(3289,484){\makebox(0,0){\strut{}$4$}}%
      \put(3875,484){\makebox(0,0){\strut{}$5$}}%
      \put(4460,484){\makebox(0,0){\strut{}$6$}}%
      \put(5046,484){\makebox(0,0){\strut{}$7$}}%
      \put(5632,484){\makebox(0,0){\strut{}$8$}}%
      \put(6217,484){\makebox(0,0){\strut{}$9$}}%
      \put(6803,484){\makebox(0,0){\strut{}$10$}}%
    }%
    \gplgaddtomacro\gplfronttext{%
      \csname LTb\endcsname%
      \put(176,2739){\rotatebox{-270}{\makebox(0,0){\strut{}\Large $\langle \varphi(x) \rangle$ \normalsize}}}%
      \put(3874,154){\makebox(0,0){\strut{}\Large $x$ \normalsize}}%
    }%
    \gplbacktext
    \put(0,0){\includegraphics{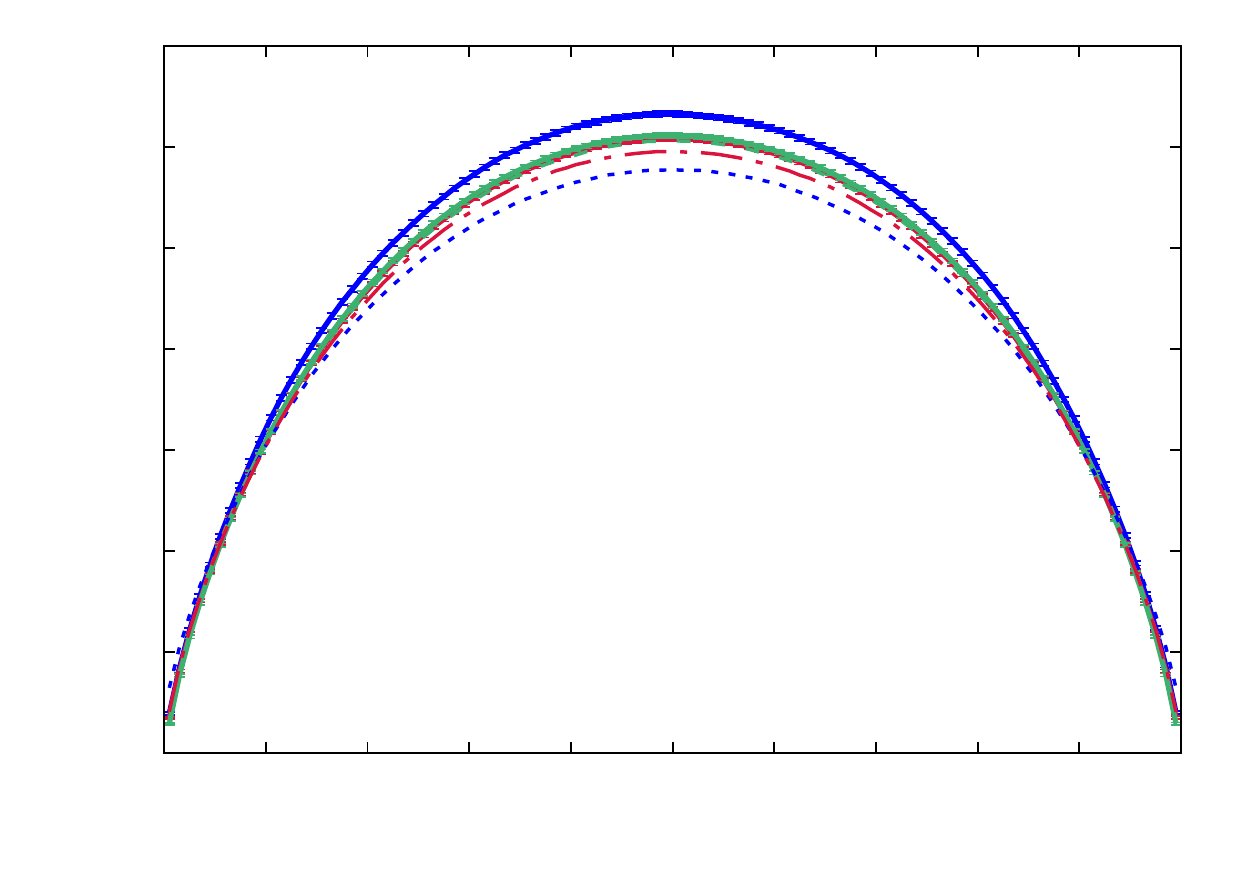}}%
    \gplfronttext
  \end{picture}%
\endgroup

%% file: flux_1B_I.tex
% GNUPLOT: LaTeX picture with Postscript
\begingroup
  \makeatletter
  \providecommand\color[2][]{%
    \GenericError{(gnuplot) \space\space\space\@spaces}{%
      Package color not loaded in conjunction with
      terminal option `colourtext'%
    }{See the gnuplot documentation for explanation.%
    }{Either use 'blacktext' in gnuplot or load the package
      color.sty in LaTeX.}%
    \renewcommand\color[2][]{}%
  }%
  \providecommand\includegraphics[2][]{%
    \GenericError{(gnuplot) \space\space\space\@spaces}{%
      Package graphicx or graphics not loaded%
    }{See the gnuplot documentation for explanation.%
    }{The gnuplot epslatex terminal needs graphicx.sty or graphics.sty.}%
    \renewcommand\includegraphics[2][]{}%
  }%
  \providecommand\rotatebox[2]{#2}%
  \@ifundefined{ifGPcolor}{%
    \newif\ifGPcolor
    \GPcolorfalse
  }{}%
  \@ifundefined{ifGPblacktext}{%
    \newif\ifGPblacktext
    \GPblacktexttrue
  }{}%
  % define a \g@addto@macro without @ in the name:
  \let\gplgaddtomacro\g@addto@macro
  % define empty templates for all commands taking text:
  \gdef\gplbacktext{}%
  \gdef\gplfronttext{}%
  \makeatother
  \ifGPblacktext
    % no textcolor at all
    \def\colorrgb#1{}%
    \def\colorgray#1{}%
  \else
    % gray or color?
    \ifGPcolor
      \def\colorrgb#1{\color[rgb]{#1}}%
      \def\colorgray#1{\color[gray]{#1}}%
      \expandafter\def\csname LTw\endcsname{\color{white}}%
      \expandafter\def\csname LTb\endcsname{\color{black}}%
      \expandafter\def\csname LTa\endcsname{\color{black}}%
      \expandafter\def\csname LT0\endcsname{\color[rgb]{1,0,0}}%
      \expandafter\def\csname LT1\endcsname{\color[rgb]{0,1,0}}%
      \expandafter\def\csname LT2\endcsname{\color[rgb]{0,0,1}}%
      \expandafter\def\csname LT3\endcsname{\color[rgb]{1,0,1}}%
      \expandafter\def\csname LT4\endcsname{\color[rgb]{0,1,1}}%
      \expandafter\def\csname LT5\endcsname{\color[rgb]{1,1,0}}%
      \expandafter\def\csname LT6\endcsname{\color[rgb]{0,0,0}}%
      \expandafter\def\csname LT7\endcsname{\color[rgb]{1,0.3,0}}%
      \expandafter\def\csname LT8\endcsname{\color[rgb]{0.5,0.5,0.5}}%
    \else
      % gray
      \def\colorrgb#1{\color{black}}%
      \def\colorgray#1{\color[gray]{#1}}%
      \expandafter\def\csname LTw\endcsname{\color{white}}%
      \expandafter\def\csname LTb\endcsname{\color{black}}%
      \expandafter\def\csname LTa\endcsname{\color{black}}%
      \expandafter\def\csname LT0\endcsname{\color{black}}%
      \expandafter\def\csname LT1\endcsname{\color{black}}%
      \expandafter\def\csname LT2\endcsname{\color{black}}%
      \expandafter\def\csname LT3\endcsname{\color{black}}%
      \expandafter\def\csname LT4\endcsname{\color{black}}%
      \expandafter\def\csname LT5\endcsname{\color{black}}%
      \expandafter\def\csname LT6\endcsname{\color{black}}%
      \expandafter\def\csname LT7\endcsname{\color{black}}%
      \expandafter\def\csname LT8\endcsname{\color{black}}%
    \fi
  \fi
    \setlength{\unitlength}{0.0500bp}%
    \ifx\gptboxheight\undefined%
      \newlength{\gptboxheight}%
      \newlength{\gptboxwidth}%
      \newsavebox{\gptboxtext}%
    \fi%
    \setlength{\fboxrule}{0.5pt}%
    \setlength{\fboxsep}{1pt}%
\begin{picture}(7200.00,5040.00)%
    \gplgaddtomacro\gplbacktext{%
      \csname LTb\endcsname%
      \put(1078,704){\makebox(0,0)[r]{\strut{}$0.001$}}%
      \put(1078,1722){\makebox(0,0)[r]{\strut{}$0.01$}}%
      \put(1078,2740){\makebox(0,0)[r]{\strut{}$0.1$}}%
      \put(1078,3757){\makebox(0,0)[r]{\strut{}$1$}}%
      \put(1078,4775){\makebox(0,0)[r]{\strut{}$10$}}%
      \put(1210,484){\makebox(0,0){\strut{}$0$}}%
      \put(1769,484){\makebox(0,0){\strut{}$1$}}%
      \put(2329,484){\makebox(0,0){\strut{}$2$}}%
      \put(2888,484){\makebox(0,0){\strut{}$3$}}%
      \put(3447,484){\makebox(0,0){\strut{}$4$}}%
      \put(4007,484){\makebox(0,0){\strut{}$5$}}%
      \put(4566,484){\makebox(0,0){\strut{}$6$}}%
      \put(5125,484){\makebox(0,0){\strut{}$7$}}%
      \put(5684,484){\makebox(0,0){\strut{}$8$}}%
      \put(6244,484){\makebox(0,0){\strut{}$9$}}%
      \put(6803,484){\makebox(0,0){\strut{}$10$}}%
    }%
    \gplgaddtomacro\gplfronttext{%
      \csname LTb\endcsname%
      \put(176,2739){\rotatebox{-270}{\makebox(0,0){\strut{}\Large $\langle \varphi(x) \rangle$ \normalsize}}}%
      \put(4006,154){\makebox(0,0){\strut{}\Large $x$ \normalsize}}%
    }%
    \gplbacktext
    \put(0,0){\includegraphics{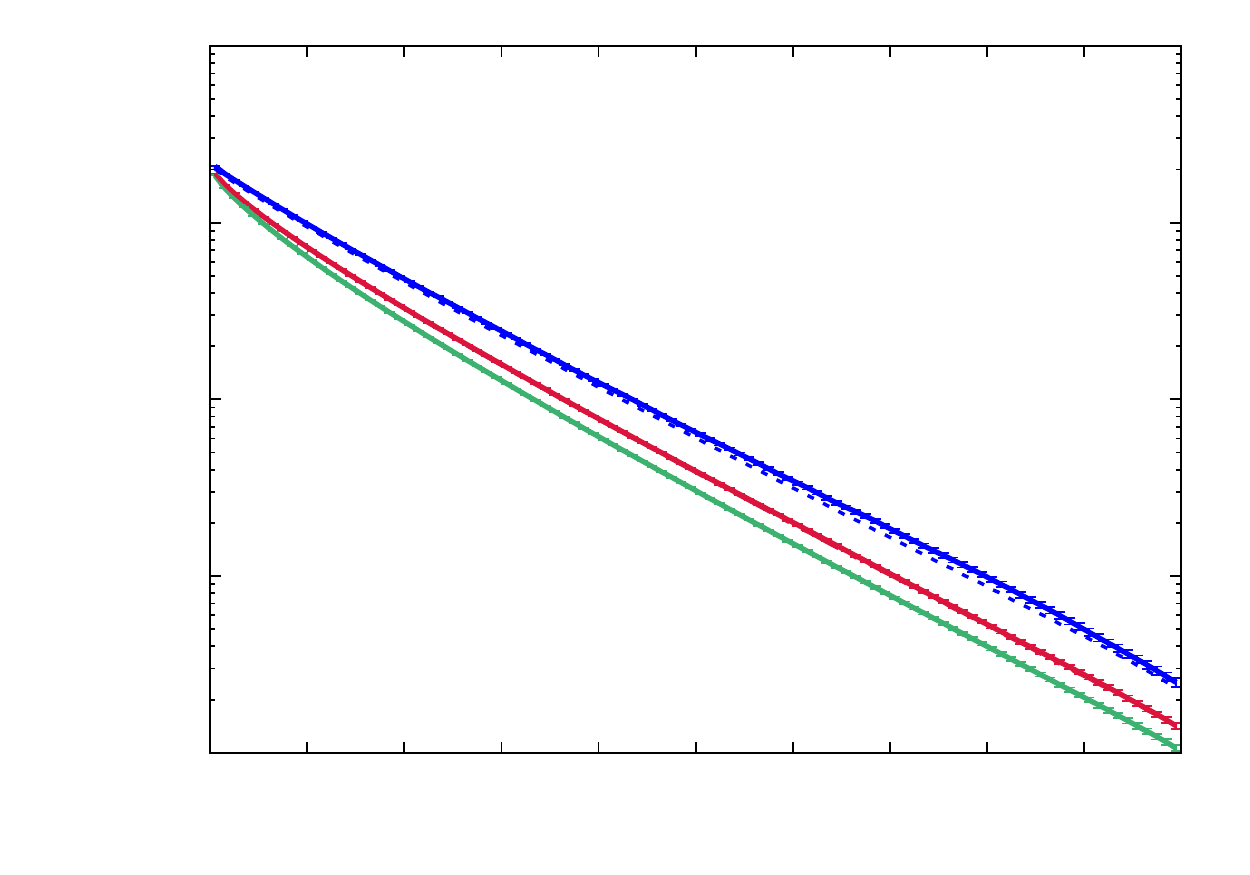}}%
    \gplfronttext
  \end{picture}%
\endgroup

%% file: flux_1B_II.tex
% GNUPLOT: LaTeX picture with Postscript
\begingroup
  \makeatletter
  \providecommand\color[2][]{%
    \GenericError{(gnuplot) \space\space\space\@spaces}{%
      Package color not loaded in conjunction with
      terminal option `colourtext'%
    }{See the gnuplot documentation for explanation.%
    }{Either use 'blacktext' in gnuplot or load the package
      color.sty in LaTeX.}%
    \renewcommand\color[2][]{}%
  }%
  \providecommand\includegraphics[2][]{%
    \GenericError{(gnuplot) \space\space\space\@spaces}{%
      Package graphicx or graphics not loaded%
    }{See the gnuplot documentation for explanation.%
    }{The gnuplot epslatex terminal needs graphicx.sty or graphics.sty.}%
    \renewcommand\includegraphics[2][]{}%
  }%
  \providecommand\rotatebox[2]{#2}%
  \@ifundefined{ifGPcolor}{%
    \newif\ifGPcolor
    \GPcolorfalse
  }{}%
  \@ifundefined{ifGPblacktext}{%
    \newif\ifGPblacktext
    \GPblacktexttrue
  }{}%
  % define a \g@addto@macro without @ in the name:
  \let\gplgaddtomacro\g@addto@macro
  % define empty templates for all commands taking text:
  \gdef\gplbacktext{}%
  \gdef\gplfronttext{}%
  \makeatother
  \ifGPblacktext
    % no textcolor at all
    \def\colorrgb#1{}%
    \def\colorgray#1{}%
  \else
    % gray or color?
    \ifGPcolor
      \def\colorrgb#1{\color[rgb]{#1}}%
      \def\colorgray#1{\color[gray]{#1}}%
      \expandafter\def\csname LTw\endcsname{\color{white}}%
      \expandafter\def\csname LTb\endcsname{\color{black}}%
      \expandafter\def\csname LTa\endcsname{\color{black}}%
      \expandafter\def\csname LT0\endcsname{\color[rgb]{1,0,0}}%
      \expandafter\def\csname LT1\endcsname{\color[rgb]{0,1,0}}%
      \expandafter\def\csname LT2\endcsname{\color[rgb]{0,0,1}}%
      \expandafter\def\csname LT3\endcsname{\color[rgb]{1,0,1}}%
      \expandafter\def\csname LT4\endcsname{\color[rgb]{0,1,1}}%
      \expandafter\def\csname LT5\endcsname{\color[rgb]{1,1,0}}%
      \expandafter\def\csname LT6\endcsname{\color[rgb]{0,0,0}}%
      \expandafter\def\csname LT7\endcsname{\color[rgb]{1,0.3,0}}%
      \expandafter\def\csname LT8\endcsname{\color[rgb]{0.5,0.5,0.5}}%
    \else
      % gray
      \def\colorrgb#1{\color{black}}%
      \def\colorgray#1{\color[gray]{#1}}%
      \expandafter\def\csname LTw\endcsname{\color{white}}%
      \expandafter\def\csname LTb\endcsname{\color{black}}%
      \expandafter\def\csname LTa\endcsname{\color{black}}%
      \expandafter\def\csname LT0\endcsname{\color{black}}%
      \expandafter\def\csname LT1\endcsname{\color{black}}%
      \expandafter\def\csname LT2\endcsname{\color{black}}%
      \expandafter\def\csname LT3\endcsname{\color{black}}%
      \expandafter\def\csname LT4\endcsname{\color{black}}%
      \expandafter\def\csname LT5\endcsname{\color{black}}%
      \expandafter\def\csname LT6\endcsname{\color{black}}%
      \expandafter\def\csname LT7\endcsname{\color{black}}%
      \expandafter\def\csname LT8\endcsname{\color{black}}%
    \fi
  \fi
    \setlength{\unitlength}{0.0500bp}%
    \ifx\gptboxheight\undefined%
      \newlength{\gptboxheight}%
      \newlength{\gptboxwidth}%
      \newsavebox{\gptboxtext}%
    \fi%
    \setlength{\fboxrule}{0.5pt}%
    \setlength{\fboxsep}{1pt}%
\begin{picture}(7200.00,5040.00)%
    \gplgaddtomacro\gplbacktext{%
      \csname LTb\endcsname%
      \put(946,704){\makebox(0,0)[r]{\strut{}$0.1$}}%
      \put(946,1286){\makebox(0,0)[r]{\strut{}$0.15$}}%
      \put(946,1867){\makebox(0,0)[r]{\strut{}$0.2$}}%
      \put(946,2449){\makebox(0,0)[r]{\strut{}$0.25$}}%
      \put(946,3030){\makebox(0,0)[r]{\strut{}$0.3$}}%
      \put(946,3612){\makebox(0,0)[r]{\strut{}$0.35$}}%
      \put(946,4193){\makebox(0,0)[r]{\strut{}$0.4$}}%
      \put(946,4775){\makebox(0,0)[r]{\strut{}$0.45$}}%
      \put(1078,484){\makebox(0,0){\strut{}$0$}}%
      \put(1651,484){\makebox(0,0){\strut{}$1$}}%
      \put(2223,484){\makebox(0,0){\strut{}$2$}}%
      \put(2796,484){\makebox(0,0){\strut{}$3$}}%
      \put(3368,484){\makebox(0,0){\strut{}$4$}}%
      \put(3941,484){\makebox(0,0){\strut{}$5$}}%
      \put(4513,484){\makebox(0,0){\strut{}$6$}}%
      \put(5086,484){\makebox(0,0){\strut{}$7$}}%
      \put(5658,484){\makebox(0,0){\strut{}$8$}}%
      \put(6231,484){\makebox(0,0){\strut{}$9$}}%
      \put(6803,484){\makebox(0,0){\strut{}$10$}}%
    }%
    \gplgaddtomacro\gplfronttext{%
      \csname LTb\endcsname%
      \put(176,2739){\rotatebox{-270}{\makebox(0,0){\strut{}\Large $\langle \varphi(x) \rangle$ \normalsize}}}%
      \put(3940,154){\makebox(0,0){\strut{}\Large $x$ \normalsize}}%
    }%
    \gplbacktext
    \put(0,0){\includegraphics{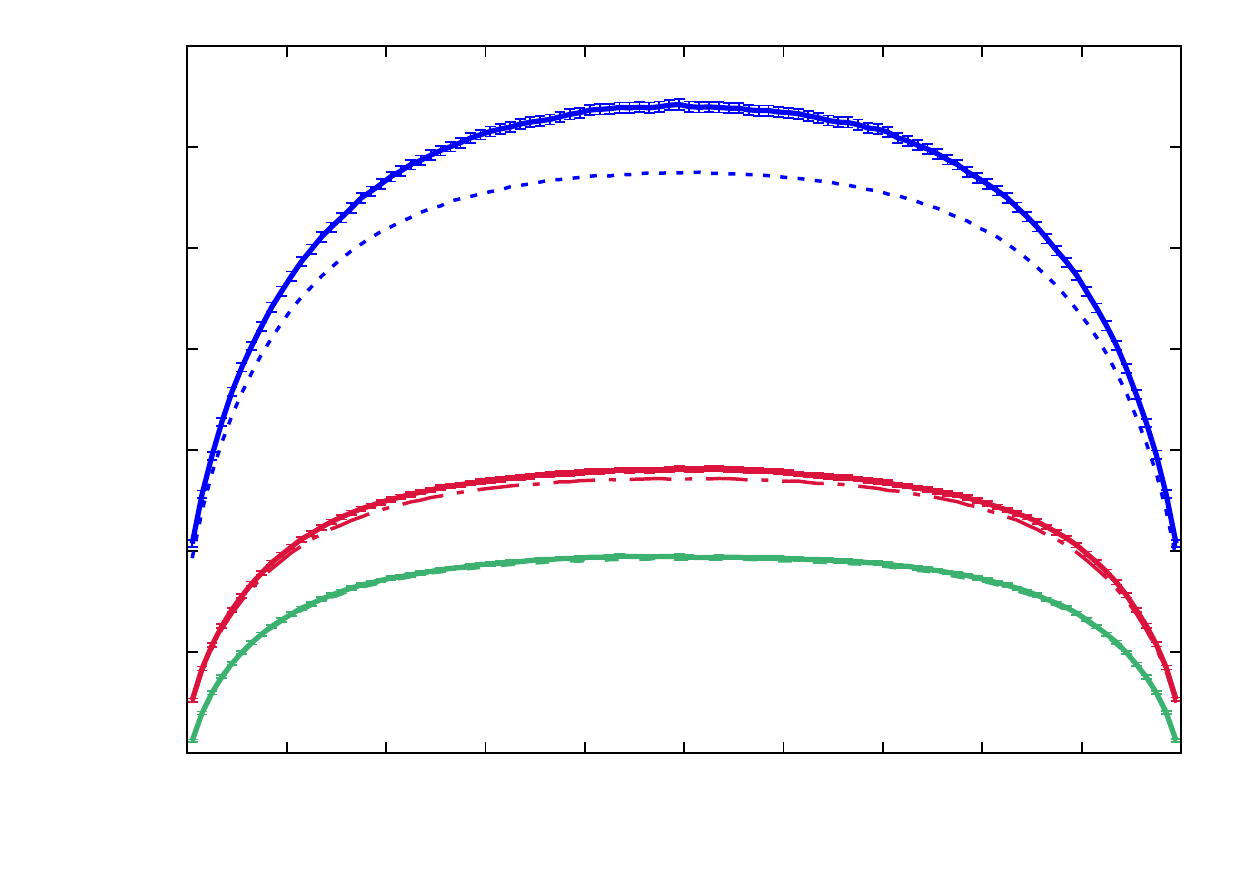}}%
    \gplfronttext
  \end{picture}%
\endgroup

%% file: flux_1C_I.tex
% GNUPLOT: LaTeX picture with Postscript
\begingroup
  \makeatletter
  \providecommand\color[2][]{%
    \GenericError{(gnuplot) \space\space\space\@spaces}{%
      Package color not loaded in conjunction with
      terminal option `colourtext'%
    }{See the gnuplot documentation for explanation.%
    }{Either use 'blacktext' in gnuplot or load the package
      color.sty in LaTeX.}%
    \renewcommand\color[2][]{}%
  }%
  \providecommand\includegraphics[2][]{%
    \GenericError{(gnuplot) \space\space\space\@spaces}{%
      Package graphicx or graphics not loaded%
    }{See the gnuplot documentation for explanation.%
    }{The gnuplot epslatex terminal needs graphicx.sty or graphics.sty.}%
    \renewcommand\includegraphics[2][]{}%
  }%
  \providecommand\rotatebox[2]{#2}%
  \@ifundefined{ifGPcolor}{%
    \newif\ifGPcolor
    \GPcolorfalse
  }{}%
  \@ifundefined{ifGPblacktext}{%
    \newif\ifGPblacktext
    \GPblacktexttrue
  }{}%
  % define a \g@addto@macro without @ in the name:
  \let\gplgaddtomacro\g@addto@macro
  % define empty templates for all commands taking text:
  \gdef\gplbacktext{}%
  \gdef\gplfronttext{}%
  \makeatother
  \ifGPblacktext
    % no textcolor at all
    \def\colorrgb#1{}%
    \def\colorgray#1{}%
  \else
    % gray or color?
    \ifGPcolor
      \def\colorrgb#1{\color[rgb]{#1}}%
      \def\colorgray#1{\color[gray]{#1}}%
      \expandafter\def\csname LTw\endcsname{\color{white}}%
      \expandafter\def\csname LTb\endcsname{\color{black}}%
      \expandafter\def\csname LTa\endcsname{\color{black}}%
      \expandafter\def\csname LT0\endcsname{\color[rgb]{1,0,0}}%
      \expandafter\def\csname LT1\endcsname{\color[rgb]{0,1,0}}%
      \expandafter\def\csname LT2\endcsname{\color[rgb]{0,0,1}}%
      \expandafter\def\csname LT3\endcsname{\color[rgb]{1,0,1}}%
      \expandafter\def\csname LT4\endcsname{\color[rgb]{0,1,1}}%
      \expandafter\def\csname LT5\endcsname{\color[rgb]{1,1,0}}%
      \expandafter\def\csname LT6\endcsname{\color[rgb]{0,0,0}}%
      \expandafter\def\csname LT7\endcsname{\color[rgb]{1,0.3,0}}%
      \expandafter\def\csname LT8\endcsname{\color[rgb]{0.5,0.5,0.5}}%
    \else
      % gray
      \def\colorrgb#1{\color{black}}%
      \def\colorgray#1{\color[gray]{#1}}%
      \expandafter\def\csname LTw\endcsname{\color{white}}%
      \expandafter\def\csname LTb\endcsname{\color{black}}%
      \expandafter\def\csname LTa\endcsname{\color{black}}%
      \expandafter\def\csname LT0\endcsname{\color{black}}%
      \expandafter\def\csname LT1\endcsname{\color{black}}%
      \expandafter\def\csname LT2\endcsname{\color{black}}%
      \expandafter\def\csname LT3\endcsname{\color{black}}%
      \expandafter\def\csname LT4\endcsname{\color{black}}%
      \expandafter\def\csname LT5\endcsname{\color{black}}%
      \expandafter\def\csname LT6\endcsname{\color{black}}%
      \expandafter\def\csname LT7\endcsname{\color{black}}%
      \expandafter\def\csname LT8\endcsname{\color{black}}%
    \fi
  \fi
    \setlength{\unitlength}{0.0500bp}%
    \ifx\gptboxheight\undefined%
      \newlength{\gptboxheight}%
      \newlength{\gptboxwidth}%
      \newsavebox{\gptboxtext}%
    \fi%
    \setlength{\fboxrule}{0.5pt}%
    \setlength{\fboxsep}{1pt}%
\begin{picture}(7200.00,5040.00)%
    \gplgaddtomacro\gplbacktext{%
      \csname LTb\endcsname%
      \put(946,704){\makebox(0,0)[r]{\strut{}$0.01$}}%
      \put(946,2061){\makebox(0,0)[r]{\strut{}$0.1$}}%
      \put(946,3418){\makebox(0,0)[r]{\strut{}$1$}}%
      \put(946,4775){\makebox(0,0)[r]{\strut{}$10$}}%
      \put(1078,484){\makebox(0,0){\strut{}$0$}}%
      \put(1651,484){\makebox(0,0){\strut{}$1$}}%
      \put(2223,484){\makebox(0,0){\strut{}$2$}}%
      \put(2796,484){\makebox(0,0){\strut{}$3$}}%
      \put(3368,484){\makebox(0,0){\strut{}$4$}}%
      \put(3941,484){\makebox(0,0){\strut{}$5$}}%
      \put(4513,484){\makebox(0,0){\strut{}$6$}}%
      \put(5086,484){\makebox(0,0){\strut{}$7$}}%
      \put(5658,484){\makebox(0,0){\strut{}$8$}}%
      \put(6231,484){\makebox(0,0){\strut{}$9$}}%
      \put(6803,484){\makebox(0,0){\strut{}$10$}}%
    }%
    \gplgaddtomacro\gplfronttext{%
      \csname LTb\endcsname%
      \put(176,2739){\rotatebox{-270}{\makebox(0,0){\strut{}\Large $\langle \varphi(x) \rangle$ \normalsize}}}%
      \put(3940,154){\makebox(0,0){\strut{}\Large $x$ \normalsize}}%
    }%
    \gplbacktext
    \put(0,0){\includegraphics{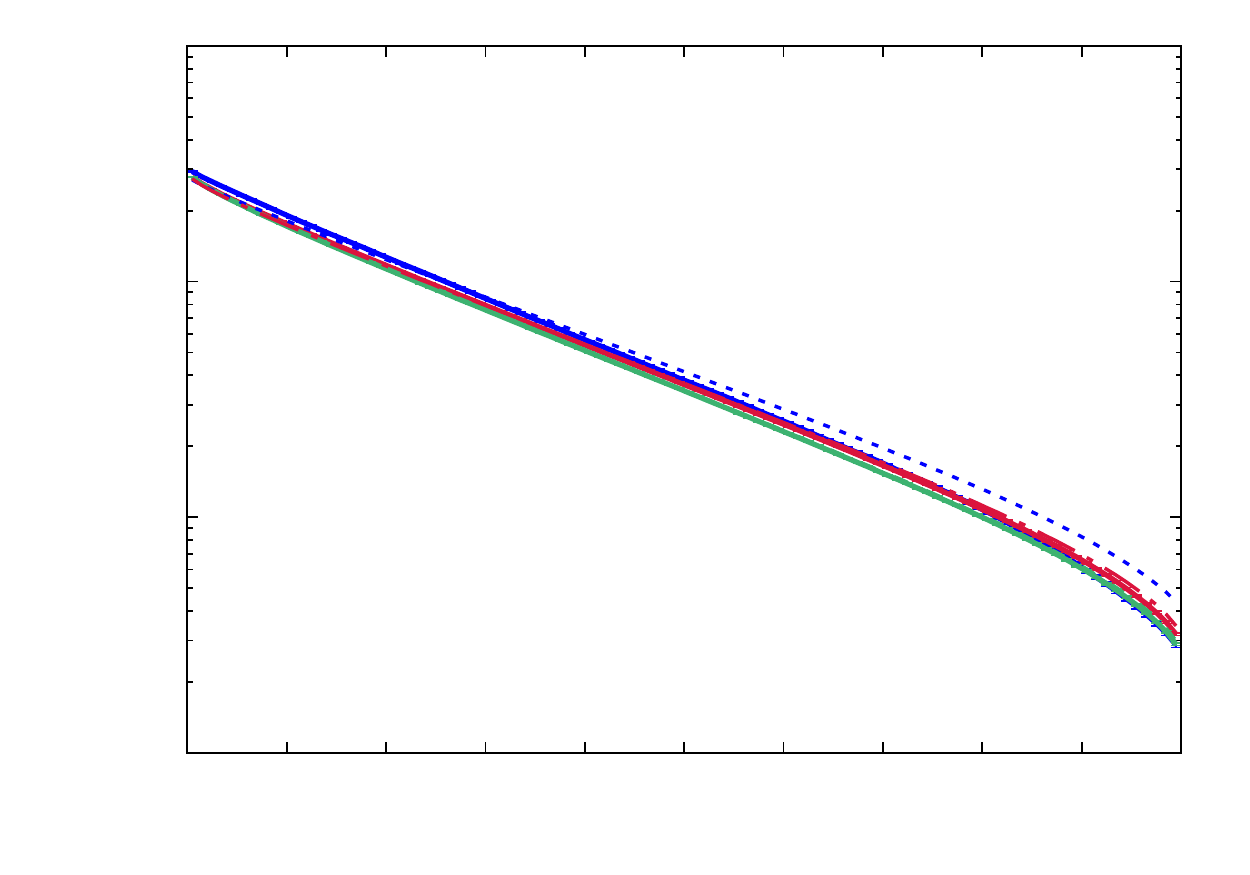}}%
    \gplfronttext
  \end{picture}%
\endgroup

%% file: flux_1C_II.tex
% GNUPLOT: LaTeX picture with Postscript
\begingroup
  \makeatletter
  \providecommand\color[2][]{%
    \GenericError{(gnuplot) \space\space\space\@spaces}{%
      Package color not loaded in conjunction with
      terminal option `colourtext'%
    }{See the gnuplot documentation for explanation.%
    }{Either use 'blacktext' in gnuplot or load the package
      color.sty in LaTeX.}%
    \renewcommand\color[2][]{}%
  }%
  \providecommand\includegraphics[2][]{%
    \GenericError{(gnuplot) \space\space\space\@spaces}{%
      Package graphicx or graphics not loaded%
    }{See the gnuplot documentation for explanation.%
    }{The gnuplot epslatex terminal needs graphicx.sty or graphics.sty.}%
    \renewcommand\includegraphics[2][]{}%
  }%
  \providecommand\rotatebox[2]{#2}%
  \@ifundefined{ifGPcolor}{%
    \newif\ifGPcolor
    \GPcolorfalse
  }{}%
  \@ifundefined{ifGPblacktext}{%
    \newif\ifGPblacktext
    \GPblacktexttrue
  }{}%
  % define a \g@addto@macro without @ in the name:
  \let\gplgaddtomacro\g@addto@macro
  % define empty templates for all commands taking text:
  \gdef\gplbacktext{}%
  \gdef\gplfronttext{}%
  \makeatother
  \ifGPblacktext
    % no textcolor at all
    \def\colorrgb#1{}%
    \def\colorgray#1{}%
  \else
    % gray or color?
    \ifGPcolor
      \def\colorrgb#1{\color[rgb]{#1}}%
      \def\colorgray#1{\color[gray]{#1}}%
      \expandafter\def\csname LTw\endcsname{\color{white}}%
      \expandafter\def\csname LTb\endcsname{\color{black}}%
      \expandafter\def\csname LTa\endcsname{\color{black}}%
      \expandafter\def\csname LT0\endcsname{\color[rgb]{1,0,0}}%
      \expandafter\def\csname LT1\endcsname{\color[rgb]{0,1,0}}%
      \expandafter\def\csname LT2\endcsname{\color[rgb]{0,0,1}}%
      \expandafter\def\csname LT3\endcsname{\color[rgb]{1,0,1}}%
      \expandafter\def\csname LT4\endcsname{\color[rgb]{0,1,1}}%
      \expandafter\def\csname LT5\endcsname{\color[rgb]{1,1,0}}%
      \expandafter\def\csname LT6\endcsname{\color[rgb]{0,0,0}}%
      \expandafter\def\csname LT7\endcsname{\color[rgb]{1,0.3,0}}%
      \expandafter\def\csname LT8\endcsname{\color[rgb]{0.5,0.5,0.5}}%
    \else
      % gray
      \def\colorrgb#1{\color{black}}%
      \def\colorgray#1{\color[gray]{#1}}%
      \expandafter\def\csname LTw\endcsname{\color{white}}%
      \expandafter\def\csname LTb\endcsname{\color{black}}%
      \expandafter\def\csname LTa\endcsname{\color{black}}%
      \expandafter\def\csname LT0\endcsname{\color{black}}%
      \expandafter\def\csname LT1\endcsname{\color{black}}%
      \expandafter\def\csname LT2\endcsname{\color{black}}%
      \expandafter\def\csname LT3\endcsname{\color{black}}%
      \expandafter\def\csname LT4\endcsname{\color{black}}%
      \expandafter\def\csname LT5\endcsname{\color{black}}%
      \expandafter\def\csname LT6\endcsname{\color{black}}%
      \expandafter\def\csname LT7\endcsname{\color{black}}%
      \expandafter\def\csname LT8\endcsname{\color{black}}%
    \fi
  \fi
    \setlength{\unitlength}{0.0500bp}%
    \ifx\gptboxheight\undefined%
      \newlength{\gptboxheight}%
      \newlength{\gptboxwidth}%
      \newsavebox{\gptboxtext}%
    \fi%
    \setlength{\fboxrule}{0.5pt}%
    \setlength{\fboxsep}{1pt}%
\begin{picture}(7200.00,5040.00)%
    \gplgaddtomacro\gplbacktext{%
      \csname LTb\endcsname%
      \put(814,704){\makebox(0,0)[r]{\strut{}$0.3$}}%
      \put(814,1156){\makebox(0,0)[r]{\strut{}$0.4$}}%
      \put(814,1609){\makebox(0,0)[r]{\strut{}$0.5$}}%
      \put(814,2061){\makebox(0,0)[r]{\strut{}$0.6$}}%
      \put(814,2513){\makebox(0,0)[r]{\strut{}$0.7$}}%
      \put(814,2966){\makebox(0,0)[r]{\strut{}$0.8$}}%
      \put(814,3418){\makebox(0,0)[r]{\strut{}$0.9$}}%
      \put(814,3870){\makebox(0,0)[r]{\strut{}$1$}}%
      \put(814,4323){\makebox(0,0)[r]{\strut{}$1.1$}}%
      \put(814,4775){\makebox(0,0)[r]{\strut{}$1.2$}}%
      \put(946,484){\makebox(0,0){\strut{}$0$}}%
      \put(1532,484){\makebox(0,0){\strut{}$1$}}%
      \put(2117,484){\makebox(0,0){\strut{}$2$}}%
      \put(2703,484){\makebox(0,0){\strut{}$3$}}%
      \put(3289,484){\makebox(0,0){\strut{}$4$}}%
      \put(3875,484){\makebox(0,0){\strut{}$5$}}%
      \put(4460,484){\makebox(0,0){\strut{}$6$}}%
      \put(5046,484){\makebox(0,0){\strut{}$7$}}%
      \put(5632,484){\makebox(0,0){\strut{}$8$}}%
      \put(6217,484){\makebox(0,0){\strut{}$9$}}%
      \put(6803,484){\makebox(0,0){\strut{}$10$}}%
    }%
    \gplgaddtomacro\gplfronttext{%
      \csname LTb\endcsname%
      \put(176,2739){\rotatebox{-270}{\makebox(0,0){\strut{}\Large $\langle \varphi(x) \rangle$ \normalsize}}}%
      \put(3874,154){\makebox(0,0){\strut{}\Large $x$ \normalsize}}%
    }%
    \gplbacktext
    \put(0,0){\includegraphics{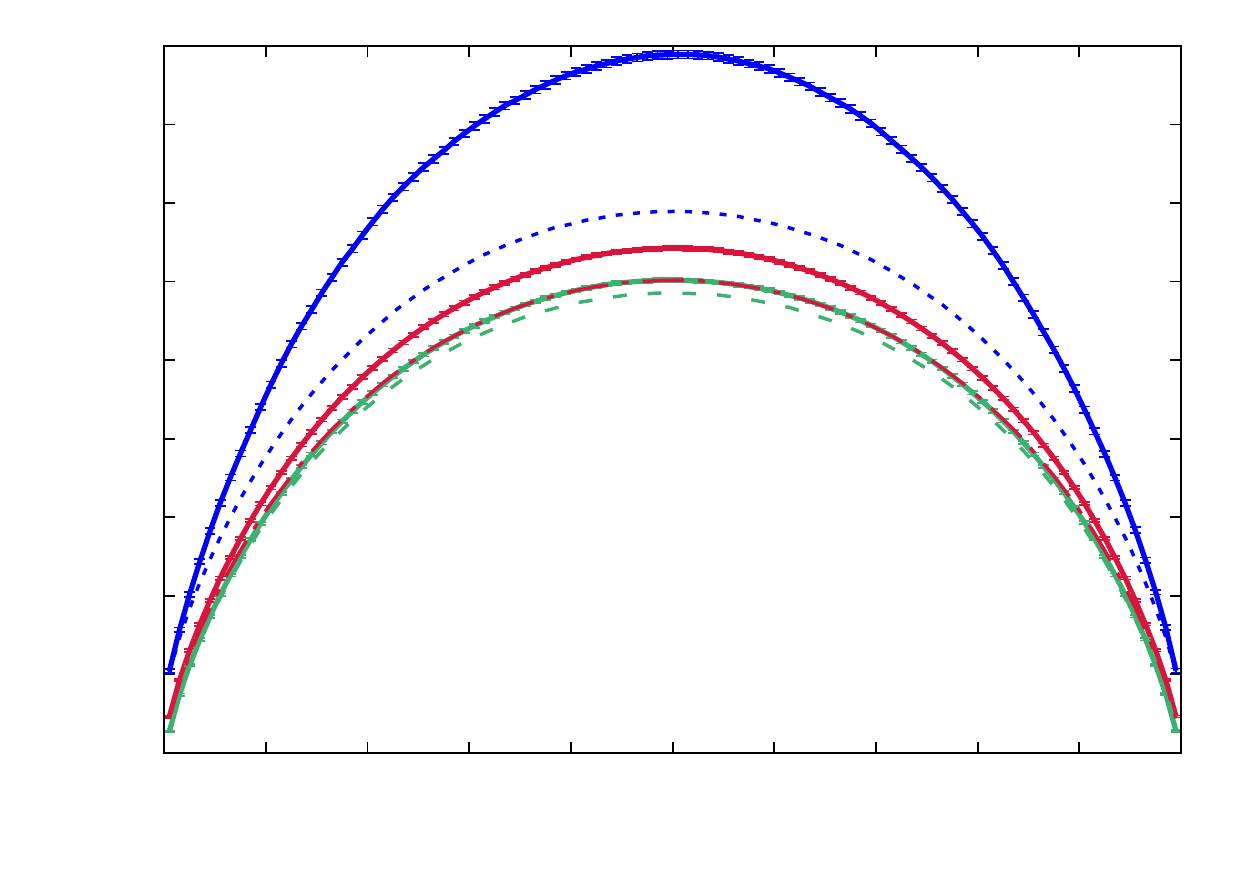}}%
    \gplfronttext
  \end{picture}%
\endgroup

%% file: flux_2A_I.tex
% GNUPLOT: LaTeX picture with Postscript
\begingroup
  \makeatletter
  \providecommand\color[2][]{%
    \GenericError{(gnuplot) \space\space\space\@spaces}{%
      Package color not loaded in conjunction with
      terminal option `colourtext'%
    }{See the gnuplot documentation for explanation.%
    }{Either use 'blacktext' in gnuplot or load the package
      color.sty in LaTeX.}%
    \renewcommand\color[2][]{}%
  }%
  \providecommand\includegraphics[2][]{%
    \GenericError{(gnuplot) \space\space\space\@spaces}{%
      Package graphicx or graphics not loaded%
    }{See the gnuplot documentation for explanation.%
    }{The gnuplot epslatex terminal needs graphicx.sty or graphics.sty.}%
    \renewcommand\includegraphics[2][]{}%
  }%
  \providecommand\rotatebox[2]{#2}%
  \@ifundefined{ifGPcolor}{%
    \newif\ifGPcolor
    \GPcolorfalse
  }{}%
  \@ifundefined{ifGPblacktext}{%
    \newif\ifGPblacktext
    \GPblacktexttrue
  }{}%
  % define a \g@addto@macro without @ in the name:
  \let\gplgaddtomacro\g@addto@macro
  % define empty templates for all commands taking text:
  \gdef\gplbacktext{}%
  \gdef\gplfronttext{}%
  \makeatother
  \ifGPblacktext
    % no textcolor at all
    \def\colorrgb#1{}%
    \def\colorgray#1{}%
  \else
    % gray or color?
    \ifGPcolor
      \def\colorrgb#1{\color[rgb]{#1}}%
      \def\colorgray#1{\color[gray]{#1}}%
      \expandafter\def\csname LTw\endcsname{\color{white}}%
      \expandafter\def\csname LTb\endcsname{\color{black}}%
      \expandafter\def\csname LTa\endcsname{\color{black}}%
      \expandafter\def\csname LT0\endcsname{\color[rgb]{1,0,0}}%
      \expandafter\def\csname LT1\endcsname{\color[rgb]{0,1,0}}%
      \expandafter\def\csname LT2\endcsname{\color[rgb]{0,0,1}}%
      \expandafter\def\csname LT3\endcsname{\color[rgb]{1,0,1}}%
      \expandafter\def\csname LT4\endcsname{\color[rgb]{0,1,1}}%
      \expandafter\def\csname LT5\endcsname{\color[rgb]{1,1,0}}%
      \expandafter\def\csname LT6\endcsname{\color[rgb]{0,0,0}}%
      \expandafter\def\csname LT7\endcsname{\color[rgb]{1,0.3,0}}%
      \expandafter\def\csname LT8\endcsname{\color[rgb]{0.5,0.5,0.5}}%
    \else
      % gray
      \def\colorrgb#1{\color{black}}%
      \def\colorgray#1{\color[gray]{#1}}%
      \expandafter\def\csname LTw\endcsname{\color{white}}%
      \expandafter\def\csname LTb\endcsname{\color{black}}%
      \expandafter\def\csname LTa\endcsname{\color{black}}%
      \expandafter\def\csname LT0\endcsname{\color{black}}%
      \expandafter\def\csname LT1\endcsname{\color{black}}%
      \expandafter\def\csname LT2\endcsname{\color{black}}%
      \expandafter\def\csname LT3\endcsname{\color{black}}%
      \expandafter\def\csname LT4\endcsname{\color{black}}%
      \expandafter\def\csname LT5\endcsname{\color{black}}%
      \expandafter\def\csname LT6\endcsname{\color{black}}%
      \expandafter\def\csname LT7\endcsname{\color{black}}%
      \expandafter\def\csname LT8\endcsname{\color{black}}%
    \fi
  \fi
    \setlength{\unitlength}{0.0500bp}%
    \ifx\gptboxheight\undefined%
      \newlength{\gptboxheight}%
      \newlength{\gptboxwidth}%
      \newsavebox{\gptboxtext}%
    \fi%
    \setlength{\fboxrule}{0.5pt}%
    \setlength{\fboxsep}{1pt}%
\begin{picture}(7200.00,5040.00)%
    \gplgaddtomacro\gplbacktext{%
      \csname LTb\endcsname%
      \put(814,704){\makebox(0,0)[r]{\strut{}$0.1$}}%
      \put(814,2740){\makebox(0,0)[r]{\strut{}$1$}}%
      \put(814,4775){\makebox(0,0)[r]{\strut{}$10$}}%
      \put(946,484){\makebox(0,0){\strut{}$0$}}%
      \put(1532,484){\makebox(0,0){\strut{}$1$}}%
      \put(2117,484){\makebox(0,0){\strut{}$2$}}%
      \put(2703,484){\makebox(0,0){\strut{}$3$}}%
      \put(3289,484){\makebox(0,0){\strut{}$4$}}%
      \put(3875,484){\makebox(0,0){\strut{}$5$}}%
      \put(4460,484){\makebox(0,0){\strut{}$6$}}%
      \put(5046,484){\makebox(0,0){\strut{}$7$}}%
      \put(5632,484){\makebox(0,0){\strut{}$8$}}%
      \put(6217,484){\makebox(0,0){\strut{}$9$}}%
      \put(6803,484){\makebox(0,0){\strut{}$10$}}%
    }%
    \gplgaddtomacro\gplfronttext{%
      \csname LTb\endcsname%
      \put(176,2739){\rotatebox{-270}{\makebox(0,0){\strut{}\Large $\langle \varphi(x) \rangle$ \normalsize}}}%
      \put(3874,154){\makebox(0,0){\strut{}\Large $x$ \normalsize}}%
    }%
    \gplbacktext
    \put(0,0){\includegraphics{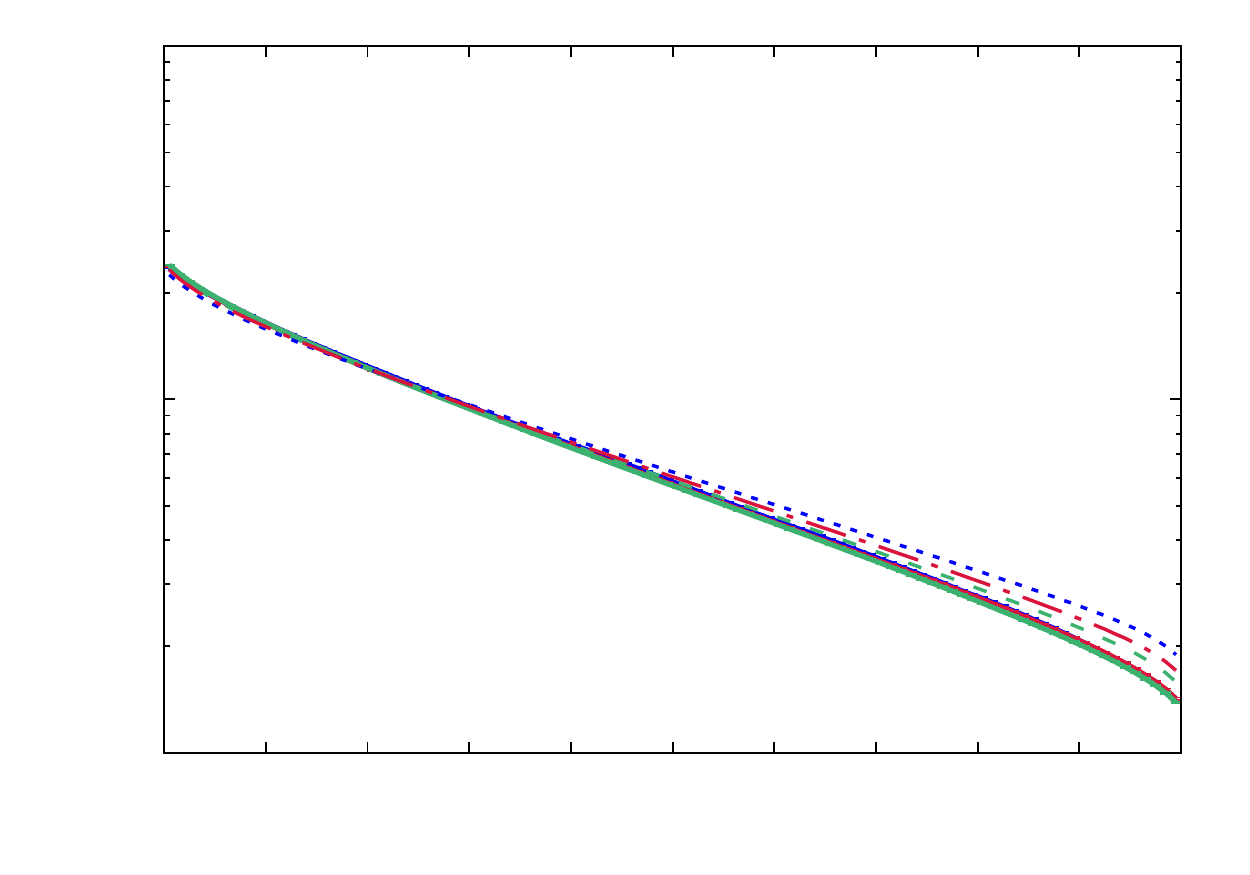}}%
    \gplfronttext
  \end{picture}%
\endgroup

%% file: flux_2A_II.tex
% GNUPLOT: LaTeX picture with Postscript
\begingroup
  \makeatletter
  \providecommand\color[2][]{%
    \GenericError{(gnuplot) \space\space\space\@spaces}{%
      Package color not loaded in conjunction with
      terminal option `colourtext'%
    }{See the gnuplot documentation for explanation.%
    }{Either use 'blacktext' in gnuplot or load the package
      color.sty in LaTeX.}%
    \renewcommand\color[2][]{}%
  }%
  \providecommand\includegraphics[2][]{%
    \GenericError{(gnuplot) \space\space\space\@spaces}{%
      Package graphicx or graphics not loaded%
    }{See the gnuplot documentation for explanation.%
    }{The gnuplot epslatex terminal needs graphicx.sty or graphics.sty.}%
    \renewcommand\includegraphics[2][]{}%
  }%
  \providecommand\rotatebox[2]{#2}%
  \@ifundefined{ifGPcolor}{%
    \newif\ifGPcolor
    \GPcolorfalse
  }{}%
  \@ifundefined{ifGPblacktext}{%
    \newif\ifGPblacktext
    \GPblacktexttrue
  }{}%
  % define a \g@addto@macro without @ in the name:
  \let\gplgaddtomacro\g@addto@macro
  % define empty templates for all commands taking text:
  \gdef\gplbacktext{}%
  \gdef\gplfronttext{}%
  \makeatother
  \ifGPblacktext
    % no textcolor at all
    \def\colorrgb#1{}%
    \def\colorgray#1{}%
  \else
    % gray or color?
    \ifGPcolor
      \def\colorrgb#1{\color[rgb]{#1}}%
      \def\colorgray#1{\color[gray]{#1}}%
      \expandafter\def\csname LTw\endcsname{\color{white}}%
      \expandafter\def\csname LTb\endcsname{\color{black}}%
      \expandafter\def\csname LTa\endcsname{\color{black}}%
      \expandafter\def\csname LT0\endcsname{\color[rgb]{1,0,0}}%
      \expandafter\def\csname LT1\endcsname{\color[rgb]{0,1,0}}%
      \expandafter\def\csname LT2\endcsname{\color[rgb]{0,0,1}}%
      \expandafter\def\csname LT3\endcsname{\color[rgb]{1,0,1}}%
      \expandafter\def\csname LT4\endcsname{\color[rgb]{0,1,1}}%
      \expandafter\def\csname LT5\endcsname{\color[rgb]{1,1,0}}%
      \expandafter\def\csname LT6\endcsname{\color[rgb]{0,0,0}}%
      \expandafter\def\csname LT7\endcsname{\color[rgb]{1,0.3,0}}%
      \expandafter\def\csname LT8\endcsname{\color[rgb]{0.5,0.5,0.5}}%
    \else
      % gray
      \def\colorrgb#1{\color{black}}%
      \def\colorgray#1{\color[gray]{#1}}%
      \expandafter\def\csname LTw\endcsname{\color{white}}%
      \expandafter\def\csname LTb\endcsname{\color{black}}%
      \expandafter\def\csname LTa\endcsname{\color{black}}%
      \expandafter\def\csname LT0\endcsname{\color{black}}%
      \expandafter\def\csname LT1\endcsname{\color{black}}%
      \expandafter\def\csname LT2\endcsname{\color{black}}%
      \expandafter\def\csname LT3\endcsname{\color{black}}%
      \expandafter\def\csname LT4\endcsname{\color{black}}%
      \expandafter\def\csname LT5\endcsname{\color{black}}%
      \expandafter\def\csname LT6\endcsname{\color{black}}%
      \expandafter\def\csname LT7\endcsname{\color{black}}%
      \expandafter\def\csname LT8\endcsname{\color{black}}%
    \fi
  \fi
    \setlength{\unitlength}{0.0500bp}%
    \ifx\gptboxheight\undefined%
      \newlength{\gptboxheight}%
      \newlength{\gptboxwidth}%
      \newsavebox{\gptboxtext}%
    \fi%
    \setlength{\fboxrule}{0.5pt}%
    \setlength{\fboxsep}{1pt}%
\begin{picture}(7200.00,5040.00)%
    \gplgaddtomacro\gplbacktext{%
      \csname LTb\endcsname%
      \put(814,704){\makebox(0,0)[r]{\strut{}$0.4$}}%
      \put(814,1383){\makebox(0,0)[r]{\strut{}$0.5$}}%
      \put(814,2061){\makebox(0,0)[r]{\strut{}$0.6$}}%
      \put(814,2739){\makebox(0,0)[r]{\strut{}$0.7$}}%
      \put(814,3418){\makebox(0,0)[r]{\strut{}$0.8$}}%
      \put(814,4096){\makebox(0,0)[r]{\strut{}$0.9$}}%
      \put(814,4775){\makebox(0,0)[r]{\strut{}$1$}}%
      \put(946,484){\makebox(0,0){\strut{}$0$}}%
      \put(1532,484){\makebox(0,0){\strut{}$1$}}%
      \put(2117,484){\makebox(0,0){\strut{}$2$}}%
      \put(2703,484){\makebox(0,0){\strut{}$3$}}%
      \put(3289,484){\makebox(0,0){\strut{}$4$}}%
      \put(3875,484){\makebox(0,0){\strut{}$5$}}%
      \put(4460,484){\makebox(0,0){\strut{}$6$}}%
      \put(5046,484){\makebox(0,0){\strut{}$7$}}%
      \put(5632,484){\makebox(0,0){\strut{}$8$}}%
      \put(6217,484){\makebox(0,0){\strut{}$9$}}%
      \put(6803,484){\makebox(0,0){\strut{}$10$}}%
    }%
    \gplgaddtomacro\gplfronttext{%
      \csname LTb\endcsname%
      \put(176,2739){\rotatebox{-270}{\makebox(0,0){\strut{}\Large $\langle \varphi(x) \rangle$ \normalsize}}}%
      \put(3874,154){\makebox(0,0){\strut{}\Large $x$ \normalsize}}%
    }%
    \gplbacktext
    \put(0,0){\includegraphics{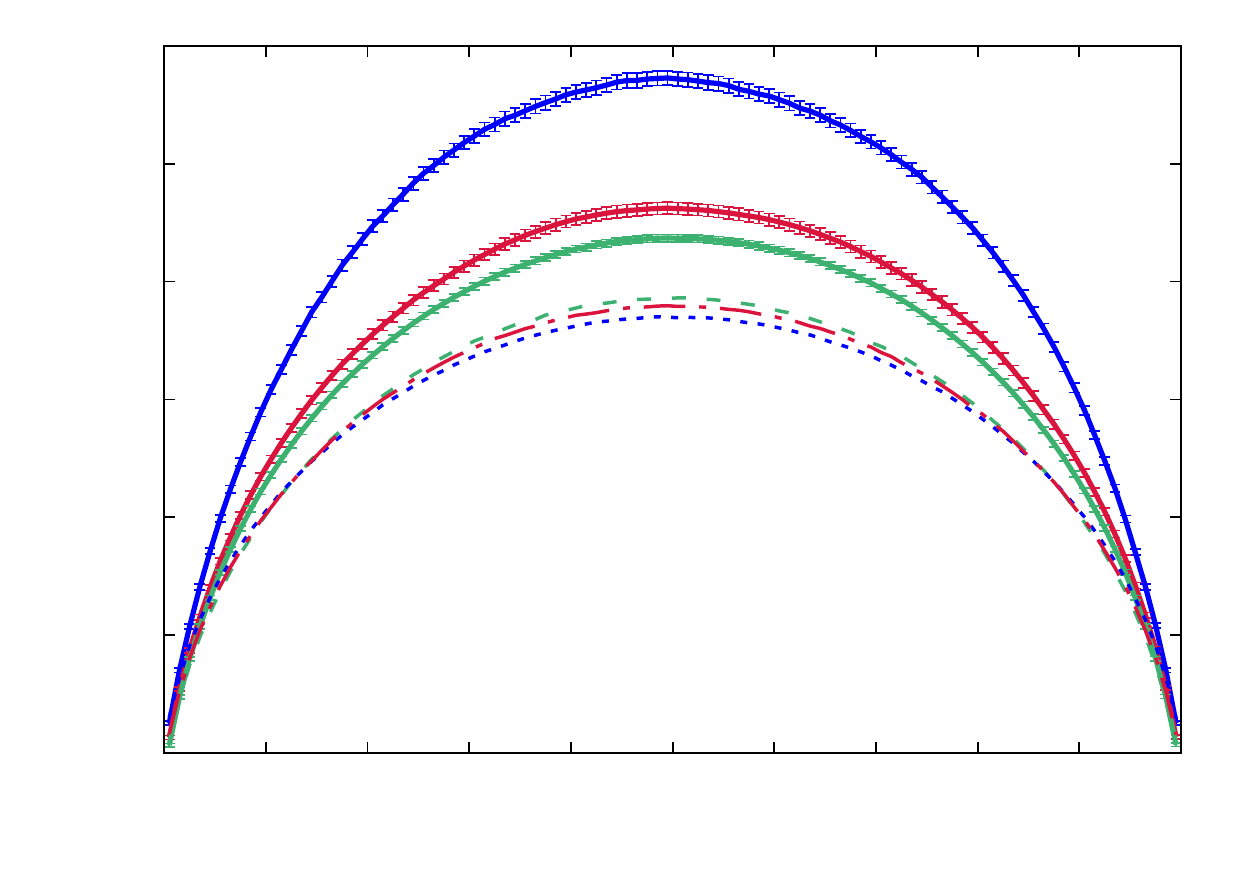}}%
    \gplfronttext
  \end{picture}%
\endgroup

%% file: flux_2B_I.tex
% GNUPLOT: LaTeX picture with Postscript
\begingroup
  \makeatletter
  \providecommand\color[2][]{%
    \GenericError{(gnuplot) \space\space\space\@spaces}{%
      Package color not loaded in conjunction with
      terminal option `colourtext'%
    }{See the gnuplot documentation for explanation.%
    }{Either use 'blacktext' in gnuplot or load the package
      color.sty in LaTeX.}%
    \renewcommand\color[2][]{}%
  }%
  \providecommand\includegraphics[2][]{%
    \GenericError{(gnuplot) \space\space\space\@spaces}{%
      Package graphicx or graphics not loaded%
    }{See the gnuplot documentation for explanation.%
    }{The gnuplot epslatex terminal needs graphicx.sty or graphics.sty.}%
    \renewcommand\includegraphics[2][]{}%
  }%
  \providecommand\rotatebox[2]{#2}%
  \@ifundefined{ifGPcolor}{%
    \newif\ifGPcolor
    \GPcolorfalse
  }{}%
  \@ifundefined{ifGPblacktext}{%
    \newif\ifGPblacktext
    \GPblacktexttrue
  }{}%
  % define a \g@addto@macro without @ in the name:
  \let\gplgaddtomacro\g@addto@macro
  % define empty templates for all commands taking text:
  \gdef\gplbacktext{}%
  \gdef\gplfronttext{}%
  \makeatother
  \ifGPblacktext
    % no textcolor at all
    \def\colorrgb#1{}%
    \def\colorgray#1{}%
  \else
    % gray or color?
    \ifGPcolor
      \def\colorrgb#1{\color[rgb]{#1}}%
      \def\colorgray#1{\color[gray]{#1}}%
      \expandafter\def\csname LTw\endcsname{\color{white}}%
      \expandafter\def\csname LTb\endcsname{\color{black}}%
      \expandafter\def\csname LTa\endcsname{\color{black}}%
      \expandafter\def\csname LT0\endcsname{\color[rgb]{1,0,0}}%
      \expandafter\def\csname LT1\endcsname{\color[rgb]{0,1,0}}%
      \expandafter\def\csname LT2\endcsname{\color[rgb]{0,0,1}}%
      \expandafter\def\csname LT3\endcsname{\color[rgb]{1,0,1}}%
      \expandafter\def\csname LT4\endcsname{\color[rgb]{0,1,1}}%
      \expandafter\def\csname LT5\endcsname{\color[rgb]{1,1,0}}%
      \expandafter\def\csname LT6\endcsname{\color[rgb]{0,0,0}}%
      \expandafter\def\csname LT7\endcsname{\color[rgb]{1,0.3,0}}%
      \expandafter\def\csname LT8\endcsname{\color[rgb]{0.5,0.5,0.5}}%
    \else
      % gray
      \def\colorrgb#1{\color{black}}%
      \def\colorgray#1{\color[gray]{#1}}%
      \expandafter\def\csname LTw\endcsname{\color{white}}%
      \expandafter\def\csname LTb\endcsname{\color{black}}%
      \expandafter\def\csname LTa\endcsname{\color{black}}%
      \expandafter\def\csname LT0\endcsname{\color{black}}%
      \expandafter\def\csname LT1\endcsname{\color{black}}%
      \expandafter\def\csname LT2\endcsname{\color{black}}%
      \expandafter\def\csname LT3\endcsname{\color{black}}%
      \expandafter\def\csname LT4\endcsname{\color{black}}%
      \expandafter\def\csname LT5\endcsname{\color{black}}%
      \expandafter\def\csname LT6\endcsname{\color{black}}%
      \expandafter\def\csname LT7\endcsname{\color{black}}%
      \expandafter\def\csname LT8\endcsname{\color{black}}%
    \fi
  \fi
    \setlength{\unitlength}{0.0500bp}%
    \ifx\gptboxheight\undefined%
      \newlength{\gptboxheight}%
      \newlength{\gptboxwidth}%
      \newsavebox{\gptboxtext}%
    \fi%
    \setlength{\fboxrule}{0.5pt}%
    \setlength{\fboxsep}{1pt}%
\begin{picture}(7200.00,5040.00)%
    \gplgaddtomacro\gplbacktext{%
      \csname LTb\endcsname%
      \put(814,704){\makebox(0,0)[r]{\strut{}$0.1$}}%
      \put(814,2740){\makebox(0,0)[r]{\strut{}$1$}}%
      \put(814,4775){\makebox(0,0)[r]{\strut{}$10$}}%
      \put(946,484){\makebox(0,0){\strut{}$0$}}%
      \put(1532,484){\makebox(0,0){\strut{}$1$}}%
      \put(2117,484){\makebox(0,0){\strut{}$2$}}%
      \put(2703,484){\makebox(0,0){\strut{}$3$}}%
      \put(3289,484){\makebox(0,0){\strut{}$4$}}%
      \put(3875,484){\makebox(0,0){\strut{}$5$}}%
      \put(4460,484){\makebox(0,0){\strut{}$6$}}%
      \put(5046,484){\makebox(0,0){\strut{}$7$}}%
      \put(5632,484){\makebox(0,0){\strut{}$8$}}%
      \put(6217,484){\makebox(0,0){\strut{}$9$}}%
      \put(6803,484){\makebox(0,0){\strut{}$10$}}%
    }%
    \gplgaddtomacro\gplfronttext{%
      \csname LTb\endcsname%
      \put(176,2739){\rotatebox{-270}{\makebox(0,0){\strut{}\Large $\langle \varphi(x) \rangle$ \normalsize}}}%
      \put(3874,154){\makebox(0,0){\strut{}\Large $x$ \normalsize}}%
    }%
    \gplbacktext
    \put(0,0){\includegraphics{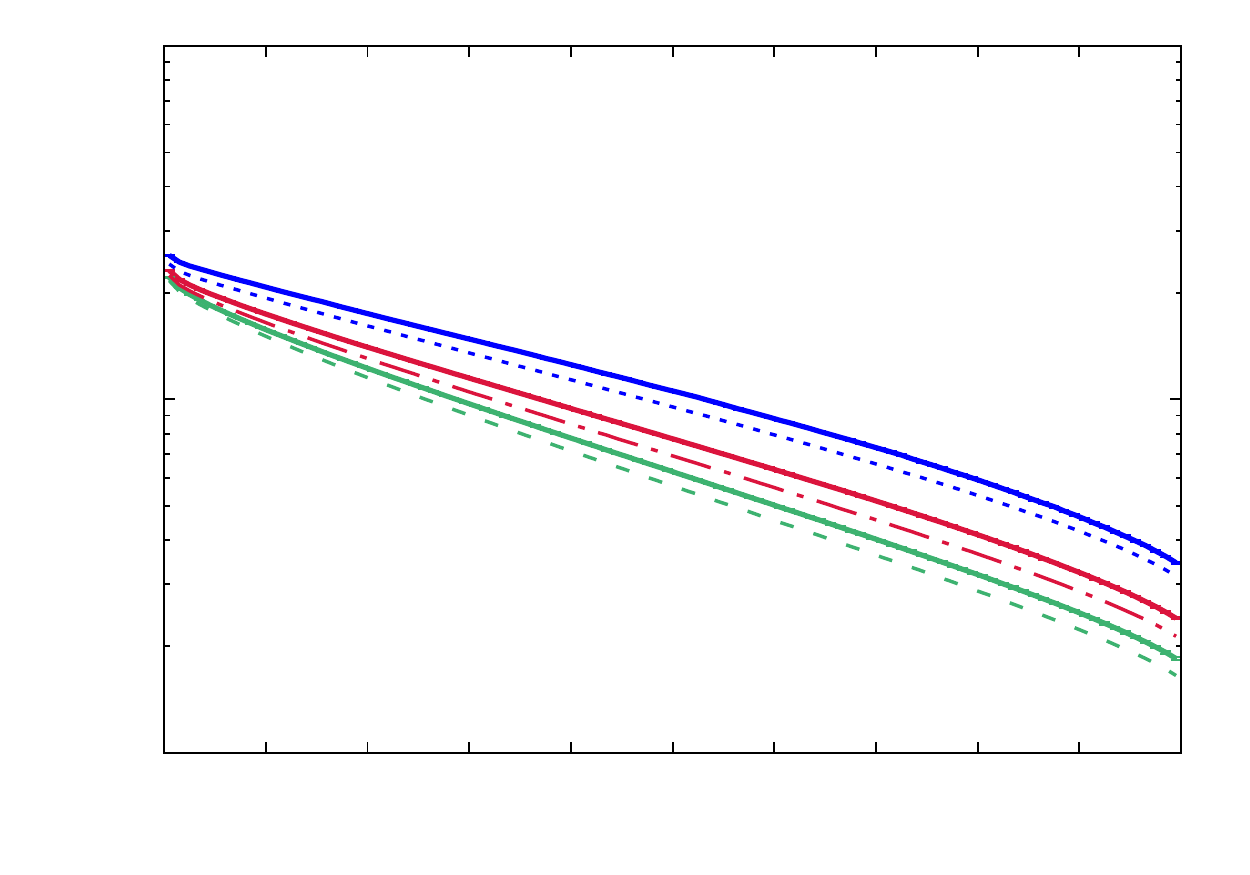}}%
    \gplfronttext
  \end{picture}%
\endgroup

%% file: flux_2B_II.tex
% GNUPLOT: LaTeX picture with Postscript
\begingroup
  \makeatletter
  \providecommand\color[2][]{%
    \GenericError{(gnuplot) \space\space\space\@spaces}{%
      Package color not loaded in conjunction with
      terminal option `colourtext'%
    }{See the gnuplot documentation for explanation.%
    }{Either use 'blacktext' in gnuplot or load the package
      color.sty in LaTeX.}%
    \renewcommand\color[2][]{}%
  }%
  \providecommand\includegraphics[2][]{%
    \GenericError{(gnuplot) \space\space\space\@spaces}{%
      Package graphicx or graphics not loaded%
    }{See the gnuplot documentation for explanation.%
    }{The gnuplot epslatex terminal needs graphicx.sty or graphics.sty.}%
    \renewcommand\includegraphics[2][]{}%
  }%
  \providecommand\rotatebox[2]{#2}%
  \@ifundefined{ifGPcolor}{%
    \newif\ifGPcolor
    \GPcolorfalse
  }{}%
  \@ifundefined{ifGPblacktext}{%
    \newif\ifGPblacktext
    \GPblacktexttrue
  }{}%
  % define a \g@addto@macro without @ in the name:
  \let\gplgaddtomacro\g@addto@macro
  % define empty templates for all commands taking text:
  \gdef\gplbacktext{}%
  \gdef\gplfronttext{}%
  \makeatother
  \ifGPblacktext
    % no textcolor at all
    \def\colorrgb#1{}%
    \def\colorgray#1{}%
  \else
    % gray or color?
    \ifGPcolor
      \def\colorrgb#1{\color[rgb]{#1}}%
      \def\colorgray#1{\color[gray]{#1}}%
      \expandafter\def\csname LTw\endcsname{\color{white}}%
      \expandafter\def\csname LTb\endcsname{\color{black}}%
      \expandafter\def\csname LTa\endcsname{\color{black}}%
      \expandafter\def\csname LT0\endcsname{\color[rgb]{1,0,0}}%
      \expandafter\def\csname LT1\endcsname{\color[rgb]{0,1,0}}%
      \expandafter\def\csname LT2\endcsname{\color[rgb]{0,0,1}}%
      \expandafter\def\csname LT3\endcsname{\color[rgb]{1,0,1}}%
      \expandafter\def\csname LT4\endcsname{\color[rgb]{0,1,1}}%
      \expandafter\def\csname LT5\endcsname{\color[rgb]{1,1,0}}%
      \expandafter\def\csname LT6\endcsname{\color[rgb]{0,0,0}}%
      \expandafter\def\csname LT7\endcsname{\color[rgb]{1,0.3,0}}%
      \expandafter\def\csname LT8\endcsname{\color[rgb]{0.5,0.5,0.5}}%
    \else
      % gray
      \def\colorrgb#1{\color{black}}%
      \def\colorgray#1{\color[gray]{#1}}%
      \expandafter\def\csname LTw\endcsname{\color{white}}%
      \expandafter\def\csname LTb\endcsname{\color{black}}%
      \expandafter\def\csname LTa\endcsname{\color{black}}%
      \expandafter\def\csname LT0\endcsname{\color{black}}%
      \expandafter\def\csname LT1\endcsname{\color{black}}%
      \expandafter\def\csname LT2\endcsname{\color{black}}%
      \expandafter\def\csname LT3\endcsname{\color{black}}%
      \expandafter\def\csname LT4\endcsname{\color{black}}%
      \expandafter\def\csname LT5\endcsname{\color{black}}%
      \expandafter\def\csname LT6\endcsname{\color{black}}%
      \expandafter\def\csname LT7\endcsname{\color{black}}%
      \expandafter\def\csname LT8\endcsname{\color{black}}%
    \fi
  \fi
    \setlength{\unitlength}{0.0500bp}%
    \ifx\gptboxheight\undefined%
      \newlength{\gptboxheight}%
      \newlength{\gptboxwidth}%
      \newsavebox{\gptboxtext}%
    \fi%
    \setlength{\fboxrule}{0.5pt}%
    \setlength{\fboxsep}{1pt}%
\begin{picture}(7200.00,5040.00)%
    \gplgaddtomacro\gplbacktext{%
      \csname LTb\endcsname%
      \put(814,704){\makebox(0,0)[r]{\strut{}$0.3$}}%
      \put(814,1111){\makebox(0,0)[r]{\strut{}$0.4$}}%
      \put(814,1518){\makebox(0,0)[r]{\strut{}$0.5$}}%
      \put(814,1925){\makebox(0,0)[r]{\strut{}$0.6$}}%
      \put(814,2332){\makebox(0,0)[r]{\strut{}$0.7$}}%
      \put(814,2739){\makebox(0,0)[r]{\strut{}$0.8$}}%
      \put(814,3147){\makebox(0,0)[r]{\strut{}$0.9$}}%
      \put(814,3554){\makebox(0,0)[r]{\strut{}$1$}}%
      \put(814,3961){\makebox(0,0)[r]{\strut{}$1.1$}}%
      \put(814,4368){\makebox(0,0)[r]{\strut{}$1.2$}}%
      \put(814,4775){\makebox(0,0)[r]{\strut{}$1.3$}}%
      \put(946,484){\makebox(0,0){\strut{}$0$}}%
      \put(1532,484){\makebox(0,0){\strut{}$1$}}%
      \put(2117,484){\makebox(0,0){\strut{}$2$}}%
      \put(2703,484){\makebox(0,0){\strut{}$3$}}%
      \put(3289,484){\makebox(0,0){\strut{}$4$}}%
      \put(3875,484){\makebox(0,0){\strut{}$5$}}%
      \put(4460,484){\makebox(0,0){\strut{}$6$}}%
      \put(5046,484){\makebox(0,0){\strut{}$7$}}%
      \put(5632,484){\makebox(0,0){\strut{}$8$}}%
      \put(6217,484){\makebox(0,0){\strut{}$9$}}%
      \put(6803,484){\makebox(0,0){\strut{}$10$}}%
    }%
    \gplgaddtomacro\gplfronttext{%
      \csname LTb\endcsname%
      \put(176,2739){\rotatebox{-270}{\makebox(0,0){\strut{}\Large $\langle \varphi(x) \rangle$ \normalsize}}}%
      \put(3874,154){\makebox(0,0){\strut{}\Large $x$ \normalsize}}%
    }%
    \gplbacktext
    \put(0,0){\includegraphics{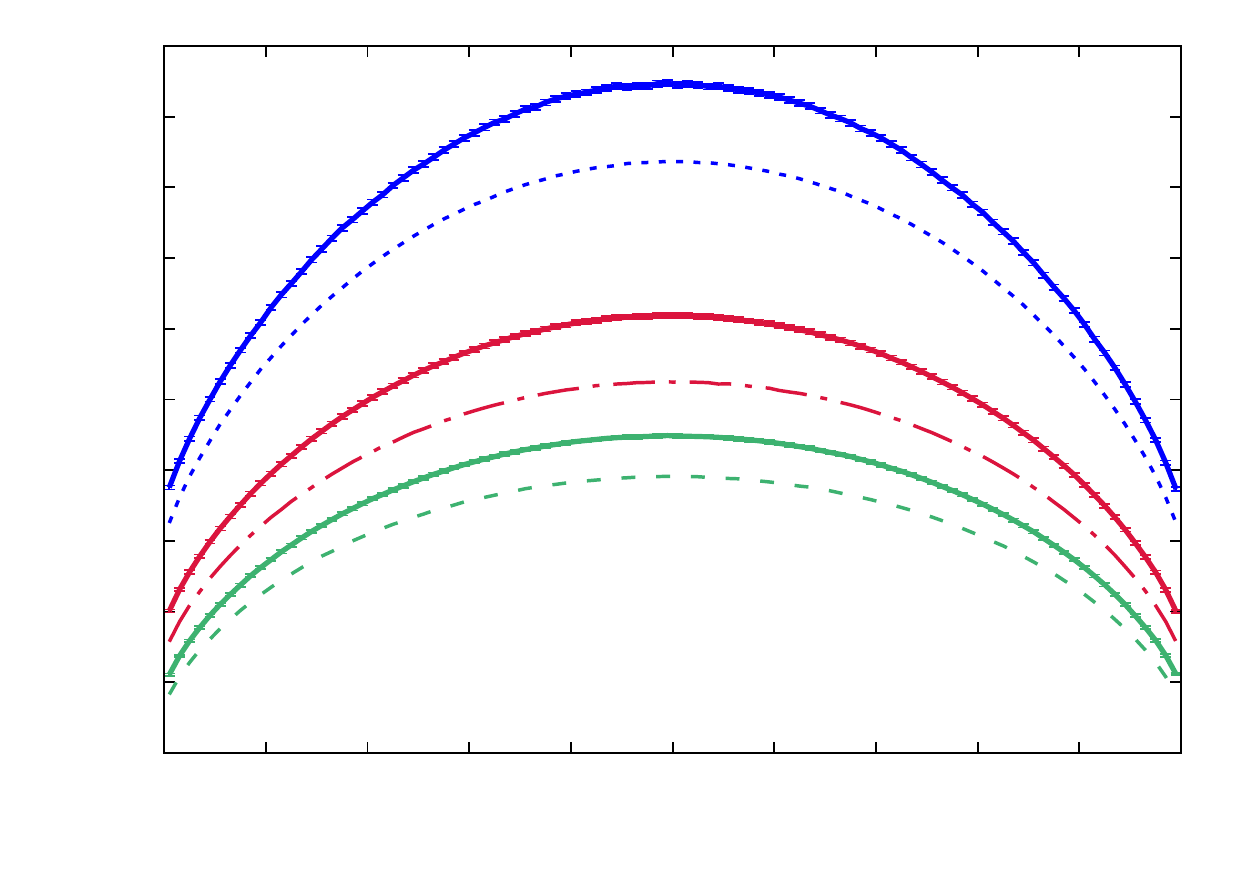}}%
    \gplfronttext
  \end{picture}%
\endgroup

%% file: flux_2C_I.tex
% GNUPLOT: LaTeX picture with Postscript
\begingroup
  \makeatletter
  \providecommand\color[2][]{%
    \GenericError{(gnuplot) \space\space\space\@spaces}{%
      Package color not loaded in conjunction with
      terminal option `colourtext'%
    }{See the gnuplot documentation for explanation.%
    }{Either use 'blacktext' in gnuplot or load the package
      color.sty in LaTeX.}%
    \renewcommand\color[2][]{}%
  }%
  \providecommand\includegraphics[2][]{%
    \GenericError{(gnuplot) \space\space\space\@spaces}{%
      Package graphicx or graphics not loaded%
    }{See the gnuplot documentation for explanation.%
    }{The gnuplot epslatex terminal needs graphicx.sty or graphics.sty.}%
    \renewcommand\includegraphics[2][]{}%
  }%
  \providecommand\rotatebox[2]{#2}%
  \@ifundefined{ifGPcolor}{%
    \newif\ifGPcolor
    \GPcolorfalse
  }{}%
  \@ifundefined{ifGPblacktext}{%
    \newif\ifGPblacktext
    \GPblacktexttrue
  }{}%
  % define a \g@addto@macro without @ in the name:
  \let\gplgaddtomacro\g@addto@macro
  % define empty templates for all commands taking text:
  \gdef\gplbacktext{}%
  \gdef\gplfronttext{}%
  \makeatother
  \ifGPblacktext
    % no textcolor at all
    \def\colorrgb#1{}%
    \def\colorgray#1{}%
  \else
    % gray or color?
    \ifGPcolor
      \def\colorrgb#1{\color[rgb]{#1}}%
      \def\colorgray#1{\color[gray]{#1}}%
      \expandafter\def\csname LTw\endcsname{\color{white}}%
      \expandafter\def\csname LTb\endcsname{\color{black}}%
      \expandafter\def\csname LTa\endcsname{\color{black}}%
      \expandafter\def\csname LT0\endcsname{\color[rgb]{1,0,0}}%
      \expandafter\def\csname LT1\endcsname{\color[rgb]{0,1,0}}%
      \expandafter\def\csname LT2\endcsname{\color[rgb]{0,0,1}}%
      \expandafter\def\csname LT3\endcsname{\color[rgb]{1,0,1}}%
      \expandafter\def\csname LT4\endcsname{\color[rgb]{0,1,1}}%
      \expandafter\def\csname LT5\endcsname{\color[rgb]{1,1,0}}%
      \expandafter\def\csname LT6\endcsname{\color[rgb]{0,0,0}}%
      \expandafter\def\csname LT7\endcsname{\color[rgb]{1,0.3,0}}%
      \expandafter\def\csname LT8\endcsname{\color[rgb]{0.5,0.5,0.5}}%
    \else
      % gray
      \def\colorrgb#1{\color{black}}%
      \def\colorgray#1{\color[gray]{#1}}%
      \expandafter\def\csname LTw\endcsname{\color{white}}%
      \expandafter\def\csname LTb\endcsname{\color{black}}%
      \expandafter\def\csname LTa\endcsname{\color{black}}%
      \expandafter\def\csname LT0\endcsname{\color{black}}%
      \expandafter\def\csname LT1\endcsname{\color{black}}%
      \expandafter\def\csname LT2\endcsname{\color{black}}%
      \expandafter\def\csname LT3\endcsname{\color{black}}%
      \expandafter\def\csname LT4\endcsname{\color{black}}%
      \expandafter\def\csname LT5\endcsname{\color{black}}%
      \expandafter\def\csname LT6\endcsname{\color{black}}%
      \expandafter\def\csname LT7\endcsname{\color{black}}%
      \expandafter\def\csname LT8\endcsname{\color{black}}%
    \fi
  \fi
    \setlength{\unitlength}{0.0500bp}%
    \ifx\gptboxheight\undefined%
      \newlength{\gptboxheight}%
      \newlength{\gptboxwidth}%
      \newsavebox{\gptboxtext}%
    \fi%
    \setlength{\fboxrule}{0.5pt}%
    \setlength{\fboxsep}{1pt}%
\begin{picture}(7200.00,5040.00)%
    \gplgaddtomacro\gplbacktext{%
      \csname LTb\endcsname%
      \put(814,704){\makebox(0,0)[r]{\strut{}$0.1$}}%
      \put(814,2740){\makebox(0,0)[r]{\strut{}$1$}}%
      \put(814,4775){\makebox(0,0)[r]{\strut{}$10$}}%
      \put(946,484){\makebox(0,0){\strut{}$0$}}%
      \put(1532,484){\makebox(0,0){\strut{}$1$}}%
      \put(2117,484){\makebox(0,0){\strut{}$2$}}%
      \put(2703,484){\makebox(0,0){\strut{}$3$}}%
      \put(3289,484){\makebox(0,0){\strut{}$4$}}%
      \put(3875,484){\makebox(0,0){\strut{}$5$}}%
      \put(4460,484){\makebox(0,0){\strut{}$6$}}%
      \put(5046,484){\makebox(0,0){\strut{}$7$}}%
      \put(5632,484){\makebox(0,0){\strut{}$8$}}%
      \put(6217,484){\makebox(0,0){\strut{}$9$}}%
      \put(6803,484){\makebox(0,0){\strut{}$10$}}%
    }%
    \gplgaddtomacro\gplfronttext{%
      \csname LTb\endcsname%
      \put(176,2739){\rotatebox{-270}{\makebox(0,0){\strut{}\Large $\langle \varphi(x) \rangle$ \normalsize}}}%
      \put(3874,154){\makebox(0,0){\strut{}\Large $x$ \normalsize}}%
    }%
    \gplbacktext
    \put(0,0){\includegraphics{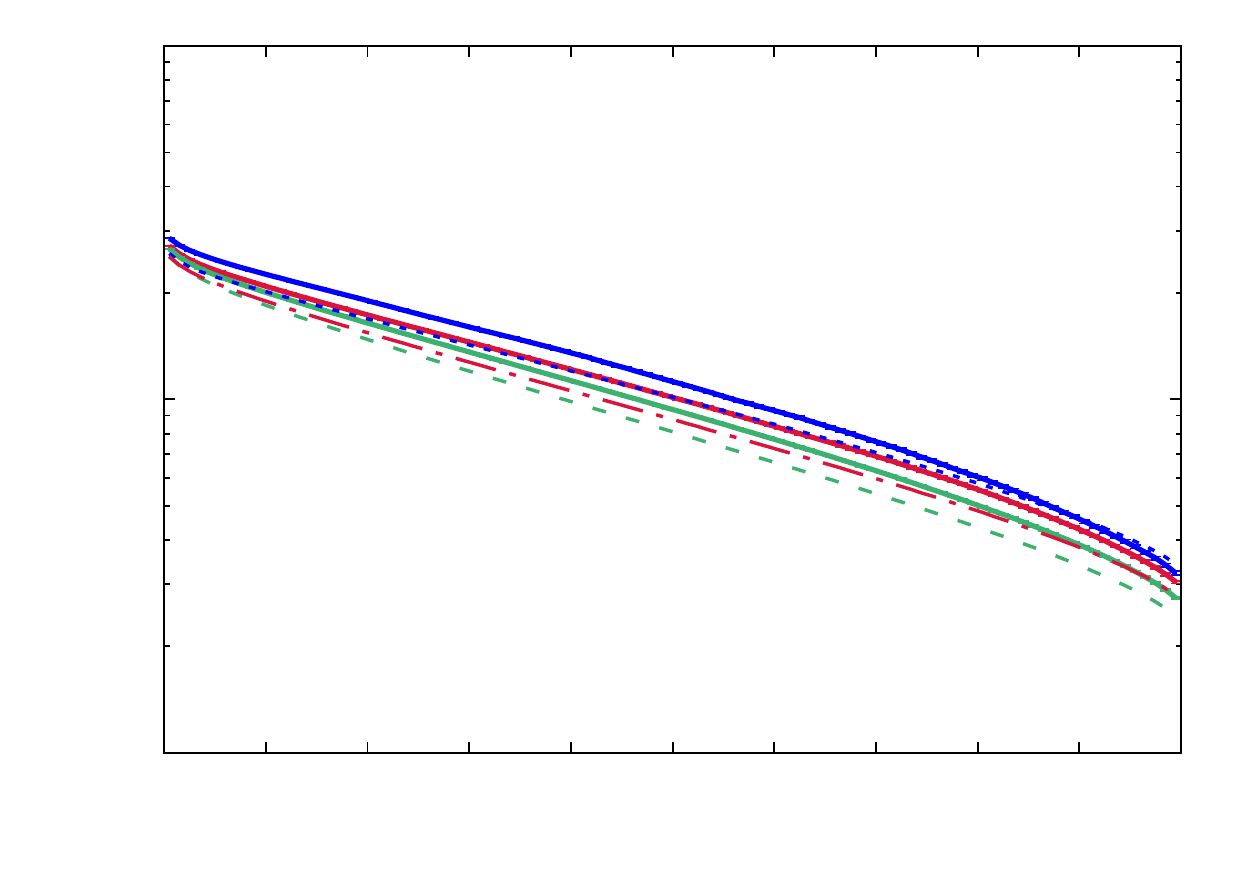}}%
    \gplfronttext
  \end{picture}%
\endgroup

%% file: flux_2C_II.tex
% GNUPLOT: LaTeX picture with Postscript
\begingroup
  \makeatletter
  \providecommand\color[2][]{%
    \GenericError{(gnuplot) \space\space\space\@spaces}{%
      Package color not loaded in conjunction with
      terminal option `colourtext'%
    }{See the gnuplot documentation for explanation.%
    }{Either use 'blacktext' in gnuplot or load the package
      color.sty in LaTeX.}%
    \renewcommand\color[2][]{}%
  }%
  \providecommand\includegraphics[2][]{%
    \GenericError{(gnuplot) \space\space\space\@spaces}{%
      Package graphicx or graphics not loaded%
    }{See the gnuplot documentation for explanation.%
    }{The gnuplot epslatex terminal needs graphicx.sty or graphics.sty.}%
    \renewcommand\includegraphics[2][]{}%
  }%
  \providecommand\rotatebox[2]{#2}%
  \@ifundefined{ifGPcolor}{%
    \newif\ifGPcolor
    \GPcolorfalse
  }{}%
  \@ifundefined{ifGPblacktext}{%
    \newif\ifGPblacktext
    \GPblacktexttrue
  }{}%
  % define a \g@addto@macro without @ in the name:
  \let\gplgaddtomacro\g@addto@macro
  % define empty templates for all commands taking text:
  \gdef\gplbacktext{}%
  \gdef\gplfronttext{}%
  \makeatother
  \ifGPblacktext
    % no textcolor at all
    \def\colorrgb#1{}%
    \def\colorgray#1{}%
  \else
    % gray or color?
    \ifGPcolor
      \def\colorrgb#1{\color[rgb]{#1}}%
      \def\colorgray#1{\color[gray]{#1}}%
      \expandafter\def\csname LTw\endcsname{\color{white}}%
      \expandafter\def\csname LTb\endcsname{\color{black}}%
      \expandafter\def\csname LTa\endcsname{\color{black}}%
      \expandafter\def\csname LT0\endcsname{\color[rgb]{1,0,0}}%
      \expandafter\def\csname LT1\endcsname{\color[rgb]{0,1,0}}%
      \expandafter\def\csname LT2\endcsname{\color[rgb]{0,0,1}}%
      \expandafter\def\csname LT3\endcsname{\color[rgb]{1,0,1}}%
      \expandafter\def\csname LT4\endcsname{\color[rgb]{0,1,1}}%
      \expandafter\def\csname LT5\endcsname{\color[rgb]{1,1,0}}%
      \expandafter\def\csname LT6\endcsname{\color[rgb]{0,0,0}}%
      \expandafter\def\csname LT7\endcsname{\color[rgb]{1,0.3,0}}%
      \expandafter\def\csname LT8\endcsname{\color[rgb]{0.5,0.5,0.5}}%
    \else
      % gray
      \def\colorrgb#1{\color{black}}%
      \def\colorgray#1{\color[gray]{#1}}%
      \expandafter\def\csname LTw\endcsname{\color{white}}%
      \expandafter\def\csname LTb\endcsname{\color{black}}%
      \expandafter\def\csname LTa\endcsname{\color{black}}%
      \expandafter\def\csname LT0\endcsname{\color{black}}%
      \expandafter\def\csname LT1\endcsname{\color{black}}%
      \expandafter\def\csname LT2\endcsname{\color{black}}%
      \expandafter\def\csname LT3\endcsname{\color{black}}%
      \expandafter\def\csname LT4\endcsname{\color{black}}%
      \expandafter\def\csname LT5\endcsname{\color{black}}%
      \expandafter\def\csname LT6\endcsname{\color{black}}%
      \expandafter\def\csname LT7\endcsname{\color{black}}%
      \expandafter\def\csname LT8\endcsname{\color{black}}%
    \fi
  \fi
    \setlength{\unitlength}{0.0500bp}%
    \ifx\gptboxheight\undefined%
      \newlength{\gptboxheight}%
      \newlength{\gptboxwidth}%
      \newsavebox{\gptboxtext}%
    \fi%
    \setlength{\fboxrule}{0.5pt}%
    \setlength{\fboxsep}{1pt}%
\begin{picture}(7200.00,5040.00)%
    \gplgaddtomacro\gplbacktext{%
      \csname LTb\endcsname%
      \put(814,704){\makebox(0,0)[r]{\strut{}$0.5$}}%
      \put(814,1111){\makebox(0,0)[r]{\strut{}$0.6$}}%
      \put(814,1518){\makebox(0,0)[r]{\strut{}$0.7$}}%
      \put(814,1925){\makebox(0,0)[r]{\strut{}$0.8$}}%
      \put(814,2332){\makebox(0,0)[r]{\strut{}$0.9$}}%
      \put(814,2739){\makebox(0,0)[r]{\strut{}$1$}}%
      \put(814,3147){\makebox(0,0)[r]{\strut{}$1.1$}}%
      \put(814,3554){\makebox(0,0)[r]{\strut{}$1.2$}}%
      \put(814,3961){\makebox(0,0)[r]{\strut{}$1.3$}}%
      \put(814,4368){\makebox(0,0)[r]{\strut{}$1.4$}}%
      \put(814,4775){\makebox(0,0)[r]{\strut{}$1.5$}}%
      \put(946,484){\makebox(0,0){\strut{}$0$}}%
      \put(1532,484){\makebox(0,0){\strut{}$1$}}%
      \put(2117,484){\makebox(0,0){\strut{}$2$}}%
      \put(2703,484){\makebox(0,0){\strut{}$3$}}%
      \put(3289,484){\makebox(0,0){\strut{}$4$}}%
      \put(3875,484){\makebox(0,0){\strut{}$5$}}%
      \put(4460,484){\makebox(0,0){\strut{}$6$}}%
      \put(5046,484){\makebox(0,0){\strut{}$7$}}%
      \put(5632,484){\makebox(0,0){\strut{}$8$}}%
      \put(6217,484){\makebox(0,0){\strut{}$9$}}%
      \put(6803,484){\makebox(0,0){\strut{}$10$}}%
    }%
    \gplgaddtomacro\gplfronttext{%
      \csname LTb\endcsname%
      \put(176,2739){\rotatebox{-270}{\makebox(0,0){\strut{}\Large $\langle \varphi(x) \rangle$ \normalsize}}}%
      \put(3874,154){\makebox(0,0){\strut{}\Large $x$ \normalsize}}%
    }%
    \gplbacktext
    \put(0,0){\includegraphics{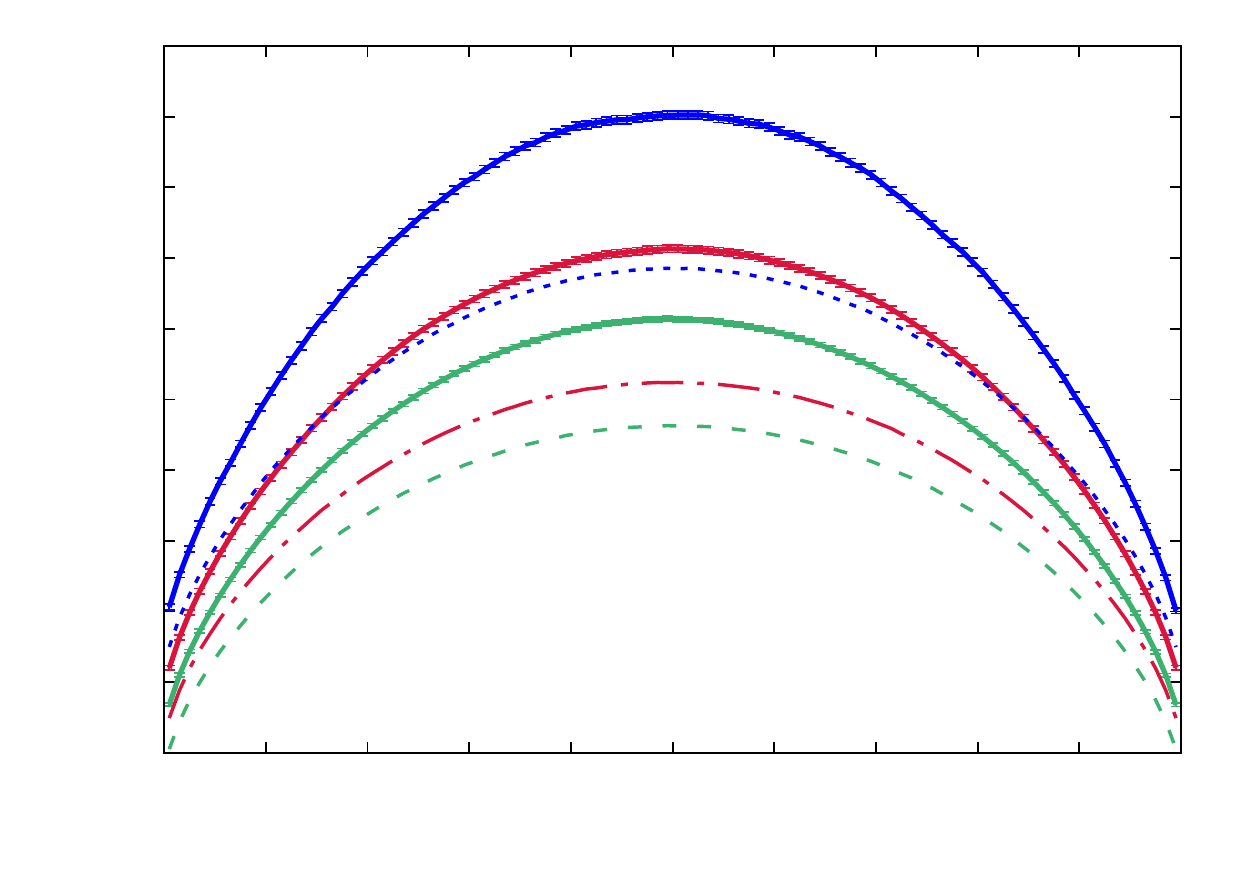}}%
    \gplfronttext
  \end{picture}%
\endgroup

%% file: flux_3A_I.tex
% GNUPLOT: LaTeX picture with Postscript
\begingroup
  \makeatletter
  \providecommand\color[2][]{%
    \GenericError{(gnuplot) \space\space\space\@spaces}{%
      Package color not loaded in conjunction with
      terminal option `colourtext'%
    }{See the gnuplot documentation for explanation.%
    }{Either use 'blacktext' in gnuplot or load the package
      color.sty in LaTeX.}%
    \renewcommand\color[2][]{}%
  }%
  \providecommand\includegraphics[2][]{%
    \GenericError{(gnuplot) \space\space\space\@spaces}{%
      Package graphicx or graphics not loaded%
    }{See the gnuplot documentation for explanation.%
    }{The gnuplot epslatex terminal needs graphicx.sty or graphics.sty.}%
    \renewcommand\includegraphics[2][]{}%
  }%
  \providecommand\rotatebox[2]{#2}%
  \@ifundefined{ifGPcolor}{%
    \newif\ifGPcolor
    \GPcolorfalse
  }{}%
  \@ifundefined{ifGPblacktext}{%
    \newif\ifGPblacktext
    \GPblacktexttrue
  }{}%
  % define a \g@addto@macro without @ in the name:
  \let\gplgaddtomacro\g@addto@macro
  % define empty templates for all commands taking text:
  \gdef\gplbacktext{}%
  \gdef\gplfronttext{}%
  \makeatother
  \ifGPblacktext
    % no textcolor at all
    \def\colorrgb#1{}%
    \def\colorgray#1{}%
  \else
    % gray or color?
    \ifGPcolor
      \def\colorrgb#1{\color[rgb]{#1}}%
      \def\colorgray#1{\color[gray]{#1}}%
      \expandafter\def\csname LTw\endcsname{\color{white}}%
      \expandafter\def\csname LTb\endcsname{\color{black}}%
      \expandafter\def\csname LTa\endcsname{\color{black}}%
      \expandafter\def\csname LT0\endcsname{\color[rgb]{1,0,0}}%
      \expandafter\def\csname LT1\endcsname{\color[rgb]{0,1,0}}%
      \expandafter\def\csname LT2\endcsname{\color[rgb]{0,0,1}}%
      \expandafter\def\csname LT3\endcsname{\color[rgb]{1,0,1}}%
      \expandafter\def\csname LT4\endcsname{\color[rgb]{0,1,1}}%
      \expandafter\def\csname LT5\endcsname{\color[rgb]{1,1,0}}%
      \expandafter\def\csname LT6\endcsname{\color[rgb]{0,0,0}}%
      \expandafter\def\csname LT7\endcsname{\color[rgb]{1,0.3,0}}%
      \expandafter\def\csname LT8\endcsname{\color[rgb]{0.5,0.5,0.5}}%
    \else
      % gray
      \def\colorrgb#1{\color{black}}%
      \def\colorgray#1{\color[gray]{#1}}%
      \expandafter\def\csname LTw\endcsname{\color{white}}%
      \expandafter\def\csname LTb\endcsname{\color{black}}%
      \expandafter\def\csname LTa\endcsname{\color{black}}%
      \expandafter\def\csname LT0\endcsname{\color{black}}%
      \expandafter\def\csname LT1\endcsname{\color{black}}%
      \expandafter\def\csname LT2\endcsname{\color{black}}%
      \expandafter\def\csname LT3\endcsname{\color{black}}%
      \expandafter\def\csname LT4\endcsname{\color{black}}%
      \expandafter\def\csname LT5\endcsname{\color{black}}%
      \expandafter\def\csname LT6\endcsname{\color{black}}%
      \expandafter\def\csname LT7\endcsname{\color{black}}%
      \expandafter\def\csname LT8\endcsname{\color{black}}%
    \fi
  \fi
    \setlength{\unitlength}{0.0500bp}%
    \ifx\gptboxheight\undefined%
      \newlength{\gptboxheight}%
      \newlength{\gptboxwidth}%
      \newsavebox{\gptboxtext}%
    \fi%
    \setlength{\fboxrule}{0.5pt}%
    \setlength{\fboxsep}{1pt}%
\begin{picture}(7200.00,5040.00)%
    \gplgaddtomacro\gplbacktext{%
      \csname LTb\endcsname%
      \put(814,704){\makebox(0,0)[r]{\strut{}$0.1$}}%
      \put(814,2740){\makebox(0,0)[r]{\strut{}$1$}}%
      \put(814,4775){\makebox(0,0)[r]{\strut{}$10$}}%
      \put(946,484){\makebox(0,0){\strut{}$0$}}%
      \put(1532,484){\makebox(0,0){\strut{}$1$}}%
      \put(2117,484){\makebox(0,0){\strut{}$2$}}%
      \put(2703,484){\makebox(0,0){\strut{}$3$}}%
      \put(3289,484){\makebox(0,0){\strut{}$4$}}%
      \put(3875,484){\makebox(0,0){\strut{}$5$}}%
      \put(4460,484){\makebox(0,0){\strut{}$6$}}%
      \put(5046,484){\makebox(0,0){\strut{}$7$}}%
      \put(5632,484){\makebox(0,0){\strut{}$8$}}%
      \put(6217,484){\makebox(0,0){\strut{}$9$}}%
      \put(6803,484){\makebox(0,0){\strut{}$10$}}%
    }%
    \gplgaddtomacro\gplfronttext{%
      \csname LTb\endcsname%
      \put(176,2739){\rotatebox{-270}{\makebox(0,0){\strut{}\Large $\langle \varphi(x) \rangle$ \normalsize}}}%
      \put(3874,154){\makebox(0,0){\strut{}\Large $x$ \normalsize}}%
    }%
    \gplbacktext
    \put(0,0){\includegraphics{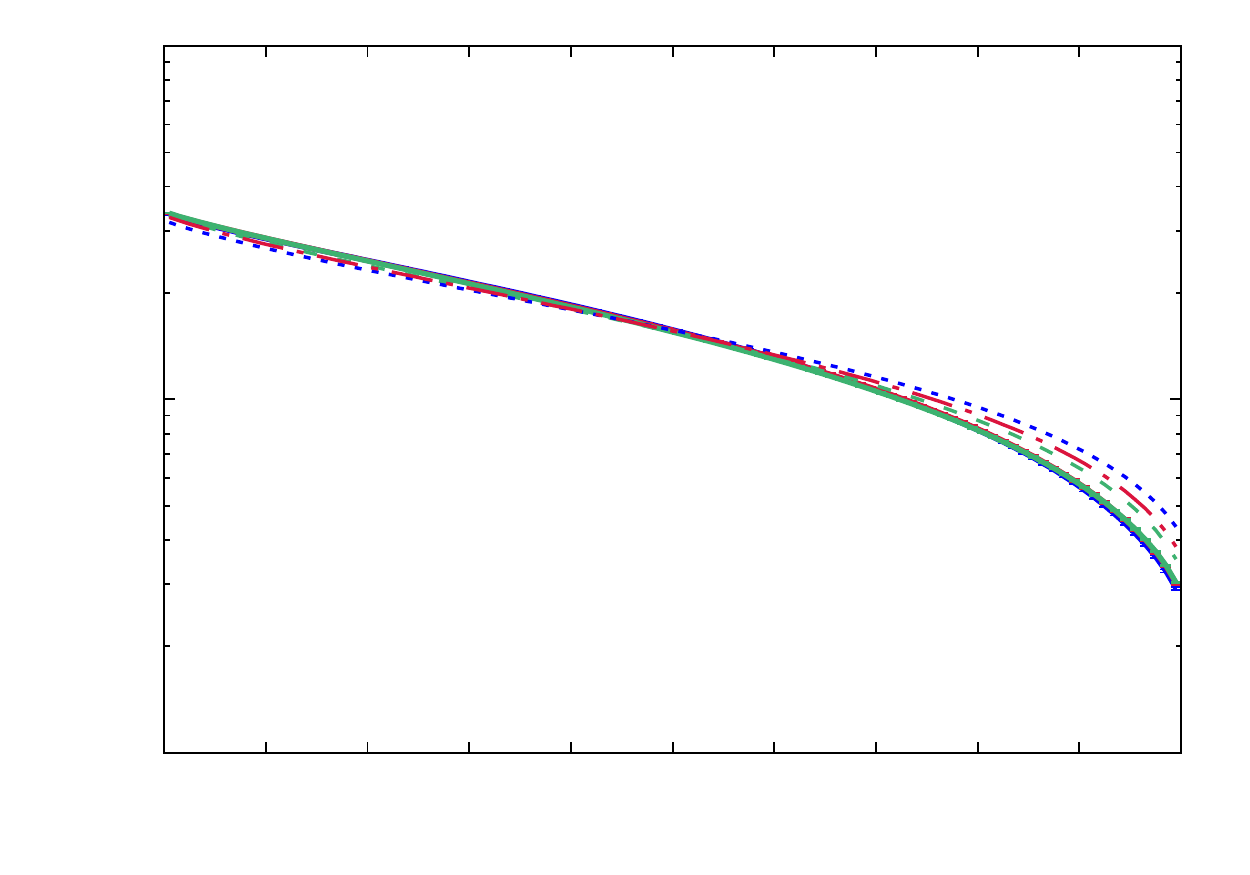}}%
    \gplfronttext
  \end{picture}%
\endgroup

%% file: flux_3A_II.tex
% GNUPLOT: LaTeX picture with Postscript
\begingroup
  \makeatletter
  \providecommand\color[2][]{%
    \GenericError{(gnuplot) \space\space\space\@spaces}{%
      Package color not loaded in conjunction with
      terminal option `colourtext'%
    }{See the gnuplot documentation for explanation.%
    }{Either use 'blacktext' in gnuplot or load the package
      color.sty in LaTeX.}%
    \renewcommand\color[2][]{}%
  }%
  \providecommand\includegraphics[2][]{%
    \GenericError{(gnuplot) \space\space\space\@spaces}{%
      Package graphicx or graphics not loaded%
    }{See the gnuplot documentation for explanation.%
    }{The gnuplot epslatex terminal needs graphicx.sty or graphics.sty.}%
    \renewcommand\includegraphics[2][]{}%
  }%
  \providecommand\rotatebox[2]{#2}%
  \@ifundefined{ifGPcolor}{%
    \newif\ifGPcolor
    \GPcolorfalse
  }{}%
  \@ifundefined{ifGPblacktext}{%
    \newif\ifGPblacktext
    \GPblacktexttrue
  }{}%
  % define a \g@addto@macro without @ in the name:
  \let\gplgaddtomacro\g@addto@macro
  % define empty templates for all commands taking text:
  \gdef\gplbacktext{}%
  \gdef\gplfronttext{}%
  \makeatother
  \ifGPblacktext
    % no textcolor at all
    \def\colorrgb#1{}%
    \def\colorgray#1{}%
  \else
    % gray or color?
    \ifGPcolor
      \def\colorrgb#1{\color[rgb]{#1}}%
      \def\colorgray#1{\color[gray]{#1}}%
      \expandafter\def\csname LTw\endcsname{\color{white}}%
      \expandafter\def\csname LTb\endcsname{\color{black}}%
      \expandafter\def\csname LTa\endcsname{\color{black}}%
      \expandafter\def\csname LT0\endcsname{\color[rgb]{1,0,0}}%
      \expandafter\def\csname LT1\endcsname{\color[rgb]{0,1,0}}%
      \expandafter\def\csname LT2\endcsname{\color[rgb]{0,0,1}}%
      \expandafter\def\csname LT3\endcsname{\color[rgb]{1,0,1}}%
      \expandafter\def\csname LT4\endcsname{\color[rgb]{0,1,1}}%
      \expandafter\def\csname LT5\endcsname{\color[rgb]{1,1,0}}%
      \expandafter\def\csname LT6\endcsname{\color[rgb]{0,0,0}}%
      \expandafter\def\csname LT7\endcsname{\color[rgb]{1,0.3,0}}%
      \expandafter\def\csname LT8\endcsname{\color[rgb]{0.5,0.5,0.5}}%
    \else
      % gray
      \def\colorrgb#1{\color{black}}%
      \def\colorgray#1{\color[gray]{#1}}%
      \expandafter\def\csname LTw\endcsname{\color{white}}%
      \expandafter\def\csname LTb\endcsname{\color{black}}%
      \expandafter\def\csname LTa\endcsname{\color{black}}%
      \expandafter\def\csname LT0\endcsname{\color{black}}%
      \expandafter\def\csname LT1\endcsname{\color{black}}%
      \expandafter\def\csname LT2\endcsname{\color{black}}%
      \expandafter\def\csname LT3\endcsname{\color{black}}%
      \expandafter\def\csname LT4\endcsname{\color{black}}%
      \expandafter\def\csname LT5\endcsname{\color{black}}%
      \expandafter\def\csname LT6\endcsname{\color{black}}%
      \expandafter\def\csname LT7\endcsname{\color{black}}%
      \expandafter\def\csname LT8\endcsname{\color{black}}%
    \fi
  \fi
    \setlength{\unitlength}{0.0500bp}%
    \ifx\gptboxheight\undefined%
      \newlength{\gptboxheight}%
      \newlength{\gptboxwidth}%
      \newsavebox{\gptboxtext}%
    \fi%
    \setlength{\fboxrule}{0.5pt}%
    \setlength{\fboxsep}{1pt}%
\begin{picture}(7200.00,5040.00)%
    \gplgaddtomacro\gplbacktext{%
      \csname LTb\endcsname%
      \put(814,704){\makebox(0,0)[r]{\strut{}$0.5$}}%
      \put(814,1286){\makebox(0,0)[r]{\strut{}$1$}}%
      \put(814,1867){\makebox(0,0)[r]{\strut{}$1.5$}}%
      \put(814,2449){\makebox(0,0)[r]{\strut{}$2$}}%
      \put(814,3030){\makebox(0,0)[r]{\strut{}$2.5$}}%
      \put(814,3612){\makebox(0,0)[r]{\strut{}$3$}}%
      \put(814,4193){\makebox(0,0)[r]{\strut{}$3.5$}}%
      \put(814,4775){\makebox(0,0)[r]{\strut{}$4$}}%
      \put(946,484){\makebox(0,0){\strut{}$0$}}%
      \put(1532,484){\makebox(0,0){\strut{}$1$}}%
      \put(2117,484){\makebox(0,0){\strut{}$2$}}%
      \put(2703,484){\makebox(0,0){\strut{}$3$}}%
      \put(3289,484){\makebox(0,0){\strut{}$4$}}%
      \put(3875,484){\makebox(0,0){\strut{}$5$}}%
      \put(4460,484){\makebox(0,0){\strut{}$6$}}%
      \put(5046,484){\makebox(0,0){\strut{}$7$}}%
      \put(5632,484){\makebox(0,0){\strut{}$8$}}%
      \put(6217,484){\makebox(0,0){\strut{}$9$}}%
      \put(6803,484){\makebox(0,0){\strut{}$10$}}%
    }%
    \gplgaddtomacro\gplfronttext{%
      \csname LTb\endcsname%
      \put(176,2739){\rotatebox{-270}{\makebox(0,0){\strut{}\Large $\langle \varphi(x) \rangle$ \normalsize}}}%
      \put(3874,154){\makebox(0,0){\strut{}\Large $x$ \normalsize}}%
    }%
    \gplbacktext
    \put(0,0){\includegraphics{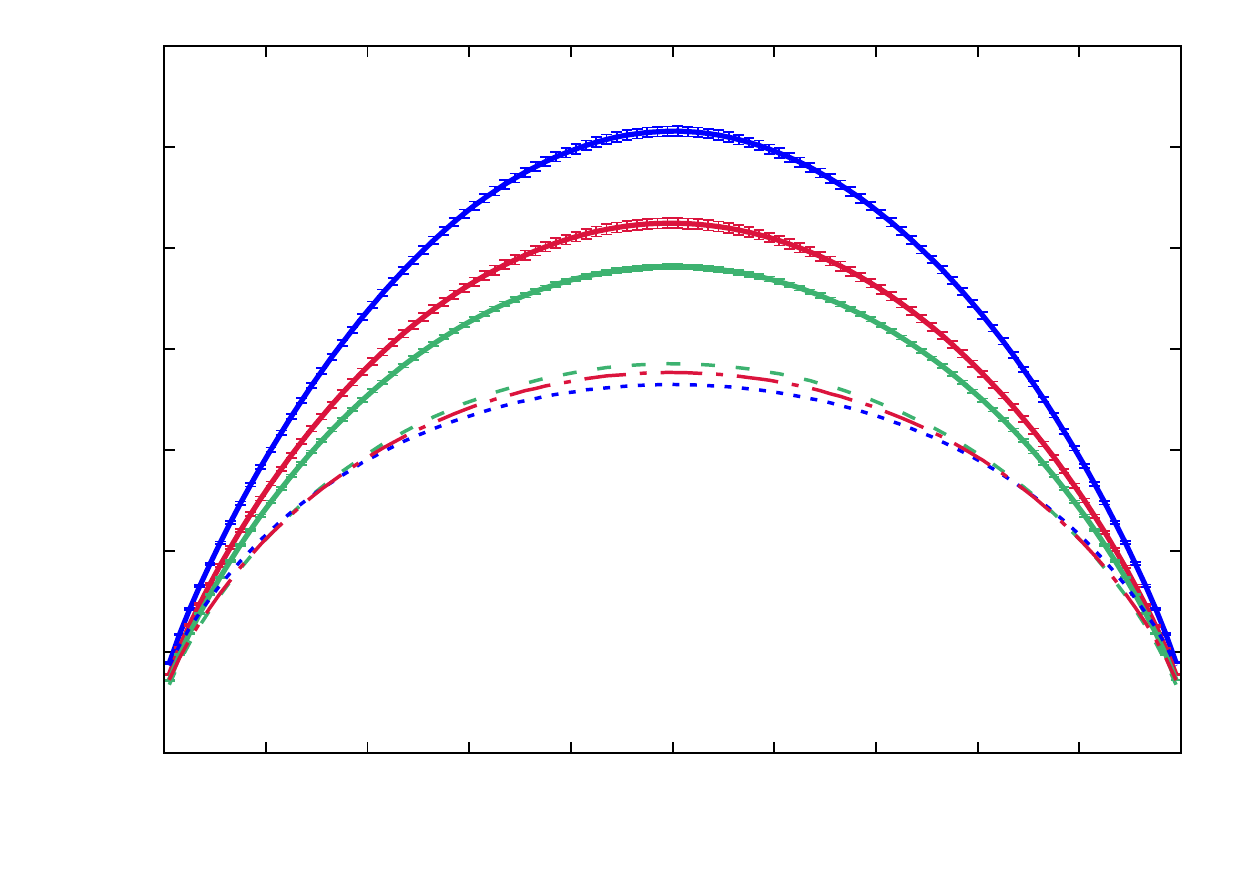}}%
    \gplfronttext
  \end{picture}%
\endgroup

%% file: flux_3B_I.tex
% GNUPLOT: LaTeX picture with Postscript
\begingroup
  \makeatletter
  \providecommand\color[2][]{%
    \GenericError{(gnuplot) \space\space\space\@spaces}{%
      Package color not loaded in conjunction with
      terminal option `colourtext'%
    }{See the gnuplot documentation for explanation.%
    }{Either use 'blacktext' in gnuplot or load the package
      color.sty in LaTeX.}%
    \renewcommand\color[2][]{}%
  }%
  \providecommand\includegraphics[2][]{%
    \GenericError{(gnuplot) \space\space\space\@spaces}{%
      Package graphicx or graphics not loaded%
    }{See the gnuplot documentation for explanation.%
    }{The gnuplot epslatex terminal needs graphicx.sty or graphics.sty.}%
    \renewcommand\includegraphics[2][]{}%
  }%
  \providecommand\rotatebox[2]{#2}%
  \@ifundefined{ifGPcolor}{%
    \newif\ifGPcolor
    \GPcolorfalse
  }{}%
  \@ifundefined{ifGPblacktext}{%
    \newif\ifGPblacktext
    \GPblacktexttrue
  }{}%
  % define a \g@addto@macro without @ in the name:
  \let\gplgaddtomacro\g@addto@macro
  % define empty templates for all commands taking text:
  \gdef\gplbacktext{}%
  \gdef\gplfronttext{}%
  \makeatother
  \ifGPblacktext
    % no textcolor at all
    \def\colorrgb#1{}%
    \def\colorgray#1{}%
  \else
    % gray or color?
    \ifGPcolor
      \def\colorrgb#1{\color[rgb]{#1}}%
      \def\colorgray#1{\color[gray]{#1}}%
      \expandafter\def\csname LTw\endcsname{\color{white}}%
      \expandafter\def\csname LTb\endcsname{\color{black}}%
      \expandafter\def\csname LTa\endcsname{\color{black}}%
      \expandafter\def\csname LT0\endcsname{\color[rgb]{1,0,0}}%
      \expandafter\def\csname LT1\endcsname{\color[rgb]{0,1,0}}%
      \expandafter\def\csname LT2\endcsname{\color[rgb]{0,0,1}}%
      \expandafter\def\csname LT3\endcsname{\color[rgb]{1,0,1}}%
      \expandafter\def\csname LT4\endcsname{\color[rgb]{0,1,1}}%
      \expandafter\def\csname LT5\endcsname{\color[rgb]{1,1,0}}%
      \expandafter\def\csname LT6\endcsname{\color[rgb]{0,0,0}}%
      \expandafter\def\csname LT7\endcsname{\color[rgb]{1,0.3,0}}%
      \expandafter\def\csname LT8\endcsname{\color[rgb]{0.5,0.5,0.5}}%
    \else
      % gray
      \def\colorrgb#1{\color{black}}%
      \def\colorgray#1{\color[gray]{#1}}%
      \expandafter\def\csname LTw\endcsname{\color{white}}%
      \expandafter\def\csname LTb\endcsname{\color{black}}%
      \expandafter\def\csname LTa\endcsname{\color{black}}%
      \expandafter\def\csname LT0\endcsname{\color{black}}%
      \expandafter\def\csname LT1\endcsname{\color{black}}%
      \expandafter\def\csname LT2\endcsname{\color{black}}%
      \expandafter\def\csname LT3\endcsname{\color{black}}%
      \expandafter\def\csname LT4\endcsname{\color{black}}%
      \expandafter\def\csname LT5\endcsname{\color{black}}%
      \expandafter\def\csname LT6\endcsname{\color{black}}%
      \expandafter\def\csname LT7\endcsname{\color{black}}%
      \expandafter\def\csname LT8\endcsname{\color{black}}%
    \fi
  \fi
    \setlength{\unitlength}{0.0500bp}%
    \ifx\gptboxheight\undefined%
      \newlength{\gptboxheight}%
      \newlength{\gptboxwidth}%
      \newsavebox{\gptboxtext}%
    \fi%
    \setlength{\fboxrule}{0.5pt}%
    \setlength{\fboxsep}{1pt}%
\begin{picture}(7200.00,5040.00)%
    \gplgaddtomacro\gplbacktext{%
      \csname LTb\endcsname%
      \put(946,704){\makebox(0,0)[r]{\strut{}$0.01$}}%
      \put(946,2061){\makebox(0,0)[r]{\strut{}$0.1$}}%
      \put(946,3418){\makebox(0,0)[r]{\strut{}$1$}}%
      \put(946,4775){\makebox(0,0)[r]{\strut{}$10$}}%
      \put(1078,484){\makebox(0,0){\strut{}$0$}}%
      \put(1651,484){\makebox(0,0){\strut{}$1$}}%
      \put(2223,484){\makebox(0,0){\strut{}$2$}}%
      \put(2796,484){\makebox(0,0){\strut{}$3$}}%
      \put(3368,484){\makebox(0,0){\strut{}$4$}}%
      \put(3941,484){\makebox(0,0){\strut{}$5$}}%
      \put(4513,484){\makebox(0,0){\strut{}$6$}}%
      \put(5086,484){\makebox(0,0){\strut{}$7$}}%
      \put(5658,484){\makebox(0,0){\strut{}$8$}}%
      \put(6231,484){\makebox(0,0){\strut{}$9$}}%
      \put(6803,484){\makebox(0,0){\strut{}$10$}}%
    }%
    \gplgaddtomacro\gplfronttext{%
      \csname LTb\endcsname%
      \put(176,2739){\rotatebox{-270}{\makebox(0,0){\strut{}\Large $\langle \varphi(x) \rangle$ \normalsize}}}%
      \put(3940,154){\makebox(0,0){\strut{}\Large $x$ \normalsize}}%
    }%
    \gplbacktext
    \put(0,0){\includegraphics{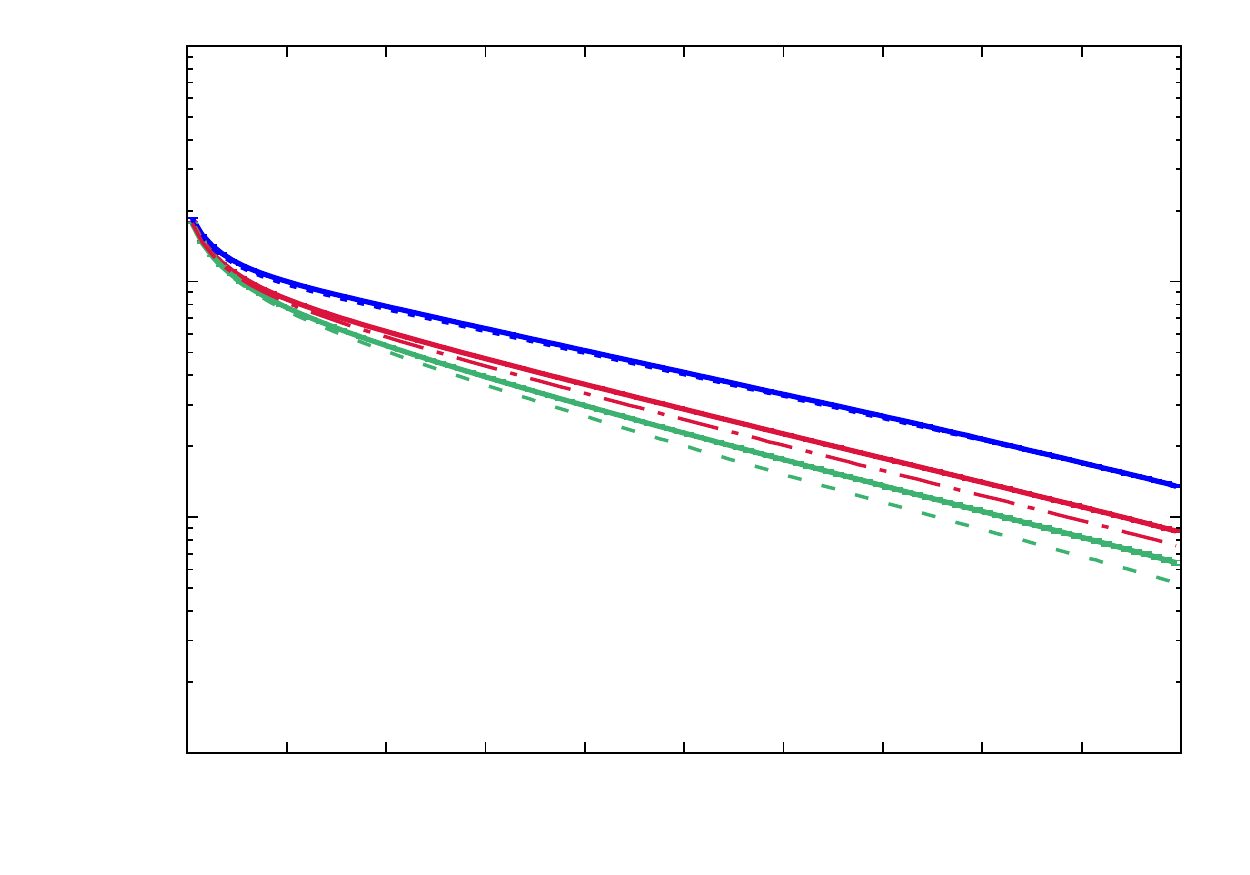}}%
    \gplfronttext
  \end{picture}%
\endgroup

%% file: flux_3B_II.tex
% GNUPLOT: LaTeX picture with Postscript
\begingroup
  \makeatletter
  \providecommand\color[2][]{%
    \GenericError{(gnuplot) \space\space\space\@spaces}{%
      Package color not loaded in conjunction with
      terminal option `colourtext'%
    }{See the gnuplot documentation for explanation.%
    }{Either use 'blacktext' in gnuplot or load the package
      color.sty in LaTeX.}%
    \renewcommand\color[2][]{}%
  }%
  \providecommand\includegraphics[2][]{%
    \GenericError{(gnuplot) \space\space\space\@spaces}{%
      Package graphicx or graphics not loaded%
    }{See the gnuplot documentation for explanation.%
    }{The gnuplot epslatex terminal needs graphicx.sty or graphics.sty.}%
    \renewcommand\includegraphics[2][]{}%
  }%
  \providecommand\rotatebox[2]{#2}%
  \@ifundefined{ifGPcolor}{%
    \newif\ifGPcolor
    \GPcolorfalse
  }{}%
  \@ifundefined{ifGPblacktext}{%
    \newif\ifGPblacktext
    \GPblacktexttrue
  }{}%
  % define a \g@addto@macro without @ in the name:
  \let\gplgaddtomacro\g@addto@macro
  % define empty templates for all commands taking text:
  \gdef\gplbacktext{}%
  \gdef\gplfronttext{}%
  \makeatother
  \ifGPblacktext
    % no textcolor at all
    \def\colorrgb#1{}%
    \def\colorgray#1{}%
  \else
    % gray or color?
    \ifGPcolor
      \def\colorrgb#1{\color[rgb]{#1}}%
      \def\colorgray#1{\color[gray]{#1}}%
      \expandafter\def\csname LTw\endcsname{\color{white}}%
      \expandafter\def\csname LTb\endcsname{\color{black}}%
      \expandafter\def\csname LTa\endcsname{\color{black}}%
      \expandafter\def\csname LT0\endcsname{\color[rgb]{1,0,0}}%
      \expandafter\def\csname LT1\endcsname{\color[rgb]{0,1,0}}%
      \expandafter\def\csname LT2\endcsname{\color[rgb]{0,0,1}}%
      \expandafter\def\csname LT3\endcsname{\color[rgb]{1,0,1}}%
      \expandafter\def\csname LT4\endcsname{\color[rgb]{0,1,1}}%
      \expandafter\def\csname LT5\endcsname{\color[rgb]{1,1,0}}%
      \expandafter\def\csname LT6\endcsname{\color[rgb]{0,0,0}}%
      \expandafter\def\csname LT7\endcsname{\color[rgb]{1,0.3,0}}%
      \expandafter\def\csname LT8\endcsname{\color[rgb]{0.5,0.5,0.5}}%
    \else
      % gray
      \def\colorrgb#1{\color{black}}%
      \def\colorgray#1{\color[gray]{#1}}%
      \expandafter\def\csname LTw\endcsname{\color{white}}%
      \expandafter\def\csname LTb\endcsname{\color{black}}%
      \expandafter\def\csname LTa\endcsname{\color{black}}%
      \expandafter\def\csname LT0\endcsname{\color{black}}%
      \expandafter\def\csname LT1\endcsname{\color{black}}%
      \expandafter\def\csname LT2\endcsname{\color{black}}%
      \expandafter\def\csname LT3\endcsname{\color{black}}%
      \expandafter\def\csname LT4\endcsname{\color{black}}%
      \expandafter\def\csname LT5\endcsname{\color{black}}%
      \expandafter\def\csname LT6\endcsname{\color{black}}%
      \expandafter\def\csname LT7\endcsname{\color{black}}%
      \expandafter\def\csname LT8\endcsname{\color{black}}%
    \fi
  \fi
    \setlength{\unitlength}{0.0500bp}%
    \ifx\gptboxheight\undefined%
      \newlength{\gptboxheight}%
      \newlength{\gptboxwidth}%
      \newsavebox{\gptboxtext}%
    \fi%
    \setlength{\fboxrule}{0.5pt}%
    \setlength{\fboxsep}{1pt}%
\begin{picture}(7200.00,5040.00)%
    \gplgaddtomacro\gplbacktext{%
      \csname LTb\endcsname%
      \put(814,704){\makebox(0,0)[r]{\strut{}$0.1$}}%
      \put(814,1383){\makebox(0,0)[r]{\strut{}$0.2$}}%
      \put(814,2061){\makebox(0,0)[r]{\strut{}$0.3$}}%
      \put(814,2740){\makebox(0,0)[r]{\strut{}$0.4$}}%
      \put(814,3418){\makebox(0,0)[r]{\strut{}$0.5$}}%
      \put(814,4097){\makebox(0,0)[r]{\strut{}$0.6$}}%
      \put(814,4775){\makebox(0,0)[r]{\strut{}$0.7$}}%
      \put(946,484){\makebox(0,0){\strut{}$0$}}%
      \put(1532,484){\makebox(0,0){\strut{}$1$}}%
      \put(2117,484){\makebox(0,0){\strut{}$2$}}%
      \put(2703,484){\makebox(0,0){\strut{}$3$}}%
      \put(3289,484){\makebox(0,0){\strut{}$4$}}%
      \put(3875,484){\makebox(0,0){\strut{}$5$}}%
      \put(4460,484){\makebox(0,0){\strut{}$6$}}%
      \put(5046,484){\makebox(0,0){\strut{}$7$}}%
      \put(5632,484){\makebox(0,0){\strut{}$8$}}%
      \put(6217,484){\makebox(0,0){\strut{}$9$}}%
      \put(6803,484){\makebox(0,0){\strut{}$10$}}%
    }%
    \gplgaddtomacro\gplfronttext{%
      \csname LTb\endcsname%
      \put(176,2739){\rotatebox{-270}{\makebox(0,0){\strut{}\Large $\langle \varphi(x) \rangle$ \normalsize}}}%
      \put(3874,154){\makebox(0,0){\strut{}\Large $x$ \normalsize}}%
    }%
    \gplbacktext
    \put(0,0){\includegraphics{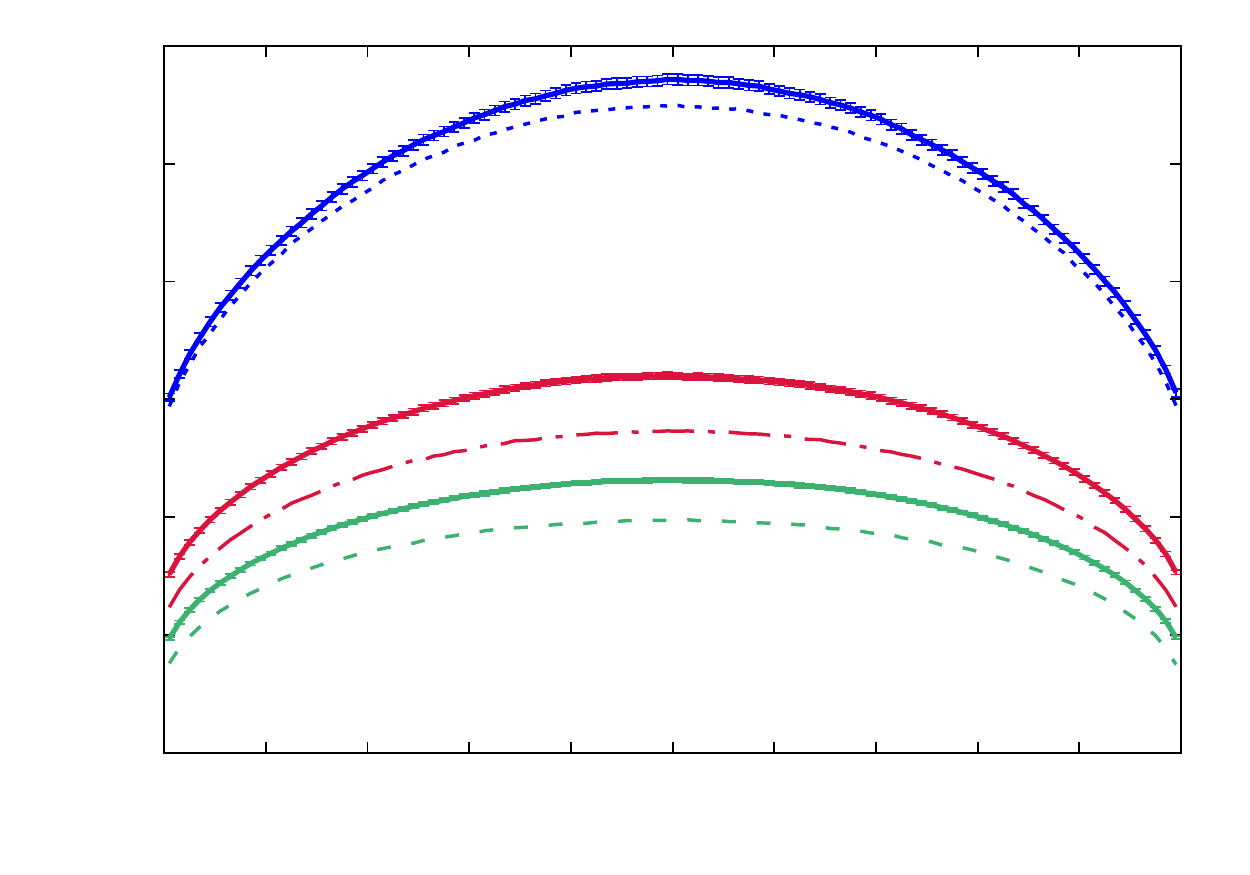}}%
    \gplfronttext
  \end{picture}%
\endgroup

%% file: flux_3C_I.tex
% GNUPLOT: LaTeX picture with Postscript
\begingroup
  \makeatletter
  \providecommand\color[2][]{%
    \GenericError{(gnuplot) \space\space\space\@spaces}{%
      Package color not loaded in conjunction with
      terminal option `colourtext'%
    }{See the gnuplot documentation for explanation.%
    }{Either use 'blacktext' in gnuplot or load the package
      color.sty in LaTeX.}%
    \renewcommand\color[2][]{}%
  }%
  \providecommand\includegraphics[2][]{%
    \GenericError{(gnuplot) \space\space\space\@spaces}{%
      Package graphicx or graphics not loaded%
    }{See the gnuplot documentation for explanation.%
    }{The gnuplot epslatex terminal needs graphicx.sty or graphics.sty.}%
    \renewcommand\includegraphics[2][]{}%
  }%
  \providecommand\rotatebox[2]{#2}%
  \@ifundefined{ifGPcolor}{%
    \newif\ifGPcolor
    \GPcolorfalse
  }{}%
  \@ifundefined{ifGPblacktext}{%
    \newif\ifGPblacktext
    \GPblacktexttrue
  }{}%
  % define a \g@addto@macro without @ in the name:
  \let\gplgaddtomacro\g@addto@macro
  % define empty templates for all commands taking text:
  \gdef\gplbacktext{}%
  \gdef\gplfronttext{}%
  \makeatother
  \ifGPblacktext
    % no textcolor at all
    \def\colorrgb#1{}%
    \def\colorgray#1{}%
  \else
    % gray or color?
    \ifGPcolor
      \def\colorrgb#1{\color[rgb]{#1}}%
      \def\colorgray#1{\color[gray]{#1}}%
      \expandafter\def\csname LTw\endcsname{\color{white}}%
      \expandafter\def\csname LTb\endcsname{\color{black}}%
      \expandafter\def\csname LTa\endcsname{\color{black}}%
      \expandafter\def\csname LT0\endcsname{\color[rgb]{1,0,0}}%
      \expandafter\def\csname LT1\endcsname{\color[rgb]{0,1,0}}%
      \expandafter\def\csname LT2\endcsname{\color[rgb]{0,0,1}}%
      \expandafter\def\csname LT3\endcsname{\color[rgb]{1,0,1}}%
      \expandafter\def\csname LT4\endcsname{\color[rgb]{0,1,1}}%
      \expandafter\def\csname LT5\endcsname{\color[rgb]{1,1,0}}%
      \expandafter\def\csname LT6\endcsname{\color[rgb]{0,0,0}}%
      \expandafter\def\csname LT7\endcsname{\color[rgb]{1,0.3,0}}%
      \expandafter\def\csname LT8\endcsname{\color[rgb]{0.5,0.5,0.5}}%
    \else
      % gray
      \def\colorrgb#1{\color{black}}%
      \def\colorgray#1{\color[gray]{#1}}%
      \expandafter\def\csname LTw\endcsname{\color{white}}%
      \expandafter\def\csname LTb\endcsname{\color{black}}%
      \expandafter\def\csname LTa\endcsname{\color{black}}%
      \expandafter\def\csname LT0\endcsname{\color{black}}%
      \expandafter\def\csname LT1\endcsname{\color{black}}%
      \expandafter\def\csname LT2\endcsname{\color{black}}%
      \expandafter\def\csname LT3\endcsname{\color{black}}%
      \expandafter\def\csname LT4\endcsname{\color{black}}%
      \expandafter\def\csname LT5\endcsname{\color{black}}%
      \expandafter\def\csname LT6\endcsname{\color{black}}%
      \expandafter\def\csname LT7\endcsname{\color{black}}%
      \expandafter\def\csname LT8\endcsname{\color{black}}%
    \fi
  \fi
    \setlength{\unitlength}{0.0500bp}%
    \ifx\gptboxheight\undefined%
      \newlength{\gptboxheight}%
      \newlength{\gptboxwidth}%
      \newsavebox{\gptboxtext}%
    \fi%
    \setlength{\fboxrule}{0.5pt}%
    \setlength{\fboxsep}{1pt}%
\begin{picture}(7200.00,5040.00)%
    \gplgaddtomacro\gplbacktext{%
      \csname LTb\endcsname%
      \put(814,704){\makebox(0,0)[r]{\strut{}$0.1$}}%
      \put(814,2740){\makebox(0,0)[r]{\strut{}$1$}}%
      \put(814,4775){\makebox(0,0)[r]{\strut{}$10$}}%
      \put(946,484){\makebox(0,0){\strut{}$0$}}%
      \put(1532,484){\makebox(0,0){\strut{}$1$}}%
      \put(2117,484){\makebox(0,0){\strut{}$2$}}%
      \put(2703,484){\makebox(0,0){\strut{}$3$}}%
      \put(3289,484){\makebox(0,0){\strut{}$4$}}%
      \put(3875,484){\makebox(0,0){\strut{}$5$}}%
      \put(4460,484){\makebox(0,0){\strut{}$6$}}%
      \put(5046,484){\makebox(0,0){\strut{}$7$}}%
      \put(5632,484){\makebox(0,0){\strut{}$8$}}%
      \put(6217,484){\makebox(0,0){\strut{}$9$}}%
      \put(6803,484){\makebox(0,0){\strut{}$10$}}%
    }%
    \gplgaddtomacro\gplfronttext{%
      \csname LTb\endcsname%
      \put(176,2739){\rotatebox{-270}{\makebox(0,0){\strut{}\Large $\langle \varphi(x) \rangle$ \normalsize}}}%
      \put(3874,154){\makebox(0,0){\strut{}\Large $x$ \normalsize}}%
    }%
    \gplbacktext
    \put(0,0){\includegraphics{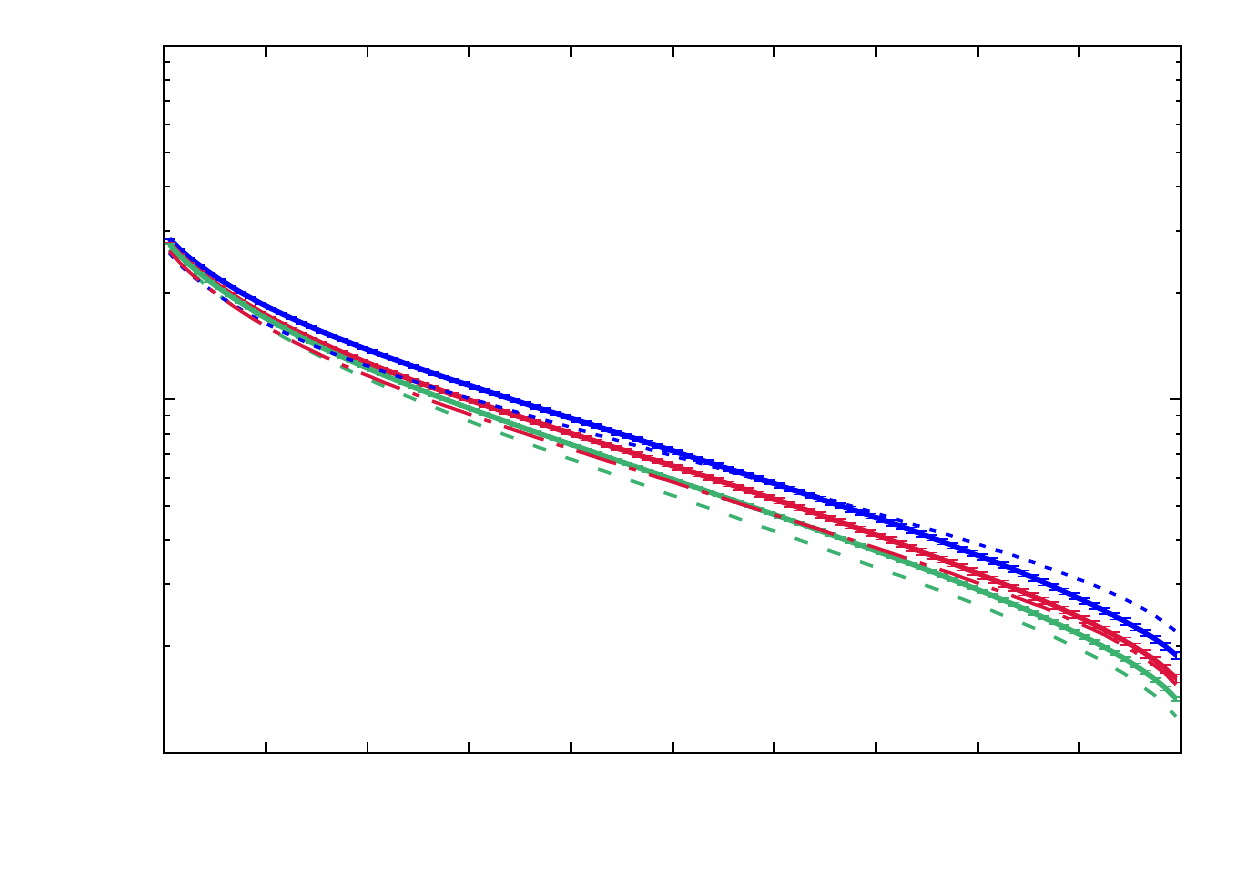}}%
    \gplfronttext
  \end{picture}%
\endgroup

%% file: flux_3C_II.tex
% GNUPLOT: LaTeX picture with Postscript
\begingroup
  \makeatletter
  \providecommand\color[2][]{%
    \GenericError{(gnuplot) \space\space\space\@spaces}{%
      Package color not loaded in conjunction with
      terminal option `colourtext'%
    }{See the gnuplot documentation for explanation.%
    }{Either use 'blacktext' in gnuplot or load the package
      color.sty in LaTeX.}%
    \renewcommand\color[2][]{}%
  }%
  \providecommand\includegraphics[2][]{%
    \GenericError{(gnuplot) \space\space\space\@spaces}{%
      Package graphicx or graphics not loaded%
    }{See the gnuplot documentation for explanation.%
    }{The gnuplot epslatex terminal needs graphicx.sty or graphics.sty.}%
    \renewcommand\includegraphics[2][]{}%
  }%
  \providecommand\rotatebox[2]{#2}%
  \@ifundefined{ifGPcolor}{%
    \newif\ifGPcolor
    \GPcolorfalse
  }{}%
  \@ifundefined{ifGPblacktext}{%
    \newif\ifGPblacktext
    \GPblacktexttrue
  }{}%
  % define a \g@addto@macro without @ in the name:
  \let\gplgaddtomacro\g@addto@macro
  % define empty templates for all commands taking text:
  \gdef\gplbacktext{}%
  \gdef\gplfronttext{}%
  \makeatother
  \ifGPblacktext
    % no textcolor at all
    \def\colorrgb#1{}%
    \def\colorgray#1{}%
  \else
    % gray or color?
    \ifGPcolor
      \def\colorrgb#1{\color[rgb]{#1}}%
      \def\colorgray#1{\color[gray]{#1}}%
      \expandafter\def\csname LTw\endcsname{\color{white}}%
      \expandafter\def\csname LTb\endcsname{\color{black}}%
      \expandafter\def\csname LTa\endcsname{\color{black}}%
      \expandafter\def\csname LT0\endcsname{\color[rgb]{1,0,0}}%
      \expandafter\def\csname LT1\endcsname{\color[rgb]{0,1,0}}%
      \expandafter\def\csname LT2\endcsname{\color[rgb]{0,0,1}}%
      \expandafter\def\csname LT3\endcsname{\color[rgb]{1,0,1}}%
      \expandafter\def\csname LT4\endcsname{\color[rgb]{0,1,1}}%
      \expandafter\def\csname LT5\endcsname{\color[rgb]{1,1,0}}%
      \expandafter\def\csname LT6\endcsname{\color[rgb]{0,0,0}}%
      \expandafter\def\csname LT7\endcsname{\color[rgb]{1,0.3,0}}%
      \expandafter\def\csname LT8\endcsname{\color[rgb]{0.5,0.5,0.5}}%
    \else
      % gray
      \def\colorrgb#1{\color{black}}%
      \def\colorgray#1{\color[gray]{#1}}%
      \expandafter\def\csname LTw\endcsname{\color{white}}%
      \expandafter\def\csname LTb\endcsname{\color{black}}%
      \expandafter\def\csname LTa\endcsname{\color{black}}%
      \expandafter\def\csname LT0\endcsname{\color{black}}%
      \expandafter\def\csname LT1\endcsname{\color{black}}%
      \expandafter\def\csname LT2\endcsname{\color{black}}%
      \expandafter\def\csname LT3\endcsname{\color{black}}%
      \expandafter\def\csname LT4\endcsname{\color{black}}%
      \expandafter\def\csname LT5\endcsname{\color{black}}%
      \expandafter\def\csname LT6\endcsname{\color{black}}%
      \expandafter\def\csname LT7\endcsname{\color{black}}%
      \expandafter\def\csname LT8\endcsname{\color{black}}%
    \fi
  \fi
    \setlength{\unitlength}{0.0500bp}%
    \ifx\gptboxheight\undefined%
      \newlength{\gptboxheight}%
      \newlength{\gptboxwidth}%
      \newsavebox{\gptboxtext}%
    \fi%
    \setlength{\fboxrule}{0.5pt}%
    \setlength{\fboxsep}{1pt}%
\begin{picture}(7200.00,5040.00)%
    \gplgaddtomacro\gplbacktext{%
      \csname LTb\endcsname%
      \put(814,704){\makebox(0,0)[r]{\strut{}$0.3$}}%
      \put(814,1111){\makebox(0,0)[r]{\strut{}$0.4$}}%
      \put(814,1518){\makebox(0,0)[r]{\strut{}$0.5$}}%
      \put(814,1925){\makebox(0,0)[r]{\strut{}$0.6$}}%
      \put(814,2332){\makebox(0,0)[r]{\strut{}$0.7$}}%
      \put(814,2739){\makebox(0,0)[r]{\strut{}$0.8$}}%
      \put(814,3147){\makebox(0,0)[r]{\strut{}$0.9$}}%
      \put(814,3554){\makebox(0,0)[r]{\strut{}$1$}}%
      \put(814,3961){\makebox(0,0)[r]{\strut{}$1.1$}}%
      \put(814,4368){\makebox(0,0)[r]{\strut{}$1.2$}}%
      \put(814,4775){\makebox(0,0)[r]{\strut{}$1.3$}}%
      \put(946,484){\makebox(0,0){\strut{}$0$}}%
      \put(1532,484){\makebox(0,0){\strut{}$1$}}%
      \put(2117,484){\makebox(0,0){\strut{}$2$}}%
      \put(2703,484){\makebox(0,0){\strut{}$3$}}%
      \put(3289,484){\makebox(0,0){\strut{}$4$}}%
      \put(3875,484){\makebox(0,0){\strut{}$5$}}%
      \put(4460,484){\makebox(0,0){\strut{}$6$}}%
      \put(5046,484){\makebox(0,0){\strut{}$7$}}%
      \put(5632,484){\makebox(0,0){\strut{}$8$}}%
      \put(6217,484){\makebox(0,0){\strut{}$9$}}%
      \put(6803,484){\makebox(0,0){\strut{}$10$}}%
    }%
    \gplgaddtomacro\gplfronttext{%
      \csname LTb\endcsname%
      \put(176,2739){\rotatebox{-270}{\makebox(0,0){\strut{}\Large $\langle \varphi(x) \rangle$ \normalsize}}}%
      \put(3874,154){\makebox(0,0){\strut{}\Large $x$ \normalsize}}%
    }%
    \gplbacktext
    \put(0,0){\includegraphics{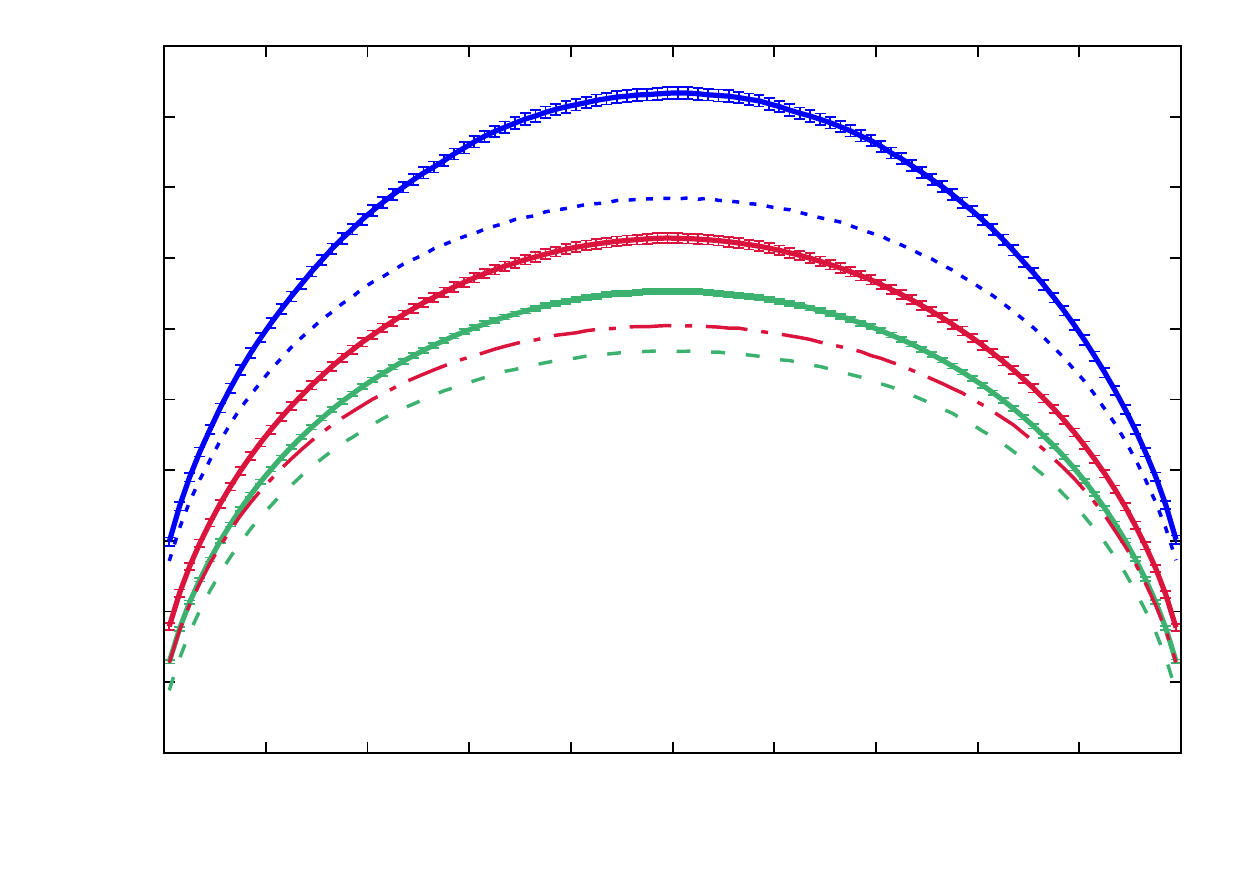}}%
    \gplfronttext
  \end{picture}%
\endgroup

%% file: flux_1A_I_ecart.tex
% GNUPLOT: LaTeX picture with Postscript
\begingroup
  \makeatletter
  \providecommand\color[2][]{%
    \GenericError{(gnuplot) \space\space\space\@spaces}{%
      Package color not loaded in conjunction with
      terminal option `colourtext'%
    }{See the gnuplot documentation for explanation.%
    }{Either use 'blacktext' in gnuplot or load the package
      color.sty in LaTeX.}%
    \renewcommand\color[2][]{}%
  }%
  \providecommand\includegraphics[2][]{%
    \GenericError{(gnuplot) \space\space\space\@spaces}{%
      Package graphicx or graphics not loaded%
    }{See the gnuplot documentation for explanation.%
    }{The gnuplot epslatex terminal needs graphicx.sty or graphics.sty.}%
    \renewcommand\includegraphics[2][]{}%
  }%
  \providecommand\rotatebox[2]{#2}%
  \@ifundefined{ifGPcolor}{%
    \newif\ifGPcolor
    \GPcolorfalse
  }{}%
  \@ifundefined{ifGPblacktext}{%
    \newif\ifGPblacktext
    \GPblacktexttrue
  }{}%
  % define a \g@addto@macro without @ in the name:
  \let\gplgaddtomacro\g@addto@macro
  % define empty templates for all commands taking text:
  \gdef\gplbacktext{}%
  \gdef\gplfronttext{}%
  \makeatother
  \ifGPblacktext
    % no textcolor at all
    \def\colorrgb#1{}%
    \def\colorgray#1{}%
  \else
    % gray or color?
    \ifGPcolor
      \def\colorrgb#1{\color[rgb]{#1}}%
      \def\colorgray#1{\color[gray]{#1}}%
      \expandafter\def\csname LTw\endcsname{\color{white}}%
      \expandafter\def\csname LTb\endcsname{\color{black}}%
      \expandafter\def\csname LTa\endcsname{\color{black}}%
      \expandafter\def\csname LT0\endcsname{\color[rgb]{1,0,0}}%
      \expandafter\def\csname LT1\endcsname{\color[rgb]{0,1,0}}%
      \expandafter\def\csname LT2\endcsname{\color[rgb]{0,0,1}}%
      \expandafter\def\csname LT3\endcsname{\color[rgb]{1,0,1}}%
      \expandafter\def\csname LT4\endcsname{\color[rgb]{0,1,1}}%
      \expandafter\def\csname LT5\endcsname{\color[rgb]{1,1,0}}%
      \expandafter\def\csname LT6\endcsname{\color[rgb]{0,0,0}}%
      \expandafter\def\csname LT7\endcsname{\color[rgb]{1,0.3,0}}%
      \expandafter\def\csname LT8\endcsname{\color[rgb]{0.5,0.5,0.5}}%
    \else
      % gray
      \def\colorrgb#1{\color{black}}%
      \def\colorgray#1{\color[gray]{#1}}%
      \expandafter\def\csname LTw\endcsname{\color{white}}%
      \expandafter\def\csname LTb\endcsname{\color{black}}%
      \expandafter\def\csname LTa\endcsname{\color{black}}%
      \expandafter\def\csname LT0\endcsname{\color{black}}%
      \expandafter\def\csname LT1\endcsname{\color{black}}%
      \expandafter\def\csname LT2\endcsname{\color{black}}%
      \expandafter\def\csname LT3\endcsname{\color{black}}%
      \expandafter\def\csname LT4\endcsname{\color{black}}%
      \expandafter\def\csname LT5\endcsname{\color{black}}%
      \expandafter\def\csname LT6\endcsname{\color{black}}%
      \expandafter\def\csname LT7\endcsname{\color{black}}%
      \expandafter\def\csname LT8\endcsname{\color{black}}%
    \fi
  \fi
    \setlength{\unitlength}{0.0500bp}%
    \ifx\gptboxheight\undefined%
      \newlength{\gptboxheight}%
      \newlength{\gptboxwidth}%
      \newsavebox{\gptboxtext}%
    \fi%
    \setlength{\fboxrule}{0.5pt}%
    \setlength{\fboxsep}{1pt}%
\begin{picture}(7200.00,5040.00)%
    \gplgaddtomacro\gplbacktext{%
      \csname LTb\endcsname%
      \put(1078,704){\makebox(0,0)[r]{\strut{}$-0.15$}}%
      \put(1078,1383){\makebox(0,0)[r]{\strut{}$-0.1$}}%
      \put(1078,2061){\makebox(0,0)[r]{\strut{}$-0.05$}}%
      \put(1078,2740){\makebox(0,0)[r]{\strut{}$0$}}%
      \put(1078,3418){\makebox(0,0)[r]{\strut{}$0.05$}}%
      \put(1078,4097){\makebox(0,0)[r]{\strut{}$0.1$}}%
      \put(1078,4775){\makebox(0,0)[r]{\strut{}$0.15$}}%
      \put(1210,484){\makebox(0,0){\strut{}$0$}}%
      \put(1769,484){\makebox(0,0){\strut{}$1$}}%
      \put(2329,484){\makebox(0,0){\strut{}$2$}}%
      \put(2888,484){\makebox(0,0){\strut{}$3$}}%
      \put(3447,484){\makebox(0,0){\strut{}$4$}}%
      \put(4007,484){\makebox(0,0){\strut{}$5$}}%
      \put(4566,484){\makebox(0,0){\strut{}$6$}}%
      \put(5125,484){\makebox(0,0){\strut{}$7$}}%
      \put(5684,484){\makebox(0,0){\strut{}$8$}}%
      \put(6244,484){\makebox(0,0){\strut{}$9$}}%
      \put(6803,484){\makebox(0,0){\strut{}$10$}}%
    }%
    \gplgaddtomacro\gplfronttext{%
      \csname LTb\endcsname%
      \put(176,2739){\rotatebox{-270}{\makebox(0,0){\strut{}\Large $\Delta [\langle \varphi(x) \rangle$]\normalsize}}}%
      \put(4006,154){\makebox(0,0){\strut{}\Large $x$ \normalsize}}%
    }%
    \gplbacktext
    \put(0,0){\includegraphics{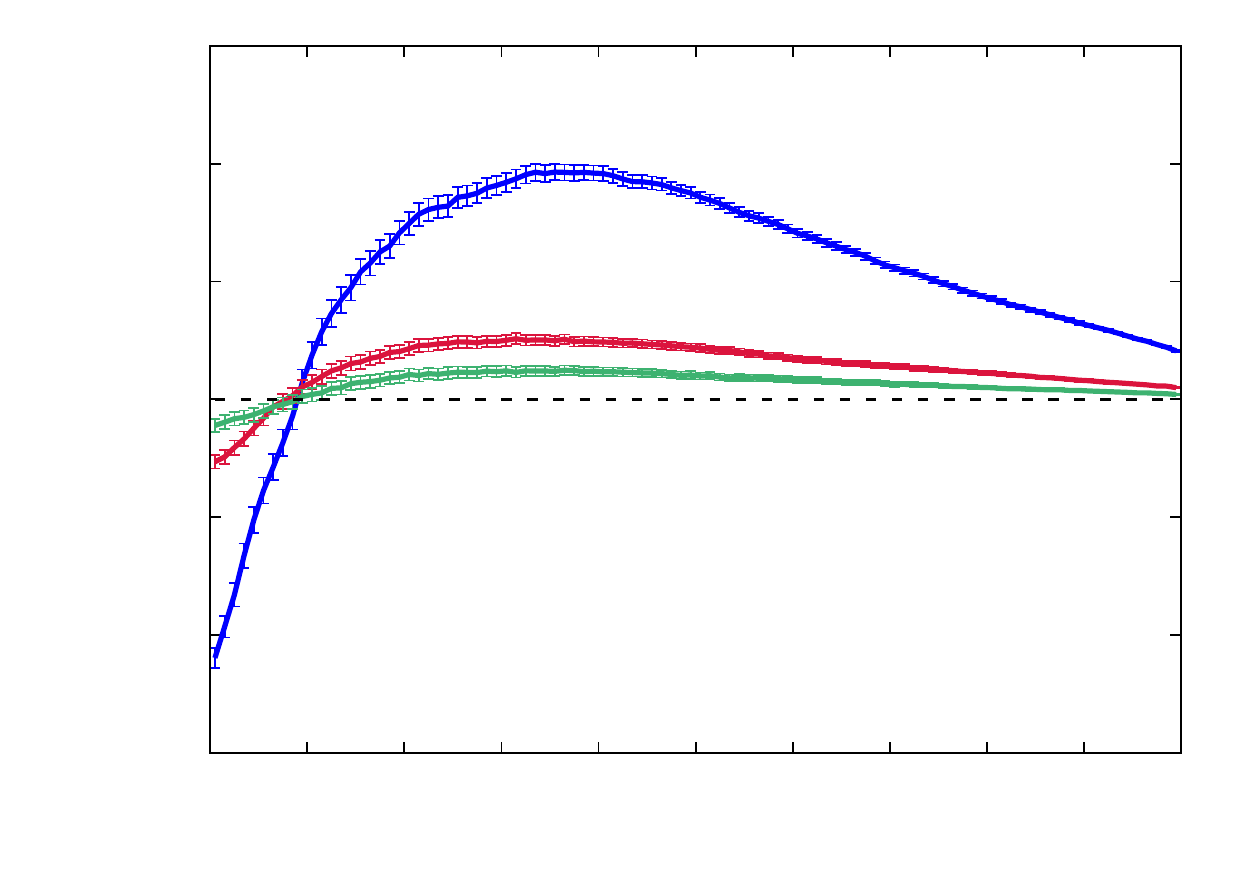}}%
    \gplfronttext
  \end{picture}%
\endgroup

%% file: flux_1A_II_ecart.tex
% GNUPLOT: LaTeX picture with Postscript
\begingroup
  \makeatletter
  \providecommand\color[2][]{%
    \GenericError{(gnuplot) \space\space\space\@spaces}{%
      Package color not loaded in conjunction with
      terminal option `colourtext'%
    }{See the gnuplot documentation for explanation.%
    }{Either use 'blacktext' in gnuplot or load the package
      color.sty in LaTeX.}%
    \renewcommand\color[2][]{}%
  }%
  \providecommand\includegraphics[2][]{%
    \GenericError{(gnuplot) \space\space\space\@spaces}{%
      Package graphicx or graphics not loaded%
    }{See the gnuplot documentation for explanation.%
    }{The gnuplot epslatex terminal needs graphicx.sty or graphics.sty.}%
    \renewcommand\includegraphics[2][]{}%
  }%
  \providecommand\rotatebox[2]{#2}%
  \@ifundefined{ifGPcolor}{%
    \newif\ifGPcolor
    \GPcolorfalse
  }{}%
  \@ifundefined{ifGPblacktext}{%
    \newif\ifGPblacktext
    \GPblacktexttrue
  }{}%
  % define a \g@addto@macro without @ in the name:
  \let\gplgaddtomacro\g@addto@macro
  % define empty templates for all commands taking text:
  \gdef\gplbacktext{}%
  \gdef\gplfronttext{}%
  \makeatother
  \ifGPblacktext
    % no textcolor at all
    \def\colorrgb#1{}%
    \def\colorgray#1{}%
  \else
    % gray or color?
    \ifGPcolor
      \def\colorrgb#1{\color[rgb]{#1}}%
      \def\colorgray#1{\color[gray]{#1}}%
      \expandafter\def\csname LTw\endcsname{\color{white}}%
      \expandafter\def\csname LTb\endcsname{\color{black}}%
      \expandafter\def\csname LTa\endcsname{\color{black}}%
      \expandafter\def\csname LT0\endcsname{\color[rgb]{1,0,0}}%
      \expandafter\def\csname LT1\endcsname{\color[rgb]{0,1,0}}%
      \expandafter\def\csname LT2\endcsname{\color[rgb]{0,0,1}}%
      \expandafter\def\csname LT3\endcsname{\color[rgb]{1,0,1}}%
      \expandafter\def\csname LT4\endcsname{\color[rgb]{0,1,1}}%
      \expandafter\def\csname LT5\endcsname{\color[rgb]{1,1,0}}%
      \expandafter\def\csname LT6\endcsname{\color[rgb]{0,0,0}}%
      \expandafter\def\csname LT7\endcsname{\color[rgb]{1,0.3,0}}%
      \expandafter\def\csname LT8\endcsname{\color[rgb]{0.5,0.5,0.5}}%
    \else
      % gray
      \def\colorrgb#1{\color{black}}%
      \def\colorgray#1{\color[gray]{#1}}%
      \expandafter\def\csname LTw\endcsname{\color{white}}%
      \expandafter\def\csname LTb\endcsname{\color{black}}%
      \expandafter\def\csname LTa\endcsname{\color{black}}%
      \expandafter\def\csname LT0\endcsname{\color{black}}%
      \expandafter\def\csname LT1\endcsname{\color{black}}%
      \expandafter\def\csname LT2\endcsname{\color{black}}%
      \expandafter\def\csname LT3\endcsname{\color{black}}%
      \expandafter\def\csname LT4\endcsname{\color{black}}%
      \expandafter\def\csname LT5\endcsname{\color{black}}%
      \expandafter\def\csname LT6\endcsname{\color{black}}%
      \expandafter\def\csname LT7\endcsname{\color{black}}%
      \expandafter\def\csname LT8\endcsname{\color{black}}%
    \fi
  \fi
    \setlength{\unitlength}{0.0500bp}%
    \ifx\gptboxheight\undefined%
      \newlength{\gptboxheight}%
      \newlength{\gptboxwidth}%
      \newsavebox{\gptboxtext}%
    \fi%
    \setlength{\fboxrule}{0.5pt}%
    \setlength{\fboxsep}{1pt}%
\begin{picture}(7200.00,5040.00)%
    \gplgaddtomacro\gplbacktext{%
      \csname LTb\endcsname%
      \put(1078,704){\makebox(0,0)[r]{\strut{}$-0.06$}}%
      \put(1078,1156){\makebox(0,0)[r]{\strut{}$-0.05$}}%
      \put(1078,1609){\makebox(0,0)[r]{\strut{}$-0.04$}}%
      \put(1078,2061){\makebox(0,0)[r]{\strut{}$-0.03$}}%
      \put(1078,2513){\makebox(0,0)[r]{\strut{}$-0.02$}}%
      \put(1078,2966){\makebox(0,0)[r]{\strut{}$-0.01$}}%
      \put(1078,3418){\makebox(0,0)[r]{\strut{}$0$}}%
      \put(1078,3870){\makebox(0,0)[r]{\strut{}$0.01$}}%
      \put(1078,4323){\makebox(0,0)[r]{\strut{}$0.02$}}%
      \put(1078,4775){\makebox(0,0)[r]{\strut{}$0.03$}}%
      \put(1210,484){\makebox(0,0){\strut{}$0$}}%
      \put(1769,484){\makebox(0,0){\strut{}$1$}}%
      \put(2329,484){\makebox(0,0){\strut{}$2$}}%
      \put(2888,484){\makebox(0,0){\strut{}$3$}}%
      \put(3447,484){\makebox(0,0){\strut{}$4$}}%
      \put(4007,484){\makebox(0,0){\strut{}$5$}}%
      \put(4566,484){\makebox(0,0){\strut{}$6$}}%
      \put(5125,484){\makebox(0,0){\strut{}$7$}}%
      \put(5684,484){\makebox(0,0){\strut{}$8$}}%
      \put(6244,484){\makebox(0,0){\strut{}$9$}}%
      \put(6803,484){\makebox(0,0){\strut{}$10$}}%
    }%
    \gplgaddtomacro\gplfronttext{%
      \csname LTb\endcsname%
      \put(176,2739){\rotatebox{-270}{\makebox(0,0){\strut{}\Large $\Delta [\langle \varphi(x) \rangle$]\normalsize}}}%
      \put(4006,154){\makebox(0,0){\strut{}\Large $x$ \normalsize}}%
    }%
    \gplbacktext
    \put(0,0){\includegraphics{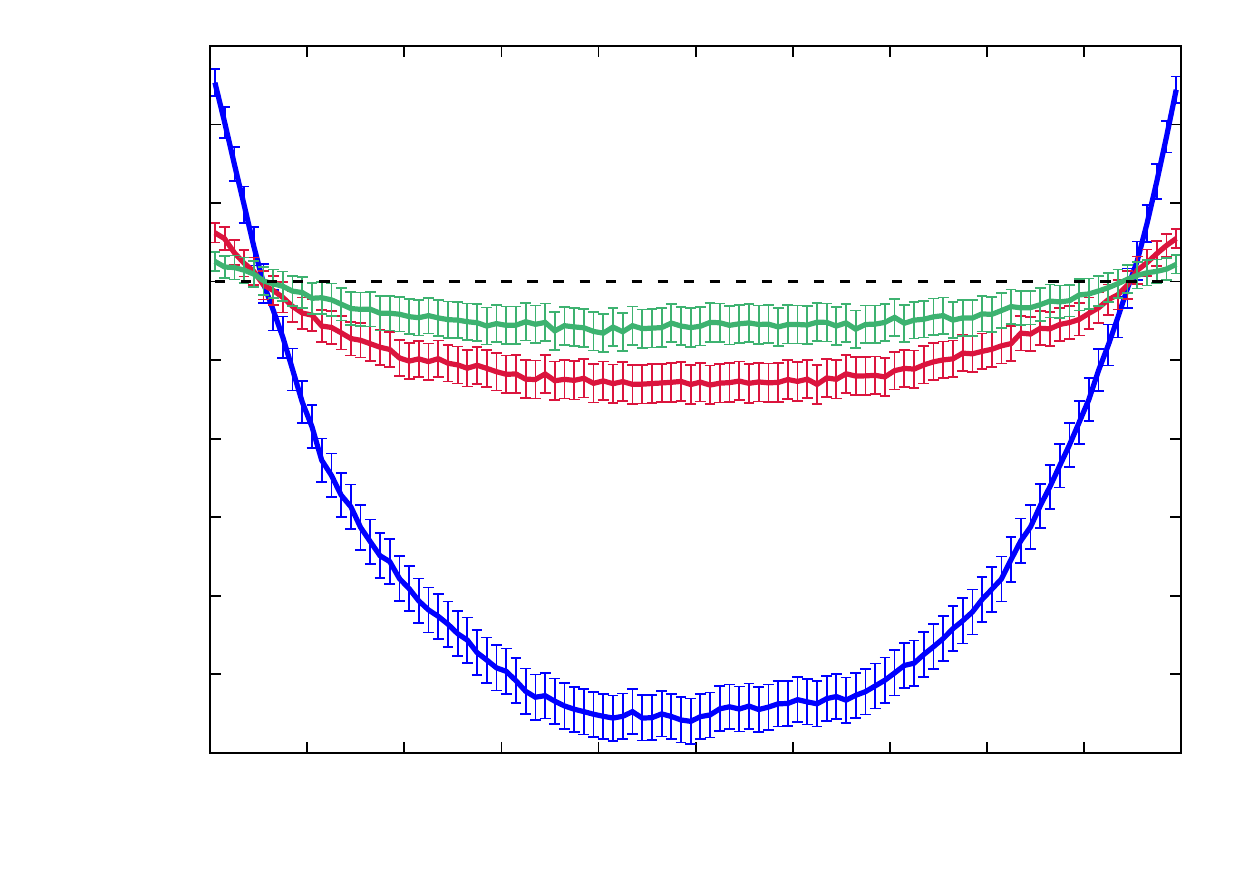}}%
    \gplfronttext
  \end{picture}%
\endgroup

%% file: flux_1B_I_ecart.tex
% GNUPLOT: LaTeX picture with Postscript
\begingroup
  \makeatletter
  \providecommand\color[2][]{%
    \GenericError{(gnuplot) \space\space\space\@spaces}{%
      Package color not loaded in conjunction with
      terminal option `colourtext'%
    }{See the gnuplot documentation for explanation.%
    }{Either use 'blacktext' in gnuplot or load the package
      color.sty in LaTeX.}%
    \renewcommand\color[2][]{}%
  }%
  \providecommand\includegraphics[2][]{%
    \GenericError{(gnuplot) \space\space\space\@spaces}{%
      Package graphicx or graphics not loaded%
    }{See the gnuplot documentation for explanation.%
    }{The gnuplot epslatex terminal needs graphicx.sty or graphics.sty.}%
    \renewcommand\includegraphics[2][]{}%
  }%
  \providecommand\rotatebox[2]{#2}%
  \@ifundefined{ifGPcolor}{%
    \newif\ifGPcolor
    \GPcolorfalse
  }{}%
  \@ifundefined{ifGPblacktext}{%
    \newif\ifGPblacktext
    \GPblacktexttrue
  }{}%
  % define a \g@addto@macro without @ in the name:
  \let\gplgaddtomacro\g@addto@macro
  % define empty templates for all commands taking text:
  \gdef\gplbacktext{}%
  \gdef\gplfronttext{}%
  \makeatother
  \ifGPblacktext
    % no textcolor at all
    \def\colorrgb#1{}%
    \def\colorgray#1{}%
  \else
    % gray or color?
    \ifGPcolor
      \def\colorrgb#1{\color[rgb]{#1}}%
      \def\colorgray#1{\color[gray]{#1}}%
      \expandafter\def\csname LTw\endcsname{\color{white}}%
      \expandafter\def\csname LTb\endcsname{\color{black}}%
      \expandafter\def\csname LTa\endcsname{\color{black}}%
      \expandafter\def\csname LT0\endcsname{\color[rgb]{1,0,0}}%
      \expandafter\def\csname LT1\endcsname{\color[rgb]{0,1,0}}%
      \expandafter\def\csname LT2\endcsname{\color[rgb]{0,0,1}}%
      \expandafter\def\csname LT3\endcsname{\color[rgb]{1,0,1}}%
      \expandafter\def\csname LT4\endcsname{\color[rgb]{0,1,1}}%
      \expandafter\def\csname LT5\endcsname{\color[rgb]{1,1,0}}%
      \expandafter\def\csname LT6\endcsname{\color[rgb]{0,0,0}}%
      \expandafter\def\csname LT7\endcsname{\color[rgb]{1,0.3,0}}%
      \expandafter\def\csname LT8\endcsname{\color[rgb]{0.5,0.5,0.5}}%
    \else
      % gray
      \def\colorrgb#1{\color{black}}%
      \def\colorgray#1{\color[gray]{#1}}%
      \expandafter\def\csname LTw\endcsname{\color{white}}%
      \expandafter\def\csname LTb\endcsname{\color{black}}%
      \expandafter\def\csname LTa\endcsname{\color{black}}%
      \expandafter\def\csname LT0\endcsname{\color{black}}%
      \expandafter\def\csname LT1\endcsname{\color{black}}%
      \expandafter\def\csname LT2\endcsname{\color{black}}%
      \expandafter\def\csname LT3\endcsname{\color{black}}%
      \expandafter\def\csname LT4\endcsname{\color{black}}%
      \expandafter\def\csname LT5\endcsname{\color{black}}%
      \expandafter\def\csname LT6\endcsname{\color{black}}%
      \expandafter\def\csname LT7\endcsname{\color{black}}%
      \expandafter\def\csname LT8\endcsname{\color{black}}%
    \fi
  \fi
    \setlength{\unitlength}{0.0500bp}%
    \ifx\gptboxheight\undefined%
      \newlength{\gptboxheight}%
      \newlength{\gptboxwidth}%
      \newsavebox{\gptboxtext}%
    \fi%
    \setlength{\fboxrule}{0.5pt}%
    \setlength{\fboxsep}{1pt}%
\begin{picture}(7200.00,5040.00)%
    \gplgaddtomacro\gplbacktext{%
      \csname LTb\endcsname%
      \put(1078,704){\makebox(0,0)[r]{\strut{}$-0.08$}}%
      \put(1078,1156){\makebox(0,0)[r]{\strut{}$-0.07$}}%
      \put(1078,1609){\makebox(0,0)[r]{\strut{}$-0.06$}}%
      \put(1078,2061){\makebox(0,0)[r]{\strut{}$-0.05$}}%
      \put(1078,2513){\makebox(0,0)[r]{\strut{}$-0.04$}}%
      \put(1078,2966){\makebox(0,0)[r]{\strut{}$-0.03$}}%
      \put(1078,3418){\makebox(0,0)[r]{\strut{}$-0.02$}}%
      \put(1078,3870){\makebox(0,0)[r]{\strut{}$-0.01$}}%
      \put(1078,4323){\makebox(0,0)[r]{\strut{}$0$}}%
      \put(1078,4775){\makebox(0,0)[r]{\strut{}$0.01$}}%
      \put(1210,484){\makebox(0,0){\strut{}$0$}}%
      \put(1769,484){\makebox(0,0){\strut{}$1$}}%
      \put(2329,484){\makebox(0,0){\strut{}$2$}}%
      \put(2888,484){\makebox(0,0){\strut{}$3$}}%
      \put(3447,484){\makebox(0,0){\strut{}$4$}}%
      \put(4007,484){\makebox(0,0){\strut{}$5$}}%
      \put(4566,484){\makebox(0,0){\strut{}$6$}}%
      \put(5125,484){\makebox(0,0){\strut{}$7$}}%
      \put(5684,484){\makebox(0,0){\strut{}$8$}}%
      \put(6244,484){\makebox(0,0){\strut{}$9$}}%
      \put(6803,484){\makebox(0,0){\strut{}$10$}}%
    }%
    \gplgaddtomacro\gplfronttext{%
      \csname LTb\endcsname%
      \put(176,2739){\rotatebox{-270}{\makebox(0,0){\strut{}\Large $\Delta [\langle \varphi(x) \rangle$]\normalsize}}}%
      \put(4006,154){\makebox(0,0){\strut{}\Large $x$ \normalsize}}%
    }%
    \gplbacktext
    \put(0,0){\includegraphics{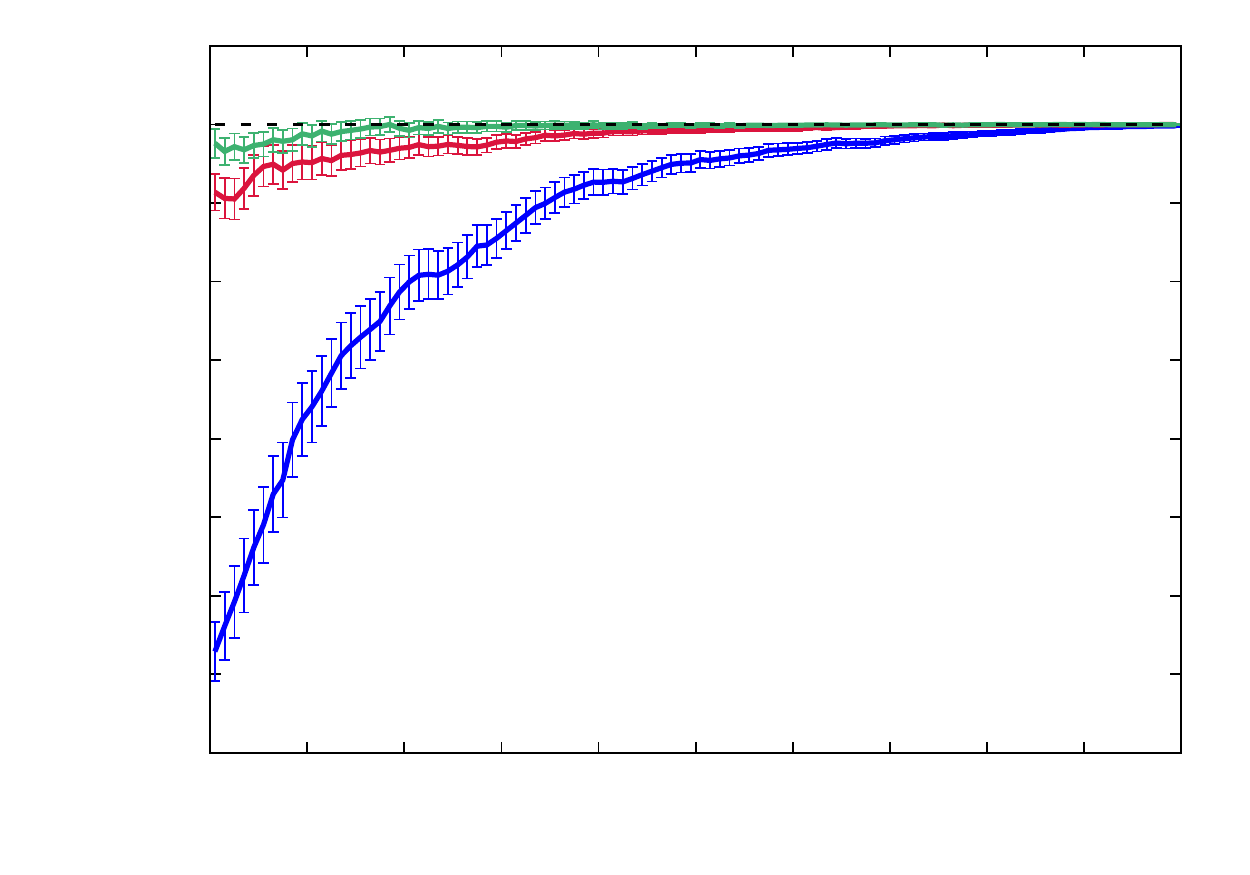}}%
    \gplfronttext
  \end{picture}%
\endgroup

%% file: flux_1B_II_ecart.tex
% GNUPLOT: LaTeX picture with Postscript
\begingroup
  \makeatletter
  \providecommand\color[2][]{%
    \GenericError{(gnuplot) \space\space\space\@spaces}{%
      Package color not loaded in conjunction with
      terminal option `colourtext'%
    }{See the gnuplot documentation for explanation.%
    }{Either use 'blacktext' in gnuplot or load the package
      color.sty in LaTeX.}%
    \renewcommand\color[2][]{}%
  }%
  \providecommand\includegraphics[2][]{%
    \GenericError{(gnuplot) \space\space\space\@spaces}{%
      Package graphicx or graphics not loaded%
    }{See the gnuplot documentation for explanation.%
    }{The gnuplot epslatex terminal needs graphicx.sty or graphics.sty.}%
    \renewcommand\includegraphics[2][]{}%
  }%
  \providecommand\rotatebox[2]{#2}%
  \@ifundefined{ifGPcolor}{%
    \newif\ifGPcolor
    \GPcolorfalse
  }{}%
  \@ifundefined{ifGPblacktext}{%
    \newif\ifGPblacktext
    \GPblacktexttrue
  }{}%
  % define a \g@addto@macro without @ in the name:
  \let\gplgaddtomacro\g@addto@macro
  % define empty templates for all commands taking text:
  \gdef\gplbacktext{}%
  \gdef\gplfronttext{}%
  \makeatother
  \ifGPblacktext
    % no textcolor at all
    \def\colorrgb#1{}%
    \def\colorgray#1{}%
  \else
    % gray or color?
    \ifGPcolor
      \def\colorrgb#1{\color[rgb]{#1}}%
      \def\colorgray#1{\color[gray]{#1}}%
      \expandafter\def\csname LTw\endcsname{\color{white}}%
      \expandafter\def\csname LTb\endcsname{\color{black}}%
      \expandafter\def\csname LTa\endcsname{\color{black}}%
      \expandafter\def\csname LT0\endcsname{\color[rgb]{1,0,0}}%
      \expandafter\def\csname LT1\endcsname{\color[rgb]{0,1,0}}%
      \expandafter\def\csname LT2\endcsname{\color[rgb]{0,0,1}}%
      \expandafter\def\csname LT3\endcsname{\color[rgb]{1,0,1}}%
      \expandafter\def\csname LT4\endcsname{\color[rgb]{0,1,1}}%
      \expandafter\def\csname LT5\endcsname{\color[rgb]{1,1,0}}%
      \expandafter\def\csname LT6\endcsname{\color[rgb]{0,0,0}}%
      \expandafter\def\csname LT7\endcsname{\color[rgb]{1,0.3,0}}%
      \expandafter\def\csname LT8\endcsname{\color[rgb]{0.5,0.5,0.5}}%
    \else
      % gray
      \def\colorrgb#1{\color{black}}%
      \def\colorgray#1{\color[gray]{#1}}%
      \expandafter\def\csname LTw\endcsname{\color{white}}%
      \expandafter\def\csname LTb\endcsname{\color{black}}%
      \expandafter\def\csname LTa\endcsname{\color{black}}%
      \expandafter\def\csname LT0\endcsname{\color{black}}%
      \expandafter\def\csname LT1\endcsname{\color{black}}%
      \expandafter\def\csname LT2\endcsname{\color{black}}%
      \expandafter\def\csname LT3\endcsname{\color{black}}%
      \expandafter\def\csname LT4\endcsname{\color{black}}%
      \expandafter\def\csname LT5\endcsname{\color{black}}%
      \expandafter\def\csname LT6\endcsname{\color{black}}%
      \expandafter\def\csname LT7\endcsname{\color{black}}%
      \expandafter\def\csname LT8\endcsname{\color{black}}%
    \fi
  \fi
    \setlength{\unitlength}{0.0500bp}%
    \ifx\gptboxheight\undefined%
      \newlength{\gptboxheight}%
      \newlength{\gptboxwidth}%
      \newsavebox{\gptboxtext}%
    \fi%
    \setlength{\fboxrule}{0.5pt}%
    \setlength{\fboxsep}{1pt}%
\begin{picture}(7200.00,5040.00)%
    \gplgaddtomacro\gplbacktext{%
      \csname LTb\endcsname%
      \put(1210,704){\makebox(0,0)[r]{\strut{}$-0.04$}}%
      \put(1210,1156){\makebox(0,0)[r]{\strut{}$-0.035$}}%
      \put(1210,1609){\makebox(0,0)[r]{\strut{}$-0.03$}}%
      \put(1210,2061){\makebox(0,0)[r]{\strut{}$-0.025$}}%
      \put(1210,2513){\makebox(0,0)[r]{\strut{}$-0.02$}}%
      \put(1210,2966){\makebox(0,0)[r]{\strut{}$-0.015$}}%
      \put(1210,3418){\makebox(0,0)[r]{\strut{}$-0.01$}}%
      \put(1210,3870){\makebox(0,0)[r]{\strut{}$-0.005$}}%
      \put(1210,4323){\makebox(0,0)[r]{\strut{}$0$}}%
      \put(1210,4775){\makebox(0,0)[r]{\strut{}$0.005$}}%
      \put(1342,484){\makebox(0,0){\strut{}$0$}}%
      \put(1888,484){\makebox(0,0){\strut{}$1$}}%
      \put(2434,484){\makebox(0,0){\strut{}$2$}}%
      \put(2980,484){\makebox(0,0){\strut{}$3$}}%
      \put(3526,484){\makebox(0,0){\strut{}$4$}}%
      \put(4073,484){\makebox(0,0){\strut{}$5$}}%
      \put(4619,484){\makebox(0,0){\strut{}$6$}}%
      \put(5165,484){\makebox(0,0){\strut{}$7$}}%
      \put(5711,484){\makebox(0,0){\strut{}$8$}}%
      \put(6257,484){\makebox(0,0){\strut{}$9$}}%
      \put(6803,484){\makebox(0,0){\strut{}$10$}}%
    }%
    \gplgaddtomacro\gplfronttext{%
      \csname LTb\endcsname%
      \put(176,2739){\rotatebox{-270}{\makebox(0,0){\strut{}\Large $\Delta [\langle \varphi(x) \rangle$]\normalsize}}}%
      \put(4072,154){\makebox(0,0){\strut{}\Large $x$ \normalsize}}%
    }%
    \gplbacktext
    \put(0,0){\includegraphics{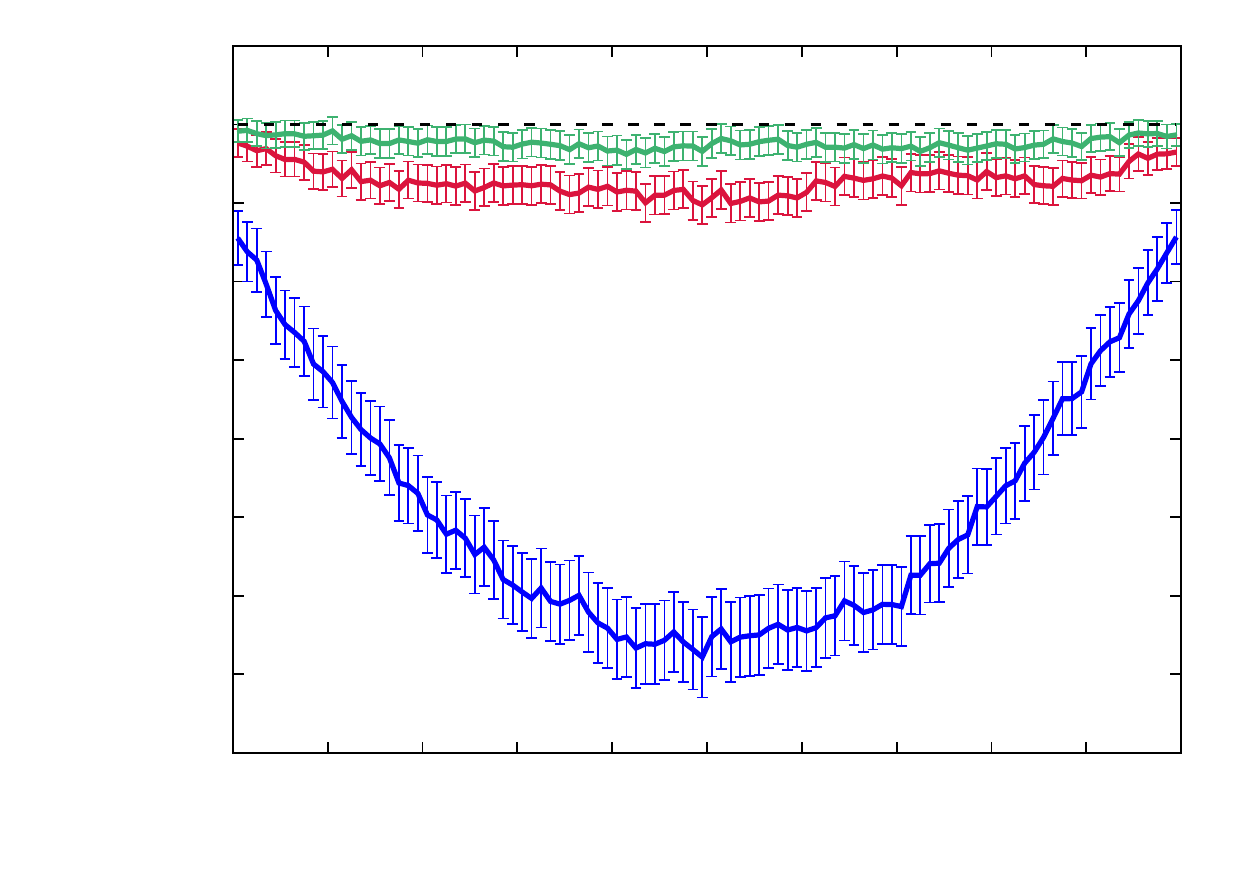}}%
    \gplfronttext
  \end{picture}%
\endgroup

%% file: flux_1C_I_ecart.tex
% GNUPLOT: LaTeX picture with Postscript
\begingroup
  \makeatletter
  \providecommand\color[2][]{%
    \GenericError{(gnuplot) \space\space\space\@spaces}{%
      Package color not loaded in conjunction with
      terminal option `colourtext'%
    }{See the gnuplot documentation for explanation.%
    }{Either use 'blacktext' in gnuplot or load the package
      color.sty in LaTeX.}%
    \renewcommand\color[2][]{}%
  }%
  \providecommand\includegraphics[2][]{%
    \GenericError{(gnuplot) \space\space\space\@spaces}{%
      Package graphicx or graphics not loaded%
    }{See the gnuplot documentation for explanation.%
    }{The gnuplot epslatex terminal needs graphicx.sty or graphics.sty.}%
    \renewcommand\includegraphics[2][]{}%
  }%
  \providecommand\rotatebox[2]{#2}%
  \@ifundefined{ifGPcolor}{%
    \newif\ifGPcolor
    \GPcolorfalse
  }{}%
  \@ifundefined{ifGPblacktext}{%
    \newif\ifGPblacktext
    \GPblacktexttrue
  }{}%
  % define a \g@addto@macro without @ in the name:
  \let\gplgaddtomacro\g@addto@macro
  % define empty templates for all commands taking text:
  \gdef\gplbacktext{}%
  \gdef\gplfronttext{}%
  \makeatother
  \ifGPblacktext
    % no textcolor at all
    \def\colorrgb#1{}%
    \def\colorgray#1{}%
  \else
    % gray or color?
    \ifGPcolor
      \def\colorrgb#1{\color[rgb]{#1}}%
      \def\colorgray#1{\color[gray]{#1}}%
      \expandafter\def\csname LTw\endcsname{\color{white}}%
      \expandafter\def\csname LTb\endcsname{\color{black}}%
      \expandafter\def\csname LTa\endcsname{\color{black}}%
      \expandafter\def\csname LT0\endcsname{\color[rgb]{1,0,0}}%
      \expandafter\def\csname LT1\endcsname{\color[rgb]{0,1,0}}%
      \expandafter\def\csname LT2\endcsname{\color[rgb]{0,0,1}}%
      \expandafter\def\csname LT3\endcsname{\color[rgb]{1,0,1}}%
      \expandafter\def\csname LT4\endcsname{\color[rgb]{0,1,1}}%
      \expandafter\def\csname LT5\endcsname{\color[rgb]{1,1,0}}%
      \expandafter\def\csname LT6\endcsname{\color[rgb]{0,0,0}}%
      \expandafter\def\csname LT7\endcsname{\color[rgb]{1,0.3,0}}%
      \expandafter\def\csname LT8\endcsname{\color[rgb]{0.5,0.5,0.5}}%
    \else
      % gray
      \def\colorrgb#1{\color{black}}%
      \def\colorgray#1{\color[gray]{#1}}%
      \expandafter\def\csname LTw\endcsname{\color{white}}%
      \expandafter\def\csname LTb\endcsname{\color{black}}%
      \expandafter\def\csname LTa\endcsname{\color{black}}%
      \expandafter\def\csname LT0\endcsname{\color{black}}%
      \expandafter\def\csname LT1\endcsname{\color{black}}%
      \expandafter\def\csname LT2\endcsname{\color{black}}%
      \expandafter\def\csname LT3\endcsname{\color{black}}%
      \expandafter\def\csname LT4\endcsname{\color{black}}%
      \expandafter\def\csname LT5\endcsname{\color{black}}%
      \expandafter\def\csname LT6\endcsname{\color{black}}%
      \expandafter\def\csname LT7\endcsname{\color{black}}%
      \expandafter\def\csname LT8\endcsname{\color{black}}%
    \fi
  \fi
    \setlength{\unitlength}{0.0500bp}%
    \ifx\gptboxheight\undefined%
      \newlength{\gptboxheight}%
      \newlength{\gptboxwidth}%
      \newsavebox{\gptboxtext}%
    \fi%
    \setlength{\fboxrule}{0.5pt}%
    \setlength{\fboxsep}{1pt}%
\begin{picture}(7200.00,5040.00)%
    \gplgaddtomacro\gplbacktext{%
      \csname LTb\endcsname%
      \put(1078,704){\makebox(0,0)[r]{\strut{}$-0.25$}}%
      \put(1078,1383){\makebox(0,0)[r]{\strut{}$-0.2$}}%
      \put(1078,2061){\makebox(0,0)[r]{\strut{}$-0.15$}}%
      \put(1078,2739){\makebox(0,0)[r]{\strut{}$-0.1$}}%
      \put(1078,3418){\makebox(0,0)[r]{\strut{}$-0.05$}}%
      \put(1078,4097){\makebox(0,0)[r]{\strut{}$0$}}%
      \put(1078,4775){\makebox(0,0)[r]{\strut{}$0.05$}}%
      \put(1210,484){\makebox(0,0){\strut{}$0$}}%
      \put(1769,484){\makebox(0,0){\strut{}$1$}}%
      \put(2329,484){\makebox(0,0){\strut{}$2$}}%
      \put(2888,484){\makebox(0,0){\strut{}$3$}}%
      \put(3447,484){\makebox(0,0){\strut{}$4$}}%
      \put(4007,484){\makebox(0,0){\strut{}$5$}}%
      \put(4566,484){\makebox(0,0){\strut{}$6$}}%
      \put(5125,484){\makebox(0,0){\strut{}$7$}}%
      \put(5684,484){\makebox(0,0){\strut{}$8$}}%
      \put(6244,484){\makebox(0,0){\strut{}$9$}}%
      \put(6803,484){\makebox(0,0){\strut{}$10$}}%
    }%
    \gplgaddtomacro\gplfronttext{%
      \csname LTb\endcsname%
      \put(176,2739){\rotatebox{-270}{\makebox(0,0){\strut{}\Large $\Delta [\langle \varphi(x) \rangle$]\normalsize}}}%
      \put(4006,154){\makebox(0,0){\strut{}\Large $x$ \normalsize}}%
    }%
    \gplbacktext
    \put(0,0){\includegraphics{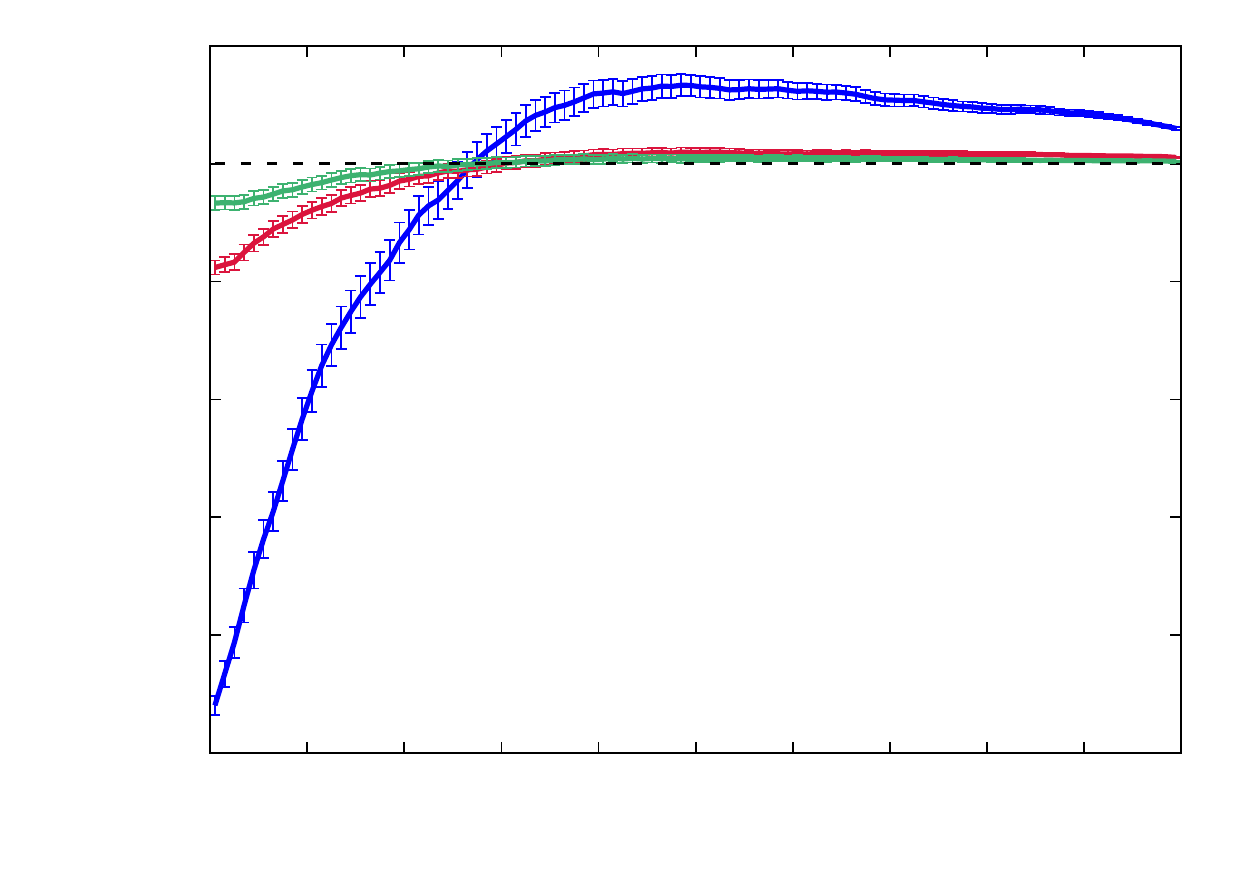}}%
    \gplfronttext
  \end{picture}%
\endgroup

%% file: flux_1C_II_ecart.tex
% GNUPLOT: LaTeX picture with Postscript
\begingroup
  \makeatletter
  \providecommand\color[2][]{%
    \GenericError{(gnuplot) \space\space\space\@spaces}{%
      Package color not loaded in conjunction with
      terminal option `colourtext'%
    }{See the gnuplot documentation for explanation.%
    }{Either use 'blacktext' in gnuplot or load the package
      color.sty in LaTeX.}%
    \renewcommand\color[2][]{}%
  }%
  \providecommand\includegraphics[2][]{%
    \GenericError{(gnuplot) \space\space\space\@spaces}{%
      Package graphicx or graphics not loaded%
    }{See the gnuplot documentation for explanation.%
    }{The gnuplot epslatex terminal needs graphicx.sty or graphics.sty.}%
    \renewcommand\includegraphics[2][]{}%
  }%
  \providecommand\rotatebox[2]{#2}%
  \@ifundefined{ifGPcolor}{%
    \newif\ifGPcolor
    \GPcolorfalse
  }{}%
  \@ifundefined{ifGPblacktext}{%
    \newif\ifGPblacktext
    \GPblacktexttrue
  }{}%
  % define a \g@addto@macro without @ in the name:
  \let\gplgaddtomacro\g@addto@macro
  % define empty templates for all commands taking text:
  \gdef\gplbacktext{}%
  \gdef\gplfronttext{}%
  \makeatother
  \ifGPblacktext
    % no textcolor at all
    \def\colorrgb#1{}%
    \def\colorgray#1{}%
  \else
    % gray or color?
    \ifGPcolor
      \def\colorrgb#1{\color[rgb]{#1}}%
      \def\colorgray#1{\color[gray]{#1}}%
      \expandafter\def\csname LTw\endcsname{\color{white}}%
      \expandafter\def\csname LTb\endcsname{\color{black}}%
      \expandafter\def\csname LTa\endcsname{\color{black}}%
      \expandafter\def\csname LT0\endcsname{\color[rgb]{1,0,0}}%
      \expandafter\def\csname LT1\endcsname{\color[rgb]{0,1,0}}%
      \expandafter\def\csname LT2\endcsname{\color[rgb]{0,0,1}}%
      \expandafter\def\csname LT3\endcsname{\color[rgb]{1,0,1}}%
      \expandafter\def\csname LT4\endcsname{\color[rgb]{0,1,1}}%
      \expandafter\def\csname LT5\endcsname{\color[rgb]{1,1,0}}%
      \expandafter\def\csname LT6\endcsname{\color[rgb]{0,0,0}}%
      \expandafter\def\csname LT7\endcsname{\color[rgb]{1,0.3,0}}%
      \expandafter\def\csname LT8\endcsname{\color[rgb]{0.5,0.5,0.5}}%
    \else
      % gray
      \def\colorrgb#1{\color{black}}%
      \def\colorgray#1{\color[gray]{#1}}%
      \expandafter\def\csname LTw\endcsname{\color{white}}%
      \expandafter\def\csname LTb\endcsname{\color{black}}%
      \expandafter\def\csname LTa\endcsname{\color{black}}%
      \expandafter\def\csname LT0\endcsname{\color{black}}%
      \expandafter\def\csname LT1\endcsname{\color{black}}%
      \expandafter\def\csname LT2\endcsname{\color{black}}%
      \expandafter\def\csname LT3\endcsname{\color{black}}%
      \expandafter\def\csname LT4\endcsname{\color{black}}%
      \expandafter\def\csname LT5\endcsname{\color{black}}%
      \expandafter\def\csname LT6\endcsname{\color{black}}%
      \expandafter\def\csname LT7\endcsname{\color{black}}%
      \expandafter\def\csname LT8\endcsname{\color{black}}%
    \fi
  \fi
    \setlength{\unitlength}{0.0500bp}%
    \ifx\gptboxheight\undefined%
      \newlength{\gptboxheight}%
      \newlength{\gptboxwidth}%
      \newsavebox{\gptboxtext}%
    \fi%
    \setlength{\fboxrule}{0.5pt}%
    \setlength{\fboxsep}{1pt}%
\begin{picture}(7200.00,5040.00)%
    \gplgaddtomacro\gplbacktext{%
      \csname LTb\endcsname%
      \put(1078,704){\makebox(0,0)[r]{\strut{}$-0.25$}}%
      \put(1078,1383){\makebox(0,0)[r]{\strut{}$-0.2$}}%
      \put(1078,2061){\makebox(0,0)[r]{\strut{}$-0.15$}}%
      \put(1078,2739){\makebox(0,0)[r]{\strut{}$-0.1$}}%
      \put(1078,3418){\makebox(0,0)[r]{\strut{}$-0.05$}}%
      \put(1078,4097){\makebox(0,0)[r]{\strut{}$0$}}%
      \put(1078,4775){\makebox(0,0)[r]{\strut{}$0.05$}}%
      \put(1210,484){\makebox(0,0){\strut{}$0$}}%
      \put(1769,484){\makebox(0,0){\strut{}$1$}}%
      \put(2329,484){\makebox(0,0){\strut{}$2$}}%
      \put(2888,484){\makebox(0,0){\strut{}$3$}}%
      \put(3447,484){\makebox(0,0){\strut{}$4$}}%
      \put(4007,484){\makebox(0,0){\strut{}$5$}}%
      \put(4566,484){\makebox(0,0){\strut{}$6$}}%
      \put(5125,484){\makebox(0,0){\strut{}$7$}}%
      \put(5684,484){\makebox(0,0){\strut{}$8$}}%
      \put(6244,484){\makebox(0,0){\strut{}$9$}}%
      \put(6803,484){\makebox(0,0){\strut{}$10$}}%
    }%
    \gplgaddtomacro\gplfronttext{%
      \csname LTb\endcsname%
      \put(176,2739){\rotatebox{-270}{\makebox(0,0){\strut{}\Large $\Delta [\langle \varphi(x) \rangle$]\normalsize}}}%
      \put(4006,154){\makebox(0,0){\strut{}\Large $x$ \normalsize}}%
    }%
    \gplbacktext
    \put(0,0){\includegraphics{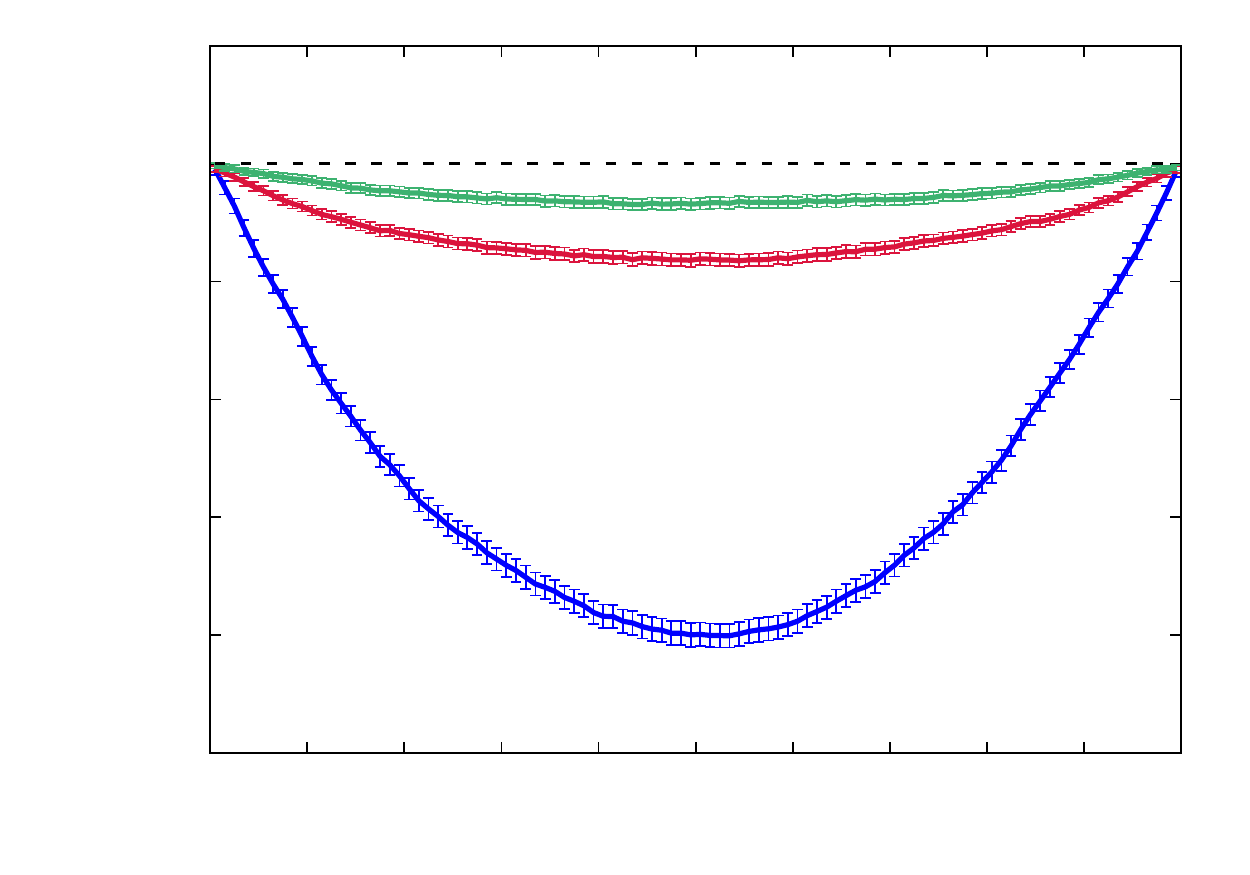}}%
    \gplfronttext
  \end{picture}%
\endgroup

%% file: flux_2A_I_ecart.tex
% GNUPLOT: LaTeX picture with Postscript
\begingroup
  \makeatletter
  \providecommand\color[2][]{%
    \GenericError{(gnuplot) \space\space\space\@spaces}{%
      Package color not loaded in conjunction with
      terminal option `colourtext'%
    }{See the gnuplot documentation for explanation.%
    }{Either use 'blacktext' in gnuplot or load the package
      color.sty in LaTeX.}%
    \renewcommand\color[2][]{}%
  }%
  \providecommand\includegraphics[2][]{%
    \GenericError{(gnuplot) \space\space\space\@spaces}{%
      Package graphicx or graphics not loaded%
    }{See the gnuplot documentation for explanation.%
    }{The gnuplot epslatex terminal needs graphicx.sty or graphics.sty.}%
    \renewcommand\includegraphics[2][]{}%
  }%
  \providecommand\rotatebox[2]{#2}%
  \@ifundefined{ifGPcolor}{%
    \newif\ifGPcolor
    \GPcolorfalse
  }{}%
  \@ifundefined{ifGPblacktext}{%
    \newif\ifGPblacktext
    \GPblacktexttrue
  }{}%
  % define a \g@addto@macro without @ in the name:
  \let\gplgaddtomacro\g@addto@macro
  % define empty templates for all commands taking text:
  \gdef\gplbacktext{}%
  \gdef\gplfronttext{}%
  \makeatother
  \ifGPblacktext
    % no textcolor at all
    \def\colorrgb#1{}%
    \def\colorgray#1{}%
  \else
    % gray or color?
    \ifGPcolor
      \def\colorrgb#1{\color[rgb]{#1}}%
      \def\colorgray#1{\color[gray]{#1}}%
      \expandafter\def\csname LTw\endcsname{\color{white}}%
      \expandafter\def\csname LTb\endcsname{\color{black}}%
      \expandafter\def\csname LTa\endcsname{\color{black}}%
      \expandafter\def\csname LT0\endcsname{\color[rgb]{1,0,0}}%
      \expandafter\def\csname LT1\endcsname{\color[rgb]{0,1,0}}%
      \expandafter\def\csname LT2\endcsname{\color[rgb]{0,0,1}}%
      \expandafter\def\csname LT3\endcsname{\color[rgb]{1,0,1}}%
      \expandafter\def\csname LT4\endcsname{\color[rgb]{0,1,1}}%
      \expandafter\def\csname LT5\endcsname{\color[rgb]{1,1,0}}%
      \expandafter\def\csname LT6\endcsname{\color[rgb]{0,0,0}}%
      \expandafter\def\csname LT7\endcsname{\color[rgb]{1,0.3,0}}%
      \expandafter\def\csname LT8\endcsname{\color[rgb]{0.5,0.5,0.5}}%
    \else
      % gray
      \def\colorrgb#1{\color{black}}%
      \def\colorgray#1{\color[gray]{#1}}%
      \expandafter\def\csname LTw\endcsname{\color{white}}%
      \expandafter\def\csname LTb\endcsname{\color{black}}%
      \expandafter\def\csname LTa\endcsname{\color{black}}%
      \expandafter\def\csname LT0\endcsname{\color{black}}%
      \expandafter\def\csname LT1\endcsname{\color{black}}%
      \expandafter\def\csname LT2\endcsname{\color{black}}%
      \expandafter\def\csname LT3\endcsname{\color{black}}%
      \expandafter\def\csname LT4\endcsname{\color{black}}%
      \expandafter\def\csname LT5\endcsname{\color{black}}%
      \expandafter\def\csname LT6\endcsname{\color{black}}%
      \expandafter\def\csname LT7\endcsname{\color{black}}%
      \expandafter\def\csname LT8\endcsname{\color{black}}%
    \fi
  \fi
    \setlength{\unitlength}{0.0500bp}%
    \ifx\gptboxheight\undefined%
      \newlength{\gptboxheight}%
      \newlength{\gptboxwidth}%
      \newsavebox{\gptboxtext}%
    \fi%
    \setlength{\fboxrule}{0.5pt}%
    \setlength{\fboxsep}{1pt}%
\begin{picture}(7200.00,5040.00)%
    \gplgaddtomacro\gplbacktext{%
      \csname LTb\endcsname%
      \put(1078,704){\makebox(0,0)[r]{\strut{}$-0.12$}}%
      \put(1078,1156){\makebox(0,0)[r]{\strut{}$-0.1$}}%
      \put(1078,1609){\makebox(0,0)[r]{\strut{}$-0.08$}}%
      \put(1078,2061){\makebox(0,0)[r]{\strut{}$-0.06$}}%
      \put(1078,2513){\makebox(0,0)[r]{\strut{}$-0.04$}}%
      \put(1078,2966){\makebox(0,0)[r]{\strut{}$-0.02$}}%
      \put(1078,3418){\makebox(0,0)[r]{\strut{}$0$}}%
      \put(1078,3870){\makebox(0,0)[r]{\strut{}$0.02$}}%
      \put(1078,4323){\makebox(0,0)[r]{\strut{}$0.04$}}%
      \put(1078,4775){\makebox(0,0)[r]{\strut{}$0.06$}}%
      \put(1210,484){\makebox(0,0){\strut{}$0$}}%
      \put(1769,484){\makebox(0,0){\strut{}$1$}}%
      \put(2329,484){\makebox(0,0){\strut{}$2$}}%
      \put(2888,484){\makebox(0,0){\strut{}$3$}}%
      \put(3447,484){\makebox(0,0){\strut{}$4$}}%
      \put(4007,484){\makebox(0,0){\strut{}$5$}}%
      \put(4566,484){\makebox(0,0){\strut{}$6$}}%
      \put(5125,484){\makebox(0,0){\strut{}$7$}}%
      \put(5684,484){\makebox(0,0){\strut{}$8$}}%
      \put(6244,484){\makebox(0,0){\strut{}$9$}}%
      \put(6803,484){\makebox(0,0){\strut{}$10$}}%
    }%
    \gplgaddtomacro\gplfronttext{%
      \csname LTb\endcsname%
      \put(176,2739){\rotatebox{-270}{\makebox(0,0){\strut{}\Large $\Delta [\langle \varphi(x) \rangle$]\normalsize}}}%
      \put(4006,154){\makebox(0,0){\strut{}\Large $x$ \normalsize}}%
    }%
    \gplbacktext
    \put(0,0){\includegraphics{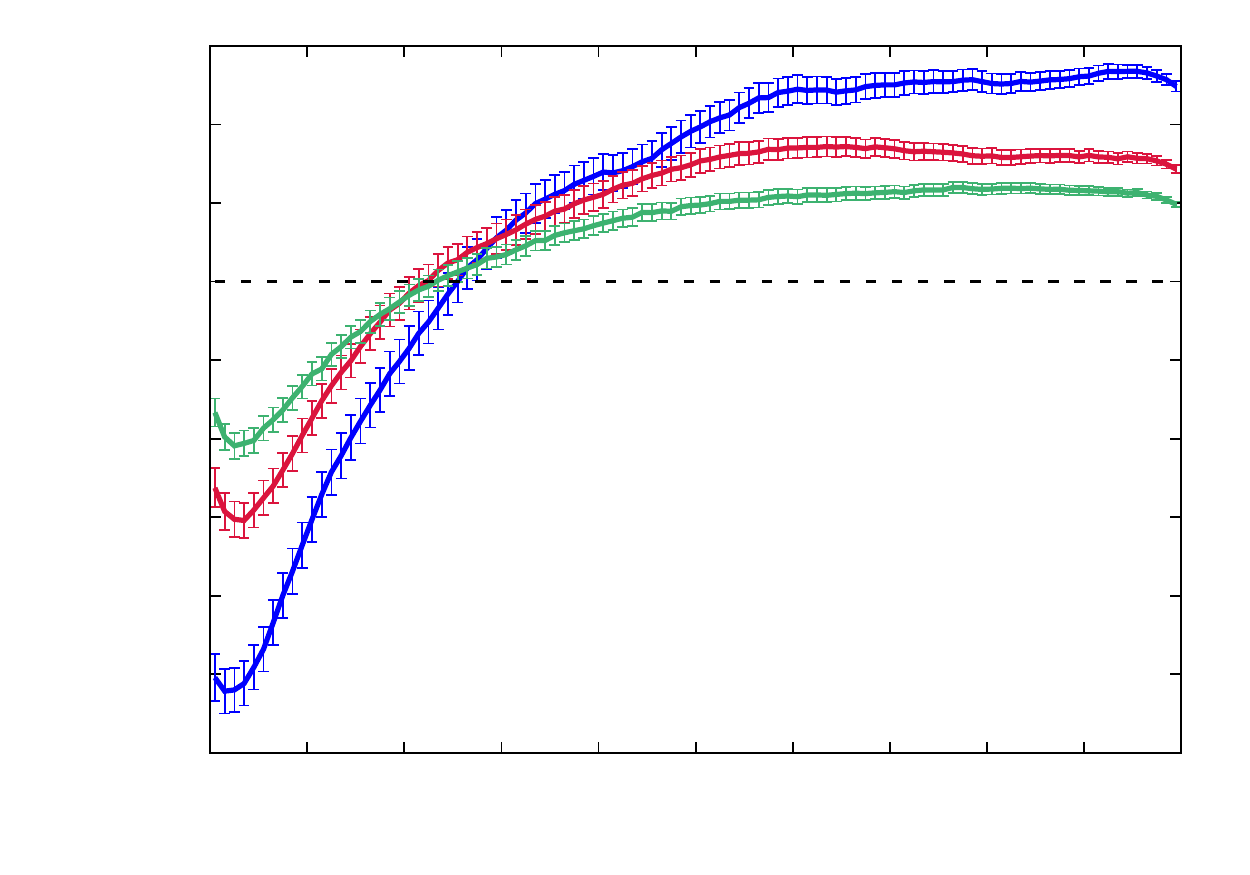}}%
    \gplfronttext
  \end{picture}%
\endgroup

%% file: flux_2A_II_ecart.tex
% GNUPLOT: LaTeX picture with Postscript
\begingroup
  \makeatletter
  \providecommand\color[2][]{%
    \GenericError{(gnuplot) \space\space\space\@spaces}{%
      Package color not loaded in conjunction with
      terminal option `colourtext'%
    }{See the gnuplot documentation for explanation.%
    }{Either use 'blacktext' in gnuplot or load the package
      color.sty in LaTeX.}%
    \renewcommand\color[2][]{}%
  }%
  \providecommand\includegraphics[2][]{%
    \GenericError{(gnuplot) \space\space\space\@spaces}{%
      Package graphicx or graphics not loaded%
    }{See the gnuplot documentation for explanation.%
    }{The gnuplot epslatex terminal needs graphicx.sty or graphics.sty.}%
    \renewcommand\includegraphics[2][]{}%
  }%
  \providecommand\rotatebox[2]{#2}%
  \@ifundefined{ifGPcolor}{%
    \newif\ifGPcolor
    \GPcolorfalse
  }{}%
  \@ifundefined{ifGPblacktext}{%
    \newif\ifGPblacktext
    \GPblacktexttrue
  }{}%
  % define a \g@addto@macro without @ in the name:
  \let\gplgaddtomacro\g@addto@macro
  % define empty templates for all commands taking text:
  \gdef\gplbacktext{}%
  \gdef\gplfronttext{}%
  \makeatother
  \ifGPblacktext
    % no textcolor at all
    \def\colorrgb#1{}%
    \def\colorgray#1{}%
  \else
    % gray or color?
    \ifGPcolor
      \def\colorrgb#1{\color[rgb]{#1}}%
      \def\colorgray#1{\color[gray]{#1}}%
      \expandafter\def\csname LTw\endcsname{\color{white}}%
      \expandafter\def\csname LTb\endcsname{\color{black}}%
      \expandafter\def\csname LTa\endcsname{\color{black}}%
      \expandafter\def\csname LT0\endcsname{\color[rgb]{1,0,0}}%
      \expandafter\def\csname LT1\endcsname{\color[rgb]{0,1,0}}%
      \expandafter\def\csname LT2\endcsname{\color[rgb]{0,0,1}}%
      \expandafter\def\csname LT3\endcsname{\color[rgb]{1,0,1}}%
      \expandafter\def\csname LT4\endcsname{\color[rgb]{0,1,1}}%
      \expandafter\def\csname LT5\endcsname{\color[rgb]{1,1,0}}%
      \expandafter\def\csname LT6\endcsname{\color[rgb]{0,0,0}}%
      \expandafter\def\csname LT7\endcsname{\color[rgb]{1,0.3,0}}%
      \expandafter\def\csname LT8\endcsname{\color[rgb]{0.5,0.5,0.5}}%
    \else
      % gray
      \def\colorrgb#1{\color{black}}%
      \def\colorgray#1{\color[gray]{#1}}%
      \expandafter\def\csname LTw\endcsname{\color{white}}%
      \expandafter\def\csname LTb\endcsname{\color{black}}%
      \expandafter\def\csname LTa\endcsname{\color{black}}%
      \expandafter\def\csname LT0\endcsname{\color{black}}%
      \expandafter\def\csname LT1\endcsname{\color{black}}%
      \expandafter\def\csname LT2\endcsname{\color{black}}%
      \expandafter\def\csname LT3\endcsname{\color{black}}%
      \expandafter\def\csname LT4\endcsname{\color{black}}%
      \expandafter\def\csname LT5\endcsname{\color{black}}%
      \expandafter\def\csname LT6\endcsname{\color{black}}%
      \expandafter\def\csname LT7\endcsname{\color{black}}%
      \expandafter\def\csname LT8\endcsname{\color{black}}%
    \fi
  \fi
    \setlength{\unitlength}{0.0500bp}%
    \ifx\gptboxheight\undefined%
      \newlength{\gptboxheight}%
      \newlength{\gptboxwidth}%
      \newsavebox{\gptboxtext}%
    \fi%
    \setlength{\fboxrule}{0.5pt}%
    \setlength{\fboxsep}{1pt}%
\begin{picture}(7200.00,5040.00)%
    \gplgaddtomacro\gplbacktext{%
      \csname LTb\endcsname%
      \put(1078,704){\makebox(0,0)[r]{\strut{}$-0.25$}}%
      \put(1078,1383){\makebox(0,0)[r]{\strut{}$-0.2$}}%
      \put(1078,2061){\makebox(0,0)[r]{\strut{}$-0.15$}}%
      \put(1078,2739){\makebox(0,0)[r]{\strut{}$-0.1$}}%
      \put(1078,3418){\makebox(0,0)[r]{\strut{}$-0.05$}}%
      \put(1078,4097){\makebox(0,0)[r]{\strut{}$0$}}%
      \put(1078,4775){\makebox(0,0)[r]{\strut{}$0.05$}}%
      \put(1210,484){\makebox(0,0){\strut{}$0$}}%
      \put(1769,484){\makebox(0,0){\strut{}$1$}}%
      \put(2329,484){\makebox(0,0){\strut{}$2$}}%
      \put(2888,484){\makebox(0,0){\strut{}$3$}}%
      \put(3447,484){\makebox(0,0){\strut{}$4$}}%
      \put(4007,484){\makebox(0,0){\strut{}$5$}}%
      \put(4566,484){\makebox(0,0){\strut{}$6$}}%
      \put(5125,484){\makebox(0,0){\strut{}$7$}}%
      \put(5684,484){\makebox(0,0){\strut{}$8$}}%
      \put(6244,484){\makebox(0,0){\strut{}$9$}}%
      \put(6803,484){\makebox(0,0){\strut{}$10$}}%
    }%
    \gplgaddtomacro\gplfronttext{%
      \csname LTb\endcsname%
      \put(176,2739){\rotatebox{-270}{\makebox(0,0){\strut{}\Large $\Delta [\langle \varphi(x) \rangle$]\normalsize}}}%
      \put(4006,154){\makebox(0,0){\strut{}\Large $x$ \normalsize}}%
    }%
    \gplbacktext
    \put(0,0){\includegraphics{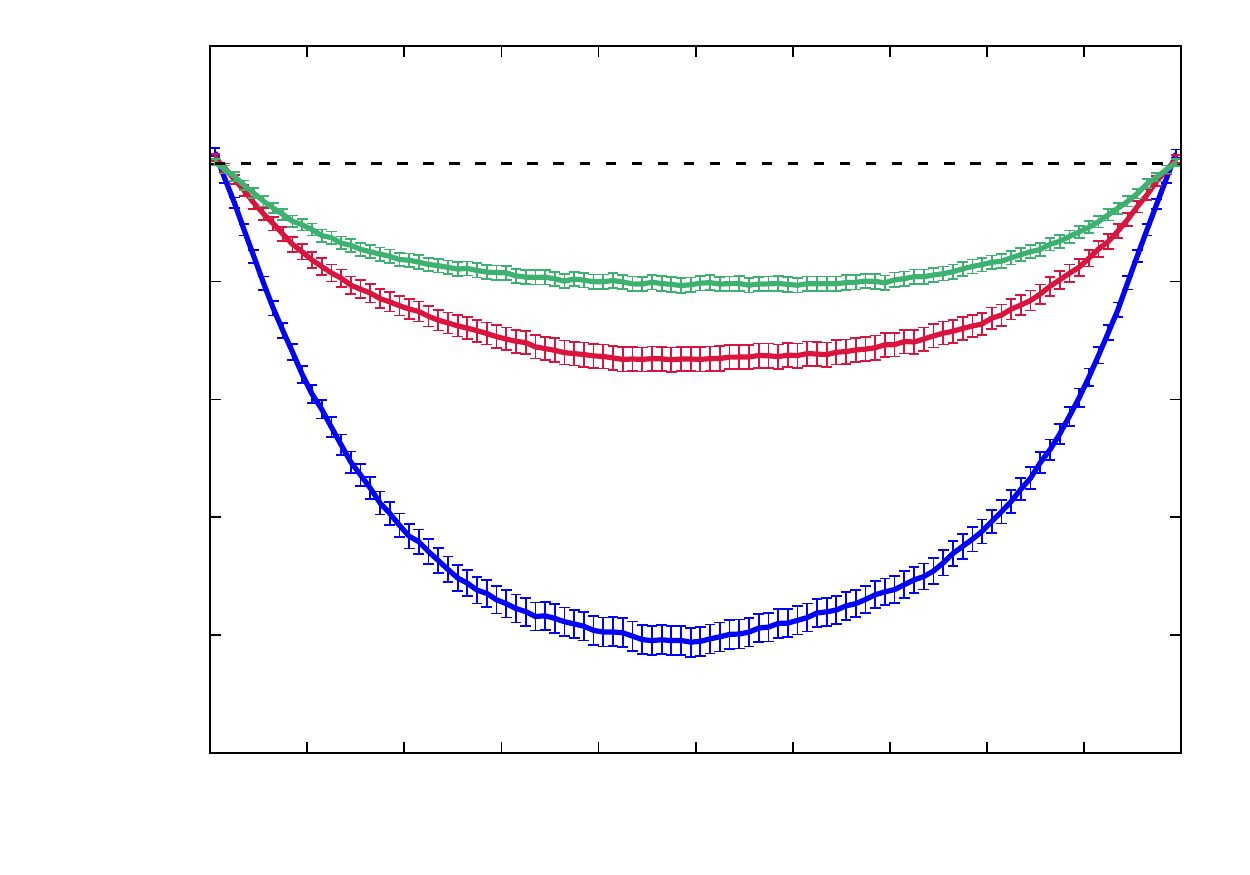}}%
    \gplfronttext
  \end{picture}%
\endgroup

%% file: flux_2B_I_ecart.tex
% GNUPLOT: LaTeX picture with Postscript
\begingroup
  \makeatletter
  \providecommand\color[2][]{%
    \GenericError{(gnuplot) \space\space\space\@spaces}{%
      Package color not loaded in conjunction with
      terminal option `colourtext'%
    }{See the gnuplot documentation for explanation.%
    }{Either use 'blacktext' in gnuplot or load the package
      color.sty in LaTeX.}%
    \renewcommand\color[2][]{}%
  }%
  \providecommand\includegraphics[2][]{%
    \GenericError{(gnuplot) \space\space\space\@spaces}{%
      Package graphicx or graphics not loaded%
    }{See the gnuplot documentation for explanation.%
    }{The gnuplot epslatex terminal needs graphicx.sty or graphics.sty.}%
    \renewcommand\includegraphics[2][]{}%
  }%
  \providecommand\rotatebox[2]{#2}%
  \@ifundefined{ifGPcolor}{%
    \newif\ifGPcolor
    \GPcolorfalse
  }{}%
  \@ifundefined{ifGPblacktext}{%
    \newif\ifGPblacktext
    \GPblacktexttrue
  }{}%
  % define a \g@addto@macro without @ in the name:
  \let\gplgaddtomacro\g@addto@macro
  % define empty templates for all commands taking text:
  \gdef\gplbacktext{}%
  \gdef\gplfronttext{}%
  \makeatother
  \ifGPblacktext
    % no textcolor at all
    \def\colorrgb#1{}%
    \def\colorgray#1{}%
  \else
    % gray or color?
    \ifGPcolor
      \def\colorrgb#1{\color[rgb]{#1}}%
      \def\colorgray#1{\color[gray]{#1}}%
      \expandafter\def\csname LTw\endcsname{\color{white}}%
      \expandafter\def\csname LTb\endcsname{\color{black}}%
      \expandafter\def\csname LTa\endcsname{\color{black}}%
      \expandafter\def\csname LT0\endcsname{\color[rgb]{1,0,0}}%
      \expandafter\def\csname LT1\endcsname{\color[rgb]{0,1,0}}%
      \expandafter\def\csname LT2\endcsname{\color[rgb]{0,0,1}}%
      \expandafter\def\csname LT3\endcsname{\color[rgb]{1,0,1}}%
      \expandafter\def\csname LT4\endcsname{\color[rgb]{0,1,1}}%
      \expandafter\def\csname LT5\endcsname{\color[rgb]{1,1,0}}%
      \expandafter\def\csname LT6\endcsname{\color[rgb]{0,0,0}}%
      \expandafter\def\csname LT7\endcsname{\color[rgb]{1,0.3,0}}%
      \expandafter\def\csname LT8\endcsname{\color[rgb]{0.5,0.5,0.5}}%
    \else
      % gray
      \def\colorrgb#1{\color{black}}%
      \def\colorgray#1{\color[gray]{#1}}%
      \expandafter\def\csname LTw\endcsname{\color{white}}%
      \expandafter\def\csname LTb\endcsname{\color{black}}%
      \expandafter\def\csname LTa\endcsname{\color{black}}%
      \expandafter\def\csname LT0\endcsname{\color{black}}%
      \expandafter\def\csname LT1\endcsname{\color{black}}%
      \expandafter\def\csname LT2\endcsname{\color{black}}%
      \expandafter\def\csname LT3\endcsname{\color{black}}%
      \expandafter\def\csname LT4\endcsname{\color{black}}%
      \expandafter\def\csname LT5\endcsname{\color{black}}%
      \expandafter\def\csname LT6\endcsname{\color{black}}%
      \expandafter\def\csname LT7\endcsname{\color{black}}%
      \expandafter\def\csname LT8\endcsname{\color{black}}%
    \fi
  \fi
    \setlength{\unitlength}{0.0500bp}%
    \ifx\gptboxheight\undefined%
      \newlength{\gptboxheight}%
      \newlength{\gptboxwidth}%
      \newsavebox{\gptboxtext}%
    \fi%
    \setlength{\fboxrule}{0.5pt}%
    \setlength{\fboxsep}{1pt}%
\begin{picture}(7200.00,5040.00)%
    \gplgaddtomacro\gplbacktext{%
      \csname LTb\endcsname%
      \put(1078,704){\makebox(0,0)[r]{\strut{}$-0.16$}}%
      \put(1078,1213){\makebox(0,0)[r]{\strut{}$-0.14$}}%
      \put(1078,1722){\makebox(0,0)[r]{\strut{}$-0.12$}}%
      \put(1078,2231){\makebox(0,0)[r]{\strut{}$-0.1$}}%
      \put(1078,2740){\makebox(0,0)[r]{\strut{}$-0.08$}}%
      \put(1078,3248){\makebox(0,0)[r]{\strut{}$-0.06$}}%
      \put(1078,3757){\makebox(0,0)[r]{\strut{}$-0.04$}}%
      \put(1078,4266){\makebox(0,0)[r]{\strut{}$-0.02$}}%
      \put(1078,4775){\makebox(0,0)[r]{\strut{}$0$}}%
      \put(1210,484){\makebox(0,0){\strut{}$0$}}%
      \put(1769,484){\makebox(0,0){\strut{}$1$}}%
      \put(2329,484){\makebox(0,0){\strut{}$2$}}%
      \put(2888,484){\makebox(0,0){\strut{}$3$}}%
      \put(3447,484){\makebox(0,0){\strut{}$4$}}%
      \put(4007,484){\makebox(0,0){\strut{}$5$}}%
      \put(4566,484){\makebox(0,0){\strut{}$6$}}%
      \put(5125,484){\makebox(0,0){\strut{}$7$}}%
      \put(5684,484){\makebox(0,0){\strut{}$8$}}%
      \put(6244,484){\makebox(0,0){\strut{}$9$}}%
      \put(6803,484){\makebox(0,0){\strut{}$10$}}%
    }%
    \gplgaddtomacro\gplfronttext{%
      \csname LTb\endcsname%
      \put(176,2739){\rotatebox{-270}{\makebox(0,0){\strut{}\Large $\Delta [\langle \varphi(x) \rangle$]\normalsize}}}%
      \put(4006,154){\makebox(0,0){\strut{}\Large $x$ \normalsize}}%
    }%
    \gplbacktext
    \put(0,0){\includegraphics{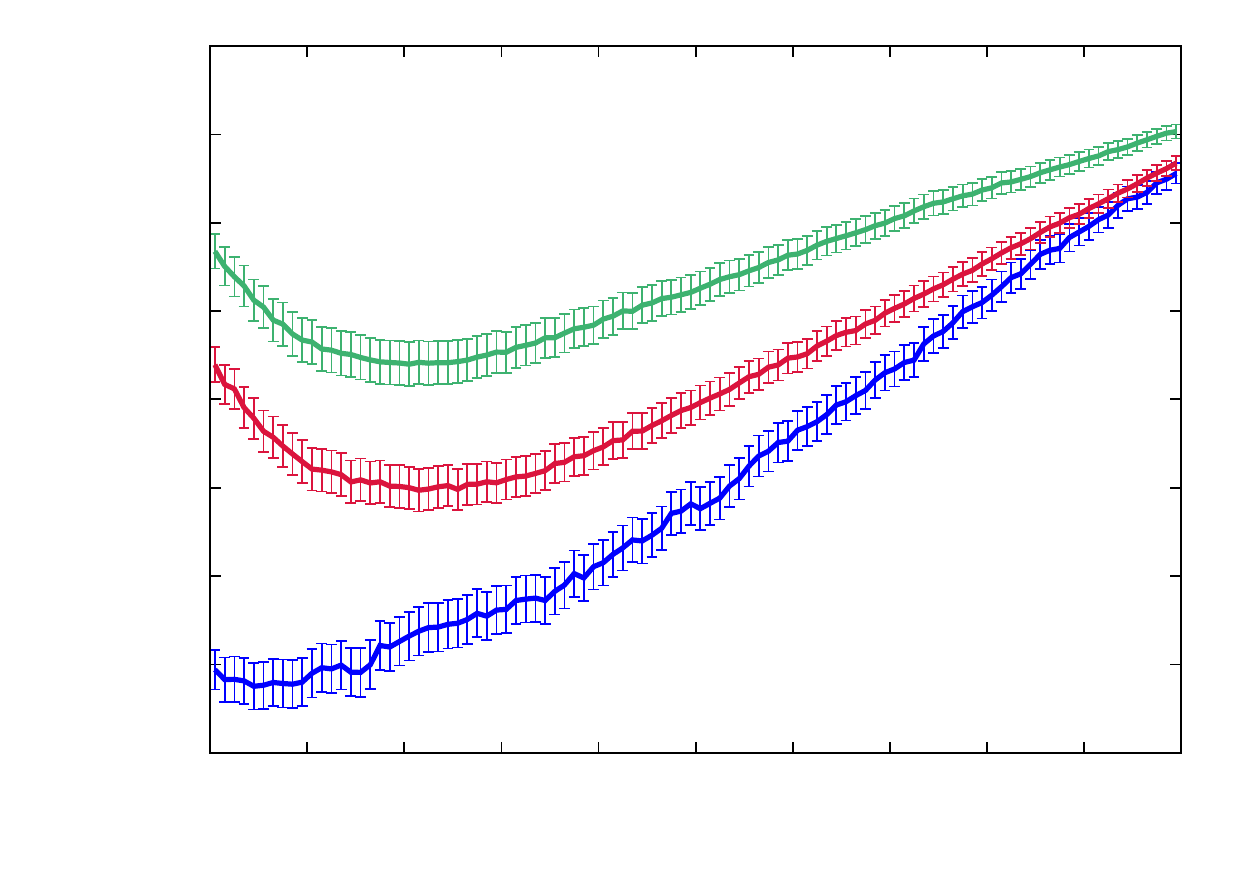}}%
    \gplfronttext
  \end{picture}%
\endgroup

%% file: flux_2B_II_ecart.tex
% GNUPLOT: LaTeX picture with Postscript
\begingroup
  \makeatletter
  \providecommand\color[2][]{%
    \GenericError{(gnuplot) \space\space\space\@spaces}{%
      Package color not loaded in conjunction with
      terminal option `colourtext'%
    }{See the gnuplot documentation for explanation.%
    }{Either use 'blacktext' in gnuplot or load the package
      color.sty in LaTeX.}%
    \renewcommand\color[2][]{}%
  }%
  \providecommand\includegraphics[2][]{%
    \GenericError{(gnuplot) \space\space\space\@spaces}{%
      Package graphicx or graphics not loaded%
    }{See the gnuplot documentation for explanation.%
    }{The gnuplot epslatex terminal needs graphicx.sty or graphics.sty.}%
    \renewcommand\includegraphics[2][]{}%
  }%
  \providecommand\rotatebox[2]{#2}%
  \@ifundefined{ifGPcolor}{%
    \newif\ifGPcolor
    \GPcolorfalse
  }{}%
  \@ifundefined{ifGPblacktext}{%
    \newif\ifGPblacktext
    \GPblacktexttrue
  }{}%
  % define a \g@addto@macro without @ in the name:
  \let\gplgaddtomacro\g@addto@macro
  % define empty templates for all commands taking text:
  \gdef\gplbacktext{}%
  \gdef\gplfronttext{}%
  \makeatother
  \ifGPblacktext
    % no textcolor at all
    \def\colorrgb#1{}%
    \def\colorgray#1{}%
  \else
    % gray or color?
    \ifGPcolor
      \def\colorrgb#1{\color[rgb]{#1}}%
      \def\colorgray#1{\color[gray]{#1}}%
      \expandafter\def\csname LTw\endcsname{\color{white}}%
      \expandafter\def\csname LTb\endcsname{\color{black}}%
      \expandafter\def\csname LTa\endcsname{\color{black}}%
      \expandafter\def\csname LT0\endcsname{\color[rgb]{1,0,0}}%
      \expandafter\def\csname LT1\endcsname{\color[rgb]{0,1,0}}%
      \expandafter\def\csname LT2\endcsname{\color[rgb]{0,0,1}}%
      \expandafter\def\csname LT3\endcsname{\color[rgb]{1,0,1}}%
      \expandafter\def\csname LT4\endcsname{\color[rgb]{0,1,1}}%
      \expandafter\def\csname LT5\endcsname{\color[rgb]{1,1,0}}%
      \expandafter\def\csname LT6\endcsname{\color[rgb]{0,0,0}}%
      \expandafter\def\csname LT7\endcsname{\color[rgb]{1,0.3,0}}%
      \expandafter\def\csname LT8\endcsname{\color[rgb]{0.5,0.5,0.5}}%
    \else
      % gray
      \def\colorrgb#1{\color{black}}%
      \def\colorgray#1{\color[gray]{#1}}%
      \expandafter\def\csname LTw\endcsname{\color{white}}%
      \expandafter\def\csname LTb\endcsname{\color{black}}%
      \expandafter\def\csname LTa\endcsname{\color{black}}%
      \expandafter\def\csname LT0\endcsname{\color{black}}%
      \expandafter\def\csname LT1\endcsname{\color{black}}%
      \expandafter\def\csname LT2\endcsname{\color{black}}%
      \expandafter\def\csname LT3\endcsname{\color{black}}%
      \expandafter\def\csname LT4\endcsname{\color{black}}%
      \expandafter\def\csname LT5\endcsname{\color{black}}%
      \expandafter\def\csname LT6\endcsname{\color{black}}%
      \expandafter\def\csname LT7\endcsname{\color{black}}%
      \expandafter\def\csname LT8\endcsname{\color{black}}%
    \fi
  \fi
    \setlength{\unitlength}{0.0500bp}%
    \ifx\gptboxheight\undefined%
      \newlength{\gptboxheight}%
      \newlength{\gptboxwidth}%
      \newsavebox{\gptboxtext}%
    \fi%
    \setlength{\fboxrule}{0.5pt}%
    \setlength{\fboxsep}{1pt}%
\begin{picture}(7200.00,5040.00)%
    \gplgaddtomacro\gplbacktext{%
      \csname LTb\endcsname%
      \put(1078,704){\makebox(0,0)[r]{\strut{}$-0.12$}}%
      \put(1078,1383){\makebox(0,0)[r]{\strut{}$-0.1$}}%
      \put(1078,2061){\makebox(0,0)[r]{\strut{}$-0.08$}}%
      \put(1078,2740){\makebox(0,0)[r]{\strut{}$-0.06$}}%
      \put(1078,3418){\makebox(0,0)[r]{\strut{}$-0.04$}}%
      \put(1078,4097){\makebox(0,0)[r]{\strut{}$-0.02$}}%
      \put(1078,4775){\makebox(0,0)[r]{\strut{}$0$}}%
      \put(1210,484){\makebox(0,0){\strut{}$0$}}%
      \put(1769,484){\makebox(0,0){\strut{}$1$}}%
      \put(2329,484){\makebox(0,0){\strut{}$2$}}%
      \put(2888,484){\makebox(0,0){\strut{}$3$}}%
      \put(3447,484){\makebox(0,0){\strut{}$4$}}%
      \put(4007,484){\makebox(0,0){\strut{}$5$}}%
      \put(4566,484){\makebox(0,0){\strut{}$6$}}%
      \put(5125,484){\makebox(0,0){\strut{}$7$}}%
      \put(5684,484){\makebox(0,0){\strut{}$8$}}%
      \put(6244,484){\makebox(0,0){\strut{}$9$}}%
      \put(6803,484){\makebox(0,0){\strut{}$10$}}%
    }%
    \gplgaddtomacro\gplfronttext{%
      \csname LTb\endcsname%
      \put(176,2739){\rotatebox{-270}{\makebox(0,0){\strut{}\Large $\Delta [\langle \varphi(x) \rangle$]\normalsize}}}%
      \put(4006,154){\makebox(0,0){\strut{}\Large $x$ \normalsize}}%
    }%
    \gplbacktext
    \put(0,0){\includegraphics{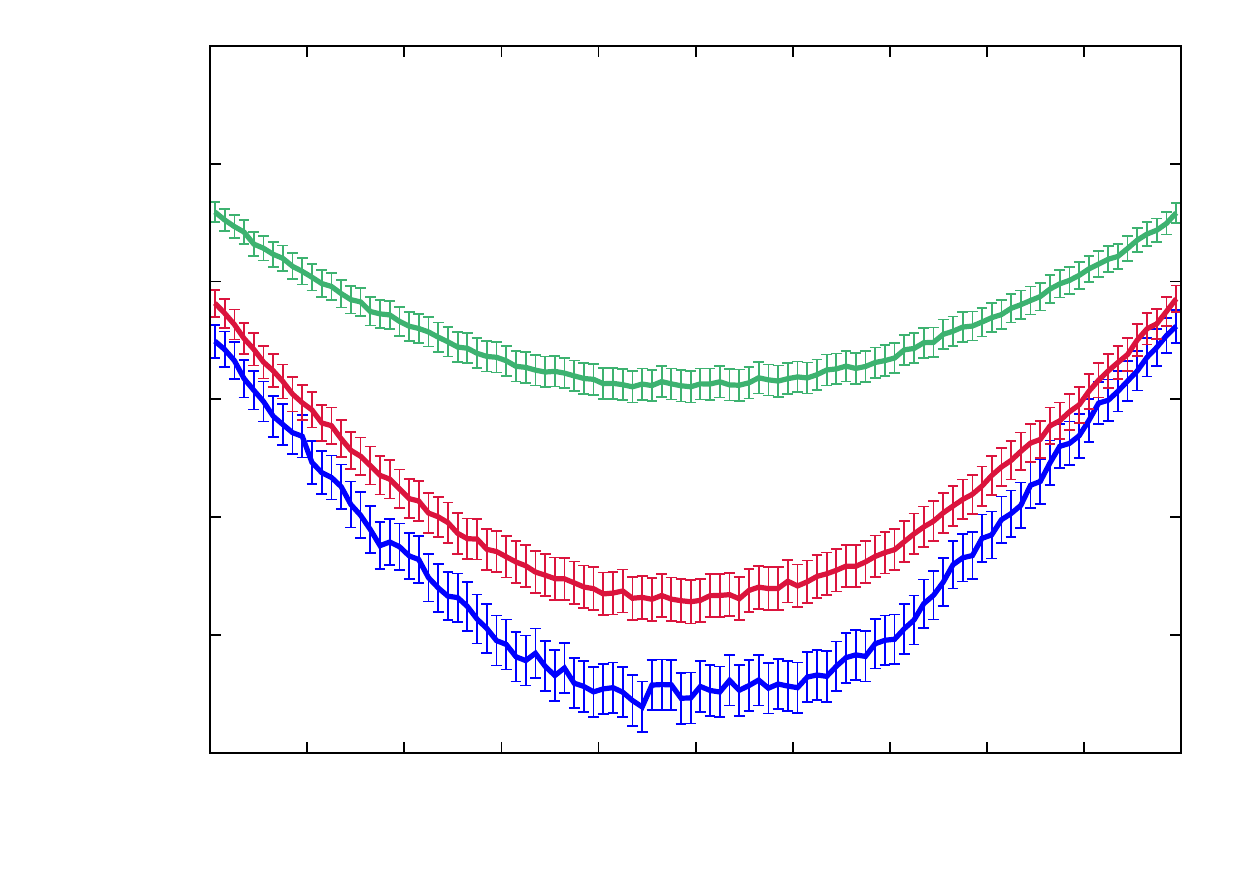}}%
    \gplfronttext
  \end{picture}%
\endgroup

%% file: flux_2C_I_ecart.tex
% GNUPLOT: LaTeX picture with Postscript
\begingroup
  \makeatletter
  \providecommand\color[2][]{%
    \GenericError{(gnuplot) \space\space\space\@spaces}{%
      Package color not loaded in conjunction with
      terminal option `colourtext'%
    }{See the gnuplot documentation for explanation.%
    }{Either use 'blacktext' in gnuplot or load the package
      color.sty in LaTeX.}%
    \renewcommand\color[2][]{}%
  }%
  \providecommand\includegraphics[2][]{%
    \GenericError{(gnuplot) \space\space\space\@spaces}{%
      Package graphicx or graphics not loaded%
    }{See the gnuplot documentation for explanation.%
    }{The gnuplot epslatex terminal needs graphicx.sty or graphics.sty.}%
    \renewcommand\includegraphics[2][]{}%
  }%
  \providecommand\rotatebox[2]{#2}%
  \@ifundefined{ifGPcolor}{%
    \newif\ifGPcolor
    \GPcolorfalse
  }{}%
  \@ifundefined{ifGPblacktext}{%
    \newif\ifGPblacktext
    \GPblacktexttrue
  }{}%
  % define a \g@addto@macro without @ in the name:
  \let\gplgaddtomacro\g@addto@macro
  % define empty templates for all commands taking text:
  \gdef\gplbacktext{}%
  \gdef\gplfronttext{}%
  \makeatother
  \ifGPblacktext
    % no textcolor at all
    \def\colorrgb#1{}%
    \def\colorgray#1{}%
  \else
    % gray or color?
    \ifGPcolor
      \def\colorrgb#1{\color[rgb]{#1}}%
      \def\colorgray#1{\color[gray]{#1}}%
      \expandafter\def\csname LTw\endcsname{\color{white}}%
      \expandafter\def\csname LTb\endcsname{\color{black}}%
      \expandafter\def\csname LTa\endcsname{\color{black}}%
      \expandafter\def\csname LT0\endcsname{\color[rgb]{1,0,0}}%
      \expandafter\def\csname LT1\endcsname{\color[rgb]{0,1,0}}%
      \expandafter\def\csname LT2\endcsname{\color[rgb]{0,0,1}}%
      \expandafter\def\csname LT3\endcsname{\color[rgb]{1,0,1}}%
      \expandafter\def\csname LT4\endcsname{\color[rgb]{0,1,1}}%
      \expandafter\def\csname LT5\endcsname{\color[rgb]{1,1,0}}%
      \expandafter\def\csname LT6\endcsname{\color[rgb]{0,0,0}}%
      \expandafter\def\csname LT7\endcsname{\color[rgb]{1,0.3,0}}%
      \expandafter\def\csname LT8\endcsname{\color[rgb]{0.5,0.5,0.5}}%
    \else
      % gray
      \def\colorrgb#1{\color{black}}%
      \def\colorgray#1{\color[gray]{#1}}%
      \expandafter\def\csname LTw\endcsname{\color{white}}%
      \expandafter\def\csname LTb\endcsname{\color{black}}%
      \expandafter\def\csname LTa\endcsname{\color{black}}%
      \expandafter\def\csname LT0\endcsname{\color{black}}%
      \expandafter\def\csname LT1\endcsname{\color{black}}%
      \expandafter\def\csname LT2\endcsname{\color{black}}%
      \expandafter\def\csname LT3\endcsname{\color{black}}%
      \expandafter\def\csname LT4\endcsname{\color{black}}%
      \expandafter\def\csname LT5\endcsname{\color{black}}%
      \expandafter\def\csname LT6\endcsname{\color{black}}%
      \expandafter\def\csname LT7\endcsname{\color{black}}%
      \expandafter\def\csname LT8\endcsname{\color{black}}%
    \fi
  \fi
    \setlength{\unitlength}{0.0500bp}%
    \ifx\gptboxheight\undefined%
      \newlength{\gptboxheight}%
      \newlength{\gptboxwidth}%
      \newsavebox{\gptboxtext}%
    \fi%
    \setlength{\fboxrule}{0.5pt}%
    \setlength{\fboxsep}{1pt}%
\begin{picture}(7200.00,5040.00)%
    \gplgaddtomacro\gplbacktext{%
      \csname LTb\endcsname%
      \put(1078,704){\makebox(0,0)[r]{\strut{}$-0.3$}}%
      \put(1078,1286){\makebox(0,0)[r]{\strut{}$-0.25$}}%
      \put(1078,1867){\makebox(0,0)[r]{\strut{}$-0.2$}}%
      \put(1078,2449){\makebox(0,0)[r]{\strut{}$-0.15$}}%
      \put(1078,3030){\makebox(0,0)[r]{\strut{}$-0.1$}}%
      \put(1078,3612){\makebox(0,0)[r]{\strut{}$-0.05$}}%
      \put(1078,4193){\makebox(0,0)[r]{\strut{}$0$}}%
      \put(1078,4775){\makebox(0,0)[r]{\strut{}$0.05$}}%
      \put(1210,484){\makebox(0,0){\strut{}$0$}}%
      \put(1769,484){\makebox(0,0){\strut{}$1$}}%
      \put(2329,484){\makebox(0,0){\strut{}$2$}}%
      \put(2888,484){\makebox(0,0){\strut{}$3$}}%
      \put(3447,484){\makebox(0,0){\strut{}$4$}}%
      \put(4007,484){\makebox(0,0){\strut{}$5$}}%
      \put(4566,484){\makebox(0,0){\strut{}$6$}}%
      \put(5125,484){\makebox(0,0){\strut{}$7$}}%
      \put(5684,484){\makebox(0,0){\strut{}$8$}}%
      \put(6244,484){\makebox(0,0){\strut{}$9$}}%
      \put(6803,484){\makebox(0,0){\strut{}$10$}}%
    }%
    \gplgaddtomacro\gplfronttext{%
      \csname LTb\endcsname%
      \put(176,2739){\rotatebox{-270}{\makebox(0,0){\strut{}\Large $\Delta [\langle \varphi(x) \rangle$]\normalsize}}}%
      \put(4006,154){\makebox(0,0){\strut{}\Large $x$ \normalsize}}%
    }%
    \gplbacktext
    \put(0,0){\includegraphics{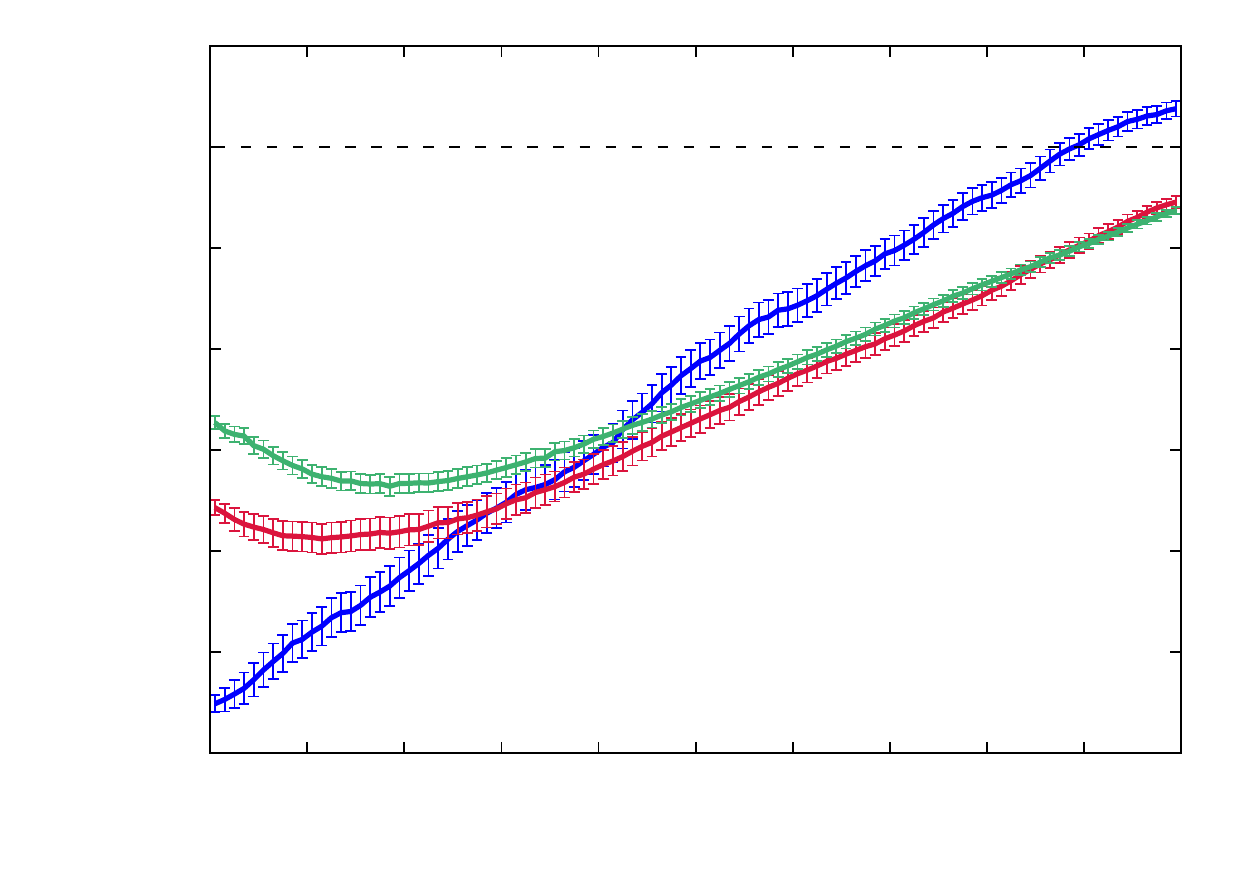}}%
    \gplfronttext
  \end{picture}%
\endgroup

%% file: flux_2C_II_ecart.tex
% GNUPLOT: LaTeX picture with Postscript
\begingroup
  \makeatletter
  \providecommand\color[2][]{%
    \GenericError{(gnuplot) \space\space\space\@spaces}{%
      Package color not loaded in conjunction with
      terminal option `colourtext'%
    }{See the gnuplot documentation for explanation.%
    }{Either use 'blacktext' in gnuplot or load the package
      color.sty in LaTeX.}%
    \renewcommand\color[2][]{}%
  }%
  \providecommand\includegraphics[2][]{%
    \GenericError{(gnuplot) \space\space\space\@spaces}{%
      Package graphicx or graphics not loaded%
    }{See the gnuplot documentation for explanation.%
    }{The gnuplot epslatex terminal needs graphicx.sty or graphics.sty.}%
    \renewcommand\includegraphics[2][]{}%
  }%
  \providecommand\rotatebox[2]{#2}%
  \@ifundefined{ifGPcolor}{%
    \newif\ifGPcolor
    \GPcolorfalse
  }{}%
  \@ifundefined{ifGPblacktext}{%
    \newif\ifGPblacktext
    \GPblacktexttrue
  }{}%
  % define a \g@addto@macro without @ in the name:
  \let\gplgaddtomacro\g@addto@macro
  % define empty templates for all commands taking text:
  \gdef\gplbacktext{}%
  \gdef\gplfronttext{}%
  \makeatother
  \ifGPblacktext
    % no textcolor at all
    \def\colorrgb#1{}%
    \def\colorgray#1{}%
  \else
    % gray or color?
    \ifGPcolor
      \def\colorrgb#1{\color[rgb]{#1}}%
      \def\colorgray#1{\color[gray]{#1}}%
      \expandafter\def\csname LTw\endcsname{\color{white}}%
      \expandafter\def\csname LTb\endcsname{\color{black}}%
      \expandafter\def\csname LTa\endcsname{\color{black}}%
      \expandafter\def\csname LT0\endcsname{\color[rgb]{1,0,0}}%
      \expandafter\def\csname LT1\endcsname{\color[rgb]{0,1,0}}%
      \expandafter\def\csname LT2\endcsname{\color[rgb]{0,0,1}}%
      \expandafter\def\csname LT3\endcsname{\color[rgb]{1,0,1}}%
      \expandafter\def\csname LT4\endcsname{\color[rgb]{0,1,1}}%
      \expandafter\def\csname LT5\endcsname{\color[rgb]{1,1,0}}%
      \expandafter\def\csname LT6\endcsname{\color[rgb]{0,0,0}}%
      \expandafter\def\csname LT7\endcsname{\color[rgb]{1,0.3,0}}%
      \expandafter\def\csname LT8\endcsname{\color[rgb]{0.5,0.5,0.5}}%
    \else
      % gray
      \def\colorrgb#1{\color{black}}%
      \def\colorgray#1{\color[gray]{#1}}%
      \expandafter\def\csname LTw\endcsname{\color{white}}%
      \expandafter\def\csname LTb\endcsname{\color{black}}%
      \expandafter\def\csname LTa\endcsname{\color{black}}%
      \expandafter\def\csname LT0\endcsname{\color{black}}%
      \expandafter\def\csname LT1\endcsname{\color{black}}%
      \expandafter\def\csname LT2\endcsname{\color{black}}%
      \expandafter\def\csname LT3\endcsname{\color{black}}%
      \expandafter\def\csname LT4\endcsname{\color{black}}%
      \expandafter\def\csname LT5\endcsname{\color{black}}%
      \expandafter\def\csname LT6\endcsname{\color{black}}%
      \expandafter\def\csname LT7\endcsname{\color{black}}%
      \expandafter\def\csname LT8\endcsname{\color{black}}%
    \fi
  \fi
    \setlength{\unitlength}{0.0500bp}%
    \ifx\gptboxheight\undefined%
      \newlength{\gptboxheight}%
      \newlength{\gptboxwidth}%
      \newsavebox{\gptboxtext}%
    \fi%
    \setlength{\fboxrule}{0.5pt}%
    \setlength{\fboxsep}{1pt}%
\begin{picture}(7200.00,5040.00)%
    \gplgaddtomacro\gplbacktext{%
      \csname LTb\endcsname%
      \put(1078,704){\makebox(0,0)[r]{\strut{}$-0.25$}}%
      \put(1078,1518){\makebox(0,0)[r]{\strut{}$-0.2$}}%
      \put(1078,2332){\makebox(0,0)[r]{\strut{}$-0.15$}}%
      \put(1078,3147){\makebox(0,0)[r]{\strut{}$-0.1$}}%
      \put(1078,3961){\makebox(0,0)[r]{\strut{}$-0.05$}}%
      \put(1078,4775){\makebox(0,0)[r]{\strut{}$0$}}%
      \put(1210,484){\makebox(0,0){\strut{}$0$}}%
      \put(1769,484){\makebox(0,0){\strut{}$1$}}%
      \put(2329,484){\makebox(0,0){\strut{}$2$}}%
      \put(2888,484){\makebox(0,0){\strut{}$3$}}%
      \put(3447,484){\makebox(0,0){\strut{}$4$}}%
      \put(4007,484){\makebox(0,0){\strut{}$5$}}%
      \put(4566,484){\makebox(0,0){\strut{}$6$}}%
      \put(5125,484){\makebox(0,0){\strut{}$7$}}%
      \put(5684,484){\makebox(0,0){\strut{}$8$}}%
      \put(6244,484){\makebox(0,0){\strut{}$9$}}%
      \put(6803,484){\makebox(0,0){\strut{}$10$}}%
    }%
    \gplgaddtomacro\gplfronttext{%
      \csname LTb\endcsname%
      \put(176,2739){\rotatebox{-270}{\makebox(0,0){\strut{}\Large $\Delta [\langle \varphi(x) \rangle$]\normalsize}}}%
      \put(4006,154){\makebox(0,0){\strut{}\Large $x$ \normalsize}}%
    }%
    \gplbacktext
    \put(0,0){\includegraphics{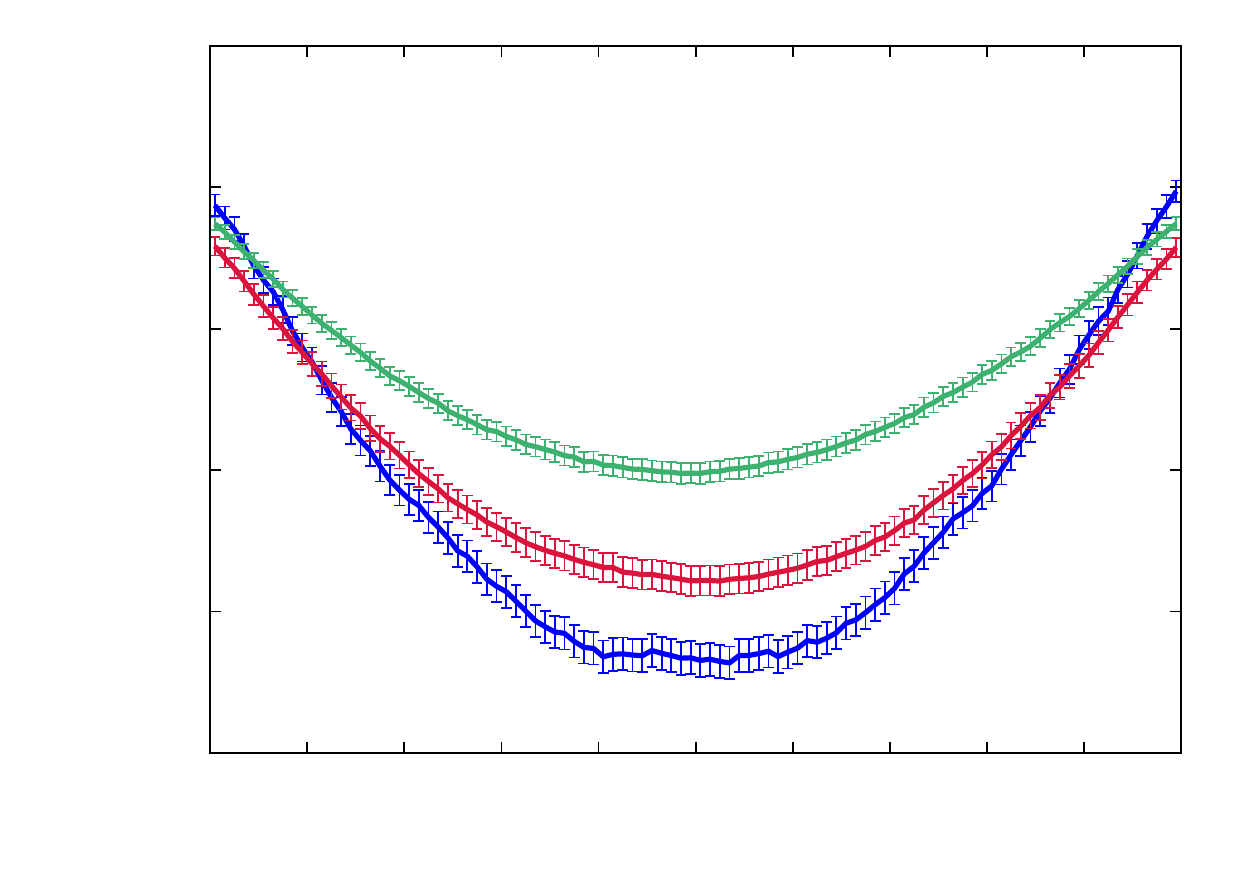}}%
    \gplfronttext
  \end{picture}%
\endgroup

%% file: flux_3A_I_ecart.tex
% GNUPLOT: LaTeX picture with Postscript
\begingroup
  \makeatletter
  \providecommand\color[2][]{%
    \GenericError{(gnuplot) \space\space\space\@spaces}{%
      Package color not loaded in conjunction with
      terminal option `colourtext'%
    }{See the gnuplot documentation for explanation.%
    }{Either use 'blacktext' in gnuplot or load the package
      color.sty in LaTeX.}%
    \renewcommand\color[2][]{}%
  }%
  \providecommand\includegraphics[2][]{%
    \GenericError{(gnuplot) \space\space\space\@spaces}{%
      Package graphicx or graphics not loaded%
    }{See the gnuplot documentation for explanation.%
    }{The gnuplot epslatex terminal needs graphicx.sty or graphics.sty.}%
    \renewcommand\includegraphics[2][]{}%
  }%
  \providecommand\rotatebox[2]{#2}%
  \@ifundefined{ifGPcolor}{%
    \newif\ifGPcolor
    \GPcolorfalse
  }{}%
  \@ifundefined{ifGPblacktext}{%
    \newif\ifGPblacktext
    \GPblacktexttrue
  }{}%
  % define a \g@addto@macro without @ in the name:
  \let\gplgaddtomacro\g@addto@macro
  % define empty templates for all commands taking text:
  \gdef\gplbacktext{}%
  \gdef\gplfronttext{}%
  \makeatother
  \ifGPblacktext
    % no textcolor at all
    \def\colorrgb#1{}%
    \def\colorgray#1{}%
  \else
    % gray or color?
    \ifGPcolor
      \def\colorrgb#1{\color[rgb]{#1}}%
      \def\colorgray#1{\color[gray]{#1}}%
      \expandafter\def\csname LTw\endcsname{\color{white}}%
      \expandafter\def\csname LTb\endcsname{\color{black}}%
      \expandafter\def\csname LTa\endcsname{\color{black}}%
      \expandafter\def\csname LT0\endcsname{\color[rgb]{1,0,0}}%
      \expandafter\def\csname LT1\endcsname{\color[rgb]{0,1,0}}%
      \expandafter\def\csname LT2\endcsname{\color[rgb]{0,0,1}}%
      \expandafter\def\csname LT3\endcsname{\color[rgb]{1,0,1}}%
      \expandafter\def\csname LT4\endcsname{\color[rgb]{0,1,1}}%
      \expandafter\def\csname LT5\endcsname{\color[rgb]{1,1,0}}%
      \expandafter\def\csname LT6\endcsname{\color[rgb]{0,0,0}}%
      \expandafter\def\csname LT7\endcsname{\color[rgb]{1,0.3,0}}%
      \expandafter\def\csname LT8\endcsname{\color[rgb]{0.5,0.5,0.5}}%
    \else
      % gray
      \def\colorrgb#1{\color{black}}%
      \def\colorgray#1{\color[gray]{#1}}%
      \expandafter\def\csname LTw\endcsname{\color{white}}%
      \expandafter\def\csname LTb\endcsname{\color{black}}%
      \expandafter\def\csname LTa\endcsname{\color{black}}%
      \expandafter\def\csname LT0\endcsname{\color{black}}%
      \expandafter\def\csname LT1\endcsname{\color{black}}%
      \expandafter\def\csname LT2\endcsname{\color{black}}%
      \expandafter\def\csname LT3\endcsname{\color{black}}%
      \expandafter\def\csname LT4\endcsname{\color{black}}%
      \expandafter\def\csname LT5\endcsname{\color{black}}%
      \expandafter\def\csname LT6\endcsname{\color{black}}%
      \expandafter\def\csname LT7\endcsname{\color{black}}%
      \expandafter\def\csname LT8\endcsname{\color{black}}%
    \fi
  \fi
    \setlength{\unitlength}{0.0500bp}%
    \ifx\gptboxheight\undefined%
      \newlength{\gptboxheight}%
      \newlength{\gptboxwidth}%
      \newsavebox{\gptboxtext}%
    \fi%
    \setlength{\fboxrule}{0.5pt}%
    \setlength{\fboxsep}{1pt}%
\begin{picture}(7200.00,5040.00)%
    \gplgaddtomacro\gplbacktext{%
      \csname LTb\endcsname%
      \put(1078,704){\makebox(0,0)[r]{\strut{}$-0.2$}}%
      \put(1078,1213){\makebox(0,0)[r]{\strut{}$-0.15$}}%
      \put(1078,1722){\makebox(0,0)[r]{\strut{}$-0.1$}}%
      \put(1078,2231){\makebox(0,0)[r]{\strut{}$-0.05$}}%
      \put(1078,2740){\makebox(0,0)[r]{\strut{}$0$}}%
      \put(1078,3248){\makebox(0,0)[r]{\strut{}$0.05$}}%
      \put(1078,3757){\makebox(0,0)[r]{\strut{}$0.1$}}%
      \put(1078,4266){\makebox(0,0)[r]{\strut{}$0.15$}}%
      \put(1078,4775){\makebox(0,0)[r]{\strut{}$0.2$}}%
      \put(1210,484){\makebox(0,0){\strut{}$0$}}%
      \put(1769,484){\makebox(0,0){\strut{}$1$}}%
      \put(2329,484){\makebox(0,0){\strut{}$2$}}%
      \put(2888,484){\makebox(0,0){\strut{}$3$}}%
      \put(3447,484){\makebox(0,0){\strut{}$4$}}%
      \put(4007,484){\makebox(0,0){\strut{}$5$}}%
      \put(4566,484){\makebox(0,0){\strut{}$6$}}%
      \put(5125,484){\makebox(0,0){\strut{}$7$}}%
      \put(5684,484){\makebox(0,0){\strut{}$8$}}%
      \put(6244,484){\makebox(0,0){\strut{}$9$}}%
      \put(6803,484){\makebox(0,0){\strut{}$10$}}%
    }%
    \gplgaddtomacro\gplfronttext{%
      \csname LTb\endcsname%
      \put(176,2739){\rotatebox{-270}{\makebox(0,0){\strut{}\Large $\Delta [\langle \varphi(x) \rangle$]\normalsize}}}%
      \put(4006,154){\makebox(0,0){\strut{}\Large $x$ \normalsize}}%
    }%
    \gplbacktext
    \put(0,0){\includegraphics{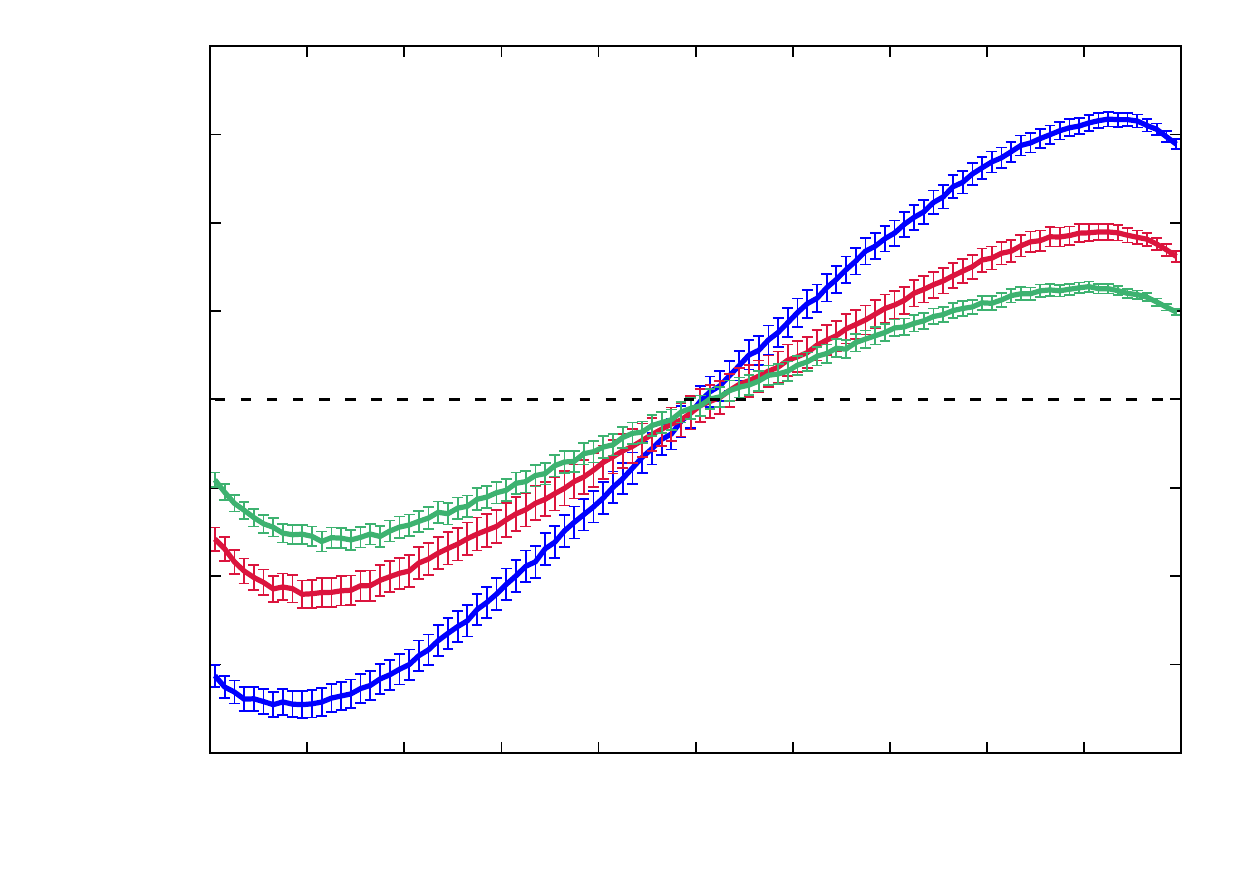}}%
    \gplfronttext
  \end{picture}%
\endgroup

%% file: flux_3A_II_ecart.tex
% GNUPLOT: LaTeX picture with Postscript
\begingroup
  \makeatletter
  \providecommand\color[2][]{%
    \GenericError{(gnuplot) \space\space\space\@spaces}{%
      Package color not loaded in conjunction with
      terminal option `colourtext'%
    }{See the gnuplot documentation for explanation.%
    }{Either use 'blacktext' in gnuplot or load the package
      color.sty in LaTeX.}%
    \renewcommand\color[2][]{}%
  }%
  \providecommand\includegraphics[2][]{%
    \GenericError{(gnuplot) \space\space\space\@spaces}{%
      Package graphicx or graphics not loaded%
    }{See the gnuplot documentation for explanation.%
    }{The gnuplot epslatex terminal needs graphicx.sty or graphics.sty.}%
    \renewcommand\includegraphics[2][]{}%
  }%
  \providecommand\rotatebox[2]{#2}%
  \@ifundefined{ifGPcolor}{%
    \newif\ifGPcolor
    \GPcolorfalse
  }{}%
  \@ifundefined{ifGPblacktext}{%
    \newif\ifGPblacktext
    \GPblacktexttrue
  }{}%
  % define a \g@addto@macro without @ in the name:
  \let\gplgaddtomacro\g@addto@macro
  % define empty templates for all commands taking text:
  \gdef\gplbacktext{}%
  \gdef\gplfronttext{}%
  \makeatother
  \ifGPblacktext
    % no textcolor at all
    \def\colorrgb#1{}%
    \def\colorgray#1{}%
  \else
    % gray or color?
    \ifGPcolor
      \def\colorrgb#1{\color[rgb]{#1}}%
      \def\colorgray#1{\color[gray]{#1}}%
      \expandafter\def\csname LTw\endcsname{\color{white}}%
      \expandafter\def\csname LTb\endcsname{\color{black}}%
      \expandafter\def\csname LTa\endcsname{\color{black}}%
      \expandafter\def\csname LT0\endcsname{\color[rgb]{1,0,0}}%
      \expandafter\def\csname LT1\endcsname{\color[rgb]{0,1,0}}%
      \expandafter\def\csname LT2\endcsname{\color[rgb]{0,0,1}}%
      \expandafter\def\csname LT3\endcsname{\color[rgb]{1,0,1}}%
      \expandafter\def\csname LT4\endcsname{\color[rgb]{0,1,1}}%
      \expandafter\def\csname LT5\endcsname{\color[rgb]{1,1,0}}%
      \expandafter\def\csname LT6\endcsname{\color[rgb]{0,0,0}}%
      \expandafter\def\csname LT7\endcsname{\color[rgb]{1,0.3,0}}%
      \expandafter\def\csname LT8\endcsname{\color[rgb]{0.5,0.5,0.5}}%
    \else
      % gray
      \def\colorrgb#1{\color{black}}%
      \def\colorgray#1{\color[gray]{#1}}%
      \expandafter\def\csname LTw\endcsname{\color{white}}%
      \expandafter\def\csname LTb\endcsname{\color{black}}%
      \expandafter\def\csname LTa\endcsname{\color{black}}%
      \expandafter\def\csname LT0\endcsname{\color{black}}%
      \expandafter\def\csname LT1\endcsname{\color{black}}%
      \expandafter\def\csname LT2\endcsname{\color{black}}%
      \expandafter\def\csname LT3\endcsname{\color{black}}%
      \expandafter\def\csname LT4\endcsname{\color{black}}%
      \expandafter\def\csname LT5\endcsname{\color{black}}%
      \expandafter\def\csname LT6\endcsname{\color{black}}%
      \expandafter\def\csname LT7\endcsname{\color{black}}%
      \expandafter\def\csname LT8\endcsname{\color{black}}%
    \fi
  \fi
    \setlength{\unitlength}{0.0500bp}%
    \ifx\gptboxheight\undefined%
      \newlength{\gptboxheight}%
      \newlength{\gptboxwidth}%
      \newsavebox{\gptboxtext}%
    \fi%
    \setlength{\fboxrule}{0.5pt}%
    \setlength{\fboxsep}{1pt}%
\begin{picture}(7200.00,5040.00)%
    \gplgaddtomacro\gplbacktext{%
      \csname LTb\endcsname%
      \put(946,704){\makebox(0,0)[r]{\strut{}$-1.4$}}%
      \put(946,1286){\makebox(0,0)[r]{\strut{}$-1.2$}}%
      \put(946,1867){\makebox(0,0)[r]{\strut{}$-1$}}%
      \put(946,2449){\makebox(0,0)[r]{\strut{}$-0.8$}}%
      \put(946,3030){\makebox(0,0)[r]{\strut{}$-0.6$}}%
      \put(946,3612){\makebox(0,0)[r]{\strut{}$-0.4$}}%
      \put(946,4193){\makebox(0,0)[r]{\strut{}$-0.2$}}%
      \put(946,4775){\makebox(0,0)[r]{\strut{}$0$}}%
      \put(1078,484){\makebox(0,0){\strut{}$0$}}%
      \put(1651,484){\makebox(0,0){\strut{}$1$}}%
      \put(2223,484){\makebox(0,0){\strut{}$2$}}%
      \put(2796,484){\makebox(0,0){\strut{}$3$}}%
      \put(3368,484){\makebox(0,0){\strut{}$4$}}%
      \put(3941,484){\makebox(0,0){\strut{}$5$}}%
      \put(4513,484){\makebox(0,0){\strut{}$6$}}%
      \put(5086,484){\makebox(0,0){\strut{}$7$}}%
      \put(5658,484){\makebox(0,0){\strut{}$8$}}%
      \put(6231,484){\makebox(0,0){\strut{}$9$}}%
      \put(6803,484){\makebox(0,0){\strut{}$10$}}%
    }%
    \gplgaddtomacro\gplfronttext{%
      \csname LTb\endcsname%
      \put(176,2739){\rotatebox{-270}{\makebox(0,0){\strut{}\Large $\Delta [\langle \varphi(x) \rangle$]\normalsize}}}%
      \put(3940,154){\makebox(0,0){\strut{}\Large $x$ \normalsize}}%
    }%
    \gplbacktext
    \put(0,0){\includegraphics{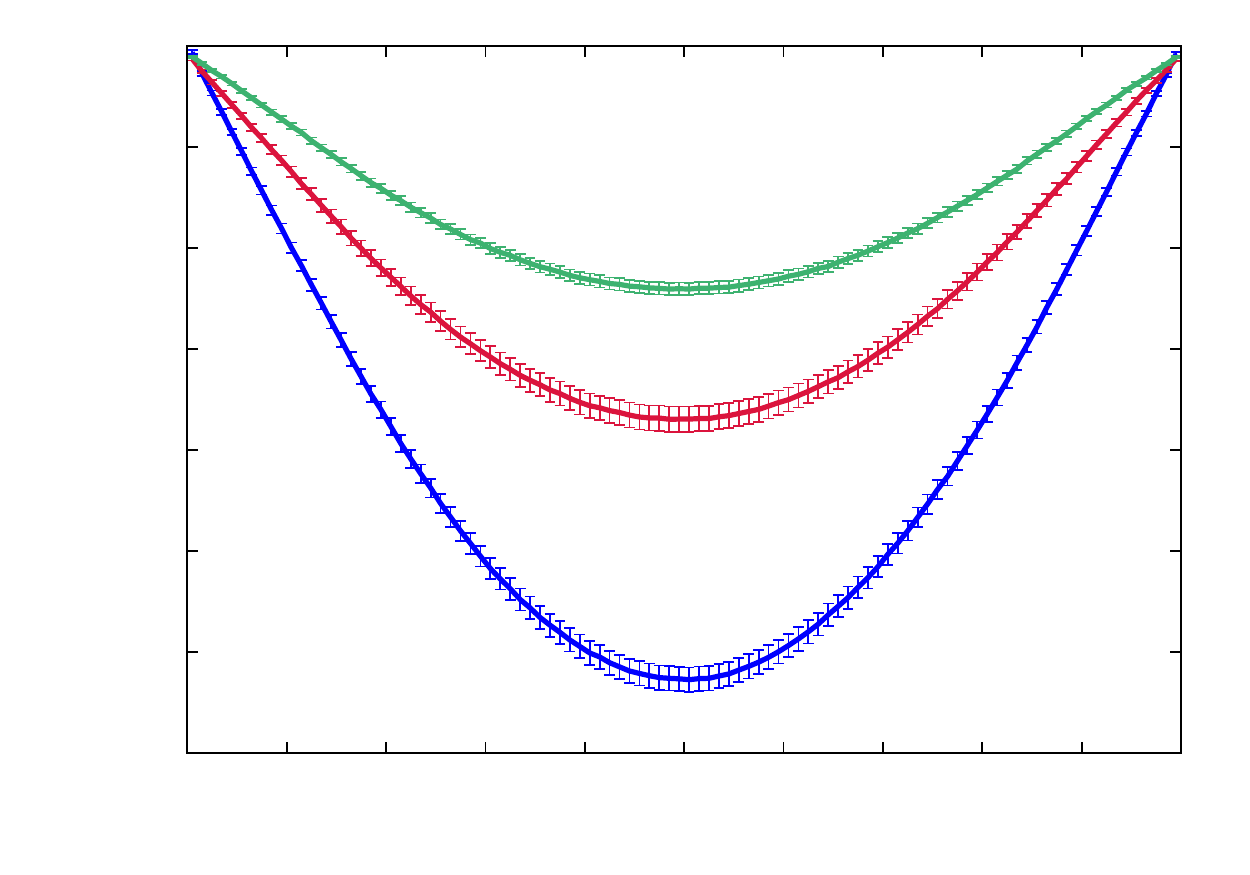}}%
    \gplfronttext
  \end{picture}%
\endgroup

%% file: flux_3B_I_ecart.tex
% GNUPLOT: LaTeX picture with Postscript
\begingroup
  \makeatletter
  \providecommand\color[2][]{%
    \GenericError{(gnuplot) \space\space\space\@spaces}{%
      Package color not loaded in conjunction with
      terminal option `colourtext'%
    }{See the gnuplot documentation for explanation.%
    }{Either use 'blacktext' in gnuplot or load the package
      color.sty in LaTeX.}%
    \renewcommand\color[2][]{}%
  }%
  \providecommand\includegraphics[2][]{%
    \GenericError{(gnuplot) \space\space\space\@spaces}{%
      Package graphicx or graphics not loaded%
    }{See the gnuplot documentation for explanation.%
    }{The gnuplot epslatex terminal needs graphicx.sty or graphics.sty.}%
    \renewcommand\includegraphics[2][]{}%
  }%
  \providecommand\rotatebox[2]{#2}%
  \@ifundefined{ifGPcolor}{%
    \newif\ifGPcolor
    \GPcolorfalse
  }{}%
  \@ifundefined{ifGPblacktext}{%
    \newif\ifGPblacktext
    \GPblacktexttrue
  }{}%
  % define a \g@addto@macro without @ in the name:
  \let\gplgaddtomacro\g@addto@macro
  % define empty templates for all commands taking text:
  \gdef\gplbacktext{}%
  \gdef\gplfronttext{}%
  \makeatother
  \ifGPblacktext
    % no textcolor at all
    \def\colorrgb#1{}%
    \def\colorgray#1{}%
  \else
    % gray or color?
    \ifGPcolor
      \def\colorrgb#1{\color[rgb]{#1}}%
      \def\colorgray#1{\color[gray]{#1}}%
      \expandafter\def\csname LTw\endcsname{\color{white}}%
      \expandafter\def\csname LTb\endcsname{\color{black}}%
      \expandafter\def\csname LTa\endcsname{\color{black}}%
      \expandafter\def\csname LT0\endcsname{\color[rgb]{1,0,0}}%
      \expandafter\def\csname LT1\endcsname{\color[rgb]{0,1,0}}%
      \expandafter\def\csname LT2\endcsname{\color[rgb]{0,0,1}}%
      \expandafter\def\csname LT3\endcsname{\color[rgb]{1,0,1}}%
      \expandafter\def\csname LT4\endcsname{\color[rgb]{0,1,1}}%
      \expandafter\def\csname LT5\endcsname{\color[rgb]{1,1,0}}%
      \expandafter\def\csname LT6\endcsname{\color[rgb]{0,0,0}}%
      \expandafter\def\csname LT7\endcsname{\color[rgb]{1,0.3,0}}%
      \expandafter\def\csname LT8\endcsname{\color[rgb]{0.5,0.5,0.5}}%
    \else
      % gray
      \def\colorrgb#1{\color{black}}%
      \def\colorgray#1{\color[gray]{#1}}%
      \expandafter\def\csname LTw\endcsname{\color{white}}%
      \expandafter\def\csname LTb\endcsname{\color{black}}%
      \expandafter\def\csname LTa\endcsname{\color{black}}%
      \expandafter\def\csname LT0\endcsname{\color{black}}%
      \expandafter\def\csname LT1\endcsname{\color{black}}%
      \expandafter\def\csname LT2\endcsname{\color{black}}%
      \expandafter\def\csname LT3\endcsname{\color{black}}%
      \expandafter\def\csname LT4\endcsname{\color{black}}%
      \expandafter\def\csname LT5\endcsname{\color{black}}%
      \expandafter\def\csname LT6\endcsname{\color{black}}%
      \expandafter\def\csname LT7\endcsname{\color{black}}%
      \expandafter\def\csname LT8\endcsname{\color{black}}%
    \fi
  \fi
    \setlength{\unitlength}{0.0500bp}%
    \ifx\gptboxheight\undefined%
      \newlength{\gptboxheight}%
      \newlength{\gptboxwidth}%
      \newsavebox{\gptboxtext}%
    \fi%
    \setlength{\fboxrule}{0.5pt}%
    \setlength{\fboxsep}{1pt}%
\begin{picture}(7200.00,5040.00)%
    \gplgaddtomacro\gplbacktext{%
      \csname LTb\endcsname%
      \put(1210,704){\makebox(0,0)[r]{\strut{}$-0.04$}}%
      \put(1210,1156){\makebox(0,0)[r]{\strut{}$-0.035$}}%
      \put(1210,1609){\makebox(0,0)[r]{\strut{}$-0.03$}}%
      \put(1210,2061){\makebox(0,0)[r]{\strut{}$-0.025$}}%
      \put(1210,2513){\makebox(0,0)[r]{\strut{}$-0.02$}}%
      \put(1210,2966){\makebox(0,0)[r]{\strut{}$-0.015$}}%
      \put(1210,3418){\makebox(0,0)[r]{\strut{}$-0.01$}}%
      \put(1210,3870){\makebox(0,0)[r]{\strut{}$-0.005$}}%
      \put(1210,4323){\makebox(0,0)[r]{\strut{}$0$}}%
      \put(1210,4775){\makebox(0,0)[r]{\strut{}$0.005$}}%
      \put(1342,484){\makebox(0,0){\strut{}$0$}}%
      \put(1888,484){\makebox(0,0){\strut{}$1$}}%
      \put(2434,484){\makebox(0,0){\strut{}$2$}}%
      \put(2980,484){\makebox(0,0){\strut{}$3$}}%
      \put(3526,484){\makebox(0,0){\strut{}$4$}}%
      \put(4073,484){\makebox(0,0){\strut{}$5$}}%
      \put(4619,484){\makebox(0,0){\strut{}$6$}}%
      \put(5165,484){\makebox(0,0){\strut{}$7$}}%
      \put(5711,484){\makebox(0,0){\strut{}$8$}}%
      \put(6257,484){\makebox(0,0){\strut{}$9$}}%
      \put(6803,484){\makebox(0,0){\strut{}$10$}}%
    }%
    \gplgaddtomacro\gplfronttext{%
      \csname LTb\endcsname%
      \put(176,2739){\rotatebox{-270}{\makebox(0,0){\strut{}\Large $\Delta [\langle \varphi(x) \rangle$]\normalsize}}}%
      \put(4072,154){\makebox(0,0){\strut{}\Large $x$ \normalsize}}%
    }%
    \gplbacktext
    \put(0,0){\includegraphics{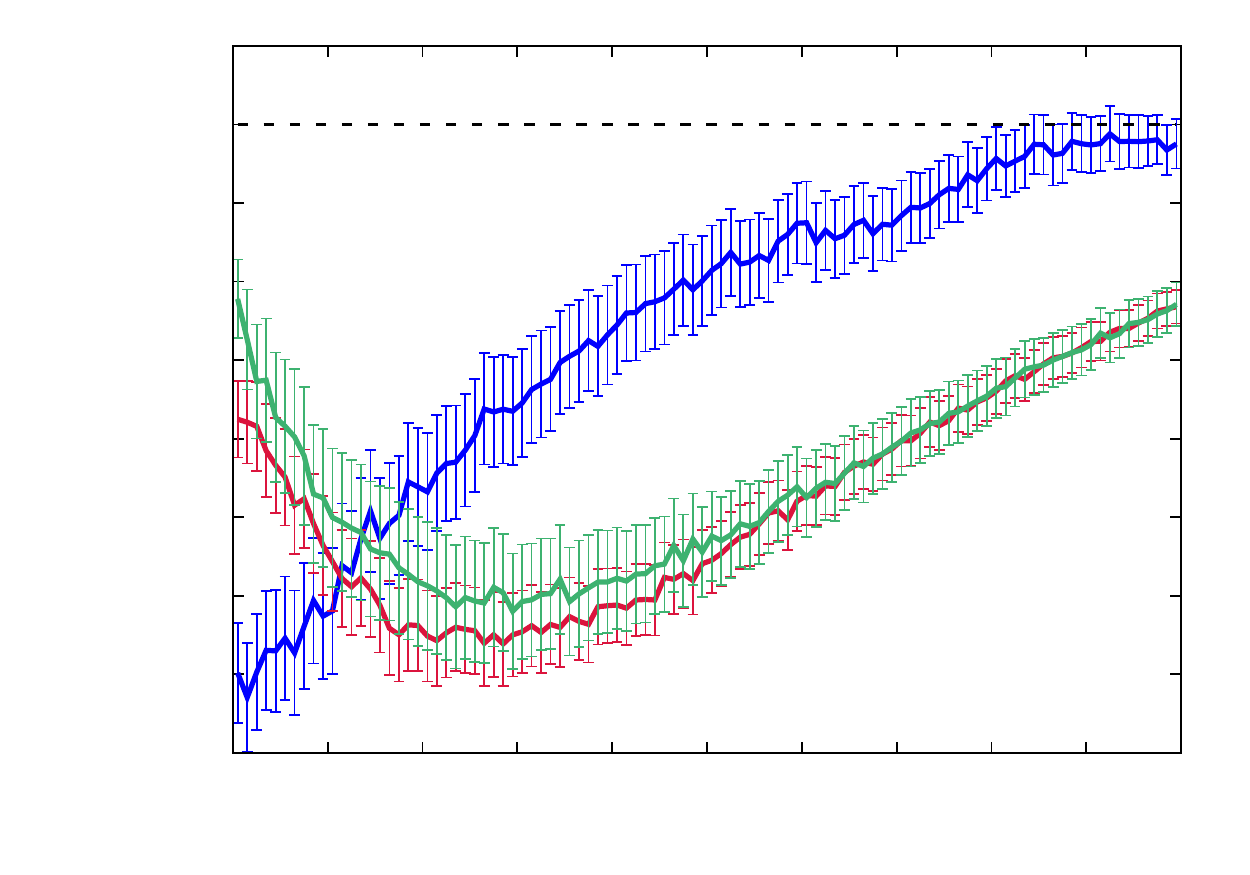}}%
    \gplfronttext
  \end{picture}%
\endgroup

%% file: flux_3B_II_ecart.tex
% GNUPLOT: LaTeX picture with Postscript
\begingroup
  \makeatletter
  \providecommand\color[2][]{%
    \GenericError{(gnuplot) \space\space\space\@spaces}{%
      Package color not loaded in conjunction with
      terminal option `colourtext'%
    }{See the gnuplot documentation for explanation.%
    }{Either use 'blacktext' in gnuplot or load the package
      color.sty in LaTeX.}%
    \renewcommand\color[2][]{}%
  }%
  \providecommand\includegraphics[2][]{%
    \GenericError{(gnuplot) \space\space\space\@spaces}{%
      Package graphicx or graphics not loaded%
    }{See the gnuplot documentation for explanation.%
    }{The gnuplot epslatex terminal needs graphicx.sty or graphics.sty.}%
    \renewcommand\includegraphics[2][]{}%
  }%
  \providecommand\rotatebox[2]{#2}%
  \@ifundefined{ifGPcolor}{%
    \newif\ifGPcolor
    \GPcolorfalse
  }{}%
  \@ifundefined{ifGPblacktext}{%
    \newif\ifGPblacktext
    \GPblacktexttrue
  }{}%
  % define a \g@addto@macro without @ in the name:
  \let\gplgaddtomacro\g@addto@macro
  % define empty templates for all commands taking text:
  \gdef\gplbacktext{}%
  \gdef\gplfronttext{}%
  \makeatother
  \ifGPblacktext
    % no textcolor at all
    \def\colorrgb#1{}%
    \def\colorgray#1{}%
  \else
    % gray or color?
    \ifGPcolor
      \def\colorrgb#1{\color[rgb]{#1}}%
      \def\colorgray#1{\color[gray]{#1}}%
      \expandafter\def\csname LTw\endcsname{\color{white}}%
      \expandafter\def\csname LTb\endcsname{\color{black}}%
      \expandafter\def\csname LTa\endcsname{\color{black}}%
      \expandafter\def\csname LT0\endcsname{\color[rgb]{1,0,0}}%
      \expandafter\def\csname LT1\endcsname{\color[rgb]{0,1,0}}%
      \expandafter\def\csname LT2\endcsname{\color[rgb]{0,0,1}}%
      \expandafter\def\csname LT3\endcsname{\color[rgb]{1,0,1}}%
      \expandafter\def\csname LT4\endcsname{\color[rgb]{0,1,1}}%
      \expandafter\def\csname LT5\endcsname{\color[rgb]{1,1,0}}%
      \expandafter\def\csname LT6\endcsname{\color[rgb]{0,0,0}}%
      \expandafter\def\csname LT7\endcsname{\color[rgb]{1,0.3,0}}%
      \expandafter\def\csname LT8\endcsname{\color[rgb]{0.5,0.5,0.5}}%
    \else
      % gray
      \def\colorrgb#1{\color{black}}%
      \def\colorgray#1{\color[gray]{#1}}%
      \expandafter\def\csname LTw\endcsname{\color{white}}%
      \expandafter\def\csname LTb\endcsname{\color{black}}%
      \expandafter\def\csname LTa\endcsname{\color{black}}%
      \expandafter\def\csname LT0\endcsname{\color{black}}%
      \expandafter\def\csname LT1\endcsname{\color{black}}%
      \expandafter\def\csname LT2\endcsname{\color{black}}%
      \expandafter\def\csname LT3\endcsname{\color{black}}%
      \expandafter\def\csname LT4\endcsname{\color{black}}%
      \expandafter\def\csname LT5\endcsname{\color{black}}%
      \expandafter\def\csname LT6\endcsname{\color{black}}%
      \expandafter\def\csname LT7\endcsname{\color{black}}%
      \expandafter\def\csname LT8\endcsname{\color{black}}%
    \fi
  \fi
    \setlength{\unitlength}{0.0500bp}%
    \ifx\gptboxheight\undefined%
      \newlength{\gptboxheight}%
      \newlength{\gptboxwidth}%
      \newsavebox{\gptboxtext}%
    \fi%
    \setlength{\fboxrule}{0.5pt}%
    \setlength{\fboxsep}{1pt}%
\begin{picture}(7200.00,5040.00)%
    \gplgaddtomacro\gplbacktext{%
      \csname LTb\endcsname%
      \put(1078,704){\makebox(0,0)[r]{\strut{}$-0.06$}}%
      \put(1078,1383){\makebox(0,0)[r]{\strut{}$-0.05$}}%
      \put(1078,2061){\makebox(0,0)[r]{\strut{}$-0.04$}}%
      \put(1078,2740){\makebox(0,0)[r]{\strut{}$-0.03$}}%
      \put(1078,3418){\makebox(0,0)[r]{\strut{}$-0.02$}}%
      \put(1078,4097){\makebox(0,0)[r]{\strut{}$-0.01$}}%
      \put(1078,4775){\makebox(0,0)[r]{\strut{}$0$}}%
      \put(1210,484){\makebox(0,0){\strut{}$0$}}%
      \put(1769,484){\makebox(0,0){\strut{}$1$}}%
      \put(2329,484){\makebox(0,0){\strut{}$2$}}%
      \put(2888,484){\makebox(0,0){\strut{}$3$}}%
      \put(3447,484){\makebox(0,0){\strut{}$4$}}%
      \put(4007,484){\makebox(0,0){\strut{}$5$}}%
      \put(4566,484){\makebox(0,0){\strut{}$6$}}%
      \put(5125,484){\makebox(0,0){\strut{}$7$}}%
      \put(5684,484){\makebox(0,0){\strut{}$8$}}%
      \put(6244,484){\makebox(0,0){\strut{}$9$}}%
      \put(6803,484){\makebox(0,0){\strut{}$10$}}%
    }%
    \gplgaddtomacro\gplfronttext{%
      \csname LTb\endcsname%
      \put(176,2739){\rotatebox{-270}{\makebox(0,0){\strut{}\Large $\Delta [\langle \varphi(x) \rangle$]\normalsize}}}%
      \put(4006,154){\makebox(0,0){\strut{}\Large $x$ \normalsize}}%
    }%
    \gplbacktext
    \put(0,0){\includegraphics{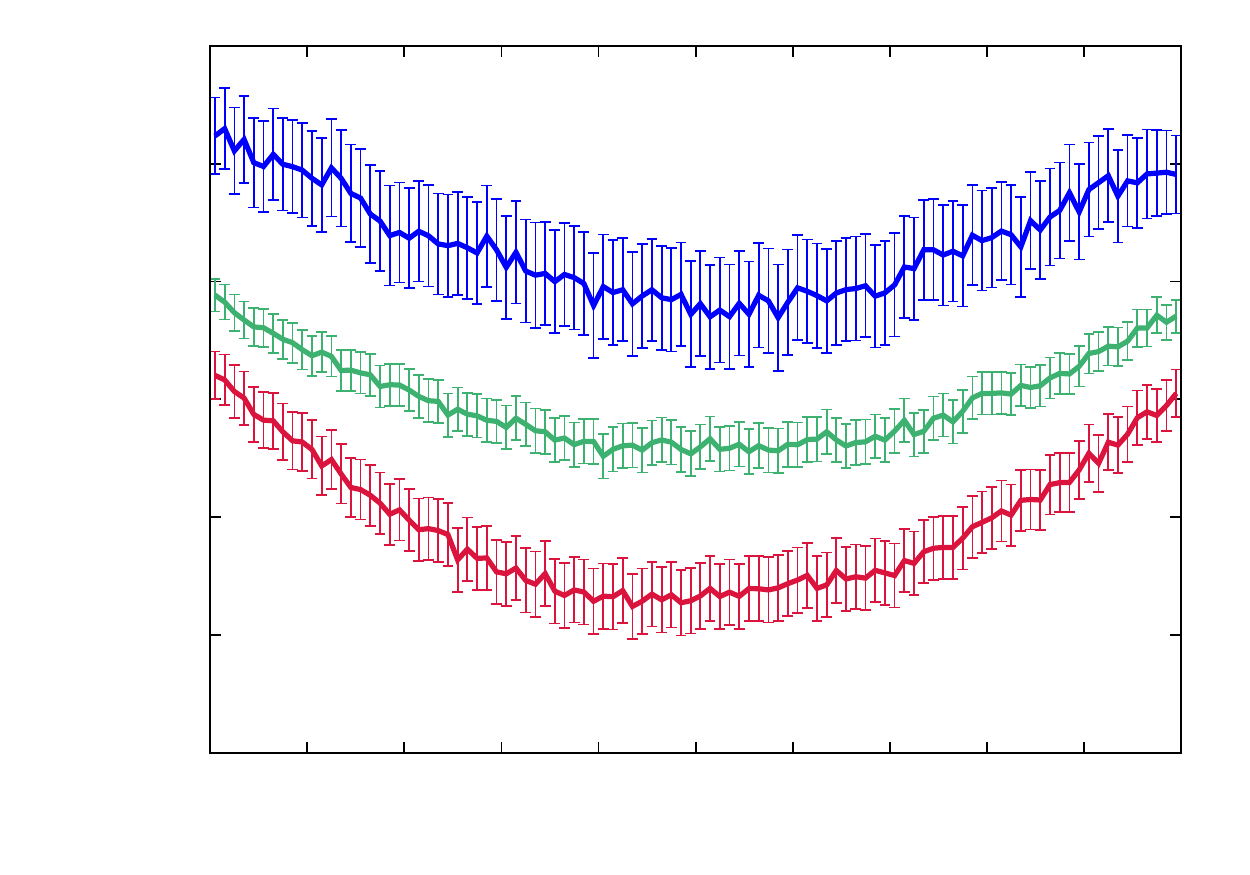}}%
    \gplfronttext
  \end{picture}%
\endgroup

%% file: flux_3C_I_ecart.tex
% GNUPLOT: LaTeX picture with Postscript
\begingroup
  \makeatletter
  \providecommand\color[2][]{%
    \GenericError{(gnuplot) \space\space\space\@spaces}{%
      Package color not loaded in conjunction with
      terminal option `colourtext'%
    }{See the gnuplot documentation for explanation.%
    }{Either use 'blacktext' in gnuplot or load the package
      color.sty in LaTeX.}%
    \renewcommand\color[2][]{}%
  }%
  \providecommand\includegraphics[2][]{%
    \GenericError{(gnuplot) \space\space\space\@spaces}{%
      Package graphicx or graphics not loaded%
    }{See the gnuplot documentation for explanation.%
    }{The gnuplot epslatex terminal needs graphicx.sty or graphics.sty.}%
    \renewcommand\includegraphics[2][]{}%
  }%
  \providecommand\rotatebox[2]{#2}%
  \@ifundefined{ifGPcolor}{%
    \newif\ifGPcolor
    \GPcolorfalse
  }{}%
  \@ifundefined{ifGPblacktext}{%
    \newif\ifGPblacktext
    \GPblacktexttrue
  }{}%
  % define a \g@addto@macro without @ in the name:
  \let\gplgaddtomacro\g@addto@macro
  % define empty templates for all commands taking text:
  \gdef\gplbacktext{}%
  \gdef\gplfronttext{}%
  \makeatother
  \ifGPblacktext
    % no textcolor at all
    \def\colorrgb#1{}%
    \def\colorgray#1{}%
  \else
    % gray or color?
    \ifGPcolor
      \def\colorrgb#1{\color[rgb]{#1}}%
      \def\colorgray#1{\color[gray]{#1}}%
      \expandafter\def\csname LTw\endcsname{\color{white}}%
      \expandafter\def\csname LTb\endcsname{\color{black}}%
      \expandafter\def\csname LTa\endcsname{\color{black}}%
      \expandafter\def\csname LT0\endcsname{\color[rgb]{1,0,0}}%
      \expandafter\def\csname LT1\endcsname{\color[rgb]{0,1,0}}%
      \expandafter\def\csname LT2\endcsname{\color[rgb]{0,0,1}}%
      \expandafter\def\csname LT3\endcsname{\color[rgb]{1,0,1}}%
      \expandafter\def\csname LT4\endcsname{\color[rgb]{0,1,1}}%
      \expandafter\def\csname LT5\endcsname{\color[rgb]{1,1,0}}%
      \expandafter\def\csname LT6\endcsname{\color[rgb]{0,0,0}}%
      \expandafter\def\csname LT7\endcsname{\color[rgb]{1,0.3,0}}%
      \expandafter\def\csname LT8\endcsname{\color[rgb]{0.5,0.5,0.5}}%
    \else
      % gray
      \def\colorrgb#1{\color{black}}%
      \def\colorgray#1{\color[gray]{#1}}%
      \expandafter\def\csname LTw\endcsname{\color{white}}%
      \expandafter\def\csname LTb\endcsname{\color{black}}%
      \expandafter\def\csname LTa\endcsname{\color{black}}%
      \expandafter\def\csname LT0\endcsname{\color{black}}%
      \expandafter\def\csname LT1\endcsname{\color{black}}%
      \expandafter\def\csname LT2\endcsname{\color{black}}%
      \expandafter\def\csname LT3\endcsname{\color{black}}%
      \expandafter\def\csname LT4\endcsname{\color{black}}%
      \expandafter\def\csname LT5\endcsname{\color{black}}%
      \expandafter\def\csname LT6\endcsname{\color{black}}%
      \expandafter\def\csname LT7\endcsname{\color{black}}%
      \expandafter\def\csname LT8\endcsname{\color{black}}%
    \fi
  \fi
    \setlength{\unitlength}{0.0500bp}%
    \ifx\gptboxheight\undefined%
      \newlength{\gptboxheight}%
      \newlength{\gptboxwidth}%
      \newsavebox{\gptboxtext}%
    \fi%
    \setlength{\fboxrule}{0.5pt}%
    \setlength{\fboxsep}{1pt}%
\begin{picture}(7200.00,5040.00)%
    \gplgaddtomacro\gplbacktext{%
      \csname LTb\endcsname%
      \put(1078,704){\makebox(0,0)[r]{\strut{}$-0.3$}}%
      \put(1078,1286){\makebox(0,0)[r]{\strut{}$-0.25$}}%
      \put(1078,1867){\makebox(0,0)[r]{\strut{}$-0.2$}}%
      \put(1078,2449){\makebox(0,0)[r]{\strut{}$-0.15$}}%
      \put(1078,3030){\makebox(0,0)[r]{\strut{}$-0.1$}}%
      \put(1078,3612){\makebox(0,0)[r]{\strut{}$-0.05$}}%
      \put(1078,4193){\makebox(0,0)[r]{\strut{}$0$}}%
      \put(1078,4775){\makebox(0,0)[r]{\strut{}$0.05$}}%
      \put(1210,484){\makebox(0,0){\strut{}$0$}}%
      \put(1769,484){\makebox(0,0){\strut{}$1$}}%
      \put(2329,484){\makebox(0,0){\strut{}$2$}}%
      \put(2888,484){\makebox(0,0){\strut{}$3$}}%
      \put(3447,484){\makebox(0,0){\strut{}$4$}}%
      \put(4007,484){\makebox(0,0){\strut{}$5$}}%
      \put(4566,484){\makebox(0,0){\strut{}$6$}}%
      \put(5125,484){\makebox(0,0){\strut{}$7$}}%
      \put(5684,484){\makebox(0,0){\strut{}$8$}}%
      \put(6244,484){\makebox(0,0){\strut{}$9$}}%
      \put(6803,484){\makebox(0,0){\strut{}$10$}}%
    }%
    \gplgaddtomacro\gplfronttext{%
      \csname LTb\endcsname%
      \put(176,2739){\rotatebox{-270}{\makebox(0,0){\strut{}\Large $\Delta [\langle \varphi(x) \rangle$]\normalsize}}}%
      \put(4006,154){\makebox(0,0){\strut{}\Large $x$ \normalsize}}%
    }%
    \gplbacktext
    \put(0,0){\includegraphics{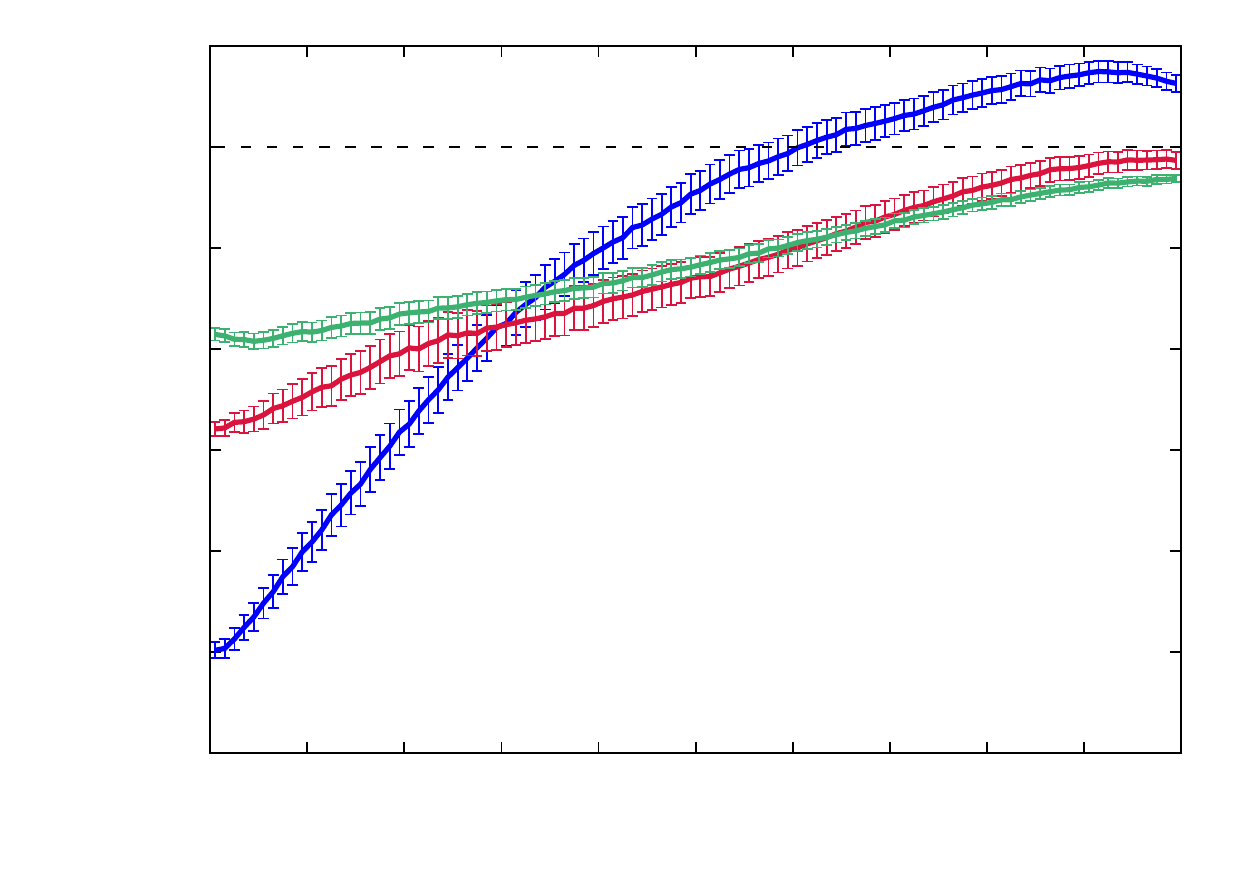}}%
    \gplfronttext
  \end{picture}%
\endgroup

%% file: flux_3C_II_ecart.tex
% GNUPLOT: LaTeX picture with Postscript
\begingroup
  \makeatletter
  \providecommand\color[2][]{%
    \GenericError{(gnuplot) \space\space\space\@spaces}{%
      Package color not loaded in conjunction with
      terminal option `colourtext'%
    }{See the gnuplot documentation for explanation.%
    }{Either use 'blacktext' in gnuplot or load the package
      color.sty in LaTeX.}%
    \renewcommand\color[2][]{}%
  }%
  \providecommand\includegraphics[2][]{%
    \GenericError{(gnuplot) \space\space\space\@spaces}{%
      Package graphicx or graphics not loaded%
    }{See the gnuplot documentation for explanation.%
    }{The gnuplot epslatex terminal needs graphicx.sty or graphics.sty.}%
    \renewcommand\includegraphics[2][]{}%
  }%
  \providecommand\rotatebox[2]{#2}%
  \@ifundefined{ifGPcolor}{%
    \newif\ifGPcolor
    \GPcolorfalse
  }{}%
  \@ifundefined{ifGPblacktext}{%
    \newif\ifGPblacktext
    \GPblacktexttrue
  }{}%
  % define a \g@addto@macro without @ in the name:
  \let\gplgaddtomacro\g@addto@macro
  % define empty templates for all commands taking text:
  \gdef\gplbacktext{}%
  \gdef\gplfronttext{}%
  \makeatother
  \ifGPblacktext
    % no textcolor at all
    \def\colorrgb#1{}%
    \def\colorgray#1{}%
  \else
    % gray or color?
    \ifGPcolor
      \def\colorrgb#1{\color[rgb]{#1}}%
      \def\colorgray#1{\color[gray]{#1}}%
      \expandafter\def\csname LTw\endcsname{\color{white}}%
      \expandafter\def\csname LTb\endcsname{\color{black}}%
      \expandafter\def\csname LTa\endcsname{\color{black}}%
      \expandafter\def\csname LT0\endcsname{\color[rgb]{1,0,0}}%
      \expandafter\def\csname LT1\endcsname{\color[rgb]{0,1,0}}%
      \expandafter\def\csname LT2\endcsname{\color[rgb]{0,0,1}}%
      \expandafter\def\csname LT3\endcsname{\color[rgb]{1,0,1}}%
      \expandafter\def\csname LT4\endcsname{\color[rgb]{0,1,1}}%
      \expandafter\def\csname LT5\endcsname{\color[rgb]{1,1,0}}%
      \expandafter\def\csname LT6\endcsname{\color[rgb]{0,0,0}}%
      \expandafter\def\csname LT7\endcsname{\color[rgb]{1,0.3,0}}%
      \expandafter\def\csname LT8\endcsname{\color[rgb]{0.5,0.5,0.5}}%
    \else
      % gray
      \def\colorrgb#1{\color{black}}%
      \def\colorgray#1{\color[gray]{#1}}%
      \expandafter\def\csname LTw\endcsname{\color{white}}%
      \expandafter\def\csname LTb\endcsname{\color{black}}%
      \expandafter\def\csname LTa\endcsname{\color{black}}%
      \expandafter\def\csname LT0\endcsname{\color{black}}%
      \expandafter\def\csname LT1\endcsname{\color{black}}%
      \expandafter\def\csname LT2\endcsname{\color{black}}%
      \expandafter\def\csname LT3\endcsname{\color{black}}%
      \expandafter\def\csname LT4\endcsname{\color{black}}%
      \expandafter\def\csname LT5\endcsname{\color{black}}%
      \expandafter\def\csname LT6\endcsname{\color{black}}%
      \expandafter\def\csname LT7\endcsname{\color{black}}%
      \expandafter\def\csname LT8\endcsname{\color{black}}%
    \fi
  \fi
    \setlength{\unitlength}{0.0500bp}%
    \ifx\gptboxheight\undefined%
      \newlength{\gptboxheight}%
      \newlength{\gptboxwidth}%
      \newsavebox{\gptboxtext}%
    \fi%
    \setlength{\fboxrule}{0.5pt}%
    \setlength{\fboxsep}{1pt}%
\begin{picture}(7200.00,5040.00)%
    \gplgaddtomacro\gplbacktext{%
      \csname LTb\endcsname%
      \put(1078,704){\makebox(0,0)[r]{\strut{}$-0.16$}}%
      \put(1078,1213){\makebox(0,0)[r]{\strut{}$-0.14$}}%
      \put(1078,1722){\makebox(0,0)[r]{\strut{}$-0.12$}}%
      \put(1078,2231){\makebox(0,0)[r]{\strut{}$-0.1$}}%
      \put(1078,2740){\makebox(0,0)[r]{\strut{}$-0.08$}}%
      \put(1078,3248){\makebox(0,0)[r]{\strut{}$-0.06$}}%
      \put(1078,3757){\makebox(0,0)[r]{\strut{}$-0.04$}}%
      \put(1078,4266){\makebox(0,0)[r]{\strut{}$-0.02$}}%
      \put(1078,4775){\makebox(0,0)[r]{\strut{}$0$}}%
      \put(1210,484){\makebox(0,0){\strut{}$0$}}%
      \put(1769,484){\makebox(0,0){\strut{}$1$}}%
      \put(2329,484){\makebox(0,0){\strut{}$2$}}%
      \put(2888,484){\makebox(0,0){\strut{}$3$}}%
      \put(3447,484){\makebox(0,0){\strut{}$4$}}%
      \put(4007,484){\makebox(0,0){\strut{}$5$}}%
      \put(4566,484){\makebox(0,0){\strut{}$6$}}%
      \put(5125,484){\makebox(0,0){\strut{}$7$}}%
      \put(5684,484){\makebox(0,0){\strut{}$8$}}%
      \put(6244,484){\makebox(0,0){\strut{}$9$}}%
      \put(6803,484){\makebox(0,0){\strut{}$10$}}%
    }%
    \gplgaddtomacro\gplfronttext{%
      \csname LTb\endcsname%
      \put(176,2739){\rotatebox{-270}{\makebox(0,0){\strut{}\Large $\Delta [\langle \varphi(x) \rangle$]\normalsize}}}%
      \put(4006,154){\makebox(0,0){\strut{}\Large $x$ \normalsize}}%
    }%
    \gplbacktext
    \put(0,0){\includegraphics{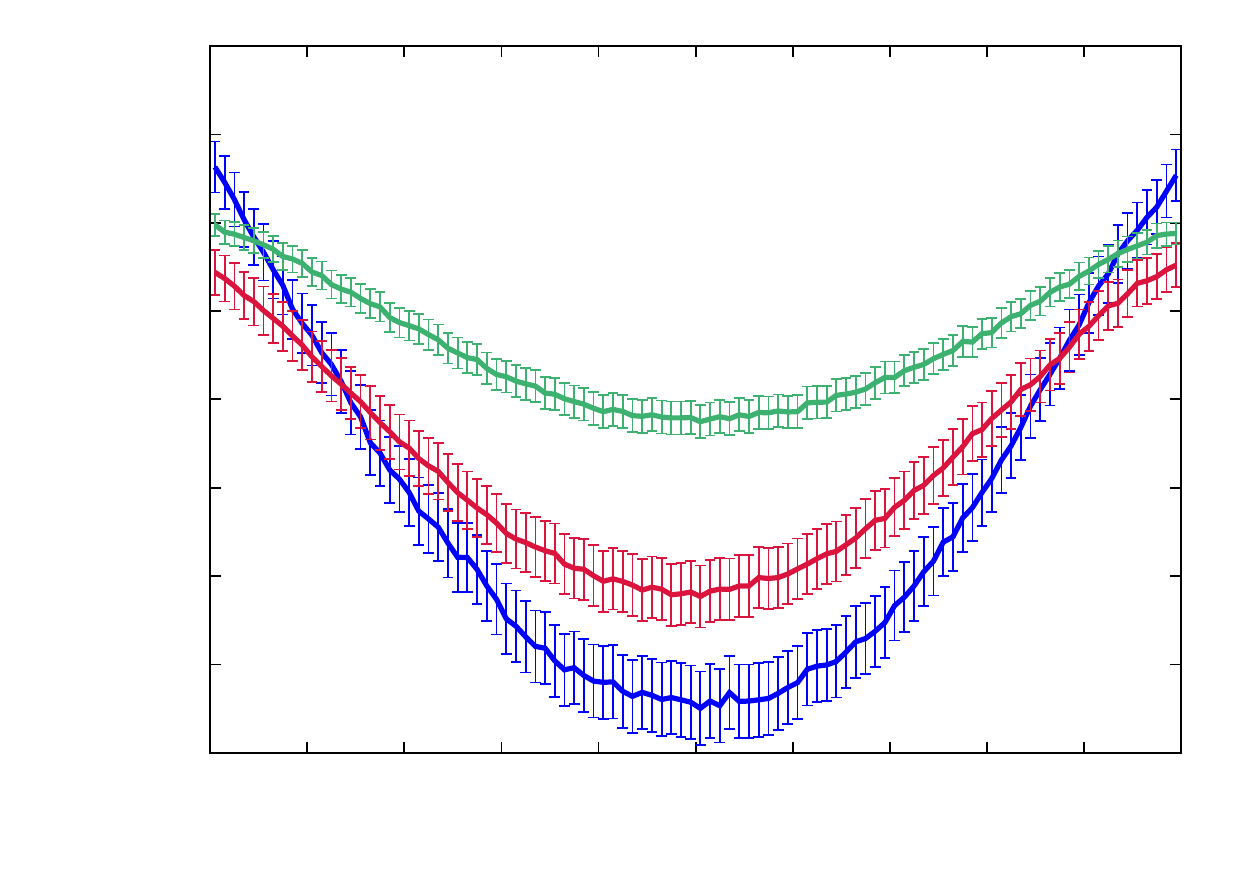}}%
    \gplfronttext
  \end{picture}%
\endgroup